\documentclass{article}
\usepackage[utf8]{inputenc}
\usepackage{graphicx}

\usepackage{amssymb}
\usepackage{amsmath}

\usepackage{dsfont}
\usepackage{csquotes}
\usepackage{ytableau}

\usepackage{enumitem}

\usepackage{relsize}
\usepackage{geometry}

\usepackage{hyperref}

\usepackage{color}

\newcommand\Op[1]{\widehat{#1}}
\newcommand\OpX{\Op{X}}
\newcommand\OpD{\Op{\Delta}}
\DeclareMathOperator\Tr{Tr}
\DeclareMathOperator\STr{STr}

\newcommand\OpDM[1]{\Op{\Delta}_{#1}}
\newcommand\MoSymOp[2]{\left[\Op{#2}\right]_{#1}}
\newcommand\MoSym[2]{\left[#2\right]_{#1}}
\newcommand\MoSymOpArg[3]{\left[\Op{#2}\right]_{#1}\hspace{-.3em}\Big(#3\Big)\;}
\newcommand\MoSymArg[3]{\left[{#2}\right]_{#1}\hspace{-.3em}\Big(#3\Big)\;}

\newcommand\sqdeti[1]{\sqrt{\det\left(i#1\right)}}
\newcommand\sqdet[1]{\sqrt{\det\left(#1\right)}}

\newcommand\OpSpH[2]{\Op{\mathcal{U}}\left[#1,\,#2\right]}
\newcommand\MoSpH[3]{\MoSym{#1}{\OpSpH{#2}{#3}}}
\newcommand\OpHc[2]{\Op{E}[#1;#2]}
\newcommand\OpH[1]{\Op{E}[#1;0]}
\newcommand\OpHW[1]{\Op{E}_0[#1;0]}
\newcommand\OpHM[2]{\Op{E}_{#1}[#2;0]}
\newcommand\OpSp[1]{\Op{\mathcal{U}}\left[#1\right]}

\newcommand\OpTDiff{\Op{\rm d}}
\newcommand\MoTDiff[1]{{\rm d}_{#1}}
\newcommand\dX{{\rm d}X}

\newcommand\au{{\underline\alpha}}
\newcommand\bu{{\underline\beta}}
\newcommand\cu{{\underline\gamma}}

\newcommand\ag{{\alpha}}
\newcommand\bg{{\beta}}
\newcommand\cg{{\gamma}}
\newcommand\dg{{\delta}}
\newcommand\ad{{\dot\alpha}}
\newcommand\bd{{\dot\beta}}

\newcommand\OpY{\Op{Y}}
\newcommand\Opy{\Op{y}}
\newcommand\Opyb{\Op{\bar{y}}}
\newcommand\yb{{\bar{y}}}
\newcommand\OpKy{{\Op\kappa_y}}
\newcommand\OpKyb{{\Op{\bar{\kappa}}_\yb}}

\newcommand\OpZ{\Op{Z}}
\newcommand\Opz{\Op{z}}
\newcommand\Opzb{\Op{\bar{z}}}
\newcommand\zb{{\bar{z}}}
\newcommand\OpKz{{\Op\kappa_z}}
\newcommand\OpKzb{{\Op{\bar{\kappa}}_\zb}}
\newcommand\OpK{{\Op\kappa}}
\newcommand\OpKb{{\Op{\bar\kappa}}}
\newcommand\dZ{{\rm d}Z}
\newcommand\dz{{\rm d}z}
\newcommand\dzb{{\rm d}\zb}
\newcommand\xb{\bar{x}}

\def\a{\alpha} 
\def\b{\beta}  

\def\d{\delta}

\def\e{\epsilon}

\def\ve{\varepsilon}

\def\m{\mu}
\def\n{\nu}
\def\k{\kappa}
\def\s{\sigma}
\def\cP{{\cal P}}
\def\vark{\varkappa}

\def\be{\begin{equation}}
\def\ee{\end{equation}}
\def\bea{\begin{eqnarray}}
\def\eea{\end{eqnarray}}
\def\ba{\begin{array}}
\def\ea{\end{array}}

\def\tn{\widetilde{\n}}

\def\tcP{\widetilde{\cP}}

\def\Log{\mbox{Log }}

\def\mso{{\mathfrak{so}}}
\def\msp{{\mathfrak{sp}}}
\def\msl{{\mathfrak{sl}}}
\def\msu{{\mathfrak{su}}}
\def\mho{{\mathfrak{ho}}}

\newcommand\OpSpHY[2]{\Op{\mathcal{U}}_{Y}\left[#1,\,#2\right]}
\newcommand\OpSpHy[2]{\Op{\mathcal{U}}_{y}\left[#1,\,#2\right]}

\newcommand\OpSpHZ[2]{\Op{\mathcal{U}}_{Z}\left[#1,\,#2\right]}
\newcommand\OpSpHz[2]{\Op{\mathcal{U}}_{z}\left[#1,\,#2\right]}

\newcommand\MoSpHY[3]{\mathcal{U}_{Y}\left[#2,\,#3\right]_{#1}}
\newcommand\MoSpHy[3]{\mathcal{U}_{y}\left[#2,\,#3\right]_{#1}}
\newcommand\MoSpHyb[3]{\mathcal{U}_{\yb}\left[#2,\,#3\right]_{#1}}

\newcommand\FTSpH[3]{\widetilde{\mathcal{U}}\left[#1,\,#2\right]_{#3}}

\DeclareMathOperator\cotan{cotan}
\newcommand\R{\mathbb{R}}
\newcommand\C{\mathbb{C}}
\newcommand\Z{\mathbb{Z}}

\DeclareMathOperator{\Pexp}{Pexp}
\DeclareMathOperator{\erf}{erf}
\DeclareMathOperator*\SeneGen{\mathlarger{\mathlarger{\mathlarger{\mathcal{O}}}}}
\newcommand\Sene{\SeneGen_{\epsilon,n,\eta}}

\renewcommand\mod{\;\text{mod}}

\newcommand\YTL[2]{{(#1,#2)}}
\newcommand\YTA[2]{{[#1,#2]}}

\advance \textheight by \topskip
\numberwithin{equation}{section}

\ytableausetup{mathmode, boxsize=2.2em}

\title{}
\date{}

\begin{document}

\newgeometry{top=3.5cm,bottom=2.5cm,left=2cm,right=2cm}
\begin{center}
\thispagestyle{empty}
{\huge Metaplectic Representation
and Ordering (In)dependence 
\\[10pt]
in Vasiliev's Higher Spin Gravity}
\\[10pt]


\vskip 1cm

David De Filippi${}^{a,}$\footnote{ \href{mailto:david.defilippi@umons.ac.be}{\texttt{david.defilippi@umons.ac.be}}},
Carlo Iazeolla${}^{b,}$\footnote{
\href{mailto:c.iazeolla@gmail.com}{\texttt{c.iazeolla@gmail.com}}},
and Per Sundell${}^{c,}$\footnote{
\href{mailto:per.sundell@unab.cl}{\texttt{per.sundell@unab.cl}
}}\\[3pt]

\vskip .4cm

${}^a$\textit{Physics of the Universe, Fields and Gravitation}\\
\textit{University of Mons -- UMONS, 20 place du Parc, 7000 Mons, Belgium}\\[3pt]
${}^b$\textit{Dipartimento di Scienze Ingegneristiche,\\ Guglielmo Marconi University -- Via Plinio 44, 00193, Roma, Italy \ \ \& }\\
\textit{Sezione INFN Roma “Tor Vergata” -- Via della Ricerca Scientifica 1, 00133, Roma, Italy}\\ [3pt]
${}^c$\textit{Centro de Ciencias Exactas, Universidad del Bío-Bío,}\\
\textit{Avda. Andrés Bello 720, 3800708, Chillán, Chile}

\end{center}

\vspace{0.8cm}

\begin{abstract}

  We investigate the formulation of Vasiliev's four-dimensional higher-spin gravity in operator form, without making reference to one specific ordering. 
  More precisely, we make use of the one-to-one mapping between operators and symbols thereof for a family of ordering prescriptions that interpolate between and go beyond Weyl and normal orderings. 
  This correspondence allows us to perturbatively integrate the Vasiliev system in operator form and in a variety of gauges.
  Expanding the master fields in inhomogenous symplectic group elements, and letting products be controlled only by the group, we specify a family of factorized gauges in which we are able to integrate the system to all orders, producing exact solutions, including but not restricted to ones presented previously in the literature; and then connect, at first order, to a family of rotated Vasiliev gauges in which the solutions can be represented in terms of Fronsdal fields. 
  The gauge function responsible for the latter transformation is explicitly constructed at first order.
  The analysis of the system in various orderings is facilitated by an analytic continuation of Gaussian symbols, by means of which one can distinguish and connect the two branches of the metaplectic double cover and give a rationale to the properties of the inner Klein operators as Gaussian delta sequences defining analytic delta densities.
  As an application of some of the techniques here developed, we evaluate twistor space Wilson line observables on our exact solutions and show their independence from auxiliary constructs up to the few first subleading orders in perturbation theory.

\end{abstract}

\maketitle
\setcounter{tocdepth}{2}
\tableofcontents

\newgeometry{top=3cm,bottom=3cm,left=2.5cm,right=2.5cm}
\section{Introduction}

\subsection{Motivations}

Recent advances in higher-spin gravity (HSG) have unveiled novel features of the theory that are unconventional from the point of view of ordinary relativistic field theories involving particles with spins less than or equal to two. 
A key feature of the theory is the presence of non-abelian higher spin symmetries, whose gauging requires infinite towers of tensor fields of arbitrarily high spin. 
While this infinite-dimensional symmetry places powerful constraints on the interactions, it also implies that the standard notion of locality needs to be replaced by some generalized notion thereof: indeed, the effective formulation of the theory in terms of perturbatively defined massless higher spin fields on spacetime backgrounds gives rise to cubic vertices in which the order of derivatives grows linearly with the spins involved, which implies some degree of non-locality already at this level of perturbation theory, only to become yet further enhanced at quartic order and beyond. 
This departure from relativistic lower-spin theory has not only blurred the (holographic re)construction of higher-spin vertices in the spacetime bulk but also led to conceptual problems, since the Noether procedure becomes trivial in the absence of any criteria that limit non-local field redefinitions  \cite{barnich,Sleight:2017pcz}.

Vasiliev's equations \cite{Vasiliev:1990en,properties,more,review99,Vasiliev:2003ev,Bekaert:2005vh,Didenko:2014dwa} provide a non-perturbative formulation of classical HSG that embeds highly complicated and spacetime non-local interactions into a compact set of first-order differential constraints for a set of differential forms, referred to as master fields, living on a fibered noncommutative extension of the spacetime manifold, sometimes referred to as \emph{correspondence space}. The evolution along the additional, noncommutative directions $\Op Z$ generates the interaction vertices among physical fields, with gauge and field-redefinition ambiguities encoded into the choice of resolution operator for the $\Op Z$-dependence (i.e., an operator that allows to integrate these evolution equations, a formal inverse of the $\Op Z$-differential). This gives some mathematical tool to control the resulting spacetime non-locality of the vertices and, possibly, to come up with a generalization of that concept adapted to HSG, allowing non-localities in some degree inherent to the system, while restricting the arbitrariness in wildly non-local field redefinitions. It is precisely by studying the consequences of this choice of resolution operator peculiar to the Vasiliev equations that progress has been recently made in addressing the problem of non-locality for some vertices, and in figuring out a proposal for this generalized concept of locality, referred to as spin-locality, which coincides with spacetime locality at the lowest order only \cite{Vasiliev:2017cae,Didenko:2018fgx,Didenko:2019xzz,Gelfond:2019tac,Didenko:2020bxd,Gelfond:2021two}.

A different but complementary approach, that uses the features of the Vasiliev system to try and bypass the difficulty associated with spacetime vertices, is to extract as much physical information as possible already from the equations in correspondence space, i.e., at the level of master fields, subject to proper boundary conditions, and of the gauge-invariant observables of the full theory. Technically, this approach is facilitated by the simple form of Vasiliev's generating system: as the master fields are subject to zero-curvature and covariant constancy conditions, to a large extent their spacetime features are stored in their dependence on fibre coordinates --- in a sort of spacetime/fibre duality much akin to a Penrose transform (see \cite{Vasiliev:2012vf} and references therein, see also \cite{neiman}). This way, non-localities of the vertices, as well as boundary conditions on spacetime fields, are translated on-shell into algebraic conditions on their generating functions, selecting specific class of functions of noncommutative fibre coordinates $\Op Y$ and auxiliary base coordinates $\Op Z$ and thereby making the problem somehow more tractable. While the exact form of such duality, in general, may be hard to exploit, specific choices of resolution operators make it especially transparent \cite{2011,2017,review,COMST}: indeed, a choice of resolution operator that effectively separates $\Op Y$ and $\Op Z$ variables (hence the corresponding gauges are referred to as \emph{factorized} gauges) not only makes it possible to push the perturbative expansion to all orders, thereby reaching to an exact solution provided certain algebraic conditions are satisfied (that put constraints on the allowed class of functions of $(\Op Y,\Op Z)$) \cite{2011,2017,cosmo,COMST}; but also singles out the fibre representative of the Weyl zero-form master field of the solution, which contains all the local data of the resulting spacetime curvatures. Extracting the corresponding potentials in general requires switching on a specific gauge function (that is, a specific finite --- possibly large --- gauge parameter) that glues the gauge fields to on-shell curvatures, and in \cite{COMST} a scheme was proposed to impose asymptotically anti-de Sitter boundary conditions at master-field level, by perturbatively adjusting gauge function and fibre representative of the Weyl zero-form at higher orders. Within this framework, an interacting theory of Fronsdal gauge fields on AdS is thereby envisaged as a perturbative branch of the Vasiliev system, reached by subjecting the master fields to specific gauge and boundary conditions that enable, among other things, to read spacetime fields as coefficient of their power series expansion in fibre coordinates. The alternative perturbative scheme just described, if successful, would have the advantage of giving a way of controlling boundary conditions at the level of master fields, thereby providing some criteria to select allowed class of functions, allowed field redefinitions, and in general in giving a better understanding of the geometry behind higher-spin fields.

One thing that these two approaches have in common is that they make use of resolution operators for the $\Op Z$-dependence that do not coincide with the simplest one, known to give rise to non-local vertices \cite{Boulanger:2015ova} and divergent boundary observables \cite{GiombiYin1}. As shown in \cite{COMST,Didenko:2019xzz} a choice of resolution operators within a certain class is equivalent to a choice of \emph{ordering prescription} for the noncommutative variables of the system. The extraction of Fronsdal fields in totally symmetric ordering of the fibre coordinates, coupled with the technical simplifications due to the real-analyticity in $\Op Y$ and $\Op Z$ of all master fields, determined the choice of a specific gauge and ``entangled'' $(\Op Y, \Op Z)$ normal-ordering prescription (with respect to the combinations $\Op Y\pm \Op Z$) for the presentation of the original Vasiliev system \cite{more,review99}. On the other hand, the physical content of a given solution, including boundary correlation functions, can be read off from classical, gauge-invariant observables of the theory \cite{Colombo:2010fu,Sezgin:2011hq,Colombo:2012jx,Didenko:2012tv,Bonezzi:2017vha,review,Sharapov:2020quq}, some of which are in fact easier to compute in factorized gauges \cite{COMST}. Besides, such classical observables are invariant under both higher spin gauge transformations and local redefinitions of the operator algebra basis (reorderings) \cite{Kontsevich:1997vb} (possibly, up to boundary terms in $(\Op Y,\Op Z)$ space), which we collectively refer to as Kontsevich gauge transformations. For the reasons collected above, it is therefore interesting to study the Vasiliev system in ordering-independent language, i.e., in operator form, and to learn to extract its physical content this way.

\subsection{Summary of our main results}

In this paper we take the first steps towards this goal, by performing the perturbative analysis of the four-dimensional Vasiliev equations at first order in operator form, thereby generalizing and extending the results of \cite{COMST}. As we shall detail in the following, the main progresses we achieved are:

\begin{enumerate}[label=\arabic*)]

 \item The integration of the $\Op Z$ dependence at the operator level, and by means of a larger family of resolution operators. 
 
 \item Such greater freedom is exploited by working in a new factorized gauge, which we shall refer to as axial holomorphic, in which a large space of solutions is first-order exact in all master fields; this also 
 
 \begin{enumerate}[label=2\alph*)]
 
 \item simplifies the construction of the gauge function enabling to extract the corresponding Fronsdal fields, as we shall show explicitly; and 
 
 \item simplifies the computation of Wilson line observables, which we push beyond the first subleading order treated in \cite{COMST}. 

\end{enumerate}
 
 \item Moreover, for the first time we frame our analytic continuation of all Gaussian functions of $\Op Y$ and $\Op Z$ (starting with the inner Klein operator present in the Vasiliev equations), recognized in previous papers to be required for kinematical as well as physical reasons (recalled in Appendix \ref{appD:motivations}), within a realization of \emph{metaplectic group} elements by means of symbol representatives of operators in any given ordering, as will be explained below. In particular, the metaplectic double-covering of the symplectic group is in our analysis extended  to the complex case within a  holomorphic, non-unitary oscillator representation. 

\end{enumerate}

One of the key technical steps to achieve point 1) above is the construction of a family of orderings that interpolate between, and in fact go beyond, normal ordering and Weyl ordering, parametrized by a Gaussian function with real symmetric matrix $M$ (with $M=0$ corresponding to Weyl, i.e., totally symmetric ordering, and $M = R$, with $R$ being a polarization of the identity, corresponding to normal orderings). Every operator in a given $M$-ordering can be mapped to a corresponding symbol, a function of commutative variables $Y$ and $Z$ --- with a star product implementing the noncommutative operator product on the resulting space of functions, and realized via a convolution. Working with symbols enables to realize $\Op Z$-commutators as $Z$-differentials, and thus to write down a concrete form of resolution operator $q^\ast$ that integrates the equations determining the $Z$-dependence of the master fields. As the operator-symbol map is one-to-one, one can promote $q^\ast \to \Op q^\ast$, and thus integrate the equations in operator form. This procedure results in the definition of a family of resolution operator $\Op q^{\ast[M,\Xi]}$, each singled out by a choice of arbitrary symmetric matrix $M$ and vector $\Xi$ that defines the homotopy contraction, thereby interpolating between various (generalized) gauges. As we shall see in a few concrete examples, such freedom in $M$ and $\Xi$ survives after one maps the so-obtained solution to a specific ordering, labelled by a matrix $M_0$, thereby encoding the freedom of realizing the source term, that $q^\ast$ acts on, and corresponding potential in different ways within a broad class of functions: in our case, the latter is built via integral transforms with kernels given by Gaussian (plus linear term) symbols, and singular limits thereof, encoding distributions.

We shall consider expansions of the local data on $SpH(4,\mathbb{C})$ group elements, where $SpH(4,\mathbb{C})$ is defined as the semi-direct product of $Sp(4,\mathbb{C})$ and the $4$-dimensional Heisenberg group: that is, exponentials of inhomogeneous quadratic polynomials in oscillators $\Op Y_{\underline{\a}}=(\Op y_\a, \Op{\bar y}_{\ad})$ satisfying a Heisenberg algebra, with complex coefficients. One of the motivations to do so is that specific Gaussian projectors have been found (in Weyl ordering) to encode some of the most relevant field configurations, appearing as solutions to either the full field equations or the linearized ones around (A)dS: among these are massless particle states \cite{fibre,2017,COMST}, bulk-to-boundary propagators \cite{GiombiYin2,Didenko:2012tv,Iazeolla:2020jee}, black-hole and black-brane-like solutions \cite{Didenko:2009td,2011,Sundell:2016mxc,2017,review,COMST,Iazeolla:2020jee,Didenko:2021vui,Didenko:2021vdb}, instantons, domain walls and FLRW-like solutions \cite{Iazeolla:2007wt,Iazeolla:2015tca,cosmo}.
In the case of scalar particle and black hole states,
we shall show that such projectors are given by evanescent pieces of $Sp(4,\mathbb{C})$ group algebra elements in singular limits.
Moreover, the Vasiliev generating system contains by construction special Gaussian elements, the so-called inner Klein operator $\Op \k$ and its hermitean conjugate $\Op{\bar \k}$, which are crucial building blocks of the source term that triggers all nonlinear corrections to the free Fronsdal equations. As an operator, $\Op{ \k}$ factorizes into $(-1)^{\Op N_y}(-1)^{\Op N_z}$, where $\Op N_X$ is a number operator counting the number of $\Op X$ oscillators. 

Products of all elements of this form are thereby controlled by the $SpH$ group algebra. As is well known (see for example \cite{CSM,Folland,Guillemin:1990ew,Woit:2017vqo} and references therein), exponentials of bilinears in oscillators with real quadratic form matrices realize the metaplectic representation of $Sp(2n,\mathbb{R})$, associating to every $Sp(2n,\mathbb{R})$ matrix two operators, distinguished by a sign, and lifting to a proper, non-projective representation of the double-covering metaplectic group $Mp(2n,\mathbb{R})$. As we shall show in this paper, the same 1:2 association can be established in the complex case, by means of a holomorphic, non-unitary oscillator representation of $Sp(2n,\mathbb{C})$ lifting to a proper representation of the double-covering group\footnote{In the literature on unitary representations, it is instead customary to identify $Mp(2n,\mathbb{C})$ and $Sp(2n;\C)$, as it follows from Bargmann's theorem that any unitary representation of a simply connected group like $Sp(2n;\C)$ is proper, i.e. non-projective.}, that we shall denote by $Mp(2n,\mathbb{C})$, and refer to as the holomorphic metaplectic group or complex metaplectic group for short. 

As the map from operators to symbols involves extracting square roots of the determinant of functions of the corresponding symplectic matrices and the ordering matrix $M$, the two branches of the metaplectic double covering can be put in correspondence with the two sheets of the Riemann surface of the square root. Thus, with the help of a specific analytic continuation, it is possible to work with symbol representatives of metaplectic group elements. In this paper, we shall give this analytic continuation only for the case of $M=0$ Weyl-ordering symbols, leaving the generic $M$-ordering case for future work\footnote{The technical reason why it is more complicated to achieve this with $M\neq 0$ symbols is the simultaneous presence of sum and products of matrices in the argument of the determinant under square root: the extension to the Riemann surface of the square root in that case is subtler, as sums are not well defined for elements living on different sheets.}. 

In any given ordering, generic symplectic matrices are realized as Gaussians in the oscillators, but there are elements whose symbol degenerates to a delta function: in particular, such is the symbol of the Kleinian in Weyl ordering. In previous papers in which we dealt with solving the equations in Weyl ordering by means of expansions of the master fields over Gaussian elements \cite{2011,2017,cosmo,COMST}, it was somehow natural (and consistent with all its properties) to realize the Kleinian via a Gaussian delta sequence, taking the singular limit after all star products had been computed. In this paper, we reframe and justify this procedure as a limit to Weyl ordering from other orderings in which the Kleinian is a regular Gaussian function (which, as we show explicitly in App. \ref{app:g hol g}, is equivalent to acting directly on the delta-function source with a reordered resolution operator). Moreover, the analytic continuation of Gaussian elements which we propose in this paper\footnote{We collect a number of motivations for considering an analytic extension of Gaussian integration formulae and delta functions in App. \ref{appD:motivations}.} provides us with a natural definition of delta one-forms, projecting onto the Riemann surface ${\cal S}_2$ of the square root as delta densities defined along a bundle of lines through the branching point. This definition generalizes straightforwardly to delta $2n$-forms, projecting to higher-dimensional delta densities $\d^{2n}(X)$ defined on the $2n$-dimensional generalization of ${\cal S}_2$.
Within this framework, it is possible to reconcile the otherwise contrasting properties of the Kleinian of behaving as a two-dimensional delta function which is however odd under the $\tau$ map (as required in order to match the corresponding property of the operator $(-1)^{{\Op N_y}}$, among other reasons), i.e., $\tau(\delta^2(y))=\delta^2(iy)=-\delta^2(y)$, by interpreting naturally the action of $\tau$ as relating delta densities defined on different surfaces embedded in ${\cal S}_2^2$.

Armed with these tools, we then proceed to analysing perturbatively the Vasiliev system at first order, in order to retrieve Fronsdal's equations in operator language.
A great advantage of the unfolded formulation, which lies at the heart of the Vasiliev system, is the reformulation of gauge field equations for arbitrary spin in terms of first-order differential constraints for a set of differential forms, allowing for all fields to be treated on equal footing and without requiring to invert the vielbein  \cite{Vasiliev:1988sa,review99,Bekaert:2005vh,Didenko:2014dwa}. Achieving that involves some amount of redundancy between the spacetime master one-form and zero-form fields.  This, in turn, implies that the embedding of the Fronsdal free equations in the Vasiliev system expanded at first order around AdS subtly depends on specific (generalized) gauge choices that enter the solution of the $\Op Z$ dependence of the master fields, and that are encoded in the choice of resolution operator $\Op q^\ast$ \cite{properties,more,Didenko:2015cwv,COMST}.
The spacetime zero-form master field $\Op \Phi$ naturally encodes the propagating degrees of freedom; according to the choice of $\Op q^\ast$ the local degrees of freedom may then be ``dualized'' into the spacetime master one-form $\Op U$ via a dynamical gluing of the two modules by means of a Chevalley-Eilenberg cocycle of the background isometry algebra.
The aforementioned factorized gauges, for instance, facilitate the dressing of linearized solutions into full ones, but, as we shall recall in Sec. \ref{sec:factor}, fail to provide the proper cocycle. This can however be achieved in other families of gauges, referred to as ``relaxed Vasiliev gauges'' in \cite{COMST,Iazeolla:2020jee}, in which it is possible to identify the Fronsdal equations within the unfolded framework \cite{more,review99,Bekaert:2005vh,Didenko:2014dwa,COMST}.
The latter result --- together with the generalized Bianchi identities contained in the equation for $\Op \Phi$ from the first-order expansion of the Vasiliev system --- is known as Central On-Mass-Shell Theorem (COMST) \cite{properties,more,review99} (see also \cite{Bychkov:2021zvd}). In a previous paper \cite{COMST}, we showed how, at first order, it is possible to construct a gauge function $\Op H$ relating solutions obtained in factorized gauges to solutions in relaxed Vasiliev gauges (equivalently, solutions obtained by means of one and the same, unshifted, resolution operator in Weyl ordering to solutions obtained in the above-mentioned entangled normal ordering): such gauge function does not alter the degrees of freedom, contained in $\Op \Phi$, and provides the Chevalley-Eilenberg cocycle, which implies that it cannot be regular. Indeed, as discussed above, this gauge transformation encodes a change of ordering, and, as the master fields are non-polynomial functions of noncommutative coordinates, ordering changes in this context are far from trivial operations, and may in general give rise to divergencies that need to be interpreted and handled. What makes it possible to do that is the fact that such divergencies in Weyl ordering manifest themselves in the singular (distributional) form of $\Op \k$, which can be dealt with via integral presentations and delta sequences \cite{COMST}. The resulting singularities are inherited by the gauge function, but turn out to give cohomologically trivial contributions to the spacetime connection, 
thereby leading to a proper, $\Op Y$-analytic Fronsdal field generating function.

In this paper, working at the level of operators, we show how it is possible to retrieve Fronsdal fields in solutions built in a new family of factorized gauges, gluing a gauge field generating function to the (unmodified) Weyl zero-form $\Op \Phi$ by means of a change in resolution operator $\Op q^\ast$, we give the interpretation of such change in terms of a change of ordering prescription, and build the corresponding first order gauge function that provides the gluing. We perform this analysis for master fields expanded over Heisenberg group elements.

More in detail, we first integrate the $\Op Z$-space equations in operator language and in a family of factorized gauges, called holomorphic gauges because the potential retains the holomorphicity of the integrated source, specified by resolution operators $\Op q^{\ast[m,\xi]}$: this allows us to interpolate between the solution with $\xi=0$ and $m$ proportional to an involutory matrix, which was used in various contexts in \cite{2011,2017,cosmo,COMST} and on which the $\Op z$-Kleinian $\Op \vark_z=(-1)^{\Op N_z}$ is realized as a regular Gaussian (only degenerating into a delta function at the boundary of the homotopy-integration domain); and the solution with $m=0$ and $\xi=\zeta$, a constant spinor, in which the homotopy integration is performed on a delta function integrand. As we shall show, the latter, new gauge, referred to as axial holomorphic, proves to be even more efficient than the former at constructing (formal) exact solution spaces, as the resulting linearized solution is already exact. Furthermore, its simplicity makes it an ideal frame to tackle the asymptotically AdS boundary condition problem along the lines envisaged in \cite{COMST}, which is our ultimate goal. For these reasons, we use it in this paper as our starting point for the extraction of Fronsdal fields.

Operating within the Heisenberg group expansion, in terms of a momentum $\Lambda$ conjugated to $\Op{Y}$, the appropriate tweak that is sufficient to achieve this is to make the shift vector momentum-dependent, $\xi=\xi(\Lambda)$: this change moves away from factorized gauge and achieves the gluing of the spacetime one-form module to the degrees of freedom in $\Op \Phi$ in operator terms. Choosing $\xi(\Lambda)=b\lambda$, where $b$ is an invertible $2\times 2$ matrix (which may be even spacetime-dependent), defines a \emph{rotated Vasiliev gauge} such that, in Weyl order, the gluing condition takes the form of the usual COMST, and any dependence on $b$ disappears. 
Note that the matrix $b$ can be spacetime-dependent, a fact that can be used to give the local Lorentz symmetry an alternative treatment to the original one \cite{properties,2011,Sezgin:2011hq}, as discussed in Sec. \ref{sec:Lorentz}.
As we show in Sec. \ref{sec:No(YZ)}, this choice of homotopy is equivalent to using an unshifted resolution operator in a particular $M$-ordering in the full $(\Op Y,\Op Z)$ oscillator space: for the simplest choice of constant $b_{\a\b}=-\check\beta \e_{\a\b}$ this family of orderings coincides with the one-parameter family considered in \cite{Didenko:2019xzz} (in terms of the parameter $\beta= 1-\check\beta$), that interpolates between the entangled normal ordering ($\check\beta=1$), Weyl ordering ($\check\beta=0$) and the limiting spin-local case ($\check\beta\to\infty$).    

Finally, we proceed to applying this formalism to the case of spherically-symmetric higher-spin black holes and massless scalar particle modes. Starting from the axial holomorphic gauge and moving to the rotated Vasiliev gauge, we first find the gauge-field generating function for black holes, and show that, as expected, it is analytic in $Y$ in Weyl order.
Then, for the case of a massless scalar particle, we check that our formalism correctly produces a trivial 1-form connection: in particular, we explicitly construct the gauge function that removes the singular yet cohomologically trivial part inherited from the axial holomorphic gauge, improving on the results of \cite{COMST} in that building such gauge function does not require the introduction of a spacetime vector field along which to perform a homotopy contraction. 

The axial holomorphic gauge introduced in this paper also allows us to push further the perturbative computation of certain gauge-invariant quantities, the Wilson line observables \cite{Engquist:2005yt,Sezgin:2005pv,Sezgin:2011hq,Colombo:2010fu,Colombo:2012jx,Bonezzi:2017vha,COMST}. As they are invariant under (large) gauge transformations, it is convenient to evaluate them on solutions in factorized holomorphic gauges, which simplify their form. In this context, the axial holomorphic gauge is especially effective, and allows us to determine their expansion coefficients (in the Weyl zero-form local datum) beyond the first subleading order that was obtained in \cite{COMST} using the so-called Gaussian holomorphic gauge. This is an important test-ground for the solution methods using factorized gauges, since, while Wilson lines are by construction unaffected by any transformation realized by a gauge function, it is less obvious that parametric integrals, intermediate-stage regulators and auxiliary spinors that may enter the concrete form of a solution in such gauges will not ruin their invariance. By this computation we have verified up to third order the independence of these coefficients from any of the auxiliary constructs that are used to write down the solution in axial holomorphic gauge. 

While ordering changes in the higher-spin context are non-trivial, due to the fact that the local symmetry algebra is infinite-dimensional, it may appear not surprising that the theory is ordering-independent at first order. Nonetheless, already the linearized analysis performed in this paper in ordering-covariant language reveals some issues that would otherwise remain hidden, and that may be crucial to address some of the open questions in the full theory. For instance, the freedom in moving in between $M$-orderings manifests itself in a greater freedom in the choice of resolution operator $q^\ast$, which includes the freedom of realizing the source terms on different functional classes. Moreover, the appearance of degenerate Gaussian symbols defining delta sequences is not limited to the intermediate steps of the computations, or to a given resolution scheme: on the contrary, it is very much tied to interesting physics, as they naturally appear, working in Weyl order on ${\cal Y}$, as the configuration that the spherically-symmetric higher-spin black holes approach at the singularity \cite{Didenko:2009td,2011,2017}, or that fluctuation master fields take at the singularity of a BTZ$_4$ black hole \cite{BTZ}. In both cases, the ill behaviour of the individual spin-$s$ Weyl tensors coalesces into a delta-function behaviour of the corresponding master field at the singularity. However, as stressed in the above-cited papers, at least some distributions in noncommutative variables can be considered smooth since they have good star product properties. In the spirit of the present paper, delta functions of noncommutative variables are in fact equivalent to bounded functions up to a change in the ordering prescription. Within this view, the resolution of, e.g., gravitational curvature singularities in HSG would amount to declaring them artifacts of the ordering choice for the infinite-dimensional symmetry algebra governing the Vasiliev system. So a better grasp of the role of orderings may prove crucial in order to properly understand the embedding of gravitational singularities into HSG --- and possible resolutions thereof, already at the classical level, due to the coupling with the higher-spin tower \cite{2011,2017,BTZ}. More generally, a full assessment of to what extent and under which conditions the theory is locally ordering-independent would provide us with crucial means to disentangle the physics from gauge artifacts and with a more powerful mathematical framework for the formulation of HSG.

Finally, we believe that the tools developed in this paper --- in particular the axial holomorphic gauge, the resolution operators related to $M$-orderings and their applications to the AdS massless UIRs --- will prove useful for the concrete implementation of the perturbative scheme proposed in \cite{COMST} (in which asymptotically AdS boundary conditions are imposed via perturbative corrections of gauge functions and master fields local data), which was one of the main motivations driving the present work. 

\subsection{Outline of the paper}

This paper is organized as follows: \\
Sec. \ref{sec:HS op} is devoted to reviewing basic properties of Vasiliev’s four-dimensional bosonic equations, linearized solution spaces, factorized vs. unfactorized gauges and perturbative schemes, and introducing the necessary formalism in operator language. \\
Sec. \ref{sec:starprod} introduces the group algebras that will be of relevance for the treatment of massless particle modes, higher-spin black hole solutions and the Klein operators: it contains the definition of the inhomogenous complex metaplectic group $MpH(2n;\C)$ and spells out the map between operators and symbols thereof in a given ordering. We introduce the notion of $M$-orderings and, finally, provide the main tool for the perturbative analysis that will follow: the family of resolution operator $\Op q^{\ast[M,\Xi]}$, interpolating between various (generalized) gauges. \\
In Sec. \ref{sec:solve Z} we  apply this formalism to integrating the $\Op Z$-space equations with various choices of $q^\ast$/gauge: holomorphic gauges, which allow to obtain full, not just linearized, solutions --- failing however to activate non-trivial Fronsdal fields; and a rotated Vasiliev gauge in which we glue, at first order, the Weyl zero-form to the spacetime connection, thereby obtaining the COMST in operator language. We show how the latter gauge entails a class of orderings that non-trivially entangle $Y$ and $Z$ variables, and in particular include the one-parameter family of orderings considered in \cite{Didenko:2019xzz}. We also construct the gauge function connecting the two types of gauge, at first order, and compute Wilson line observables, checking their invariance in the sense specified above. \\ 
Finally, in Sec. \ref{sec:PTBH}, we obtain the regular gauge-field generating function corresponding to AdS massless particles and higher spin black hole states, which are encoded by projectors that we show are obtained from rescaled limits of $Sp(4,\C)$ elements.  

The paper is completed by six appendices. In App. \ref{app:cvt} we spell out the AdS and spinor conventions used throughout the paper. App. \ref{app:anal} is especially important, as it contains the details of the analytic continuation which enables to distinguish symbol representatives of metaplectic group elements and the definition of delta densities related to the Klein operators. App. \ref{1psymp} contains a number of details on the properties of $M$-ordering symbols of $Sp(2n,\C)$, in particular deriving their general form, and showing how the analytic continuation of App. \ref{app:anal} enables to distinguish the two branches of the metaplectic double cover. In App. \ref{app:star} we give some lemmas that are instrumental for Sec. \ref{sec:starprod}. In App. \ref{app:g hol g} we show the detailed steps leading to the all-order perturbative solution in holomorphic Gaussian gauge given in Sec. \ref{sec:solve Z}, which also furnishes an example of how the freedom in $M$ of $\Op q^{\ast[M,\Xi]}$ can encode the freedom of realizing the integrated source term on different functional classes. Finally, in App. \ref{app:OWL} we collect the computations related to the results on the Wilson line observables collected in Sec. \ref{sec:solve Z}.

\section{Vasiliev's higher spin gravity}
\label{sec:HS op}

In this Section, we pass from associative higher spin algebras \cite{Eastwood:2002su}, via the COMST \cite{COMST} for unfolded Fronsdal fields, and perturbatively defined formal HSG \cite{Vasiliev:1988sa,Sharapov:2019vyd}, to Vasiliev's fully nonlinear formulation of HSG using differential forms on noncommutative spaces.

The classical moduli space of Vasiliev's theory consists of free differential subalgebras (FDA) of differential graded associative operators algebras (DGA) of horizontal forms on fibered manifolds
\be {\cal Y}\times {\cal G}_0\rightarrow {\cal C}\to {\cal X}\times {\cal Z}\ ,\ee
where ${\cal Y}$ is a noncommutative symplectic manifold; ${\cal X}$ is a commuting manifold embeddable into the universal classifying space $B{\cal G}_0$ of a group ${\cal G}_0$ acting on ${\cal Y}$ via moment maps; and ${\cal Z}$ is a noncommutative symplectic manifold equipped with two-forms attached to conical singularities in codimension two.
These two-forms combine with covariantly constant zero-forms into sources of one-form connections interpretable, in a dual fashion, either as  deformations of the symplectic structure on ${\cal Y}\rightarrow {\cal T}_{p_0}\to \{p_0\}\times \cal Z$ positioned at a base point $p_0\in {\cal X}$, referred to as the unfolding point, or as self-interacting Fronsdal fields in asymptotic regions ${\cal X}_\infty$ of ${\cal X}$ created by i) switching on gauge functions from ${\cal X}\times{\cal Z}$ to ${\cal G}$ so as to create noncommutative twistor spaces ${\cal Y}\rightarrow {\cal T}_{p}\to  \{p\}\times{}\cal Z$ over base points $p\in {\cal X}$; and ii) imposing suitable boundary conditions as $p$ approaches ${\cal X}_\infty$.

The horizontal forms making up the classical moduli spaces are thus quantum mechanical composite operators that can be represented by symbols given by distributions on ${\cal T}_p$, thought of as a classical symplectic manifold, composed by means of twisted convolution formulae given by integrals over ${\cal T}_p$.
The functional form of these distributions thus depend not only on the choice of local coordinates on ${\cal C}$ but also on the choice of basis for the operator algebra, that is, on the choice of ordering prescription as operators are sent to classical symbols.
Indeed, the observables of the theory are intrinsically defined functionals of the horizontal forms obtained by integrating over subspaces of ${\cal C}$ so as to achieve invariance under diffeomorphisms of ${\cal X}$ and isomorphisms of the operator algebra on ${\cal T}_p$, that is, changes of its basis, which thus include re-orderings as well as symplectomorphisms \cite{Kontsevich:1997vb}.
Thus, just as in ordinary gauge theory, the construction of classical moduli spaces in HSG consists of first exhibiting the local degrees of freedom residing in gauge functions as well as ordering schemes, and then examine which of these remain inherent at the level of classical observables.

Thus, in order to provide a cohesive framework for the sequel of this paper, we shall present Vasiliev's theory using an operator language that does not refer to any specific ordering scheme.
Such schemes will then be introduced separately in the following section, and then studied in more detail in the remainder of the paper.

Finally, our line of presentation will highlight a difference between formal HSG and Vasiliev's theory vis a vis going off-shell, as Vasiliev's formalism lends itself relatively straightforwardly to the construction of noncommutative AKSZ sigma models with BV algebras consisting of single-trace functionals, which provide a natural generalization of the BV algebras of local functionals of a ordinary commutative AKSZ sigma model.

\subsection{Higher spin algebras from conformal particles}
\label{sec:UEA}

The basic building block for HSG on $AdS_{1+N}$ is Dirac's conformal particle on the real cone with signature $(2,N)$, which provides an irreducible one-sided $\mso(2,N)$ module ${\cal S}$ that remain irreducible under $\mso(1,N)$\footnote{Interestingly, ${\cal S}$ constitutes the internal spin degree of freedom of the Majorana field equation in flat $(N+1)$-dimensional spacetime  \cite{Bekaert:2009pt}.}.
This module comprises a plethora of $\mso(2,N)$ representations \cite{Bars:2000qm}, including unirreps that decompose under the maximal compact $\mso(2)\oplus \mso(N)$ subalgebra of $\mso(2,N)$ into compact weight spaces $D^\pm(\tfrac12 (N-2);(0))$ with energy bounded from below and above, respectively, by $\pm\tfrac12(N-2)$, referred to as singletons and anti-singletons, as their weights occupy single lines in the energy-spin plane.

For $N\geqslant 3$, their squares\footnote{Multi-singleton and multi-anti-singleton extensions yield massive spectra arising naturally within first-quantized non-compact Wess--Zumino--Witten models with integer spectral flows \cite{Engquist:2007pr,Gaberdiel:2021qbb} related to tensionless extended objects in $AdS_{1+N}$ \cite{Engquist:2005yt} and corresponding extensions of the Vasiliev system \cite{Vasiliev:2012tv}.} decompose under $\mso(2,N)$ into massless particles and anti-particles in $AdS_{1+N}$, viz.
\be D^\pm(\tfrac12 (N-2);(0))^{\otimes 2}=\bigoplus_{s=0}^\infty D^\pm(s+N-2;(s))\ ,\ee
as first found by Flato and Fronsdal in four spacetime dimensions \cite{Flato:1978qz} and later generalized to higher $N$ in the context of HSG in \cite{Sezgin:2001zs,Sezgin:2001ij,Vasiliev:2004cm}.
Moreover, the direct product of a singleton and an anti-singleton can be rearranged \cite{fibre,Basile:2018dzi} into a compact weight space with unbounded energy equipped with a real structure and a positive definite bilinear form \cite{fibre} whose states correspond to higher spin generalizations of the Coulomb and Schwarzschild solutions \cite{2011,2017}.

The conformal particle module is characterized by its annihilator ${\cal I}[{\cal S}]$ (for example, see \cite{Engquist:2005yt,fibre}), which is the ideal of the unital enveloping algebra\footnote{For a mathematical treatise on enveloping algebras of finite-dimensional Lie algebras, see e.g. \cite{DixmierBook}; we take their elements to be polynomials in the Lie algebra generators of finite degree. } ${\rm Env}[\mso(2,N)]$ that vanishes in ${\cal S}$.
In the basis
\be \Op{J}_{A,B}=-\Op{J}_{B,A}\ ,\qquad A,B\in\{0',0,1,\dots,N\}\ ,\ee
of $\mso(2,N)$ normalized such that
\begin{align}
  \label{eq:SO [J,J]}
  \left[\Op{J}_{A,B},\,\Op{J}_{C,D}\right]
  &=
  i\left(\eta_{AD}\Op{J}_{B,C}+\eta_{BC}\Op{J}_{A,D}-\eta_{AC}\Op{J}_{B,D}-\eta_{BD}\Op{J}_{A,C}\right)
  \,,
\end{align}
where $\eta_{AB}={\rm diag}(-,-,+,\cdots,+)$, the annihilator is generated by
\be \Op{V}_{AB}:=\frac12 \Op{J}_{(A,}{}^C\Op{J}_{B),C}-\frac1{N+2} \eta_{AB}\Op{C}_2\approx 0\ ,\qquad \Op{V}_{A,B,C,D}:=\Op{J}_{[A,B}\Op{J}_{C,D]}\approx 0\ ,\ee
where $\Op{C}_2[\mso(2,N)]:=\frac12 \Op{J}^{A,B}\Op{J}_{A,B}$, which can be shown to assume the value $\Op{C}_2[\mso(2,N)|{\cal S}]\approx 1-\frac{N^2}{4}$.

The conformal particle thus induces the associative algebra
\be \mho(2,N):={\rm Env}[\mso(2,N)]/{\cal I}[{\cal S}]\ ,\ee
which acts on itself from the left and the right, and through twisted versions
\be \rho_{\alpha,\beta}(\hat{f}) \hat g:= \alpha(\hat{f}) \hat g-\hat g\beta(\hat{f})\ ,\qquad \hat{f},\hat g\in \mho(2,N)\ ,\ee
of the adjoint action ${\rm ad}\equiv\rho_{{\rm Id},{\rm Id}}$ labelled by linear $\mso(2,N)$ automorphisms $\alpha$ and $\beta$.
Such automorphisms act faithfully on any subspace of ${\rm Env}[\mso(2,N)]$ preserved under the adjoint $\mso(2,N)$ action, including the conformal particle annihilator, and hence they lift to (linear) $\mho(2,N)$ automorphisms.
It follows that
\be [\rho_{\alpha,\beta}(\hat{f}_1),\rho_{\alpha,\beta}(\hat{f}_2)]=\rho_{\alpha,\beta}([\hat{f}_1,\hat{f}_2])\ ,\ee
where $[\hat{f}_1,\hat{f}_2]:={\rm ad}(\hat{f}_1)\hat{f}_2=\hat{f}_1\hat{f}_2-\hat{f}_2\hat{f}_1$, which assigns the Lie algebra $(\mho(2,N),{\rm ad})$ a module ${\cal T}_{\alpha,\beta}:= (\mho(2,N),\rho_{\alpha,\beta})$, referred to as the $(\alpha,\beta)$-twisted $\mho(2,N)$-module.
Letting $\mathfrak{h}$ denote the maximal subalgebra of $\mso(2,N)$ stabilized by $\alpha$ and $\beta$, it follows that
    \be {\cal T}_{\alpha,\beta}\downarrow_{\rho_{\alpha,\beta}(\mso(2,N))}=\bigoplus_{\vec\lambda_0\in \Sigma_{\alpha,\beta}} {\cal T}_{\alpha,\beta}^{\vec\lambda_0}\ ,\ee
where ${\cal T}_{\alpha,\beta}^{\vec\lambda_0}$ denotes a $\mso(2,N)$ irrep generated from a finite-dimensional $\mathfrak h$ irrep $\tau^{\vec\lambda_0}$, i.e. an irreducible $\mathfrak{h}$ tensor, referred to as the reference state, by means of the $(\alpha,\beta)$-twisted $\mso(2,N)$ action; for example, $\tau^{\vec\lambda_0}$ can be the smallest $\mathfrak{h}$ irrep contained in ${\cal T}_{\alpha,\beta}^{\vec\lambda_0}$.
Thus, ${\cal T}_{\alpha,\beta}^{\vec\lambda_0}\downarrow_{{\rm ad}(\mathfrak{h})}= \bigoplus_{\vec\lambda\in \sigma_{\mathfrak{h}}^{\vec\lambda_0}}\tau^{\vec\lambda}$, where $\tau^{\vec\lambda}$ denote $\mathfrak{h}$ irreps; in what follows, we shall take $\vec\lambda$ to be highest weight labels.
In particular, under the adjoint action,
\be \mho(2,N)\downarrow_{{\rm ad(\mso(2,N))}}=\bigoplus_{n=0}{\cal T}^{[n,n]} \ ,\ee
where ${\cal T}^{[n,n]}\equiv {\cal T}_{{\rm Id},{\rm Id}}^{[n,n]}$ denote the irreducible $\mso(2,N)$ tensor of highest weight $[n,n]$, consisting of monomials in $\widehat J_{A,B}$ in degree $n$ projected onto the Young tableaux \eqref{eq:YT ambient rect}.
An alternative decomposition of $\mho(2,N)$ can be obtained by splitting $\Op{J}_{A,B}$ into  $\Op{M}_{ab}:=\lambda \Op{J}_{a,b}$, generating an $\mso(1,N)$ to be identified as the Lorentz algebra, and transvections $\Op{P}_{a}:=\lambda \Op{J}_{0',a}$ with closure relations \eqref{eq:SO(2,N) Lorentz}.
Thus, the Lorentz $\mso(1,N)$ is stabilized by the automorphism $\pi$ defined by
\begin{align}
  \label{eq:pi vect}
  \pi(\Op{M}_{ab}):=\Op{M}_{ab}\ ,\qquad \pi(\Op{P}_a):=-\Op{P}_a\ ,
\end{align}
which is outer in $\mho(2,N)$, but, as we shall see, inner in a suitable non-polynomial extension of $\mho(2,N)$.
The corresponding twisted-adjoint action $\widetilde{\rm ad}:=\rho_{{\rm Id},\pi}$, viz.
\be \widetilde{\rm ad}(\Op{f})\Op{g}=\Op{f}\Op{g}-\Op{g}\pi(\Op{f})\ ,\ee
induces a twisted-adjoint $\mho(2,N)$ module $\widetilde{\cal T}\equiv {\cal T}_{{\rm Id},\pi}$ with $\mso(2,N)$ decomposition
\be \widetilde{\cal T}\downarrow_{\widetilde{\rm ad}(\mso(2,N))}=\bigoplus_{s=0}^\infty \widetilde{\cal T}^{[s,s]}\ ,\ee
where $\widetilde{\cal T}^{[s,s]}\equiv {\cal T}_{{\rm Id},\pi}^{[s,s]}$ is the infinite-dimensional $\mso(2,N)$ irrep with smallest Lorentz tensor $\tau^{(s,s)}$ of highest weight $(s,s)$, viz.
\be \widetilde{\cal T}^{[s,s]}\downarrow_{{\rm ad}(\mso(1,N))}=\bigoplus_{k=0}^\infty \widetilde\tau^{(s+k,s)}\ ,\qquad s=0,1,2,\dots\ ,\ee
where $\widetilde\tau^{(s+k,s)}$ are thus Lorentz tensors of highest weights $(s+k,s)$ built from $s$ powers of $\widehat{M}_{a,b}$ and $k$ powers of $\widehat P_a$ projected onto the Young tableaux \eqref{eq:YT Lorentz} with $n\to s+k$ and $t\to s$.

\subsection{Weyl zero-form and Bargmann--Wigner equations}

Letting ${\cal X}$ be a real commuting manifold with charts ${\cal U}$, the DGAs  $\mathcal{E}({\cal U}):=\Omega({\cal U})\otimes \mho(2,N)$, where $\Omega({\cal U})$ denotes the space of forms on ${\cal U}$, decompose under form degree into $\mathcal{E}({\cal U})=\bigoplus_{p}\mathcal{E}_{[p]}({\cal U})$, where $\mathcal{E}_{[p]}({\cal U})=\Omega_{[p]}({\cal U})\otimes \mho(2,N)$.
A locally $AdS_{N+1}$ background is an element
\be
  \label{eq:AdS W}
  \Op\Omega
  =
  -\frac{i}2 \Omega^{AB} \Op{J}_{A,B}
  \,,\qquad
  \Omega^{AB}
  \in 
  \Omega_{[1]}({\cal U})
  \,,
\ee
submitted to the flatness condition
\begin{align}
  \label{eq:AdS R=0}
  d\Op\Omega+\Op\Omega\;\Op\Omega
  &\approx0
  \,,
\end{align}
viewed as a background equation of motion, which implies the nilpotency of the associated adjoint and twisted-adjoint covariant derivatives
\be
  \label{eq:def acd}
  \Op{D}{\Op{f}_{[p]}}
  := d{\Op{f}_{[p]}}+\Op\Omega\;\Op{f}-(-1)^p\Op{f}\;\Op\Omega\ ,\qquad \Op{D}_\pi\Op{f}_{[p]}:=
  d \Op{f}_{[p]}+\Op\Omega\;\Op{f}-(-1)^p\Op{f}_{[p]}\;\pi(\Op\Omega)\,,\ee
of $\widehat f_{[p]}\in \mathcal{E}_{[p]}({\cal U})$.

Introducing a twisted-adjoint zero-form $\widehat C\in \mathcal{E}_{[0]}({\cal U})$, the linear equation 
\be 
  \label{eq:DC=0}
  \Op{D}_\pi\Op{C}
  \approx 0\,,\ee
defines a (universally) Cartan integrable system (CIS) together with \eqref{eq:AdS R=0}, with $\mathfrak{so}(2,N)$ decomposition
\be 
  \Op{D}_\pi\Op{C}^{\YTA{s}{s}}
  :=d\Op{C}^{\YTA{s}{s}}+\Op\Omega\;\Op{C}^{\YTA{s}{s}}-\Op{C}^{\YTA{s}{s}}\pi(\Op\Omega)
  \approx 0\,,\qquad s=0,1,\dots\,,\ee
where $\Op{C}^{\YTA{s}{s}}\in \Omega_{[0]}({\cal U})\otimes\widetilde{\cal T}^{[s,s]}$.
Lorentz decomposing 
\be
  \label{eq:W0}
  \Op\Omega
  = 
  \Op{\omega}+\Op{e}
  \,,\qquad
  \Op{\omega}
  :=
  -\frac{i}2 \omega^{ab}\Op{M}_{ab}
  \,,\qquad
  \Op{e}
  :=
  -ie^a\Op{P}_a
  \,,
\ee
induces a decomposition of the twisted-adjoint covariant derivative, viz.
\be
  \Op{D}_\pi
  =
  \Op\nabla+\widetilde{\rm ad}(\Op{e})
  \,,\qquad
  \widetilde{\rm ad}(\Op{e})=\Op\sigma_\pi^{(+)}+\Op\sigma_\pi^{(-)}
  \,,
\ee
where $\Op\nabla:=d+{\rm ad}(\Op{\omega})$ is the Lorentz covariant derivative, and
\be \Op\sigma_\pi^{(\pm)}:\Omega_{[p]}({\cal U})\otimes\widetilde\tau^{(s+k,s)}\to \Omega_{[p+1]}({\cal U})\otimes\widetilde\tau^{(s+k\pm 1,s)}\,,\ee
which obey 
\begin{align}
  \Op\nabla^2 +\Op\sigma_\pi^{(+)}\Op\sigma_\pi^{(-)}+\Op\sigma_\pi^{(-)}\Op\sigma_\pi^{(+)}
  &\approx 0
  \,,&
  \Op\nabla \Op\sigma_\pi^{(\pm)}+\Op\sigma_\pi^{(\pm)}\Op\nabla 
  &\approx0
  \,,&
  (\Op\sigma_\pi^{(\pm)})^2
  &\approx 0
  \,,
\end{align}
as a consequence of $(\Op{D}_\pi)^2\approx 0$.
Thus, expanding
\be
  \label{eq:basisPhi}
  \Op{C}^{\YTA s s}
  =
  \sum_{k=0}^\infty \Op{C}^{\YTL{s+k} s}
  \,,\qquad
  \Op{C}^{\YTL{s+k} s}
  \in 
  \Omega_{[0]}({\cal U})\otimes\widetilde\tau{(s+k,s)}
  \,,
\ee
the Cartan integrable spin-$s$ subsystem assumes the manifestly Lorentz covariant form
\be
  \Op\nabla\Op{e}
  \approx 0\ ,\qquad d\Op\omega+\Op\omega\;\Op\omega+\Op{e}^2
  \approx 0
  \,,
\ee
\be 
  \Op\nabla\Op{C}^{\YTL{s+k} s}+\Op\sigma_\pi^{(-)}\Op{C}^{\YTL{s+k+1} s}+\Op\sigma_\pi^{(+)}\Op{C}^{\YTL{s+k-1} s}\approx 0
  \,,\qquad k=0,1,\dots\ .
  \label{CComst}
\ee

If $\hat e$ defines a non-degenerate frame in ${\cal U}$, then 
\be 
  \Omega_{[p]}({\cal U})
  \otimes \widetilde\tau^{(s,s+k)}\stackrel{\widehat e}{\cong}\Omega_{[0]}({\cal U})\otimes \tau^{(1,\dots,1)_p}\otimes \widetilde\tau^{(s,s+k)}
  \,,\qquad (1,\dots,1)_p\equiv \underbrace{(1,\dots,1)}_{\tiny \mbox{$p$ times}}\ .
  \label{innerderivatives}
\ee
Lorentz decomposing 
\be \tau^{(1,\dots,1)_p}\otimes \widetilde\tau^{\vec s}=\bigoplus_{\vec s'}\tau^{\vec s'}([p];\vec s)\ ,\ee
where $\vec s'$ are highest $\mso(1,N)$ weights, the cohomology $H\left({\Op\sigma}^{(-)}_\pi\big|\Omega({\cal U})\otimes \widetilde{\cal T}^{[s,s]}\right)$ contains
\be H_{[0]}\left({\Op\sigma}^{(-)}_\pi\big|\Omega({\cal U})\otimes \widetilde{\cal T}^{[s,s]}\right)\cong \tau^{(s,s)}([0];s,s)\,,\ee
while $\tau^{(s+k,s)}([0];s+k,s)$ with $k>0$ is in the co-kernel of $\Op\sigma_\pi^{(-)}$.
Thus, using $\Op{C}^{\YTL{s+k}s}:= C^{(s+k,s)}_{a_1\cdots a_{s+k},b_1\cdots b_s}\Op{T}^{a_1\dots a_{s+k},b_1\dots b_s}$, Eq. \eqref{CComst} implies that $\Op{C}^{\YTL{s+k} s}$ are given by $k$ symmetrized traceless Lorentz covariant derivatives of $\Op{C}^{\YTL{s}s}$, which in its turn obeys\footnote{Eq. \eqref{2.32}, which is not Young-projected, implies the zero-divergence condition $\Op\nabla^a C^{(s,s)}_{ab_2\dots b_{s},c_1\dots c_s}
  \approx 0
  \,.$  } 
\begin{align}
  &&
  (\Op\nabla^2 +2)C^{(0,0)}
  &\approx 0
  \,,&&
  s=0\ ,\\
   \Op\nabla^a C^{(1,1)}_{a,b}
  &\approx 0
  \,,&
  \Op\nabla^{\phantom{[}}_{[a} C^{(1,1)}_{b,c]}
  &\approx 0
  \,,&&
  s=1\ ,\\
  &&
  \Op\nabla^{\phantom{[}}_{[a} C^{(s,s)}_{b|b_2\dots b_s,|c]c_2\dots c_s}
  &\approx 0
  \,,&&
  s=2,3,\dots
  \,,\label{2.32}
\end{align}
i.e. $C^{(0,0)}$ is a Klein--Gordon scalar field with mass $-2$; $C^{(1,1)}_{a,b}$
is a Faraday tensor obeying Maxwell's equations; and $C^{(s,s)}_{a_1\cdots a_{s},b_1\cdots b_s}$ with $s\geqslant 2$ is a tower of spin-$s$ (traceless) Weyl tensors obeying the Bargmann--Wigner equations.
In this sense, $\widehat C$ is referred to as the Weyl zero-form, and $\Op{C}^{\YTA ss}$ as its spin-$s$ Weyl zero-form components, which thus serve as generating functionals for $\Op{C}^{\YTL ss}$ and all their derivatives on-shell in regions of ${\cal U}$ where the transvection gauge field $e^a$ provides a non-degenerate frame.

\subsection{Unfolded Fronsdal fields and Central On-Mass-Shell Theorem}

The spin-$s$ Weyl zero-form $\Op{C}^{\YTA ss}$ with $s\geqslant 1$, which thus sits in the twisted-adjoint representation, can be glued to an adjoint one-form $\Op{W}^{\YTA{s-1}{s-1}}\in \Omega_{[1]}({\cal U})\otimes {\cal T}^{[s-1,s-1]}$ by means of a map
\be \Sigma_{2;1}: (\mathcal{E}_{[1]}({\cal U}))^{\wedge 2} \otimes \mathcal{E}_{[0]}({\cal U})\ ,\label{cocyclemap}\ee
that obeys the cocycle condition
\be
  \Op{D}\Sigma^{\YTA{s-1}{s-1}}\approx 0\ ,\qquad \Op\Sigma^{\YTA{s-1}{s-1}}
  :=
  \Sigma_{2;1}(\Op{\Omega},\Op{\Omega};\Op{C}^{\YTA{s}{s}})
  \ ,
  \label{Dsigma}
\ee
modulo \eqref{eq:DC=0} and \eqref{eq:AdS R=0}, ensuring the Cartan integrability of
\begin{align}
  \Op{D}\Op{W}^{\YTA{s-1}{s-1}}+ \Sigma_{2;1}(\Op{\Omega},\Op{\Omega};\Op{C}^{\YTA{s}{s}})
  \approx0
  \,.\label{eq:DW=S s}
\end{align}
Introducing the generating function
\begin{align}
  \Op{W}
  :&=
  \sum_{s=1}^{\infty}\Op{W}^{[s-1,s-1]}
  \in\mathcal{E}_{[0]}({\cal U})
  \,,
\end{align}
the Cartan integrable subsystems for $s=1,\dots$ can be grouped together as
\begin{align}
  \Op{D}\Op{W}+ \Sigma_{2;1}(\Op{\Omega},\Op{\Omega};\Op{C})
  \approx0
  \,.\label{eq:DW=S}
\end{align}
Eq. \eqref{Dsigma} determines $\widehat\Sigma^{\YTA{s-1}{s-1}}$ modulo a homogenous solution in the image of $\Op{D}$ that can be removed by 
\be
  \delta_{\Op\Upsilon} \Op{W}^{\YTA{s-1}{s-1}}
  =
  -\Op{\Upsilon}^{\YTA{s-1}{s-1}}
  \,,\qquad
  \delta_{\Op\Upsilon}\Op\Sigma^{\YTA{s-1}{s-1}}
  =
  \Op{D}\Op{\Upsilon}^{\YTA{s-1}{s-1}}
  \,,
  \label{shift}
\ee
where $\Op{\Upsilon}^{\YTA{s-1}{s-1}}\in \Omega_{[1]}({\cal U})\otimes {\cal T}^{\YTA{s-1}{s-1}}$, that is, $\widehat\Sigma^{\YTA{s-1}{s-1}}$ is an element in $H_{[2]}\left(\Op{D}\big|\Omega({\cal U})\otimes {\cal T}^{[s-1,s-1]}\right)$ represented by a homogeneous tri-linear map, as indicated in \eqref{cocyclemap}, which vanishes if $\Op{C}^{\YTA{s-1}{s-1}}$ vanishes.

Treating $(\Op\Omega,\Op{C}^{\YTA ss},\Op{W}^{\YTA{s-1}{s-1}})$ as dynamical variables, there are two types of Cartan gauge symmetries: i) background gauge transformations
\begin{align}
  \label{eq:BG gt}
  \delta_{\Op{\zeta}} \Op{\Omega}
  &=
  \Op{D}\Op{\zeta}
  \,,&
  \delta_{\Op{\zeta}} \Op{C}^{\YTA ss}
  &=
  \Op{C}^{\YTA ss}\pi(\Op{\zeta})-\Op{\xi}\Op{C}^{\YTA ss}
  \,,&
  \delta_{\Op{\zeta}}\Op{W}^{\YTA{s-1}{s-1}}
  &=
  [\Op{W}^{\YTA{s-1}{s-1}},\Op{\zeta}]
  \,,
\end{align}
with $\Op{\zeta}\in \Omega_{[0]}({\cal U})\otimes \mso(2,N)$, under which 
\be \delta_{\Op{\zeta}} \Op\Sigma^{\YTA{s-1}{s-1}}=[\Op\Sigma^{\YTA{s-1}{s-1}},\Op{\zeta}]\ ,\ee
comprising Killing symmetries obeying $\delta_{\Op{\zeta}} \Op{\Omega}=0$; and ii) abelian gauge transformations
\begin{align}
  \label{eq:abelian}
  \delta_{\Op{\cal E}} \Op{\Omega}
  &=0
  \,,&
  \delta_{\Op{\cal E}}\Op{C}^{\YTA ss}
  &=0
  \,,&
  \delta_{\Op{\cal E}} \Op{W}^{\YTA{s-1}{s-1}}
  &=
  \Op{D}{\Op{\cal E}^{\YTA{s-1}{s-1}}}
\end{align}
with adjoint parameters ${\Op{\cal E}^{\YTA{s-1}{s-1}}}\in \Omega_{[0]}({\cal U})\otimes {\cal T}^{[s-1,s-1]}$, under which
\be \delta_{\Op{\cal E}}\Op\Sigma^{\YTA{s-1}{s-1}}=0\ .\ee
Lorentz decomposing 
\be
  \Op{D}
  =
  \Op\nabla +{\rm ad}(\Op{e})
  \,,\qquad
  {\rm ad}(\Op{e})
  =
  {\Op\sigma}^{(+)}+{\Op\sigma}^{(-)}
  \,,
\ee
where
\begin{align}
  {\Op\sigma}^{(\pm)}:\Omega_{[p]}({\cal U})\otimes\tau^{(s-1,t)}\to \Omega_{[p+1]}({\cal U})\otimes\tau^{(s-1,t\pm 1)}
  \,,
\end{align}
it follows from $(\Op{D})^2\approx 0$ that
\begin{align}
  \Op\nabla^2 +{\Op\sigma}^{(+)}{\Op\sigma}^{(-)}+{\Op\sigma}^{(-)}{\Op\sigma}^{(+)}\approx 0
  \,,&&
  \Op\nabla {\Op\sigma}^{(\pm)}+{\Op\sigma}^{(\pm)}\Op\nabla \approx0\ ,\qquad ({\Op\sigma}^{(\pm)})^2\approx 0
  \,.
\end{align}
Thus, using the expansions \eqref{eq:W0}-\eqref{eq:basisPhi} and
\begin{align}
  \Op{W}^{\YTA{s-1}{s-1}}
  =
  \sum_{t=0}^{s-1} \Op{W}^{\YTL{s-1}{t}}
  \,,&&
  {\Op{\cal E}^{\YTA{s-1}{s-1}}}
  =\sum_{t=0}^{s-1} \Op{\cal E}^{\YTL{s-1}{t}}
  \,,\\
  \Op\Sigma^{\YTA{s-1}{s-1}}=\sum_{t=0}^{s-1} \Op{\Sigma}^{\YTL{s-1}{t}}
  \,,&&
  \Op\Upsilon^{\YTA{s-1}{s-1}}=\sum_{t=0}^{s-1} \Op{\Upsilon}^{\YTL{s-1}{t}}
  \,,
\end{align}
where $\Op{W}^{\YTL{s-1}{t}}, \Op{\Upsilon}^{\YTL{s-1}{t}}\in \Omega_{[1]}({\cal U})\otimes \tau^{(s-1,t)}$, $\Op{\cal E}^{\YTL{s-1}{t}}\in \Omega_{[0]}({\cal U})\otimes \tau^{(s-1,t)}$ and $\Op{\Sigma}^{\YTL{s-1}{t}}\in \Omega_{[2]}({\cal U})\otimes \tau^{(s-1,t)}$, the gluing equation \eqref{eq:DW=S s} and abelian gauge transformations decompose into
\begin{align}
  \label{eq:weq}
  \Op\nabla \Op{W}^{\YTL{s-1}{t}}+{\Op\sigma}^{(-)}\Op{W}^{\YTL{s-1}{t+1}}+{\Op\sigma}^{(+)}\Op{W}^{\YTL{s-1}{t-1}}+ \Op{\Sigma}^{\YTL{s-1}{t}}\approx 0
  \,,\end{align}
and
\begin{align}  \label{eq:abellor}
  \delta_{\Op{\cal E}} \Op{W}^{\YTL{s-1}{t}}=\Op\nabla \Op{\cal E}^{\YTL{s-1}{t}}+{\Op\sigma}^{(-)}\Op{\cal E}^{\YTL{s-1}{t+1}}+{\Op\sigma}^{(+)}\Op{\cal E}^{\YTL{s-1}{t-1}} \,,
\end{align}
respectively, while the cocycle condition takes the form
\begin{align}
  \label{eq:dsigmalor}&&
  \Op{\Sigma}^{\YTL{s-1}{t}}+{\Op\sigma}^{(-)}\Op{\Sigma}^{\YTL{s-1}{t+1}}+{\Op\sigma}^{(+)}\Op{\Sigma}^{\YTL{s-1}{t-1}}\approx 0\,,\end{align}
with attendant tensorial shift symmetry 
\begin{align} 
  \label{shiftlor}
  \delta_{\Op\Upsilon} \Op{W}^{\YTL{s-1}{t}}=-\Op\Upsilon^{\YTL{s-1}{t}}\ ,\qquad \delta_{\Op\Upsilon}\Op{\Sigma}^{\YTL{s-1}{t}}=\Op\nabla\Op\eta^{\YTL{s-1}{t}}+{\Op\sigma}^{(-)}\Op{\Upsilon}^{\YTL{s-1}{t+1}}+{\Op\sigma}^{(+)}\Op{\Upsilon}^{\YTL{s-1}{t-1}}
  \,.
\end{align}

Assuming $\widehat e$ to be non-degenerate, which induces the isomorpism
\be \Omega_{[p]}({\cal U})\otimes \tau^{(s-1,t)}\stackrel{\widehat e}{\cong}\Omega_{[0]}({\cal U})\otimes\tau^{ (1,\dots,1)_p} \otimes \tau^{(s-1,t)}\ , \qquad (1,\dots,1)_p\equiv \underbrace{(1,\dots,1)}_{\tiny \mbox{$p$ times}}\ ,\label{innerderivatives2}\ee
and Lorentz decomposing 
\be \tau^{(1,\dots,1)_p}\otimes \tau^{\vec s}=\bigoplus_{\vec s'}\tau^{\vec s'}([p];\vec s)\ ,\ee
where $\vec s'$ are highest $\mso(1,N)$ weights,
it follows from the cocycle condition \eqref{eq:dsigmalor} modulo the shift symmetry \eqref{shiftlor} that
\begin{align} \Op\Sigma^{\YTA{s-1}{s-1}}{\in}&H_{[2]}\left({\Op\sigma}^{(-)}\big| \Omega({\cal U})\otimes {\cal T}^{[s-1,s-1]}\right)\\& \cong\underbrace{\tau^{(s)}([2];s-1,1)\oplus \tau^{(s-2)}([2];s-1,1)}_{\mbox{current}}\oplus \underbrace{\tau^{(s,s)}([2];s-1,s-1)}_{\mbox{Weyl}}\ ,\label{H[2]}\end{align}
where the spin-$s$ current cannot\footnote{There exists a spin-$s$ current that is linear in $C^{[0,0]}$ which can be removed by a higher-spin generalized Weyl rescaling.} be activated at the linearized level under the assumption that $\Op\Sigma^{\YTA{s-1}{s-1}}$ is linear in $\Op{C}^{\YTA ss}$, and the linearized spin-$s$ Weyl tensor \cite{Vasiliev:1988sa}
\be \Op\Sigma^{\YTL{s-1}{s-1}}= e^c\wedge e^d\wedge C^{(s,s)}_{c a_1\cdots a_{s-1},d b_1\cdots b_{s-1}}\Op{T}^{a_1\dots a_{s-1},b_1\dots b_{s-1}}\,.\label{eq:S=eeC}
\ee
From \eqref{eq:weq} and \eqref{eq:abellor}, it follows that $\Op{W}^{\YTA{s-1}{s-1}}$ decomposes into i) auxiliary gauge fields in the co-kernel of ${\Op\sigma}^{(-)}$; ii) shift-symmetry gauge fields in the image of ${\Op\sigma}^{(-)}$, which can be eliminated using the abelian gauge transformations \eqref{eq:abelian} with parameters in the co-kernel of ${\Op\sigma}^{(-)}$; and iii) an algebraically unconstrained gauge field
\be \Op{W}^{\YTL{s-1}{0}}\in H_{[1]}\left({\Op\sigma}^{(-)}\big| \Omega({\cal U})\otimes {\cal T}^{[s-1,s-1]}\right)\cong \tau^{(s)}([1];s,0)\oplus\tau^{(s-2)}([1];s,0)\,,\ee
that is, a doubly traceless symmetric tensor gauge field of rank $s\geq 1$, alias a spin-$s$ Fronsdal tensor, with equation of motion given by the current projection of \eqref{eq:weq}, which is a second-order differential operator invariant under the residual abelian gauge transformations with traceless parameter
\be \Op{\epsilon}^{(s-1,0)}\in H_{[0]}\left({\Op\sigma}^{(-)}\big| \Omega({\cal U})\otimes {\cal T}^{[s-1,s-1]}\right)\cong \tau^{(s-1)}([0];s,0)\,.\ee
In summary, Eq. \ref{eq:DW=S s} with cocycle given by \eqref{eq:S=eeC} constitutes a CIS describing a rank-$s$ Fronsdal field on-shell; this fact is referred to as the Central On-Mass-Shell Theorem \cite{properties,more,review99}.

\subsection{Classical linearized solution spaces}

Subjecting Eqs.
(\ref{eq:AdS R=0},\,\ref{eq:DC=0},\,\ref{eq:DW=S}) to appropriate boundary conditions on ${\cal X}$ yields spaces of linearized Fronsdal fields propagating on locally anti-de Sitter backgrounds\footnote{Integration on charts ${\cal U}\subset{\cal X}$ yields locally defined linearized solution spaces that can be glued together into into sections over ${\cal X}$ using transition functions, which along with the holonomies for $\Op{\Omega}$ in ${\cal G}_0(SO(2,N))$ define linearized solution spaces globally; for examples, see \cite{BTZ}.
} corresponding to Weyl zero-forms in a non-polynomial extension $\widetilde{\cal M}_{[0]}({\cal U})$ of ${\cal E}_{[0]}({\cal U})$ arising from ``stalks'' $\widetilde{\cal M}$ corresponding to adjoint modules ${\cal A}$ with associative algebra structures to be studied in Sec. \ref{sec:starprod} and constructed explicitly in the case of $N=3$ for specific boundary conditions in Secs. \ref{sec:solve Z} and \ref{sec:PTBH}.

Integrating Eq. \eqref{eq:AdS R=0} starting from a base point $x_0\in{\cal U}$ yields a gauge function
\be \Op{L}:{\cal U}\to {\cal G}_0(SO(2,N))\ ,\qquad \widehat L|_{x_0}=1\ ,\label{eq:AdS gf Op}\ee
in a representation ${\cal G}_0:SO(2,N)\to {\cal A}$, such that
\be 
 \Op{\Omega}\approx \Op{L}^{-1} d\Op{L}\ ;\label{LinversedL}\ee
hence, the (nilpotent) covariant derivatives in \eqref{eq:def acd} can be written as
\bea
\Op{D}{\Op{f}_{[p]}}&=&{\rm Ad}_{\widehat L^{-1}} (d{\rm Ad}_{\widehat L}\widehat f_{[p]})\ ,\qquad {\rm Ad}_{\widehat L}\widehat f_{[p]}:= \Op{L}\; \Op{f}_{[p]}\; \Op{L}^{-1}\ ,\\
  \Op{D}_\pi\Op{f}_{[p]}
  &=&\widetilde{{\rm Ad}}_{\widehat L^{-1}} (d\widetilde{{\rm Ad}}_{\widehat L}\widehat f_{[p]})\ ,\qquad \widetilde{{\rm Ad}}_{\widehat L}\widehat f_{[p]}:=\Op{L}\; \Op{f}_{[p]}\; \pi(\Op{L}^{-1})
  \ ,
\eea
for $\Op{f}_{[p]}\in\mathcal{E}_{[p]}({\cal U})$, where ${\rm Ad}_{\widehat L}$ acts faithfully on $\mathcal{E}_{[p]}({\cal U})$, and $\widetilde{{\rm Ad}}_{\widehat L}$ in non-polynomial  $\widetilde{\cal M}_{[p]}({\cal U})$ extensions of $\mathcal{E}_{[p]}({\cal U})$.

Thus, for $p=0$, integration of Eq. \eqref{eq:DC=0}, subject to boundary and regularity conditions on ${\cal X}$, requires an extension of $\mathcal{E}_{[0]}(\cal U)$ to the $SO(2,N)$-module
\be \widetilde{\cal M}_{[0]}({\cal U}|\widehat L;\widetilde{\cal M}):= \left\{\widetilde{\rm Ad}_{\Op{L}^{-1}}\Op{C}'\mid_{\cal U}\big|\Op{C}'\in \widetilde{\cal M}\right\}\ ,\qquad d\widehat{C}'=0\ ,\ee
where the integration constant belongs to the twisted-adjoint $\overline{\mathfrak{ho}}(2,N)$-module
\be\widetilde{\cal M}\downarrow_{\widetilde{\rm ad}(\mho(2,N)}
=\bigoplus_{\vec\lambda_0} \widetilde {\cal M}^{\vec \lambda_0}\ ,\label{stalk}\ee
with $\widetilde {\cal M}^{\vec \lambda_0}$ being spaces of operators built from $\widehat J_{A,B}$, comprising irreducible twisted-adjoint representations of $\mho(2,N)$, making up the spectrum of local degrees of freedom of the system.
The representation $\widetilde {\cal M}^{\vec \lambda_0}$ need not be an $SO(2,N)$-module, which manifests itself in possible singularities in the unfolded Weyl tensors.
Letting $\overline{\widetilde {\cal T}}$ be the space of power-series expansions in the basis of Young-projected monomials given in Eq. \eqref{eq:YT Lorentz}, there exists a monomorphism $\rho$ from $\widetilde{\cal M}_{[0]}({\cal U}'|\widehat L;\widetilde{\cal M})$ into $\overline{\widetilde {\cal T}}$ \cite{Boulanger:2013naa,Boulanger:2015uha},  both viewed as twisted-adjoint $\mho(2,N)$-modules, with domain
\be\label{2.66} {\cal U}'(\widehat L;\widehat C'):=\left\{x\in {\cal U}\,\Big|\, \widetilde{\rm Ad}_{\Op{L}^{-1}}|_x\Op{C}'\in \overline{\widetilde {\cal T}}\right\}\ ,\ee
such that 
\be \Op{C}|_{{\cal U}'(\widehat L;\widehat C')}\approx  \rho(\widetilde{\rm Ad}_{\Op{L}^{-1}}\Op{C}')\ , \label{CIF}\ee
consists of regular Weyl tensors $C^{[s,s]|\vec\lambda_0}$; these local data can then be glued together into globally defined solutions on ${\cal X}$ exhibiting various types of distinct singularities and asymptotic boundary conditions depending on $\Op{C}'$ and the holonomies in $\widehat L$ \cite{fibre,BTZ,Iazeolla:2020jee,Iazeolla:2022dal}.

An example, spelled out in more detail in Sec. \ref{sec:PTBH}, is obtained by taking ${\cal X}\cong S^1\times S^3$, and imposing $\widehat C|_{S^1\times \{s_\infty\}}=0$ at conformal infinity $S^1\times \{s_\infty\}$; particle and anti-particle modes in $D^\pm(s+N-2,(s))$ then arise from operators $\widehat C'\in \widetilde{\cal M}_{[0]}(x_0)$ such that $\widehat C$ is in the domain of $\rho$ in all of ${\cal X}$, while black-hole modes arise from operators in $\widetilde{\cal M}_{[0]}(x_0)$ such that $\widehat C$ belongs to the domain of $\rho$ only in a subset ${\cal X}'$ of ${\cal X}$, and coalesce to distributions given by Gaussian delta sequences as $x\to x_0$ where $x_0\in{\cal X}\setminus {\cal X}'$ are the points where the Weyl tensors of these modes have classical singularities \cite{2011}; the analyticity properties of these solutions are studied in Sec. \ref{sec:starprod} and Appendix \ref{app:anal}. 
In what follows, we shall most often suppress the map $\rho$.

Finally, the Cartan integration of Eq. \eqref{eq:DW=S} using the formalism developed in \cite{cosmo,COMST}, yields the generating function $\widehat W$ of the unfolded Fronsdal fields in terms of a construct built from the vacuum gauge function $\widehat L$, the zero-form integration $\widehat C'$ and a linearized gauge function.
This integration, which is algebraically more involved than that of \eqref{eq:DC=0} due to the presence of the cocycle and the need to tune the linearized gauge function such that $\widehat W\in \mathcal{E}_{[1]}(\cal{U})$, will be explored in further detail below for particle and black-hole modes.

\subsection{Formal higher spin gravity}
\label{sec:formalHS}

Higher spin gravities with gauge algebra $\mathfrak{ho}(2,N)$ and vacuum solution $\widehat\Omega$, can be described perturbatively by a CIS of the form  
\begin{align}
  \label{eq:deom W}
  &0\approx d\Op{W}+\Op{W}\;\Op{W}
  +\Sigma_{2;1}(\Op{W},\Op{W};\Op{C})
  +\Sigma_{2;2}(\Op{W},\Op{W};\Op{C},\Op{C})
  +\cdots
  \,,\\
  \label{eq:deom C}
  &0\approx d\Op{C}+\Op{W}\;\Op{C}-\Op{C}\;\Op\pi(\Op{W})
  +\Sigma_{1;2}(\Op{W};\Op{C},\Op{C})
  +\Sigma_{1;3}(\Op{W};\Op{C},\Op{C},\Op{C})
  +\cdots
  \,,
\end{align}
where $\Sigma_{p;q}$ are multi-linear maps that 
\begin{enumerate}[label=\roman*),ref=(\roman*)]
\item\label{it:maniLor} preserves manifest Lorentz-covariance; and 
\item\label{it:cocyle} obeys perturbatively defined cocycle conditions\footnote{Under the assumption that the perturbative scheme can be executed with master fields in $\mathcal{E}({\cal U})$, the higher cocycles are generated by the Chevalley-Eilenberg cohomology of $\mathfrak{ho}(2,N)$.} that ensure Cartan integrability for $\Op{C}\in {\cal E}_{[0]}({\cal U})$ and $\Op{W}\in {\cal E}_{[1]}({\cal U})$ order-by-order in $\Op{C}$, which implies invariance under the Cartan gauge transformations
\begin{align} \delta_{\widehat{\epsilon}} \widehat{W}&=d\widehat{\epsilon}+[\widehat{W},\widehat{\epsilon}]+2\Sigma_{2;1}(\Op{W},\Op{\epsilon};\Op{C})
  +2\Sigma_{2;2}(\Op{W},\Op{\epsilon};\Op{C},\Op{C})+\cdots\ ,\label{eq:dgt W}\\
 \delta_{\widehat{\epsilon}} \widehat{C}&=-\widehat{\epsilon}\widehat{C}+\widehat{C}\widehat{\pi}(\widehat{\epsilon})-\Sigma_{1;2}(\Op{\epsilon};\Op{C},\Op{C})
  -\Sigma_{1;3}(\Op{\epsilon};\Op{C},\Op{C},\Op{C})+\cdots\label{eq:dgt C}\ ,\end{align}
\end{enumerate}
If the deformations are  real-analytic functions of the coupling constant $g$ in a perturbative expansion
\begin{align}
  \Op{W}
  &=
  \Op{\Omega}+g\Op{W}^{(1)}+O(g^2)
  \,,&
  \Op{C}
  &=
  g\Op{C}^{(1)}+ O(g^2)
  \,,\qquad g\in \R\,,
\end{align}
around a background $\widehat\Omega$ obeying
\be d\widehat{\Omega}+\widehat \Omega\widehat \Omega\approx 0\ ,\qquad \widehat\Omega\in {\cal E}_{[1]}({\cal U})\ ,\label{Omegaho}\ee
then it follows on general grounds that the linearized approximation 
\begin{align}
  \label{eq:deom Wlin}
  d\Op{W}^{(1)}+\Op{\Omega}\;\Op{W}^{(1)}+\Op{W}^{(1)}\;\Op{\Omega}
  +\Sigma_{2;1}(\Op{\Omega},\Op{\Omega};\Op{C}^{(1)})
  \approx &0
  \,,\\
  \label{eq:deom Clin}
  d\Op{C}^{(1)}+\Op{\Omega}\;\Op{C}^{(1)}-\Op{C}^{(1)}\;\Op\pi(\Op{\Omega})\approx 0
  \,,
\end{align}
form a CIS together with \eqref{Omegaho}; in particular, for $\widehat\Omega\in \Omega_{[1]}({\cal U})\otimes \mso(2,N)$, the linearized system reduces to  (\ref{eq:DC=0},\,\ref{eq:DW=S}); henceforth, we shall let $\widehat C$ and $\widehat W$ stand for the full fields and $\widehat C^{(1)}$ and $\widehat W^{(1)}$ denote their linearized approximations.

The formal treatment can be streamlined \cite{Sharapov:2019vyd} by extending the ${\cal E}({\cal U})$ as a DGA to 
\be \Upsilon({\cal U}):={\cal E}({\cal U})\otimes {\rm Gr}[\psi]\otimes \mathbb{C}[\mathbb{Z}_2[\hat k]\ ,\ee
where I) the outer Klein operator $\hat k$ has degree $0$ and obeys
\be \hat k \Op{M}_{ab}=\Op{M}_{ab}\hat k\ ,\qquad \hat k\Op{P}_{a}=-\Op{P}_{a}\hat k\ ,\qquad \hat k\hat k=1\ ,\ee
which stabilizes ${\cal I}[{\cal S}]$, whereby the adjoint and twisted-adjoint $\mho(2,N)$ modules are embedded into the adjoint representation of $\mho(2,N)\otimes \mathbb{C}[\mathbb{Z}_2[\hat k]]$; and II) ${\rm Gr}[\psi]$ is the Grassmann algebra generated by an outer nilpotent closed variable $\psi$ of degree $1$, which thus anti-commutes to odd forms and $d$ using Koszul sign conventions with total degree given by form degree and $\psi$-degree.
The Quillen superconnection
\be \widehat X\in \Upsilon_{[1]}({\cal U})\ ,\ee
of total degree one, has an expansion in $(\psi,\hat k)$ that comprises $\Op{\Phi}$ and $\Op{W}$ together with an adjoint zero-form and twisted-adjoint one-form, such that \eqref{eq:deom W}--\eqref{eq:deom C} can be embedded into

\be d\Op{X}+\Op{X}^2+\Sigma_{3}(\Op{X},\Op{X},\Op{X})\
  +\Sigma_{4}(\Op{X},\Op{X},\Op{X},\Op{X})
  +\cdots\approx 0\ ,\ee
using multi-linear maps $\Sigma_n:(\Upsilon_{[1]}({\cal U}))^{\otimes n}\to \Upsilon_{[2]}({\cal U})$ of intrinsic degree $2-n$ compatible with Lorentz-covariance and Cartan integrability.

As shown in \cite{Sharapov:2019vyd}, the existence of $\{\Sigma_n\}$ is equivalent to a deformation of $\mho(2,N)\otimes \mathbb{C}[\mathbb{Z}_2[\hat k]$ as an associative algebra along a Hochschild cohomology element, which is a noncommutative counterpart of a Poisson structure on a commuting manifold.
The resulting formal HSG present conceptual challenges:
\begin{enumerate}[label=\alph*)]
\item \emph{Classical perturbation theory:} Representing $\mho(2,N)$ using symbols, the deformations $\Sigma_n$ are non-local functionals on the fiber space $\mso(2,N)$ whose evaluation on linearized solutions obeying physically desirable boundary conditions on ${\cal X}$ resulting in non-polynomial symbols is non-trivial;
\item \emph{Variational principle:} The fiber non-locality blurs what criteria controls the class of functionals making up the Batalin--Vilkovisky bracket algebra required for constructing master actions and observables in an off-shell formulation.
\item \emph{Classical moduli space:} the Hochschild cohomology obeying \ref{it:cocyle} contains an infinite set of elements, of which only a finite-dimensional subset obeys \ref{it:maniLor}, that may eventually have to be included into a partition function with a suitable path integral measure.
\end{enumerate}
One possible way to address the above issues is to employ the Frobenius--Chern--Simons (FCS) off-shell extension of Vasiliev's HSG theory \cite{Boulanger:2015kfa}, as will be outlined further in Sec. \ref{Sec2.8}.

In what follows, we shall outline Vasiliev's four-dimensional minimal bosonic HSG model, which provides a setting that is non-trivial (with local degrees of freedom) yet accessible to exact treatments.

\subsection{Twistorial fiber algebra in four dimensions}
\label{sec:OpY}

From the Lie algebra isomorphism $\mso(2,3)\cong\msp(4;\R)$ and the fact that $C_2[\mso(2,N)|{\cal S}]$ equals the quadratic Casimir of the metaplectic representation of $\msp(4;\R)$, it follows that $\mho(2,3)$ is isomorphic to the space of even polynomials in an $Sp(4;\R)$-quartet $\OpY_\au$, $\au=1,\dots,4$, obeying the Heisenberg commutation rules
\begin{align}
  \label{eq:[Y,Y]}
  \left[\OpY_\au,\,\OpY_\bu\right]
  =
  2i\epsilon_{\underline{\a\b}}\,,
\end{align}
coordinatizing a noncommutative holomorphic symplectic manifold ${\cal Y}$; for our spinorial notation, see App. \ref{app:cvt}.
Thus, $\mho(2,3)$ is spanned by even totally symmetric monomials
\begin{align}
  \label{eq:M(Y)}
  \Op{J}_{\au_1\cdots\au_{2n}}
  =
  \OpY_{(\au_1}\cdots\OpY_{\au_{2(s-1)})}
  \,,\qquad n=0,1,2,\dots\ ,
\end{align}
and from 
  $\left[\Op\ell(\ell_1),\,\Op\ell(\ell_2)\right]
  =
  \Op{\mathcal{\ell}}\left([\ell_1,\,\ell_2]\right)$, where $\Op{\ell}(\ell_1)
  :=
  -\frac{i}{4}\OpY \ell\OpY$ for bi-spinors  $\ell_\au\,^\bu$, it follows using Eq. \eqref{eq:M(Gamma)} that the normalized $\mso(2,3)$ generators 
\begin{align}
  \Op{J}_{A,B}
  =
  \frac{1}{8}\OpY\Gamma_{AB}\OpY
  \,.
\end{align}
The twisted-adjoint Weyl zero-form $\Op{C}= \sum_{s=0}^{\infty}\Op{C}^{\YTA ss}$, where
\begin{align}
  \label{eq:C Taylor}
 s=0\ :\quad C^{[0,0]}
  &=\sum_{k=0}^{\infty}C_{\ag_1\cdots\ag_{k}\ad_1\cdots\ad_k}
  \widehat{J}^{\ag_1\dots\ag_{k},\ad_1\dots\ad_{k}}\ ,\\
 s\geqslant 1\ :\quad C^{[s,s]}&=
  \sum_{k=0}^{\infty}\left(
  C_{\ag_1\cdots\ag_{2s+k}\ad_1\cdots\ad_k}
  \widehat{J}^{\ag_1\dots\ag_{2s+k},\ad_1\dots\ad_{k}}+
  C_{\ag_1\cdots\ag_k\ad_1\cdots\ad_{2s+k}}\widehat{J}^{\ag_1\dots\ag_{k},\ad_1\dots\ad_{2s+k}}\right)\ ,
\end{align}
and the adjoint one-form $\Op{W}=\sum_{s=1}^{\infty}\Op{W}^{[s-1,s-1]}$, where
\begin{align}
  \label{eq:W Taylor}
  s\geqslant 1\ :\quad \Op{W}^{[s-1,s-1]}
  &=
  \sum_{t=0}^{s-1}\left(
  W_{\ag_1\cdots\ag_{s-1+t}\ad_1\cdots\ad_{s-1-t}}
\widehat{J}^{\ag_1\dots\ag_{s-1+t},\ad_1\dots\ad_{s-1-t}}\right.\nonumber\\&\qquad\qquad \left.+\,
  W_{\ag_1\cdots\ag_{s-1-t}\ad_1\cdots\ad_{s-1+t}}
  \widehat{J}^{\ag_1\dots\ag_{s-1-t},\ad_1\dots\ad_{s-1+t}}\right)
  \,,
\end{align}
obey the integer-spin projection\footnote{The component fields are assumed to be Grassmann even.}
\be
  \label{eq:BP Y}
  C_{\ag(m)\ad(n)} = C_{\ag(m)\ad(n)}\delta_{m+n,0\,\mod\,2}
  \,,\qquad
  W_{\ag(m)\ad(n)} = W_{\ag(m)\ad(n)} 
  \delta_{m+n,0\,\mod\,2}
  \,,
\ee
referred to as the bosonic projection, ensuring perturbative expansions around (locally) $AdS_4$ backgrounds in terms of Lorentz tensorial gauge fields.
These can be made real by imposing 
\be
  \label{eq:RC Y}
  \left(C_{\ag(m)\ad(n)}\right)^\ast = (-)^m C_{\ag(n)\ad(m)}
  \,,\qquad
  \left(W_{\ag(m)\ad(n)}\right)^\ast = -W_{\ag(n)\ad(m)}
  \,.
\ee
The model can be truncated further to even Lorentz spins by imposing
\begin{align}
  \label{eq:MBP Y}
  C_{\ag(m)\ad(n)} &= C_{\ag(m)\ad(n)}\delta_{m-n\mod4,0}
  \,,&
  W_{\ag(m)\ad(n)} &= W_{\ag(m)\ad(n)} 
  \delta_{m+n\mod4,0}
  \,,
\end{align}
referred to as the minimal bosonic projection.

The twisting map \eqref{eq:pi vect} admits two chiral spinorial realisations, viz.
\be
  \label{eq:pi(Y)}
  \Op\pi (f(\Opy,\Opyb))
  =
  f(-\Opy,\Opyb)
  \,,\qquad
  \Op{\bar\pi}(f(\Opy,\Opyb))
  =
  f(\Opy,-\Opyb)
  \,,
\ee
which are involutive automorphisms of the oscillator algebra, such that 
\be \widehat\pi\widehat{\bar \pi}(\widehat C)=\widehat C\ ,\qquad \widehat\pi\widehat{\bar \pi}(\widehat W)=\widehat W\ ,\ee
for bosonic master fields.

The chiral twisting maps \eqref{eq:pi(Y)} are inner, viz.
\be
  \Op\pi (f(\OpY))
  =
  \OpKy\;f(\OpY)\;\OpKy
  \,,\qquad
  \Op{\bar\pi}(f(\OpY))
  =
  \OpKyb\;f(\OpY)\;\OpKyb
  \,,
\ee
using Klein operators $\OpKy$ and $\OpKyb$, which are holomorphic oscillator representations of elements in $Sp(4;\C)$ (not in $Sp(4;\R)$), obeying
\begin{align}
  \label{eq:def Ky}
  \OpKy^2=\OpKyb^2=1
  \,,&&
  \left\{\OpKy,\,\Opy_\ag\right\}=\left[\OpKyb,\,\Opy_\ag\right]=0
  \,,&&
  \left[\OpKy,\,\Opyb_\ad\right]=\left\{\OpKyb,\,\Opyb_\ad\right\}=0
  \,.
\end{align}
These operators can be used to define chiral adjoint Weyl zero-forms \cite{Engquist:2005yt,Gelfond:2008td,Didenko:2009td,fibre}
\be \widehat{\Psi}:= \Op{C}\,\widehat \kappa_y\ ,\qquad\widehat{\overline{\Psi}}:= \Op{C}\,\widehat{\bar\kappa}_{\bar y}\ ,\label{eq:Psi=Ck}\ee
related via $\Op\Psi
  =
  \Op{\bar\Psi}\;\OpKy\;\OpKyb$.
Thus, at the linearized level,
\be
  \Op\Psi^{(1)}
   \approx
  \Op{L}^{-1}\;\Op\Psi^{\prime(1)}\;\Op{L}
  \,,\qquad
  \Op{\bar\Psi}^{(1)}
   \approx
  \Op{L}^{-1}\;\Op{\bar\Psi}^{\prime(1)}\;\Op{L}
  \,,\label{eq:Psi}
\ee
where $\widehat L$ is the vacuum gauge function and the chiral zero-form integration constants
\be
  \label{eq:Psi'=C'k}
  \Op\Psi^{\prime(1)}
  :=
  \Op{C}^{\prime(1)}\;\OpKy
  \,,\qquad
  \Op{\bar\Psi}^{\prime(1)}
  :=
  \Op{C}^{\prime(1)}\;\OpKyb
  \,,
\ee
belong to the adjoint $SO(2,N)$-module 
\be {\cal A}:= \widetilde {\cal M}\,\widehat{\kappa}_y\cong \widetilde {\cal M}\,\widehat{\bar\kappa}_{\bar y}\ ,\label{calA}\ee
where $\widetilde{\cal M}$ is the stalk defined in \eqref{stalk}. 
As shown in \cite{Vasiliev:2012vf,neiman} (see also \cite{Didenko:2021vui}), the induced relation between $\Op\Psi^{\prime(1)}$ and $\Op{C}^{(1)}$ provides an unfolded version of the Penrose transform;
the nonlinear completion of Penrose's transform is one of the main features of Vasiliev's system, to which we turn next.

\subsection{Four-dimensional fully nonlinear field equations}
\label{sec:VE}

Vasiliev's method \cite{properties,more,review99} for generating solutions to the deformation problem in Sec. \ref{sec:formalHS}, uses fibrations ${\cal T}$ with generic fiber ${\cal Y}$ over a holomorphic symplectic twistor space ${\cal Z}$ coordinatized by a $Sp(4;\R)$ quartet
\begin{align}
  \label{eq:[Z,Z][Y,Z]}
  \OpZ_\au
  &=
  (\Opz_\ag,\,-\Opzb_\ad)
  \,,&
  \left[\OpZ_\au,\,\OpZ_\bu\right]
  &=
  -2i\epsilon_{\underline{\a\b}}
  \,,&
  \left[\OpY_\au,\,\OpZ_\bu\right]
  &=
  0
  \,;
\end{align}
${\cal T}$ is in its turn fibered over ${\cal X}$, as described on-shell by a flat horizontal superconnection generating an on-shell FDA forming a subspace of an operator DGA with finite observables, alias a noncommutative fibration\footnote{The noncommutative fibration admits semi-classical descriptions in terms of symbols that may degenerate viewed as bundles with fibre $\mathfrak{ho}(2,3)$, defined using Eq. \eqref{2.66}, which provides a natural mechanism for resolving various classical singularities in gauge theory and gravity formulated using tensor fields; for examples, see \cite{2011,BTZ}.}.

Viewing the total space as a fibration over ${\cal U}$, the corresponding superconnection comprises a one-form $\Op{U}$ and zero-forms $\Op\Phi$ and $\widehat S_{\underline\alpha}=(\widehat S_{\alpha},-\widehat {\overline S}_{\dot\alpha})$, subject to
\begin{itemize}
    \item[i)] integer-spin projection conditions
\begin{align}
  \label{eq:BP VE}
  \Op\pi\Op{\bar\pi}(\Op A)
  &=
  \Op A
  \,,&
  \Op\pi\Op{\bar\pi}(\Op\Phi)
  &=
  \Op\Phi
  \,,&
  \Op\pi\Op{\bar\pi}(\Op{S}_{\underline\alpha})
  &=
  -\Op{S}_{\underline\alpha}
  \,,
\end{align}
suitable for a non-linear bosonic HSG model consisting perturbatively of Fronsdal tensor gauge fields, where the chiral involutive automorphisms 
\begin{align}
  \label{eq:def pi}
  \Op\pi(\Opy,\Opyb,\Opz,\Opzb)
  &=
  (-\Opy,\Opyb,-\Opz,\Opzb)
  \,,&
  \Op\pi(\Op{f}\;\Op{g})
  &=
  \Op\pi(\Op{f})\;\Op\pi(\Op{g})
  \,,\\
  \label{eq:def pibar}
  \Op{\bar\pi}(\Opy,\Opyb,\Opz,\Opzb)
  &=
  (\Opy,-\Opyb,\Opz,-\Opzb)
  \,,&
  \Op{\bar\pi}(\Op{f}\;\Op{g})
  &=
  \Op{\bar\pi}(\Op{f})\;\Op{\bar\pi}(\Op{g})
  \,,
\end{align}
are inner by means of the chiral Klein operators on ${\cal Y}$ and dittos on ${\cal Z}$, viz.
\begin{align}
  \label{eq:def k}
  \OpKz^2&=1
  \,,&
  \left\{\OpKz,\,\Opz_\ag\right\}&=0
  \,,&
  \left[\OpKz,\,\Opzb_\ad\right]&=0
  \,,&
  \OpK:&=\OpKy\;\OpKz
  \,,&
  \Op\pi(f(\OpY,\,\OpZ))
  :&=
  \OpK\;f(\OpY,\,\OpZ)\;\OpK
  \,,\\
  \label{eq:def kb}
  \OpKzb^2&=1
  \,,&
  \left\{\OpKzb,\,\Opzb_\ad\right\}&=0
  \,,&
  \left[\OpKzb,\,\Opz_\ag\right]&=0
  \,,&
  \OpKb:&=\OpKyb\;\OpKzb
  \,,&
  \Op{\bar\pi}(f(\OpY,\,\OpZ))
  :&=
  \OpKb\;f(\OpY,\,\OpZ)\;\OpKb
  \,;
\end{align}
\item[ii)] flatness conditions on ${\cal U}$, viz.
\bea
  \label{eq:VEx}
  d\Op{U}+\Op{U}\;\Op{U}
  =0
  \,,\qquad
  d\Op{\Phi}+\Op{U}\;\Op{\Phi}-\Op{\Phi}\;\Op\pi(\Op{U})
  =0
  \,, \\ \label{eq:dS}
  d\Op{S}_\ag+\Op{U}\;\Op{S}_\ag-\Op{S}_\ag\;\Op{U}
  =0
  \,,\qquad
  d\Op{\bar S}_\ad+\Op{U}\;\Op{\bar S}_\ad-\Op{\bar S}_\ad\;\Op{U}
  =0\ ;
\eea
\item[iii)] algebraic zero-form constraints on the total space, viz.
\begin{align}
  \label{eq:[S,S]}
  \left[\Op{S}_\ag,\,\Op{\bar S}_\ad\right]
  &=0
  \,,&
  \left[\Op{S}_\ag,\,\Op{S}_\bg\right]
  +
  2i\epsilon_{\ag\bg}\left(
  1-e^{i\theta}\Op\Phi\;\OpK
  \right)
  &=
  0
  \,,&
  \left[\Op{\bar S}_\ad,\,\Op{\bar S}_\bd\right]
  +
  2i\epsilon_{\ad\bd}\left(
  1-e^{-i\theta}\Op\Phi\;\OpKb
  \right)
  &=
  0
  \,,\\&&\label{eq:[S,Phi]}
  \Op{S}_\ag\;\Op\Phi+\Op\Phi\;\Op\pi(\Op{S}_\ag)
  &=0
  \,,&
  \Op{\bar S}_\ad\;\Op\Phi+\Op\Phi\;\Op{\bar\pi}(\Op{\bar S}_\ad)
  &=0
  \,,
\end{align}
where $\theta$ is a real parameter.
\end{itemize}
Eqs. \eqref{eq:[S,S]} and \eqref{eq:[S,Phi]} ensure the integrability of Eqs. \eqref{eq:VEx} and \eqref{eq:dS} on ${\cal U}$, and imply that $\Op{S}_{\alpha}$ and $\Op{\overline S}_{\dot\alpha}$ are two mutually commuting $Sp(2;\C)$-quartets deformed \`a la Wigner by $\widehat \Phi$ \cite{Wigner:50,Yang:51,Vasiliev:1989re}.

The purpose of Eqs. (\ref{eq:dS}--\ref{eq:[S,Phi]}) is to solve for the $\Op Z$ dependence of all fields in terms of $\Op Z$-independent, projected fields\begin{align}
  \label{eq:C(Phi) W(U)}
  \Op{W}
  :&=
  \Op{\mathcal{P}}^{(W)}\Op{U}
  \,,&
  \Op{C}
  :&=
  \Op{\mathcal{P}}^{(C)}\Op\Phi
  \,.
\end{align} 
These fields obey deformed equations  \eqref{eq:deom W} and \eqref{eq:deom C} with non-trivial cocycles $\Sigma_{p;q}$ obtained by projecting the equations in \eqref{eq:VEx}.
The non-triviality of the latter cocycles results from the non-triviality of Wigner's deformation.
The projectors can be built perturbatively following the procedure reviewed in Sec.
\ref{sec:VE pert}.

Vasiliev's original formulation relied on symbol calculus referring to a specific normal-ordering scheme; in what follows, we shall instead proceed using the operator formulation which is manifestly  ordering independent with the aim of examining to what extent this property can be retained at the level of classical moduli spaces.

Introducing auxiliary line-elements $dZ^\au$ obeying
\begin{align}
  \label{eq:[dZ,f]}
  \dZ^\au\;\dZ^\bu
  &=
  -\dZ^\bu\;\dZ^\au
  \,,&
  \dZ^\au\;\Op{f}_{[p]}
  &=
  (-1)^p\Op{f}_{[p]}\;\dZ^\au
  \,,
\end{align}
where $p$ denotes the total form degree, and corresponding differentials
\begin{align}
  \label{eq:diff}
  \Op{q}\Op{f}
  :&=
  \frac{i}{2}\dZ^\au \left[\OpZ_\au,\,\Op{f}\right]
  \,,&
  \Op\Delta
  :&=
  d+\Op{q}
  \,,
\end{align}
with prefactor chosen for later convenience, the deformed oscillators assemble into a one-form master field
\be \Op{S}:=\dZ^\au \Op{S}_\au\ ,\ee
whose adjoint action is a covariantization of $\Op{q}$.
Thus, upon defining connections
\begin{align}
  \label{eq:A=U+V}
  \Op{V}
  :&=
  \frac{i}{2}\dZ^\au\left(\Op{S}_\au-\Op{Z}_\au\right)
  \,,&
  \Op{A}
  :&=
  \Op{U}+\Op{V}
  \,;
\end{align}
extending the chiral twisting automorphisms by declaring 
\begin{align}
  \label{eq:pidZ}
  \Op\pi(\dz_\ag)
  &=
  -\dz_\ag
  \,,&
  \Op\pi(\dzb_\ad)
  &=
  \dzb_\ad
  \,,&
  \Op{\bar\pi}(\dz_\ag)
  &=
  \dz_\ag
  \,,&
  \Op{\bar\pi}(\dzb_\ad)
  &=
  -\dzb_\ad
  \,;
\end{align}
and assembling the Klein operators into a two-form
\be
  \Op{J}
  :=
  -\frac{i}{4}\left(
  e^{i\theta}\OpKy\;\OpKz\;\dz^\ag\dz_\ag
  +e^{-i\theta}\OpKyb\;\OpKzb\;\dzb^\ad\dzb_\ad
  \right)
  \,,\ee
which is thus a closed and twisted-central two-form on ${\cal Z}$, viz.
\be \widehat{\Delta}\widehat J=0\ ,\qquad \widehat J\widehat f=\widehat{\pi}(\widehat f)\widehat J\ ,\ee
for any form $\widehat f$ obeying $\widehat{\pi}\widehat{\bar\pi}(\widehat f)=\widehat f$; Eqs. (\ref{eq:VEx}\,-\,\ref{eq:dS}) acquire the form
\begin{align}
  \label{eq:VE A}
  \Op\Delta\Op{A}+\Op{A}\;\Op{A}
  +\Op\Phi\;\Op{J}
  &=0
  \,,\\\label{eq:VE Phi}
  \Op\Delta\Op\Phi+\Op{A}\;\Op\Phi-\Op\Phi\;\Op\pi(\Op{A})
  &=0
  \,,
\end{align}
defining a CIS on ${\cal U}\times {\cal Z}$ with a closed and twisted-central element.
The system is invariant under Cartan gauge transformations
\begin{align}
  \label{eq:VE igt}
  \delta_{\Op{g}}\Op{A}
  =
  \Op\Delta\Op{g}
  +
  \left[\Op{A},\,\Op{g}\right]
  \,,&&
  \delta_{\Op{g}}\Op\Phi
  =
  -\Op{g}\;\Op\Phi
  +\Op\Phi\;\Op\pi(\Op{g})
  \,,
\end{align}
where $\widehat{\pi}\widehat{\bar\pi}(\widehat g)=\widehat g$,
whose finite form reads
\begin{align}
  \label{eq:VE fgt}
  \Op{A}\to
  \Op{G}^{-1}\;\Op\Delta\Op{G}
  +
  \Op{G}^{-1}\;\Op{A}\;\Op{G}
  \,,&&
  \Op\Phi\to
  \Op{G}^{-1}\;\Op\Phi'\;\Op\pi(\Op{G})
  \,,
\end{align}
for maps $\Op{G}:{\cal U}\times {\cal Z}\to {\cal G}$ obeying $\widehat{\pi}\Op{\bar\pi}(\widehat G)$, where ${\cal G}$ is a group of similarity transformations of the fibration generated by polynomials in $Y_{\underline\alpha}$, containing ${\cal G}_0$ as a subgroup. 

The bosonic HSG model can be truncated by means of (anti-)linear and (anti-)automorphisms preserving Eqs. (\ref{eq:VE A},\,\ref{eq:VE Phi}): Using the hermitean conjugation operation
\begin{align}
  \label{eq:def dagger}
  (\Opy,\Opyb,\Opz,\Opzb)^\dagger
  &=
  (\Opyb,\Opy,\Opzb,\Opz)
  \,,&
  \Op\Delta\left(\Op{f}\,^\dagger\right)
  &=
  \left(\Op\Delta\Op{f}\right){}^\dagger
  \,,&
  (\Op{f}\;\Op{g})
  &=
  (-)^{\deg\Op{f}\deg\Op{g}}\,
  (\Op{g})^\dagger\;(\Op{f})^\dagger
  \,,
\end{align}
reality of the Lorentz tensorial component fields is ensured by imposing
\begin{align}
  \label{eq:RC VE}
  \Op A\,^\dagger
  &=
  -\Op A
  \,,&
  \Op\Phi\,^\dagger
  &=
  \Op\pi\Op\Phi
  \,,&
  \Op G\,^\dagger
  &=
  \Op G^{-1}
  \,.
\end{align}
Using the linear anti-automorphism
\begin{align}
  \label{eq:def tau}
  \Op\tau(\Opy,\Opyb,\Opz,\Opzb)
  &=
  f(i\Opy,i\Opyb,-i\Opz,-i\Opzb)
  \,,&
  \Op\tau\Op\Delta
  &=
  \Op\Delta\Op\tau
  \,,&
  \Op\tau(\Op{f}\;\Op{g})
  &=
  (-)^{\deg\Op{f}\deg\Op{g}}\,
  \Op\tau(\Op{g})\;\Op\tau(\Op{f})
  \,,
\end{align}
the even-spin projection \eqref{eq:MBP Y} generalizes to
\begin{align}
  \label{eq:MBP VE}
  \Op\tau(\Op A)
  &=
  -\Op A
  \,,&
  \Op\tau(\Op\Phi)
  &=
  \Op\pi(\Op\Phi)
  \,,&
  \Op\tau(\Op G)
  &=
  \Op G^{-1}
  \,,
\end{align}
defining the (non-linear) minimal bosonic HSG model.

\subsection{Observables and gauge functions}

Letting $\Tr$ denote a trace operation on the undeformed oscillator algebra, the quantities \cite{Sezgin:2005pv,Engquist:2005yt,Sezgin:2011hq,COMST}
\begin{align}
  \label{eq:def OWL}
  {\cal I}_{n,p}({\cal M})
  :=
  \frac{1}{(2\pi)^2} \Tr\left(
  \left(\Op\Phi\;\OpK\right)^{n}\;
  \Pexp\left(\int_{L({\cal M})}dZ^\au\,\Op{A}_\au\right)\;
  \exp\left(\tfrac{i}{2}{\cal M}^\au \OpZ_\au\right)\;
  \OpK^p\;\OpKb^p
  \right)
  \,,
\end{align}
where $n=0,1,\dots$, $p=0,1$, and $L({\cal M})$ is a straight line\footnote{%
The open Wilson line is equivalent to a closed Wilson loop with path given by the straight line followed by the reversed straight line; inserting transition functions on ${\cal Z}$ at the cusps, the resulting holonomy, and hence associated the Wilson Line observable, is the same as for the open line \cite{COMST}.
} in ${\cal Z}$, are invariant under Cartan gauge transformations \eqref{eq:VE fgt} and closed on ${\cal X}$ on-shell (which means that they can be evaluated at any point in ${\cal X}$).
In strictly positive form degrees on ${\cal X}$, charges $\oint_\Sigma {\cal J}$ are given by integrals over topologically non-trivial surfaces $\Sigma\subset {\cal X}$ of on-shell (de Rham) closed forms ${\cal J}=\Tr \widehat{\cal J}$ that are invariant under a proper subgroup of the group of Cartan gauge transformations defining the transition functions of the HSG geometry, analogous to the group of locally defined Lorentz transformations in ordinary gravity \cite{Sezgin:2011hq,Sharapov:2020quq}.

The Vasiliev system can be integrated on ${\cal U}$ by taking
\begin{align}
  \label{eq:VE CI}
  \Op{U}=\Op{G}^{-1}\;d\Op{G}
  \,,&&
  \Op\Phi=\Op{G}^{-1}\;\Op\Phi'\;\Op{\pi}(\Op{G})
  \,,&&
  \Op{S}_\ag=\Op{G}^{-1}\;\Op{S}'_\ag\;\Op{G}
  \,,&&
  \Op{\bar S}_\ad=\Op{G}^{-1}\;\Op{\bar S}'_\ad\;\Op{\pi}(\Op{G})
  \,,
\end{align}
where the primed fields are  zero-form integration constants on ${\cal U}$, annihilated by $d$, obeying the Wigner system in Eqs. (\ref{eq:[S,S]},\,\ref{eq:[S,Phi]}) on ${\cal Z}$, which in its turn can be solved using operator methods.
The resulting classical HSG solution spaces are thus coordinatized by the moduli for the Wigner system and the gauge function $\widehat G$ modulo the subgroup of proper gauge transformations that preserve the various charges introduced above.
The identification of the resulting space of large gauge transformations is a non-trivial issue of which some aspects will be discussed in Sec. \ref{sec:Lorentz}.

\subsection{Constraints on interactions from noncommutative geometry} \label{Sec2.8}

The classical moduli spaces of
Vasiliev's four-dimensional HSG are quasi-free subalgebras ${\cal O}({\cal M})$ of horizontal DGAs $\Omega_{\rm hor}({\cal M})$ arising as cosets inside spaces $\Omega({\cal M})$ of forms on fibered noncommutative differential Poisson manifolds ${\cal M}$, given locally by Cartesian products of a commutative manifolds ${\cal X}$ and fibered noncommutative complex four-manifolds 
\be {\cal Y}\rightarrow {\cal T}\to {\cal Z}\ ,\ee
alias twistor spaces, where $\cal Y$ and $\cal Z$ are coordinatized by $Y_{\underline\alpha}$ and $Z_{\underline\alpha}$, respectively.
The algebra ${\cal O}({\cal M})$ is generated by a horizontal one-form $\widehat A=-\widehat A^\dagger$ and a zero-form $\widehat B=\widehat \Phi\widehat k=\widehat B^\dagger$ obeying 
\begin{align}\label{rVE}
\widehat{\Delta}\,\widehat A+ \widehat A \,\widehat A + {\cal F}(\widehat B) \,\widehat I - \overline{\cal F}(\widehat B)\,\widehat {\bar{I}} \approx 0~,\quad \widehat{\Delta}\,\widehat B+\widehat A\,\widehat  B-\widehat B\,\widehat  A \approx 0\ ,
\end{align}
constituting a CIS, where the juxtaposition denotes an associative noncommutative deformation of the wedge product; $\widehat \Delta$ denotes a compatible differential, that we shall assume is given by the undeformed differential in Eq. \eqref{eq:diff}; ${\cal F}$ and $\overline{\cal F}:=({\cal F})^\dagger$ are composite operators encoding on-shell interaction ambiguities; and $\widehat I:=\widehat k\widehat J=-\widehat I^\dagger$ is a special central cohomologically non-trivial holomorphic horizontal two-form on ${\cal T}$.
Requiring manifest Lorentz covariance and parity invariance implies ${\cal F} =\widehat  B$ and $\theta\in \{0,\pi\}$, referred to as the Type A and B models \cite{Sezgin:2001zs}, respectively.
These models have perturbative expansions in asymptotically locally AdS$_4$ (ALAdS$_4$) regions consisting of one real Fronsdal field for each even Lorentz spin, with the spin-two field identified, via its minimal Lorentz couplings, as the graviton, and the scalar field having parity $+1$ or $-1$ in the Type A and B models, respectively.

The moduli space admits subspaces in which ${\cal O}({\cal M})$ is represented by symbols given by classical horizontal forms on ${\cal M}$ composed using a twisted convolution product, also known as star product, representing the operator product.
This space contains different subspaces arising by imposing various boundary conditions on the symbols, whereby the projections of $\widehat A$ and $\widehat B$ onto ${\cal X}$, respectively, define a connection $\widehat W=\widehat A|_{\cal X}$ whose holonomies belong to spaces of BTZ-like higher spin vacua, and a Weyl zero-form $\widehat B|_{\cal X}$ whose integration constant $\widehat B'$ belongs to operator algebras on ${\cal Y}$ on which the holonomies act faithfully, containing different types of local degrees of freedom on ${\cal X}$. 

The linearized Weyl zero-form on $\cal X$ can be constructed by taking the vacuum gauge function $\Op{L}$ to be defined modulo a group $\Gamma\subset {\cal G}_0(SO(2,3))$ of (metaplectic) holonomies acting on the left and a structure subgroup ${\cal H}\subset {\cal G}_0(SO(2,3))$ acting on the right, and expanding $\widehat B'$ over adjoint and twisted-adjoint ${\cal G}_0(SO(2,3))$ modules that are invariant under $\Gamma$ and on which ${\cal H}$ acts tensorially, that is, without any indecomposable module substructure.
Taking ${\cal H}={\rm Stab}(k)\cong {\cal G}_0(SO(1,3))$ and $\mathcal{X}$ to contain homotopy cylinders $\mathcal{X}^{(\xi)}\cong \mathcal{X}^{(\xi)}_3\times [0,1[$ on which
\be \Op{L}|_{\mathcal{X}^{(\xi)}}= \Op{L}^{(\xi)}\Op{b}^{(\xi)}\ ,\ee
where $\Op{L}^{(\xi)}$ are three-dimensional gauge functions of conformal geometries on $\mathcal{X}^{(\xi)}_3$ (defined modulo holonomies and using an $SO(1,2)$ structure group) and $\Op{b}^{(\xi)}$ are radial gauge functions on $[0,1[$, and assuming the parallel transport of $\widehat B$ to degenerate on $\mathcal{X}^{(\xi)}_3\times \{0\}$ in the sense that
\be \Op{B}|_{\mathcal{X}^{(\xi)}_3\times \{0\}}\approx 0\ ,\label{2.12}\ee
which thus constrains $\Op{L}$ as well as $\Op{B}'$, the linearized configuration describes a set of defects surrounded by ALAdS geometries glued together into a global geometry, which need not obey the metricity condition away from the defects.

While the noncommutativity does not affect the essence of Cartan integration, it constrains the spaces boundary conditions, as the symbols must belong to special classes of distributions in order for the star product algeba to close; for a review on the construction of HSG moduli spaces containing massless particles, massive Type D modes, domain walls and boundary-to-bulk propagators, using Cartan integration, see for instance \cite{Iazeolla:2020jee}.

More precisely, the closed and central element $\widehat I$ is built from a holomorphic (finite) volume form on ${\cal Z}$ and a holomorphic Klein operator on ${\cal Y}$ induced by the discrete Weil map \cite{Vasiliev:1989qh,Vasiliev:1989re} exchanging symbols with their chiral Fourier transforms, or, equivalently, traces with supertraces.
As a result, the Vasiliev system induces a duality between deformations of, on the one hand, the noncommutative symplectic structure on ${\cal T}$, and, on the other hand, the BTZ-like higher spin vacuum on ${\cal X}$, with deformation parameter $\widehat C'$ expanded in dual bases appropriated, respectively, by the operator algebra in twistor space and the holonomies on ${\cal X}$; for example, operator algebras on ${\cal T}$ obtained from singleton representations of Dirac's conformal particle are dualized into massless particle and generalized Petrov Type D modes carried by Fronsdal fields on ALAdS$_4$ regions of ${\cal X}$.

It is worth stressing that, thought of as an extension of gravity, Vasiliev's HSG exhibits two remarkable feaures: Firstly, being based on DGAs rather than Riemannian structures, the latter arise on-shell only locally on ${\cal X}$ in the neighbourhood of defects where the vacuum gauge function $\widehat L$ blows up in accordance with Eq. \eqref{2.12}.  
Second, an observable of a CIS is an integral of an on-shell closed globally defined element in the corresponding FDA.
Remarkably, the Vasiliev system admits infinitely many charges given by integrals of top-forms on ${\cal Z}$ as in Eq. \eqref{eq:def OWL} with a dual interpretation as on-shell closed zero-forms on ${\cal X}$.
In ALAdS regions, these ``zero-form charges'' are given by strongly coupled expansions in derivatives of the Weyl tensors $C^{(s,s)}$ regularized by the twistor space map so as to provide observables that are non-local on ${\cal X}$ serving as extensive variables \cite{Colombo:2010fu}; for example, computed on multi-body solutions in HSG, they ``count'' the number of bodies making up these solutions, and when evaluated on boundary-to-bulk propagators, their perturbative expansions yield semi-classical HS amplitudes \cite{Engquist:2005yt,Colombo:2012jx,Didenko:2012tv}.

The fact that the (regularized) zero-form charges remain finite (and constant) even as the separate Weyl tensors fall off the asymptotic limit, suggests that HSG holds clues to physically desirable infra-red modifications of ordinary gravity governed by requiring the existence of zero-form charges built from the spin-two Weyl zero-form $C^{[2,2]}$, referred to as strong Cartan integrability, which one may think of as an analog of the existence of Lax-pairs for lower-dimensional field theories.

In summary so far, the classical moduli spaces of Vasiliev's HSG thus consists of noncommutative geometries, and exhibit the following key features:
\begin{itemize}
    \item[i)] \emph{Background covariance:} The classical moduli, taken from cohomological elements, holonomies and transition functions, parametrize noncommutative twistor spaces fibered over BTZ-like higher spin vacua on ${\cal X}$, whereby the deformations merge with the background into a single dynamical entity (given by a Quillen superconnection) akin to how gravitons merge with the background metric to form a dynamical metric;
    \item[ii)] \emph{Semi-classical topologies:} The solutions are built on top of semi-classical (fibered) manifolds ${\cal M}$ of distinct topologies, which facilitates the incorporation of various topologies beyond that of global AdS$_4$ \cite{BTZ};
    \item[iii)] \emph{Differential graded algebra operations:} The moduli spaces are FDA shells inside DGAs equipped with classical observables constructed using no other operations than the noncommutative wedge products, compatible differentials and trace operations of the background geometry.
\end{itemize}
Subjecting HSG to Hamilton's variational principle, property (iii) facilitates the embedding of the Vasiliev system into noncommutative topological AKSZ sigma models of FCS type, in which the fundamental fields are horizontal differential forms on fibered noncommutative differential Poisson manifolds, serving as backgrounds, and all BV functionals are constructed using the basic operations of the underlying DGA, that is, the trace, product and compatible deformation of the de Rham differential.
The resulting definition of BV algebra serves as a noncommutative counterpart to that of local BV algebra used in commuting AKSZ sigma models\footnote{Beyond the semi-classical level, the BV algebra faces potential perturbative anomalies due to infra-red-ultra-violet mixing \cite{Ferrari:2005kx}, though the cancellation of topological anomalies of even and odd forms \cite{Wu:1990ci} are encouraging.}, providing HSG with a natural substitute for the notion of locality in ordinary gravity coupled to matter, based on Riemannian structures, referred to as DGA locality, which is suitable for a path integral formalism distinct from that of the Fronsdal program \cite{Boulanger:2012bj,Boulanger:2015kfa}.
Finally, the FCS model contains a dynamical two-form whose background values may provide the moduli parameters of the Hochschild cohomology of formal higher spin gravity, thus providing a natural inclusion of these degrees of freedom into a partition function, thereby resolving issue (c) in Sec. \ref{sec:formalHS}.

In what follows, we shall explore the above framework mainly at the classical and semi-classical level by examining perturbative expansions of classical field configurations and a select set of zero-form charges.

\subsection{Perturbation theory}
\label{sec:VE pert}

Locally $AdS_4$ spacetimes are given by field configurations of the form
\begin{align}
  \label{eq:VE BG}
  \Op{U}^{(0)}&=\Op\Omega
  \,,&
  \Op\Phi^{(0)}&=0
  \,,&
  \Op{S}^{(0)}_\au&=\OpZ_\au\,,
\end{align}
where $\widehat \Omega$ is built from a vacuum gauge function $\Op{L}$ as in Eq. \eqref{LinversedL}.
Expanding the master fields around
\eqref{eq:VE BG}, viz.
\begin{align}
  \Op{U}
  &=
  \widehat\Omega+\sum_{n=1}^{+\infty}\Op{U}^{(n)}
  \,,&
  \Op{V}
  &=
  \sum_{n=1}^{+\infty}\Op{V}^{(n)}
  \,,&
  \Op\Phi
  &=
  \sum_{n=1}^{+\infty}\Op\Phi^{(n)}
  \,,
\end{align}
where the suffices refers to the order in the linearized Weyl zero-form $\widehat C$,
the fluctuations obey
\begin{align}
  \label{eq:qPhin}
  \Op{q}\Op\Phi^{(n)}
  +
  \sum_{k=1}^{n-1}\Op{V}^{(n-k)}\;\Op\Phi^{(k)}
  +
  \sum_{k=1}^{n-1}\Op\Phi^{(k)}\;\Op\pi\left(\Op{V}^{(n-k)}\right)
  &=0
  \,,\\
  \label{eq:DPhin}
  \Op{D}_\pi\Op\Phi^{(n)}
  +
  \sum_{k=1}^{n-1}\Op{U}^{(n-k)}\;\Op\Phi^{(k)}
  +
  \sum_{k=1}^{n-1}\Op\Phi^{(k)}\;\Op\pi\left(\Op{U}^{(n-k)}\right)
  &=0
  \,,\\
  \label{eq:qVn}
  \Op{q}\Op V^{(n)}
  +
  \sum_{k=1}^{n-1}\Op{V}^{(k)}\;\Op{V}^{(n-k)}
  +\Phi^{(n)}\;\Op{J}
  &=0
  \,,\\
  \label{eq:qUn}
  \Op{q}\Op U^{(n)}
  +
  \Op{D}\Op V^{(n)}
  +
  \sum_{k=1}^{n-1}\Op{U}^{(n-k)}\;\Op{V}^{(k)}
  +
  \sum_{k=1}^{n-1}\Op{V}^{(n-k)}\;\Op{U}^{(k)}
  &=0
  \,,\\
  \label{eq:DUn}
  \Op{D}\Op U^{(n)}
  +
  \sum_{k=1}^{n-1}\Op{U}^{(k)}\;\Op{U}^{(n-k)}
  &=0
  \,,
\end{align}
using the background covariant differentials $\widehat D$ and $\widehat D_\pi$ on ${\cal U}$ defined in \eqref{eq:def acd}.
Eq. \eqref{eq:qPhin} implies that $\Op\Phi^{(1)}$ is covariantly constant and $\OpZ$-independent,
that is, it can be identified with $\Op{C}^{(1)}$, and we will make no distinction between $\Op\Phi^{(1)}$ and $\Op{C}^{(1)}$ in the rest of the paper.
It follows that the perturbative expansions of the zero-form charges  \eqref{eq:def OWL} start with
\begin{align}
  {\cal I}^{(n)}_{n,p}(\mathcal{M})
  =
  \frac{1}{(2\pi)^2} \Tr\Big(
  \left(\Op\Phi^{(1)}\;\OpK\right)^{n}\;
  \exp\left(\tfrac{i}{2}\mathcal{M}^\au \OpZ_\au\right)\;
  \OpK^p\;\OpKb^p
  \Big)
  \,.
\end{align}
The dependence on $\Op{Z}$ of the remaining fluctuations can be obtained by repeatedly integrating equations of the form
\begin{align}
  \label{eq:qf=g}
  \Op{q}\Op{f}=\Op{g}
  \,,\qquad \Op{q}\Op{g}=0\ ,
\end{align}
using sources $\widehat g$ built from lower-order fluctuations.
Solutions to Eq. \eqref{eq:qf=g} are provided by \emph{resolution operators} $\Op{q}^{\ast}$
such that
\begin{align}
  \Op{\mathcal{P}}
  :=
  1
  -\Op{q}^{\ast}\Op{q}
  -\Op{q}\Op{q}^{\ast}
  \,,
\end{align}
projects onto the cohomology of $\Op{q}$ on ${\cal Z}$ treated as a topologically trivial space\footnote{Treating ${\cal Z}$ as a topologically non-trivial noncommutative space, the  cohomology contains the twisted-central one-form $\widehat J$ as well as elements in degree one \cite{Iazeolla:2007wt}.}.
In other words, $\Op{q}^{\ast}$ provides a resolution of the identity, viz.
\begin{align}
  1
  =
  \Op{q}^{\ast}\Op{q}
  +\Op{q}\Op{q}^{\ast}
  +\Op{\mathcal{P}}
  \,.\label{2.135}
\end{align}
If $\Op{g}$ in addition obeys $\Op{\mathcal{P}}\Op{g}=0$, then Eq. \eqref{2.135} implies that $\Op{q}^{\ast}\Op{g}$ is a particular solution to Eq. \eqref{eq:qf=g}, viz.
\begin{align}
  \label{eq:qq*g=g}
  \Op{q}\left(\Op{q}^{\ast}\Op{g}\right)
  =
  \Op{g}-\Op{q}^{\ast}\Op{q}\Op{g}-\Op{\mathcal{P}}\Op{g}
  =
  \Op{g}
  \,.
\end{align}
Thus, applying the resolution of the identity to a general solution $\Op{f}$ to Eq. \eqref{eq:qf=g}, yields 
\begin{align}
  \label{eq:sol qf=g}
  \Op{f} = \Op{q}^{\ast}\Op{g} + \Op{q}\Op{h} + \Op{c}\,,
\end{align}
where the \emph{gauge function}
\begin{align}
  \label{eq:h(q*)}
  \Op{h}
  &=
  \Op{q}^{\ast}\Op{f}
  \,,
\end{align}
and the \emph{integration constant}
\begin{align}
  \label{eq:c(q*)}
  \Op{c}
  &=
  \Op{\mathcal{P}}\Op{f}
  \,.
\end{align}
It follows that if $\left(\Op{q}^{\ast[i]}\right)_{i=1,2}$ are two different resolution operators, then 
\begin{align}
  \Op{f}
  &=
  \Op{q}^{\ast[1]}\Op{g} + \Op{q}\Op{h}^{[1]} + \Op{c}^{[1]}
  \nonumber\\&=
  \left(
  \Op{q}^{\ast[2]}\Op{q}
  + \Op{q}\Op{q}^{\ast[2]}
  + \Op{\mathcal{P}}^{[2]}
  \right)\Op{q}^{\ast[1]}\Op{g}
  + \Op{q}\Op{h}^{[1]} + \Op{c}^{[1]}
  \nonumber\\&=
  \Op{q}^{\ast[2]}\Op{g} + \Op{q}\Op{h}^{[2]} + \Op{c}^{[2]}
  \,,
\end{align}
are two equivalent expressions for the general solution to Eq. \eqref{eq:qf=g}, 
where 
\begin{align}
  \label{eq:h,c(q*)}
  \Op{h}^{[2]}
  &:=
  \Op{h}^{[1]}
  +
  \Op{q}^{\ast[2]}\Op{q}^{\ast[1]}\Op{g}
  \,,&
  \Op{c}^{[2]}
  &:=
  \Op{c}^{[1]}
  +
  \Op{\mathcal{P}}^{[2]}\Op{q}^{\ast[1]}\Op{g}
  \,.
\end{align}
The freedom in choosing $\Op{c}^{(n)}$ and $\Op{h}^{(n)}$ when integrating perturbatively, 
is equivalent to the that in choosing $\Op{C}^\prime$ and $\Op{H}$ in the following rewriting of \eqref{eq:VE CI}:
\begin{align}
  \label{eq:CI LHC}
  \Op{A}
  &=
  (\Op{L}\;\Op{H})^{-1}\;\Op\Delta(\Op{L}\;\Op{H})
  +(\Op{L}\;\Op{H})^{-1}\;
  \Op{\mathcal{V}}\left[\Op{C}^\prime\right]\;
  (\Op{L}\;\Op{H})
  \,,&
  \Op\Phi
  &=
  (\Op{L}\;\Op{H})^{-1}\;\Op{C}^\prime\;
  \Op\pi(\Op{L}\;\Op{H})
  \,,&
  \Op\Delta\Op{C}^\prime
  &=0
  \,,
\end{align}
where the one-form
\begin{align}
  \label{eq:pert part sol}
  \Op{\mathcal{V}}\left[\Op{C}^\prime\right]
  &=
  \sum_{k=1}^{+\infty}
  \Op{\mathcal{V}}^{(k)}\left[\Op{C}^\prime\right]
  \,,&
  \Op{\mathcal{V}}^{(k)}\left[a\;\Op{C}^\prime\right]
  &=
  a^k\;\Op{\mathcal{V}}^{(k)}\left[\Op{C}^\prime\right]
  \,,
\end{align}
is a particular solution to
\begin{align}
  \Op{q}\Op{\mathcal{V}}\left[\Op{C}^\prime\right]
  +\Op{\mathcal{V}}\left[\Op{C}^\prime\right]\;\Op{\mathcal{V}}\left[\Op{C}^\prime\right]
  +\Op{C}^\prime\;\Op{J}
  &=
  0
  \,,&
  d\Op{\mathcal{V}}\left[\Op{C}^\prime\right]
  &=0
  \,.\label{2.146}
\end{align}
Given a solution up to a given order $n-1$ in perturbations around the background
\begin{align}
  \Op{C}^{\prime(0)}
  &=0
  \,,&
  \Op{H}^{(0)}
  &=1
  \,,
\end{align}
the $n^{\rm th}$ order of \eqref{eq:CI LHC} reads
\begin{align}
  \label{eq:CI Phin}
  \Op{\Phi}^{(n)}
  &=
  \Op{\Phi}^{(n)}_{\rm l.o.}
  +
  \Op{L}^{-1}\;\Op{C}^{\prime(n)}\;
  \Op\pi(\Op{L})
  \,,\\
  \label{eq:CI Vn}
  \Op{V}^{(n)}
  &=
  \Op{V}^{(n)}_{\rm l.o.}
  +\Op{L}^{-1}\;
  \Op{\mathcal{V}}^{(1)}\left[\Op{C}^{\prime(n)}\right]\;
  \Op{L}
  +
  \Op{q}\Op{H}^{(n)}
  \,,\\
  \label{eq:CI Un}
  \Op{U}^{(n)}
  &=
  \Op{U}^{(n)}_{\rm l.o.}
  +
  \Op{D}\Op{H}^{(n)}
  \,,
\end{align}
where $(\Op{\Phi}^{(n)}_{\rm l.o.},\,\Op{V}^{(n)}_{\rm l.o.},\,\Op{U}^{(n)}_{\rm l.o.})$ is a particular solution constructed from the moduli of orders $n'<n$.
Thus:
\begin{itemize}
  \item[i)] Eq. \eqref{eq:DPhin} restricts the integration constant that appears when solving \eqref{eq:qPhin} to the form \eqref{eq:CI Phin};
  \item[ii)] Solving Eq. \eqref{eq:qVn} brings in a gauge function $\Op{H}^{(n)}$ modulo a piece $\Op{H}_0^{(n)}$ that commutes with $\OpZ$;
  \item[iii)] Working locally on ${\cal U}$, whose de Rham cohomology is trivial, the integration constant of Eq. \eqref{eq:qUn} is given by $\Op{D}\Op{H}_0^{(n)}$ in order to comply with \eqref{eq:DUn}.
\end{itemize}

\subsection{Holomorphic and factorized gauges}\label{sec:factor}

A special form of the particular solution $\mathcal{V}\left[\Op{C}^\prime\right]$ can be obtained with the help of a resolution operator $\Op{q}^{\ast[F]}$ with simple properties:
Firstly, it is \emph{factorised}, in the sense that it satisfies
\begin{align}
  \label{eq:rhoF}
  \Op{q}^{\ast[F]}\left(\Op{f}_Y\;\Op{g}\right)
  &=
  \Op{f}_Y\;\Op{q}^{\ast[F]}\Op{g}
  \,,&
  d\Op{q}^{\ast[F]}\Op{g}
  &=
  \Op{q}^{\ast[F]}d\Op{g}
  \,,
\end{align}
for any $\Op{g}$ and for any $\Op{q}$-closed $\Op{f}_Y$.
Second, it is \emph{holomorphic}, that is
\begin{align}
  \label{eq:rhoFhol}
  \left[\Opz,\,\Op{g}\right]
  &=0
  &\Rightarrow&&
  \left[\Opz,\,\Op{q}^{\ast[F]}\Op{g}\right]
  &=0
  \,,&&&&&
  \left[\Opzb,\,\Op{g}\right]
  &=0
  &\Rightarrow&&
  \left[\Opzb,\,\Op{q}^{\ast[F]}\Op{g}\right]
  &=0
  \,.
\end{align}
Using the definition \eqref{eq:Psi'=C'k} of the chiral zero-form integration constants, Eq. \eqref{2.146} can be solved to first order in $\Op{C}^\prime$ by
\begin{align}
  \label{eq:vz1}
  \mathcal{V}^{(1)}
  \left[\Op{C}^{\prime}\right]
  &=
  \Op\Psi^{\prime}\;\Op{v}^{[F]}_{1}
  +
  \Op{\bar\Psi}^{\prime}\;\Op{\bar{v}}^{[F]}_{1}
  \,,&
  \Op{v}^{[F]}_{1}
  :&=
  \frac{i}{4}e^{i\theta}\Op{q}^{\ast[F]}\OpKz\;\dz^\ag\dz_\ag
  \,,&
  \Op{\bar{v}}^{[F]}_{1}
  :&=
  \frac{i}{4}e^{-i\theta}\Op{q}^{\ast[F]}\OpKzb\;\dzb^\ad\dzb_\ad
  \,.
\end{align}
Assuming that ${\cal A}$ defined in \eqref{calA} is a non-unital associative algebra, a perturbatively exact solution can then be built recursively as 
\begin{align}
  \label{eq:fact sol}
  \Op{\mathcal{V}}^{(k)}
  &=
  \Op\Psi^{\prime k}\;\Op{v}^{[F]}_{k}
  +
  \Op{\bar\Psi}^{\prime k}\;\Op{\bar{v}}^{[F]}_{k}
  \,,&
  \Op{v}^{[F]}_{k}
  &=
  -\sum_{\ell=1}^{k-1}
  \Op{q}^{\ast[F]}\left(\Op{v}^{[F]}_{\ell}\;\Op{v}^{[F]}_{k-\ell}\right)
  \,,&
  \Op{\bar{v}}^{[F]}_{k}
  &=
  -\sum_{\ell=1}^{k-1}
  \Op{q}^{\ast[F]}\left(\Op{\bar{v}}^{[F]}_{\ell}\;\Op{\bar{v}}^{[F]}_{k-\ell}\right)
  \,,
\end{align}
for $k=1,2,3\dots$, relying on the integer-spin projection \eqref{eq:BP VE}, which translates into
\begin{align}
  \label{eq:BP holG}
  \Op\Psi'\Op{\bar\Psi}'
  &=
  \Op{\bar\Psi}'\Op\Psi'
  \,,&
  \Op\pi\left(\Op{v}_k^{[F]}\right)
  &=
  \Op{v}_k^{[F]}
  \,,&
  \Op{\bar\pi}\left(\Op{\bar{v}}_k^{[F]}\right)
  &=
  \Op{\bar{v}}_k^{[F]}
  \,,
\end{align}
as well as the lemmas
\begin{align}
  \Op{v}^{[F]}_{k}\Op{\bar{v}}^{[F]}_{\ell}
  &=-
  \Op{\bar{v}}^{[F]}_{\ell}\Op{v}^{[F]}_{k}
  \,,&
  \left[\Op\Psi^\prime,\,\Op{v}^{[F]}_{k}\right]
  &=
  \left[\Op{\bar\Psi}^\prime,\,\Op{v}^{[F]}_{k}\right]
  =
  0
  \,,&
  \left[\Op\Psi^\prime,\,\Op{\bar{v}}^{[F]}_{k}\right]
  &=
  \left[\Op{\bar\Psi}^\prime,\,\Op{\bar{v}}^{[F]}_{k}\right]
  =
  0
  \,.
\end{align}
The solution (\ref{eq:CI LHC},\,\ref{eq:pert part sol},\,\ref{eq:fact sol}) is said to be in \emph{holomorphic gauge} if $\Op{L}\;\Op{H}=1$, and in \emph{factorised $L$-gauge} if $\Op{H}=1$.

While the factorized resolution operator  $\Op{q}^{\ast[F]}$ enables the perturbative construction of $\widehat \Phi$ and $\widehat V$ to all orders, it does not dress the one-form $\widehat U$ \cite{COMST}, since the $\Op{Z}$-independence of the gauge function $\Op{L}$ and the $x$- and $\Op{Y}$-independence of $\Op{v}^{[F]}_{k}$, $\Op{\bar{v}}^{[F]}_{k}$, implies that the particular solution
\be \Op{L}^{-1}\;
  \Op{\mathcal{V}}^{(1)}\left[\Op{C}^{\prime(n)}\right]\;
  \Op{L} \ = \ \Op{L}^{-1}\;\Op\Psi^{\prime k}\;\Op{L}\;\Op{v}^{[F]}_{k}
  +
  \Op{L}^{-1}\;\Op{\bar\Psi}^{\prime k}\;\Op{L}^{-1}\;\Op{\bar{v}}^{[F]}_{k} \ =: \ \Op\Psi^{k}\;\Op{v}^{[F]}_{k}
  +
  \Op{\bar\Psi}^{k}\;\Op{\bar{v}}^{[F]}_{k}\ ,\ee
with chiral zero-forms $\widehat \Psi$ and $\widehat{\overline\Psi}$ defined in \eqref{eq:Psi}, obeys
\be \Op{D}( \Op\Psi^{k}\;\Op{v}^{[F]}_{k}
  +
  \Op{\bar\Psi}^{k}\;\Op{\bar{v}}^{[F]}_{k}) =  (\Op{D}\Op\Psi^{k})\;\Op{v}^{[F]}_{k}+
 (\Op{D}\Op{\bar\Psi}^{k})\;\Op{\bar{v}}^{[F]}_{k}\ ,\ee
as $\Op{D}\Op\Psi=0=\Op{D}\Op{\bar\Psi}$.
Thus, $\Op{D}\Op{V}^{(1)}=0$ in Eq. \eqref{eq:qUn}, which implies that $\Op{U}^{(1)}$ is independent of $\Op{\Phi}$ (in this gauge), hence removable on ${\cal U}$ by means of a gauge transformation. 
This conclusion extends iteratively to higher orders as well \cite{COMST}. 
In other words, in the factorized gauge, no nontrivial cocycle $\Sigma_{2;1}$ is activated in Eq. \eqref{eq:DW=S}, whereby the zero-form module, which does receive non-trivial corrections, is not glued to the spacetime one-form module.

However, any solution is equivalent to a factorized solution by means of a large gauge transformation (affecting charges in strictly positive form degree).
Thus, the zero-form charge  \eqref{eq:def OWL} of any solution written in the form \eqref{eq:CI LHC} can be computed in holomorphic gauge.
Assuming that the trace can be factorized as $\Tr=\Tr_{\OpY}\Tr_{\Opz}\Tr_{\Opzb}$, these observables read
\begin{align}
  \label{eq:OWL fact}
  {\cal I}_{n,p}(\mathcal{M})
  =&
  \sum_{N=n}^\infty\sum_{P=0}^1
  \alpha_{n,p;N,P}\Tr_{\OpY}\left(
  \Op\Psi^{\prime N}\OpKy^P\;\OpKyb^P\right)
  \,,\\
  \label{eq:OWL alpha}
  \alpha_{n,p;N,P}
  :=&
  \sum_{\substack{0 \leqslant\,\bar{k}\,\leqslant\, N-n\\\bar{k}\equiv P-p\mod2}}
  \beta_{(p+n){\rm mod}\ 2,N-n-\bar{k}}\,
  \bar\beta_{p,\bar{k}}
  \,,\\
  \label{eq:OWL beta}
  \beta_{p;k}
  :=&
  \frac{1}{2\pi}
  \frac{\left(\partial_\nu\right)^k}{k!}
  \left.\Tr_{\Opz}\left(
  \Pexp\left(\int_{L(\mu)}\dz^\ag\,
  \sum_{j=1}^\infty \nu^{j}\;\Op{v}^{[F]}_{\ag\,j}\right)\;
  \exp\left(\tfrac{i}{2}\mu\,\Opz\right)\;
  \OpKz^{p}
  \right)\right\vert_{\nu=0}
  \,,\\
  \label{eq:OWL bb}
  \bar\beta_{p,\bar{k}}
  :=&
  \frac{1}{2\pi}
  \frac{\left(\partial_{\bar\nu}\right)^{\bar{k}}}{\bar{k}!}
  \left.\Tr_{\Opzb}\left(
  \Pexp\left(\int_{L(\bar\mu)}dz^\ad\,
  \sum_{\bar{\jmath}=1}^\infty \bar\nu^{\bar\jmath}\;\Op{\bar{v}}^{[F]}_{\ad\,\bar\jmath}\right)\;
  \exp\left(\tfrac{i}{2}\bar\mu\,\Opzb\right)\;
  \OpKzb^{p}
  \right)\right\vert_{\bar\nu=0}
  \,.
\end{align}

We shall return to the perturbative schemes  formulated above in
Sec. \ref{sec:solve Z} ( including perturbative corrections to the twistor space Wilson line observables for $N-n=0,1,2,3$ given in Eqs. (\ref{eq:OWL alpha answer 1}-\ref{eq:OWL alpha answer 5})), armed with resolution operators built algebraic tools to be introduced next.

\section{Symbol calculus }
\label{sec:starprod}

Classical HSG geometries containing ALAdS regions connected via bulk regions filled with particle modes and extended objects, are thus described by horizontal forms on 
\be {\cal M}=\cup_{p\in{\cal X}} {\cal T}_p\ ,\ee
where ${\cal Y}\rightarrow {\cal T}_p\to \{p\}\times {\cal Z}$ are noncommutative twistor spaces, and ${\cal X}$ is a commutative manifold embedded into the base manifold $\cal B$ of a universal ${\cal G}_0$-bundle, viz. 
\be {\cal G}_0\rightarrow B{\cal G}_0\to {\cal B}\hookleftarrow {\cal X}\ .\ee
On charts ${\cal X}_\xi\subseteq {\cal X}$, vacuum gauge functions $\widehat L:{\cal X}_\xi\times{\cal Z}\to {\cal G}_0$ induce isomorphisms ${\cal T}_p \stackrel{\widehat L}{\cong}{\cal T}_{p_0}$, for $p,p_0\in {\cal X}_\xi$, where ${\cal T}_{p_0}$ gives rise to an operator algebra $\Omega_{\rm hor}({\cal T}_{p_0})$ (capturing the quantum mechanics of the underlying conformal particle).
Letting ${\rm Pr}_O:{\cal Z}\to O$ denote a projection to a base point $O\in {\cal Z}$ (which has the effect of setting $Z=Z(O)$ in symbols), the fiber algebra 
\be {\cal A}:=({\rm Pr}_O)^\ast \Omega({\cal Y})\subset \Omega({\cal T}_{p_0})\ ,\ee
contains the locally-defined Weyl zero-form integration constant $\widehat\Psi$, which is a composite operator on ${\cal Y}$ encoding local HSG degrees of freedom.
HSG geometries containing particle and black-hole modes arise by taking ${\cal A}$ to be the orbit of the algebra ${\rm End}(F)$ of operators in a Fock space $F$, generated by the group algebra of the $\mathbb{Z}_2$ subgroup of a complexified version of ${\cal G}_0$ generated by inner Klein operators $\widehat{\kappa}_y$; taking $\widehat\Psi\in {\rm End}(F)$ yields black-hole modes, while $\widehat\Psi \in {\rm End}(F)\widehat\kappa_y$ yields massless particle modes, as spelled out in Sec. \ref{sec:PTBH}, and, originally, in \cite{2017}.
For a parallel construction using different Fock spaces leading to bulk-to-boundary propagators and boundary Green's functions, see \cite{Iazeolla:2020jee}.

In this Section, we review operator algebras ${\cal A}[\C^{2n}]$ arising by quantizing the holomorphic symplectic $\C^{2n}$, that are themselves (infinite-dimensional) complex manifolds, suitable for constructing physically desirable HSG geometries, including those mentioned above.
As we shall see, these operator algebras arise within the group algebra of the holomorphic inhomogenous metaplectic group $MpH(2n;\mathbb{C})$, including analytic delta function sequences, which contains the unitary projective metaplectic representation of $Sp(2n;\mathbb{R})$ (with cocycle factors in $\mathbb{Z}_2=\{\pm 1\}$).
The complex metaplectic group $Mp(2n;\mathbb{C})$ arises from non-unitary representations of $Sp(2n;\mathbb{C})$ using holomorphic oscillators, giving rise maps from $Sp(2n;\mathbb{C})$ to ${\cal A}[\C^{2n}]$, viewed as complex manifolds, that we propose are holomorphic in a sense to be made precise below, and that we will corroborate in a number of ways.

\subsection{Holomorphic Heisenberg group}   \label{sec:3.1}

\paragraph{Weyl algebra.}

Quantizing the holomorphic symplectic $\C^{2n}$ with canonical two-form $\omega=\frac12 dX^I \wedge dX^J \theta_{IJ}$, $I=1,\dots,2n$, with constant $\theta_{IJ}$, deforms the algebra of holomorphic polynomials into a unital associative algebra ${\cal P}[\C^{2n}]$, referred as the holomorphic Weyl algebra, generated by operators $\OpX^{I}$ obeying the Heisenberg commutation relations
\be \left[\OpX^{I},\,\OpX^{J}\right]=2i\theta^{IJ}\ ,\qquad \theta^{IK}\theta_{JK}:=\delta^I_J\,,\ee
equipped with a chiral supertrace operation ${\rm STr}:{\cal P}[\C^{2n}]\to \mathbb{C}$ defined by
\be\label{eq:Str(Poly)}
\STr\left[\OpX^{(I_1}\cdots \OpX^{I_p)}\right]=\delta_{p,0}\ ,\qquad p\in\{0,1,\dots\}\ ,\ee
which is graded with respect to the monomial degree $p$.

The holomorphic Weyl algebra can be equipped with a number of hermitean conjugation operations, that is, anti-linear anti-automorphisms $\dagger$; for example, 
\be (\widehat X^I)^\dagger=\widehat X^I\ .\ee
Alternatively, by first tensoring ${\cal P}[\C^{2n}]$ by an anti-holomorphic copy ${\cal P}[\overline\C^{2n}]$ generated by $\widehat{\overline{X}}{}^{\dot I}$, $\dot I=1,\dots,2n$, obeying 
\be [\widehat{\overline{X}}{}^{\dot I},\widehat{\overline{X}}{}^{\dot J}]=2i\overline{\Theta}{}^{\dot I\dot J}\ ,\qquad \overline{\Theta}{}^{\dot I\dot J} = \Theta^{\dot I\dot J}\ ,\ee
and $[\widehat{\overline{X}}{}^{\dot I},\widehat X^{I}]=0$, the product ${\cal P}[\C^{2n}]\otimes {\cal P}[\overline\C^{2n}]$ can be equipped by a hermitean conjugation maps $\dagger_{\mathbb{C}}$; for example, 
\be
  (\OpX^{I})^{\dagger_{\mathbb{C}}}
  =\widehat{\overline{X}}{}^{\dot I}:=(\widehat{X}^I)^{\dagger}\ ,\qquad \dagger\circ\dagger={\rm Id}\ .\ee
In what follows, we shall work mainly at the level of the holomorphic algebra.

\paragraph{Symbols.}

The holomorphic Weyl algebra ${\cal P}[\C^{2n}]$ can be extended to associative algebras ${\cal A}[\C^{2n}]$ equipped with complex linear maps\footnote{In general, an associative algebra ${\cal A}$ can be topologically non-trivial in the sense that its symbolization requires several charts glued together via fusion rules, viz. ${\cal A}=\bigoplus_\xi {\cal A}_\xi$ with product rule ${\cal A}_\xi{\cal A}_\eta=\bigoplus_\zeta {\cal N}_{\xi,\eta}{}^\zeta{\cal A}_\zeta$ where ${\cal N}_{\xi,\eta}{}^\zeta\in\{0,1\}$ represent a finite-dimensional associative algebra with generators $e_\xi$ obeying $e_\xi e_\eta=\bigoplus_\zeta {\cal N}_{\xi,\eta}{}^\zeta e_\zeta$; for example, a matrix fusion rule yields ${\cal A}\cong {\rm mat}_n({A}_0)$ where $ A_0$ is the associative algebra associated to the identity matrix. } 
\be [\cdot]:{\cal A}[\C^{2n}]\to {\cal D}[\C^{2n}|\R^{2n}]\ ,\ee
referred to as Wigner maps, where ${\cal D}[\C^{2n}|\R^{2n}]$ are spaces of classical distributions, referred to as symbols, on spaces of test functions defined on $\R^{2n}$ sub-spaces of $\C^{2n}$, such that
\begin{align}\label{eq:def star}
    \left[\Op{f}_1 \Op{f}_2\right]=\left[\Op{f}_1\right]\star \left[ \Op{f}_2\right]\ ,
\end{align}
where $\star$ is a composition rule for symbols, referred to as a star product, which thus represents the operator product by letting the symbols act on themselves.
Conversely, a quantization scheme, or Weyl map, is a complex linear map
\be \widehat{\rm Op}: {\cal D}[\C^{2n}|\R^{2n}]\to {\cal A}[\C^{2n}]\ ,\ee
such that
\be \widehat{\rm Op}\left[[\Op{f}]\right]=\Op{f}\ ,\ee
for all $\Op{f}\in {\cal A}[\C^{2n}]$, or, equivalently,
\begin{align}
 [\widehat{\rm Op}\left[f\right]]=f\ ,
\end{align}
for all $f\in {\cal D}[\C^{2n}|\R^{2n}]$.
The symbols can furthermore be used to equip ${\cal A}[\C^{2n}]$ with chiral (super)trace operations and ${\cal A}[\C^{2n}]$ or ${\cal A}[\C^{2n}]\otimes {\cal A}[\overline \C^{2n}]$ with hermitean conjugation operations.

The algebra ${\cal P}[\C^{2n}]$ can be viewed as an infinite-dimensional manifold coordinatized by the Wigner maps; this manifold has a complex structure that is respected by the operator product.
We shall assume that this holomorphic structure extends to ${\cal A}[\C^{2n}]$.
This requires the star products, which are integrals over $\R^{2n}$ planes in $\C^{2n}$, to respect the complex analyticity of symbols, which is non-trivial, since if $f:\Omega\times \Omega'\to \C$ is analytic on $\Omega\subset \C^{N}$ and $\Omega'\subset \R^{N'}$ is non-compact, then the integral $I(u):=\int_{v\in \Omega'} d^{N'} v f(u,v)$ need not depend analytically on $u$; one method to achieve a holomorphic operator algebra is to compactify the integral over $\Omega'$ and possibly go to multiple covers of $\Omega$, such as the double covers that will be introduced below in order to define the metaplectic representation of the symplectic groups over $\C$.

\paragraph{Group algebra.}

We assume that ${\cal A}[\C^{2n}]$ is an irreducible module for the group algebra $\C[H_{2n+1}(\C)]$ of the holomorphic Heisenberg group,
spanned by the set $\C^{2n+1}\equiv\C^{2n}\times\C$ with product rule
\be 
(K_1;\,k_1)\,(K_2;\,k_2)
= 
(K_1+K_2; k_1+k_1+c(K_1,K_2))\,,\ee
where $c: \C^{2n}\otimes \C^{2n}\to \C$ obeys the commutation rule
\be c(K_1,K_2)-c(K_2,K_1)=2K_1 K_2\ ,\ee
and the cocycle condition 
\be c(K_1,K_2)+c(K_1+K_2,K_3)-c(K_1,K_2+K_3)-c(K_2,K_3)=0
\,.
\ee
Denoting the representation map by
\be \Op{E}:\C[H_{2n+1}]\to {\cal A}[\C^{2n}]\,,\ee
we furthermore assume that the central elements are realised as
\begin{align}
  \OpHc0{k}
  :&=
  e^{ik}
  \,,&
  \OpHc{K}{k}
  &=
  e^{ik}
  \OpH{K}
  \,,&
  \OpH{K_1}\;\OpH{K_2}
  &=
  e^{ic(K_1,K_2)}
  \OpH{K_1+K_2}
  \,.&
\end{align}
Expanding the distributions over plane waves $E_K:\R^{2n}\to \C : X\to E_K(X):=e^{iKX}$, 
we define our quantization map through the aforementioned representation of $\C[H_{2n+1}(\C)]$ as
\begin{align}
  \widehat{\rm Op}\left[E_K\right]
  :=
  \OpH{K}
  \,,
\end{align}
thereby inducing its action on $\mathcal{D}[\C^{2n}\vert\R^{2n}]$ 
\cite{Kubo,Cohen66,Cohenbook}
\begin{equation}
  \label{eq:WeylMap}
  \widehat{f}=
  \int d^{2n}X\; [\widehat f](X)\, \OpD(X)\ ,\qquad\OpD(X):= \widehat{\rm Op}\left[\delta^{2n}_X\right]\ ,
\end{equation}
where $\{\delta^{2n}_X\}_{X\in \R^{2n}}$ is a complete set of Dirac delta densities\footnote{If $\phi:\R^{2n}\to\C$ is bounded and continuous at $X$, then $\int_{\R^{2n}} d^{2n}X' \delta^{2n}_X(X') \phi(X') =\phi(X)$.} given by
\be \delta^{2n}_X(X')= \int_{\R^{2n}}\frac{d^{2n}K}{(2\pi)^{2n}}\;e^{-iKX}\:E_K(X')\ .\ee

\paragraph{Weyl quantization.} The Weyl ordering scheme amounts to representing the Heisenberg group algebra elements projectively using Weyl--Heisenberg translation operators
\be
\label{eq:reprH WO}
\OpHW{K}:=e^{iK\OpX}
\,,\ee
whose product rule is dictated by the Baker--Campbell--Hausdorff formula as
\begin{equation}
  \label{eq:BCH}
  \OpHW{K_1}\OpHW{K_2}
  =
  e^{iK_1K_2}\;\OpHW{K_1+K_2}
  \,,
\end{equation}
that is $c_0(K_1,K_2)=K_1 K_2$.
Thus, in this scheme,
\begin{align}
  \label{eq:WeylMap0}
  \widehat{f}
  &=
  \int d^{2n}X\; [\widehat f]_0(X)\, \OpD_0(X)
  \,,&
  \OpD_0(X)
  &= 
  \int_{\R^{2n}}\frac{d^{2n}K}{(2\pi)^{2n}}\;e^{-iKX}\OpHW{K}
  \,, 
\end{align}
where $[\widehat f]_0$ is referred to as the  Weyl-ordering symbol of $\Op{f}$; in a slight abuse of notation, Taylor expansion of $\OpHW{K}$ in wave numbers yields
\begin{align}
  \widehat{\rm Op}_{0}\left[X^{I_1}\cdots X^{I_p}\right]
  =
  \OpX^{(I_1}\cdots \OpX^{I_p)}
  \,,&&
  \qquad p\in\{0,1,\dots\}
  \,,
\end{align}
that is, classical monomials are mapped under Weyl quantization to totally symmetric operator monomials\footnote{The operator  $\Op\Delta_0(X)$ squares to a constant, which serves as a defining property of Weyl order \cite{Pierre Bieliavsky}.}

\paragraph{Weyl transform.} 
In Quantum Mechanics (see, e.g., chapter 4 in \cite{Wong1998}),
the mapping \eqref{eq:WeylMap0} between symbols and operators is known as the \emph{Weyl transform}, which can be presented as a map $f\mapsto W_f$ that sends a phase-space function $f$ to an integral operator $W_f$ acting on wave functions in the position basis  viz.
\bea (W_f\psi)(x): = \int \frac{d^n x' d^n p}{(2\pi)^n} \,f\left(\frac{x+x'}{2},p\right)\,e^{i(x-x')\cdot p} \psi(x') \ ,\label{Weyltr1} \eea
where $\psi(x):= \langle x|\psi\rangle$ and the completeness relation $1=\int d^n x'|x' \rangle\langle x'|$ is assumed. Eq. \eqref{Weyltr1} in its turn can be identified as the transformed wave function, in position basis,
\be 
 (W_f\psi)(x)=\langle x| \Op f| \psi\rangle   = \int d^n x' \langle x| \Op f| x'\rangle\psi(x')\ ,\label{matrel}\ee
via the operator $\hat f$ with Weyl-ordering symbol $[\hat f]_0=f$.
To show the equivalence between \eqref{eq:WeylMap0} and \eqref{Weyltr1}, one splits the symplectic coordinates into positions and momenta, say $\Op X^{I}=\sqrt{2}(\Op x^1,\Op p^1,...,\Op x^n,\Op p^n)$, and rewrites \eqref{eq:WeylMap0} as
\be \widehat{f}
  =
  \int  \frac{d^{n}x \,d^n p\,d^n y \, d^n q}{\pi^{2n}}\, [\widehat f]_0(x,p)\, e^{2iq\cdot (\Op x-x)-2iy\cdot(\Op p-p)} \ ,\label{Weyltr0}\ee
where $a\cdot b:= a^ib^j\d_{ij}$.
Plugging Eq. \eqref{eq:BCH}, $e^{2iq\cdot \Op x-2iy\cdot\Op p}=e^{-2iy\cdot\Op p}e^{2iq\cdot \Op x}e^{2iy\cdot q}$, into Eq.  \eqref{Weyltr0} gives
\be \widehat{f}
  =
  \int  \frac{d^{n}x \,d^n p\,d^n y \, d^n q}{\pi^{2n}}\, [\widehat f]_0(x,p)\, e^{-2iq\cdot x+2iy\cdot (p+q)}e^{-2iy\cdot\Op p}e^{2iq\cdot \Op x} \ .\ee
Thus, the matrix element of an operator $\Op f$ between two position eigenstates can be written as
\bea \langle z'|\Op f |z\rangle =  \int \frac{d^{n}x \,d^n p\,d^n y \, d^n q}{\pi^{2n}}[\widehat f]_0(x,p) \, e^{-2iq\cdot x+2iy\cdot (p+q)} e^{2iq\cdot z}  \langle z'|e^{-2iy\cdot\Op p}| z\rangle \ ,\eea
which, after inserting $1=\int d^n p'|p' \rangle\langle p'|$, and using $\langle z'|p' \rangle=(2\pi)^{-n/2} e^{ip'\cdot z'}$, yields 
\be \langle z'|\Op f |z\rangle = \int \frac{d^n p}{(2\pi)^n} \,[\widehat f]_0\left(\frac{z+z'}{2},p\right)\,e^{i(z'-z)\cdot p} \ . \label{Weyltrsand}\ee
Inserting \eqref{Weyltrsand} into \eqref{matrel} gives \eqref{Weyltr1}. Thus, provided that one identifies $[\hat f]_0=f$, the action of $\Op f$ on a state vector $|\psi\rangle$ can be represented in position basis as the integral operator $W_f$ defined in \eqref{Weyltr1}. 

Anticipating Section \ref{sec:Sp}, we note that the metaplectic group elements are often realized \cite{Folland,Guillemin:1990ew,deGosson} in terms of integral operators of the above type. 
However, by means of the Weyl transform, these group elements can equally well be represented in terms of symbols, which is the realization that we will exploit in this paper. In Appendix \ref{app:U(1) ex} we provide an example of the relation between the two realizations in the familiar case of the evolution operator of the harmonic oscillator.

\paragraph{$M$-ordering schemes.}

In higher spin geometries, where thus ${\cal T}_p$ are fibers over points $p\in{\cal X}$, it is natural to employ locally defined Weyl maps to describe Vasiliev's master fields in different regions of ${\cal X}$.
Requiring these maps to preserve the operator identity
\begin{equation}
  \label{eq:X=dPW}
  \OpX_I=-i\frac{\partial}{\partial K^{I}}e^{iK\OpX}\vert_{K=0}
  \,,
\end{equation}
that is, the corresponding Wigner maps to obey
\begin{align}
    X_I=-i\frac{\partial}{\partial K^{I}}\left.\left[e^{iK\OpX}\right](X)\right\vert_{K=0}\ ,
\end{align}
it is natural to consider 
\begin{align}
  \left[e^{iK\OpX}\right]_{M}
  :&=
  e^{-\tfrac{i}2KMK} E_K
  \,,&
  \widehat{\rm Op}_{M}\left[E_K\right]
  &=
  e^{iK\OpX+\tfrac{i}2KMK}
  \,,\label{eq:OpM(EK)}
\end{align}
parametrized by symmetric complex matrices $M$, that is, $M_{IJ}=M_{JI}\in\C$, which we refer to as the $M$-ordering schemes.
These schemes are tantamount to changes of parametrization for the representation \eqref{eq:reprH WO}
\begin{align}
\OpHM{M}{K}
:&=
e^{\tfrac{i}2KMK}\OpHW{K}
\,,& 
c_M(K_1,K_2)=K_1(1-M)K_2
\,.
\end{align}
In this scheme,
\begin{align}
  \OpDM{M}(X) &:= \widehat{\rm Op}_{M}\left[\delta^{2n}_X\right]=\int\frac{d^{2n}K}{(2\pi)^{2n}}\;
  e^{-iKX} \widehat{\rm Op}_{M}\left[E_K\right]
  \nonumber\\&=
  \label{eq:DeltaM}
  \int\frac{d^{2n}K}{(2\pi)^{2n}}\;
  e^{-iKX+\tfrac{i}2KMK}\OpHW{K} 
  \,,
\end{align}
in terms of which $\widehat f\in {\cal A}[\C^{2n}]$ can be expanded as
\begin{equation}
  \label{eq:WeylMapM}
  \widehat f=\int d^{2n}X [\widehat f]_M(X)\OpDM{M}(X)
  \,,
\end{equation}
using distributions $[\widehat f]_M\in{\cal D}_M[\C^{2n}|\R^{2n}]$, referred to as the $M$-ordering symbol of $\widehat f$; conversely, an $f\in {\cal D}_M[\C^{2n}|\R^{2n}]$ is sent to
\be \widehat{\rm Op}_{M}[f]
  =
  \int d^{2n}X\; f(X)\, \OpDM{M}(X)\ ,\ee
in ${\cal A}[\C^{2n}]$.
For simplicity, we assume throughout the paper that $M$ is real.
The $M$-ordering scheme reduces to the Weyl-ordering scheme for $M=0$, and normal ordering schemes for $M$ given by square roots of the unit, as explained in App. \ref{app:X+-}.

\subsection{Traces and star products}

\paragraph{Trace operation.}
The embedded Heisenberg group algebra can be equipped with the linear map
\begin{align}
  \label{eq:TrPW}
  \Tr\left[\OpHW{K}\right]
  :&=
  (2\pi)^{2n}\delta^{2n}(K)\ ,
\end{align}
which is indeed cyclic, viz. $\Tr \left[\OpHW{K_1}\;\OpHW{K_2}\right]=
  \Tr\left[\OpHW{K_2}\;\OpHW{K_1}\right]$.
It follows that
\begin{align}
  \label{eq:TrDelta}
  \Tr\left[\OpDM{M}(X)\right]
  &=
  1\ ,
\end{align}
for all $X\in \R^{2n}$; hence, if $\Op{f}\in{\cal A}[\C^{2n}]$, then
\begin{align}
  \label{eq:MoTr}
  \Tr\left[\Op{f}\right]
  &=
  \int d^{2n}X\; \left[\Op{f}\right]_{M}(X)  \,.\end{align}
The $M$-ordering symbol of an operator $\Op{f}$ can be retrieved by means of the generalization 
\begin{equation}
\label{eq:MoSym}
  \left[\widehat f\right]_M(X)
  =
  \int\frac{d^{2n}X'}{(4\pi)^n\sqdeti{M}}
  e^{-\tfrac{i}4(X-X')M^{-1}(X-X')}
  \Tr\left[\,\OpDM{M}(X')\Op{f}\,\right]
  \,,\qquad \det M\neq 0\,,
\end{equation}
of the standard Wigner map
\begin{equation}
\label{eq:WignerMap0}
  \left[\widehat f\right]_0(X)
  =
  \Tr\left[\,\OpDM{0}(X)\Op{f}\,\right]
  \,,
\end{equation}
which sends an operator to its Weyl-ordering symbol.
Composition of the generalized Wigner map \eqref{eq:MoSym} with its supposed inverse \eqref{eq:WeylMapM} using formula \eqref{eq:Tr(opD^2)} yields the transition function between the symbols of a given operator in two different $M$-ordering schemes, viz.
\begin{equation}
  \left[\widehat f\right]_{M_0}(X_0)
  =
  \int\frac{d^{2n}X_1}{(2\pi)^n\sqdeti{(M_0-M_1)}}
  e^{-\tfrac{i}2(X_0-X_1)(M_0-M_1)^{-1}(X_0-X_1)}
  \left[\widehat f\right]_{M_1}(X_1)
  \,.
\end{equation}
Taking $M_1\to M_2$ using the analytic delta sequence \eqref{eq:delta gauss}, one obtains a tautology which shows that the maps in Eq. \eqref{eq:WeylMapM} and Eq. \eqref{eq:MoSym} are indeed each other's inverses.
The trace operation induces the operator Fourier transform 
\be
  \label{eq:FTOp}
  \widetilde{f}(K)
  :=
  \frac{1}{(2\pi)^n}\Tr\left[
  \Op{f}\;\OpHW{-K}\right]
  \,,\qquad
  \Op{f}
  =
  \int\frac{d^{2n}K}{(2\pi)^n}\;
  \widetilde{f}(K)\;
  \OpHW{K}
  \,.
\ee

\paragraph{Supertrace operation.}

The linear map $\STr:{\cal A}[\C^{2n}]\to \C$ defined by
\begin{align}
  \label{eq:def STr}
  \STr\, \left[\OpHW{K}\right]
  &:=
  1
  \,,
\end{align}
enjoys the graded cyclicity property
\be \STr\left[\OpHW{K_1}\;\OpHW{K_2}\right]
=\STr\left[\OpHW{K_2}\;\OpHW{K_1}\right]\ ,\ee
and defines a non-degenerate bilinear form on ${\cal A}[\C^{2n}]$.
If $M$ is invertible, then
\begin{align}
  \STr\left[\OpDM{M}(X)\right]
  &=
  \frac{1}{(2\pi)^n\sqdeti{M}}
  \exp\left(\tfrac{i}2XM^{-1}X\right)
  \,,
\end{align}
as can be seen using Eq. \eqref{eq:gauss int}, from which it follows that
\begin{align}
  \STr[\Op{f}]
  &=
  \int\frac{d^{2n}X}{(2\pi)^n}\;\frac{e^{\tfrac{i}2XM^{-1}X}}{\sqdeti{M}}\left[\widehat f\right]_{M}(X)
  \ .
\end{align}
Taking $M\to0$ using the delta sequence \eqref{eq:delta gauss} yields
\be
  \STr\left[\OpDM{0}(X)\right]
  =
  \delta^{2n}(X)
  \,,\qquad
  \STr(\Op{f})
  =
  \left[\widehat f\right]_{0}(0)\,,
\ee
justifying the notation as it induces the supertrace \eqref{eq:Str(Poly)} of the Weyl algebra.
As we shall see below, the trace and supertrace operations are interchanged upon insertions of the inner Klein operator defined in \eqref{eq:K=U(-1)}.

\paragraph{Twisted convolution formula.}

Combining the Weyl and Wigner maps in Eqs. \eqref{eq:WeylMapM} and \eqref{eq:MoSym}, respectively, and using the lemma \eqref{eq:Tr(opD^3)}, yields the following integral formula for the star-product \eqref{eq:def star} in $M$-ordering:
\begin{align}
  &
  \left(\MoSymOp{M}{f_1}\star_M\left[\widehat f_2\right]_{M}\right)(X)
  :=
  \left[\widehat f_1 \widehat f_2\right]_M(X)
  \nonumber\\&=
  \int\frac{d^{6n}(X_0,X_1,X_2)}{(4\pi)^n} \frac{e^{-\tfrac{i}4(X-X_0)M^{-1}(X-X_0)}}{\sqdeti{M}}\Tr\left[\OpDM{M}(X_0)\OpDM{M}(X_1)\OpDM{M}(X_2)\right] \left[\widehat f_1\right]_{M}(X_1)\left[\widehat f_2\right]_{M}(X_2)
  \nonumber\\&=
  \label{eq:starM int}
  \int\frac{d^{2n}X_1^\prime\,d^{2n}X_2^\prime}{(2\pi)^{2n}}
  \,\frac{e^{-iX^\prime_1(M+1)^{-1}X^\prime_2}}{\sqdet{M^2-1}}
  \left[\widehat f_1\right]_{M}(X+X_1)\left[\widehat f_2\right]_{M}(X+X_2)
  \,,
\end{align}
which reduces in the limit $M\to 0$ to
\begin{align}
  \left(\MoSymOp{0}{f_1}\star\MoSymOp{0}{f_2}\right)(X)
  &=
  \label{eq:star0 int}
  \int\frac{d^{2n}X_1\,d^{2n}X_2}{(2\pi)^{2n}}
  \,e^{-iX_1X_2}
  \left[\widehat f_1\right]_{0}(X+X_1)\left[\widehat f_2\right]_{0}(X+X_2)
  \,.
\end{align}
Provided that the symbols are real-analytic at the origin, the auxiliary integration can be converted into auxiliary differentiation operators, referred to as Bopp shifts, viz.
\begin{align}
  \label{eq:starM bopp}
  \left(\MoSymOp{M}{f_1}\star\MoSymOp{M}{f_2}\right)(X)
  &=
  \left.
  \MoSymOpArg{M}{f_1}{X-i(M-1)\partial_{X_2}}\MoSymOpArg{M}{f_2}{X+X_2}
  \right\vert_{X_2=0}
  \\&=
  \left.
  \MoSymOpArg{M}{f_2}{X-i(M+1)\partial_{X_1}}\MoSymOpArg{M}{f_1}{X+X_1}
  \right\vert_{X_1=0}
  \,.,
\end{align} 
and even further into the exponential form
\begin{align}
  \label{eq:starM exp}
  \left(\MoSymOp{M}{f_1}\star\MoSymOp{M}{f_2}\right)(X)
  &=
  \left.
  e^{i\partial_{X_1}(M-1)\partial_{X_2}}
  \MoSymOpArg{M}{f_1}{X+X_1}\MoSymOpArg{M}{f_2}{X+X_2}
  \right\vert_{X_1=X_2=0}
  \,.
\end{align}
The star product simplifies in normal ordering schemes where $M=R$ with $R$ given by a root of the unit, for which 
\begin{align}
  &
\left(\MoSymOp{R}{f_1}\star\MoSymOp{R}{f_2}\right)(X)
  \nonumber\\&=
  \int\frac{d^{n}X_{1,-R}\,d^{n}X_{2,+R}}{\pi^{n}}
  \,e^{-\frac{i}2 X_{1,-R}X_{2,+R}}
  \MoSymOpArg{R}{f_1}{X+X_{1,-R}}\MoSymOpArg{R}{f_2}{X+X_{2,+R}}
  \,,
\end{align}
where the integration variables $X_{1,-R}$ and $X_{2,+R}$ are projected as in Eq. \eqref{eq:XpmR}.

\subsection{Metaplectic groups and inner Klein operators}
\label{sec:Sp}

The complex symplectic group $Sp(2n;\C)$ consists of the matrices $U\in {\rm mat}_{2n}(\C)$ obeying
\begin{align}
  U_I{}^{I'}\;U_J{}^{J'}\;\Theta_{I'J'}
  &=
  \Theta_{IJ}
  &\Leftrightarrow&&
  U\;U^t
  &=
  -1
  &&\text{where\ }&
  (U^t)_{IJ}&:=U_{JI}
  \,.
\end{align}
Its Lie algebra $\mathfrak{sp}(2n;\C)\cong {\rm Sym}_{2n}(\C)$, equipped with the NW-SE matrix commutator, and oscillator realization 
\be  \label{eq:repr sp}
\widehat{\cal G}: G\in \mathfrak{sp}(2n;\C)\mapsto \Op{\mathcal{G}}(G)
  :=
  -\frac{i}{4}\OpX G\OpX
  \,,\qquad
  \left[\Op{\mathcal{G}}(G_1),\,\Op{\mathcal{G}}(G_2)\right]
  =
  \Op{\mathcal{G}}\left([G_1,\,G_2]\right)
  \,.
\ee
As is well known, in the real case the oscillators provide a proper unitary representation of $\mathfrak{sp}(2n;\R)$ but only a projective unitary representation\footnote{In a four-dimensional HSG characterized by admitting a maximally symmetric spacetime vacuum with background isometry group $SO(t,s)$, where $t+s=5$, the finite-dimensional spinor is the fundamental representation of the spin group $Spin(t,s)$, which is given by a corresponding real form of $Sp(4;\mathbb{C})$; for example, $Spin(1,4)=USp(2,2)$ and $Spin(2,3)=Sp(4;\mathbb{R})$.
These symplectic groups also admit infinite-dimensional projective representations in terms of even functions of oscillators, which lift to proper representations of corresponding metaplectic groups.
Thus, it is important to distinguish between the signs appearing under symplectic rotations of canonical coordinates, thought of as spinors, corresponding to $2\pi$ rotations in $SO(t,s)$, and those arising in the multiplication of operators furnishing projective representations of symplectic group elements; this distinction is of importance in explaining the sign ambiguities observed in \cite{neiman}.} of $Sp(2n;\R)$, referred to as \emph{metaplectic representation}, or the Segal--Shale--Weil representation \cite{Folland,Guillemin:1990ew,Woit:2017vqo}. In the following, we shall introduce a parallel construction of a projective representation map
\be \widehat{\cal U}: Sp(2n;\C)|_{\rm cut}\to {\cal A}[\C^{2n}]\,,\label{350}\ee
where the domain denotes an $Sp(2n;\C)$ equipped with a branch cut, which is holomorphic\footnote{The holomorphicity refers to the holomorphic structure on ${\cal A}[\C^{2n}]$ defined by the Wigner map $\left[\cdot\right]_M$, and not to the dependence of the symbols on $X^I$, which will degenerate to delta functions on submanifolds on $Sp(2n;\C)$ determined by the choice of $M$.} and not unitary\footnote{According to Bargmann's theorem, a unitary representation of a group $G$ with trivial $\pi_1(G)$ can be de-projectivized.
Thus, as $\pi_1(Sp(2n;\C))=\{e\}$, it follows that any unitary representation of $Sp(2n;\C)$ is non-projective; for example, $U\mapsto \widehat{\mathcal{U}}(U,0) (\widehat{\mathcal{U}}(U^{-1},0))^{\dagger_\C}$ is a unitary non-projective representation of $Sp(2n;\C)$ in ${\cal A}[\C^{2n}]\otimes {\cal A}[\overline\C^{2n}]$. 
Conversely, the restriction of $\widehat{\cal R}$ to the real metaplectic group $$\mathbb{Z}_2\rightarrow Mp(2n;\R)\stackrel{Pr}{\longrightarrow} Sp(2n;\R)$$
yields the unitary projective metaplectic representation of $Sp(2n;\R)$, which cannot be de-projectivized as $\pi_1(Sp(2n;\R))=\mathbb{Z}$; indeed, the restriction of $\widehat{\cal U}$ to the topological $S^1\subset Sp(2n;\R)$ is double-valued.}, and which lifts to a proper representation
\be \widehat{\mathcal{R}}: Mp(2n;\C)\to {\cal A}[\C^{2n}]\ ,\label{3.54}\ee
of the holomorphic metaplectic double cover
\be \mathbb{Z}_2 \rightarrow Mp(2n;\C)\stackrel{Pr}{\longrightarrow} Sp(2n;\C)\ee
of $Sp(2n;\C)$, with projection map defined by 
\begin{align} \widehat{\cal R}(g)\widehat X^I (\widehat{\cal R}(g))^{-1}&= \widehat X^J (Pr(g))_J{}^I\ ,\qquad g\in Mp(2n;\C)\ .\end{align}
To construct these maps, we first apply the Wigner map $\left[\cdot \right]_M$ to the representation
\be \widehat{\cal U}[e^{-2G}]= \exp\left(-2\Op{\mathcal{G}}(G)\right)
  \equiv
  \exp\left(\tfrac{i}{2}\OpX G\OpX\right)\ ,\ee
of a symplectic group element $e^{-2G}$ in the image of the exponentiation map, which yields (c.f. App. \ref{app:stargauss})
\begin{align}
  \label{eq:stargauss}
  \left[\widehat{\cal U}[e^{-2G}]\right]_M(X)
  =
  \frac{1}{\sqdet{\cosh G-M\sinh G}}
  \exp\left(
  \tfrac{i}{2}X\left(
  (\tanh G)^{-1}-M
  \right)^{-1}X\right)
  \,,
\end{align}
using a specific branch for the square-root function.
Eq. \eqref{eq:stargauss} defines a single-valued map from $\mathfrak{sp}(2n;\C)|_{\rm cut}$ to ${\cal D}[\C^{2n}|\R^{2n}]$.
While the exponentiation map $\exp: \mathfrak{sp}(2n;\C)\to Sp(2n;\C)$ is not surjective, the $M$-ordering star product formula can be used to extend $\left[\widehat{\cal U}\right]_M$ analytically from $\exp(\mathfrak{sp}(2n;\C))$ (which is thus a proper subset of $Sp(2n;\C)$) to a single-valued function 
\be \left[\widehat{\cal U}\right]_M:Sp(2n;\C)|_{\rm cut}\to {\cal D}[\C^{2n}|\R^{2n}]\ ,\ee
given by 
\begin{align}
  \label{eq:OpSp U}
  \left[\widehat{\cal U}[U]\right]_M(X)
  =
  \frac{ 1}{\sqdet{\frac{1+U-M(1-U)}{2}}}
  \exp\left(
  \tfrac{i}{2}X\left(\tfrac{1+U}{1-U}-M\right)^{-1}X
  \right)
  \,;
\end{align}
for the case of $n=1$ and $M=0$,  see App. \ref{app:U(1) ex}.
Composing $\left[\widehat{\cal U}\right]_M$ with the Weyl map yields 
\be \widehat{\cal U}:=\widehat{\rm Op}_M\circ \left[\widehat{\cal U}\right]_M\ ,\ee
that is hence single-valued on $Sp(2n;\C)|_{\rm cut}$ discontinuity at the cut, and furnishes a projective representation, viz.
\begin{align}
  \label{eq:repr Sp}
  \OpSp{U_1}\;\OpSp{U_2}
  &=
  e^{i\varphi(U_1,\,U_2)}
  \OpSp{U_1 U_2}
  \,,\qquad \varphi: Sp(2n;\C)|_{\rm cut}\times Sp(2n;\C)|_{\rm cut}\to \{0,\pi\}\ ,
\end{align}
where $\varphi$, which thus depends on the choice of the cut, obeys the cocycle condition
\be \varphi(U_1,U_2)+\varphi(U_1 U_2,U_3)-\varphi(U_1,U_2U_3)-\varphi(U_2,U_3)=0\ .\ee
The function $[\widehat{\cal U}]_M$ can be continued analytically to a single-valued function 
\be\left[\widehat{\cal R}[g_\pm]\right]_M(X)
  =\pm \left[\widehat{\cal U}[Pr(g_\pm)]\right]_M(X)
  \,,\qquad g_\pm\in Mp(2n;\C)_\pm\ ,\ee
defined on two copies of $Sp(2n;\C)|_{\rm cut}$ glued together into 
\be Mp(2n;\C)\stackrel{M\,,\,\sqrt{\phantom1}}{=\!=} Mp(2n;\C)_+\cup Mp(2n;\C)_-\ ,\qquad Mp(2n;\C)_\pm \stackrel{\rm top}{\cong} Sp(2n;\C)|_{\rm cut}\ ,\label{Mppatches}\ee
such that $\widehat{\cal R}:= \widehat{\rm Op}_M \circ [\widehat{\cal R}]_M$ is intrinsically defined, i.e. independently of the choice of Wigner map $[\cdot]_M$ and branch cut, and holomorphic upon viewing $Mp(2n;\C)$ and ${\cal A}[\C^{2n}]$ as two complex manifolds.
Since $\widehat{\cal U}(Pr(g))$ cannot vanish, it follows that $Mp(2n;\C)$ indeed covers $Sp(2n;\C)$ twice.

In other words, the two patches of the metaplectic group corresponds to the two sheets of the Riemann surface of the square root function in the prefactor of \eqref{eq:OpSp U}.
Therefore, the choice of ${\rm arg}(\sqdet{\frac{1+U-M(1-U)}{2}})$ determines the sign distinguishing the two patches of $Mp(2n;\C)$, i.e., the two metaplectic operators corresponding to a given $Sp(2n;\C)$ element.
Thus, this argument plays a role analogous to that of the Maslov index of operators in $Mp(2n;\R)$; for the general definition and references, see \cite{deGosson}, and for details in the case of $n=1$ and $M=0$, see App. \ref{app:U(1) ex}.

Finally, let us remark on two different types of limits that can be taken inside $Mp(2n;\C)$: Firstly, the symbol $[\widehat{\cal R}(g)]_M(X)\in {\cal D}[\C^{2n}|\R^{2n}]$, with $g\in Mp(2n;\C)$, coordinatizing the operator $\widehat{\cal R}(g)\in {\cal A}[\C^{2n}]$, approaches analytic delta sequences on surfaces where $1+Pr(g)-M(1-Pr(g))$ degenerates; for example, if $g_\pm\in Mp(2n;\C)_\pm$, and  $U:=Pr(g_\pm)$ does not have any unit eigenvalue, then 
\begin{align}
  \label{eq:Sp crit order}
  [\widehat{\cal R}(g_\pm)]_{\tfrac{1+U}{1-U}}(X)=\pm \left[\OpSp{U}\right]_{\tfrac{1+U}{1-U}}(X)\ ,\qquad  \left[\OpSp{U}\right]_{\tfrac{1+U}{1-U}}(X)
  =
  \frac{(2i\pi)^n}{\sqdet{\frac{1-U}{2}}}
  \,\delta^{2n}(X)
  \,,
\end{align}
as spelled out for $M=0$ in App. \ref{app:anal}.
Secondly, the space $Mp(2n;\C)$ can be extended to a compact space $\overline{Mp}(2n;\C)$ by adding points $p_\infty$ at infinity where
\be \left[\widehat{\cal R}[p_\infty]\right]_M(X)=0\ ;\ee
these points correspond to projectors 
\be \widehat{\cal P}[p_\infty]:= \lim_{g\to p_\infty} {\cal N}(g) \widehat{\cal R}[g]\ ,\label{3.68}\ee
where ${\cal N}:Mp(2n;\C)\to \C$ diverge at $p_\infty$ so as to cancel the evanescent prefactor in $\left[\widehat{\cal U}[Pr(g)]\right]_M(X)$ leaving a unique normalization constant. 
Thus, one may view these projectors as ramification points of $\overline{Mp}(2n;\C)$ as they are not doubly replicated, and one can replace \eqref{3.68} by
\be \widehat{\cal P}[p_\infty]:= \lim_{U\to p_\infty} {\cal N}'(U) \widehat{\cal U}[g]\ ,\qquad {\cal N}(g)={\cal N}'(Pr(g))\ ,\ee
which defines a compactification $\overline{Sp}(2n;\C)$ of $Sp(2n;\C)$.
As we shall see in Sec. \ref{sec:PTBH}, scalar and spinor singleton state projectors, on which $AdS$ massless particle solutions are based, are special examples of such limits in $\overline{Sp}(2n;\C)$.

In practice, instead of working with two distinct patches for $Mp(2n;\C)$, we analytically continue Eq. \eqref{eq:OpSp U} from one patch to the entire metaplectic group, so as to obtain a projective holomorphic representation map
\be \widehat{\cal Q}: Sp(2n;\C)\to {\cal A}[\C^{2n}]\ ,\qquad \widehat{\cal Q}[U]=\{\widehat{\cal U}[U],-\widehat{\cal U}[U]\}\ .\ee 
In what follows, we shall work with the analytically continued, hence double-valued, symbol $[\widehat{\cal U}]_M$, thus in a slight abuse of notation.

\paragraph{Inner Klein operators.}
The center of $Sp(2n;\C)$ is the $\mathbb{Z}_2$ generated by $-I$.
Letting $R$ be the root of the unity as in Eq. \eqref{eq:R^2=1} and Eq. \eqref{eq:det R}, such that $e^{i(2N+1)\pi R}=-I$ for $N\in \mathbb Z$, the branch prescribed in App. \ref{app:anal} amounts to cutting $Sp(2n;\C)$ open.
In particular, we choose the cut such that
\be
\label{eq:U(-1) seq}
\OpSp{-I}
:=
\OpSp{e^{i\pi R}}
=
(-1)^n\OpSp{e^{-i\pi R}}
\,;
\ee
in Weyl order,
\begin{align}
  \label{eq:K0}
  \left[\OpSp{e^{\pm i\pi R}}\right]_0(X)=\pm (2\pi i)^n \delta^{2n}(X)\,,
\end{align}
as computed in App. \ref{app:X+-}.
Thus, the metaplectic group algebra contains the operators
\be
  \label{eq:K=U(-1)}
  \Op{K}
  :=
  (-i)^n \;\OpSp{e^{i\pi R}}
  \,,
\ee
which hence obey
\be
  \OpX^I\;\Op{K}
  =
  -\Op{K}\;\OpX^I
  \,,\qquad
  (\Op{K})^2
  =1
  \,,\qquad \MoSym0{\Op{K}}(X)=(2\pi)^n\delta^{2n}(X)\ .
\ee

\subsection{Inhomogeneous metaplectic group}
\label{sec:SpH}

The inhomogeneous metaplectic group
\be MpH(2n;\C)=Mp(2n;\C)\ltimes H_{2n+1}(\C)\ ,\ee
with product rule
\be (g_1;K_1;k_1)(g_2;K_2;k_2)=(g_1 g_2;K_1+Pr(g_2) K_2 ;k_1+k_2+c(K_1,K_2))\ ,\ee
and projection $(g;K;k)\mapsto (Pr(g),K) \in ISp(2n;\C)$; the latter is cut open and represented projectively in ${\cal A}$ by
\begin{align}
  \label{eq:def OpSpH}
  \OpSpH{U}{K}:=\OpHW{K}\;\OpSp{U}
  \,,
\end{align}
with $M$-ordered symbol
\begin{align}
  &
  \left[\OpSpH{U}{K}\right]_M(X)
  =
  \left(\left[\OpHW{K}\right]_M \star_M \left[\OpSp{U}\right]_M\right)(X)
  \nonumber\\
  \label{eq:MoSpH}
  &=
  \frac{1}{\sqdet{\frac{1+U-M(1-U)}{2}}}
  \exp\left(
  \tfrac{i}{2}X\left(\tfrac{1+U}{1-U}-M\right)^{-1}X
  +2iK\tfrac{U}{1-U}\left(\tfrac{1+U}{1-U}-M\right)^{-1}X
  \right)
  \nonumber\\&\quad\times
  \exp\left(
  \tfrac{i}{2}K\left(
  \tfrac{1+U}{1-U}
  -\tfrac{4U}{1-U}\left(\tfrac{1+U}{1-U}-M\right)^{-1}\tfrac{1}{1-U}
  \right)K
  \right)
  \,.
\end{align}
as can be seen using the formula \eqref{eq:starM int} for the $\star_M$-product.
The reverse product
\begin{align}
  &
  \left(\left[\OpSp{U}\right]_M \star_M \left[\OpHW{K}\right]_M\right)(X)
  \nonumber\\
  &=
  \frac{1}{\sqdet{\frac{1+U-M(1-U)}{2}}}
  \exp\left(
  \tfrac{i}{2}X\left(\tfrac{1+U}{1-U}-M\right)^{-1}X
  +2iK\tfrac{1}{1-U}\left(\tfrac{1+U}{1-U}-M\right)^{-1}X
  \right)
  \nonumber\\&\quad\times
  \exp\left(
  \tfrac{i}{2}K\left(
  \tfrac{1+U}{1-U}
  -\tfrac{4}{1-U}\left(\tfrac{1+U}{1-U}-M\right)^{-1}\tfrac{U}{1-U}
  \right)K
  \right)
  \\&=
  \left[\OpSpH{U}{UK}\right]_M
  \,,
\end{align}
whose combination with Eqs. \eqref{eq:BCH} and \eqref{eq:repr Sp} yields the projective closure rule
\begin{align}
  \label{eq:repr SpH}
  \OpSpH{U_1}{K_1}
  \OpSpH{U_2}{K_2}
  =e^{iK_1U_1K_2+i\varphi(U_1,\,U_2)}
  \OpSpH{U_1U_2}{K_1+U_1K_2}
  \,,
\end{align}
where $\varphi$ is the cocycle factor in \eqref{eq:repr Sp}, which was derived in \cite{Didenko:2012tv} using the Weyl-ordering order symbol
\begin{align}
  \label{eq:WoSpH}
  \left[\OpSpH{U}{K}\right]_0(X)
  =
  \frac{1}{\sqdet{\frac{1+U}{2}}}
  \exp\left(
  \tfrac{i}{2}X\tfrac{1-U}{1+U}X
  +2iK\tfrac{U}{1+U}X
  +\tfrac{i}{2}K\tfrac{1-U}{1+U}K
  \right)
  \,.
\end{align}
The formula for the inverse, viz.
\begin{align}
  \OpSpH{U}{K}^{-1}
  &=
  \OpSpH{U^{-1}}{-U^{-1}K}
  =
  \OpSpH{U^{-1}}{-KU}
  \,
\end{align}
yields the adjoint action 
\begin{align}
  &
  \OpSpH{U}{K}^{-1}
  \OpSpH{V}{L}
  \OpSpH{U}{K}
  \\\nonumber&=
  e^{iKU^{-1}(1+V^{-1})L-iKU^{-1}VUK}\,\OpSpH{U^{-1}VU}{U^{-1}L+(1-U^{-1}VU)K}
  \,.
\end{align}
In particular, the adjoint $Sp(2n;\C)$ action reads
\begin{align}
  \label{eq:Sp rot}
  \OpSpH{U}{0}^{-1}
  \OpSpH{V}{L}
  \OpSpH{U}{0}
  =
  \OpSpH{U^{-1}VU}{LU}
  \,,
\end{align}
and the action of the Heisenberg group on itself is
\begin{align}
  \OpHW{K}^{-1}
  \OpHW{L}
  \OpHW{K}
  =
  \OpHW{L}
  \exp\left(2iLK\right)
  \,.
\end{align}
Thus, if $\widehat f=f(\widehat X)$ is expandable  over plane waves, then
\begin{align}
  \label{eq:SpH on f(X)}
  \OpSpH{U}{0}^{-1}
  f(\OpX)\;
  \OpSpH{U}{0}
  &=
  f(U\OpX)
  \,,&
  \OpHW{K}^{-1}
  f(\OpX)\;
  \OpHW{K}
  &=
  f(\OpX+2K)
  \,.
\end{align}
The trace \eqref{eq:MoTr} of an $ISp(2n;\C)$ group element reads
\begin{align}
  \label{eq:Tr SpH}
  \Tr\,\OpSpH{U}{K}
  &=
  \frac{(2i\pi)^n}{\sqdet{\tfrac{1-U}{2}}}
  \exp\left(\tfrac{i}{2}K\tfrac{1+U}{1-U}K\right)
  \,,
\end{align}
which due to the phase assignment \eqref{eq:U(-1) seq} applies directly to $U=-I$.
The operator Fourier transform \eqref{eq:FTOp} of $\OpSpH{U}{K}$ is given by
\begin{align}
  \label{eq:FT OpSpH}
  \FTSpH{U}{K}{K'}
  &=
  \frac{1}{(2\pi)^n}\Tr\left(
  \OpSpH{U}{K}\;\OpSpH{I}{-K'}\right)
  \nonumber\\&=
  \frac{i^n}{\sqdet{\tfrac{1-U}{2}}}
  \exp\left(
  \tfrac{i}{2}(K-K')\tfrac{1+U}{1-U}(K-K')
  +iKK'\right)
  \,.
\end{align}
Since each operator is unambiguously defined by its Fourier transform, it follows that the delta operators \eqref{eq:DeltaM} are related to $Sp(2n)$ operators, viz.
\begin{align}
  \OpDM{M}(X) =
  \frac{(-i)^n}{(2\pi)^n\sqdet{M+1}}\;
  \OpSpH{\tfrac{M-1}{M+1}}{\tfrac{1}{M+1}X}
  \exp\left(\tfrac{i}{2}X\tfrac{M}{M^2-1}X\right)
  \,.
\end{align}
The Fourier representation also facilitates the identification of  limits of delta sequences, such as
\begin{align}
  \label{eq:U(-1,K) deltaseq}
  \lim_{\epsilon\to0^+}
  \frac{1}{\epsilon^{2n}}
  \OpSpH{-I}{\tfrac{1}{\epsilon}K}
  =
  (2i\pi)^n\delta^{2n}(K)
  \,,
\end{align}
where we note that the r.h.s. is a constant with respect to the oscillators $\OpX_I$.
In Weyl order, one can see that the sign is indeed the same as in in \eqref{eq:K0}.

The inner Klein operator \eqref{eq:K=U(-1)} can be used to twist the trace operation
\begin{align}
  \Tr\left[\Op{f}\;\Op{K}\right]
  =
  (-i)^n\Tr\left[\Op{f}\;\OpSpH{-I}0\right]
  =
  \STr\left[\Op{f}\right]
  \,,
\end{align}
where the identification with the supertrace comes from the result of its application to $MpH(2n,\C)$
\begin{align}
  \label{eq:STr SpH}
  \STr\,\OpSpH{U}{K}
  &=
  \frac{(2\pi)^n}{\sqdet{\tfrac{1+U}{2}}}
  \exp\left(\tfrac{i}{2}K\tfrac{1-U}{1+U}K\right)
  \,.
\end{align}

\subsection{Differentials}
\label{sec:dX}

Equipping $C^{2n}$ with a trivial differential Poisson structure, the quantization of $\Omega(\C^{2n})$ in the presence of boundary conditions, yields a DGA $\Omega[\C^{2n}]$ with (inner) differential $\Op{Q}$ given by
\begin{align}
  \Op{Q}\Op{f}:=\left[-\frac{i}{2}\dX\OpX,\,\Op{f}\,\right]
  \,,\qquad \Op{f}\in \Omega(\C^{2n})\ ,
\end{align}
where $dX^I$ are anti-commuting elements commuting to $\widehat X^I$, which defines a linear map $\Op{Q}:\Omega_{[p]}[\C^{2n}]\to \Omega_{[p+1]}[\C^{2n}]$ obeying the graded Leibniz rule. 
From \eqref{eq:starM bopp}, it follows that 
\begin{align}
  Q\MoSymOp{M}{f\,}
  :=
  \MoSym{M}{\Op{Q}\Op{f}\,}
  =
  \dX^{I}\partial_{X^{I}}\MoSymOp{M}{f\,}
  \,,
\end{align}
where $Q$ is thus the de Rham differential on $\C^{2n}$.
Assuming that $\widehat{\cal U}[MpH(2n;\C)]\subset \Omega_{[p]}[\C^{2n}]$, the action of $\widehat Q$ on the operator representatives of $MpH(2n;\C)$ can be computed from \eqref{eq:X=dPW}, viz.
\begin{align}
  \Op{Q}\,\OpSpH{U}{K}
  &=
  \left.
  -\frac{1}{2}\dX\partial_L
  \left(\OpSpH{1}{L}\OpSpH{U}{K}-\OpSpH{U}{K}\OpSpH{1}{L}\right)
  \right\vert_{L=0}
  \nonumber\\&=
  \frac12\dX\left(\frac{1-U}{U}\partial_K
  -i\frac{1+U}{U}K\right)
  \OpSpH{U}{K}
  \,.
\end{align}
Letting ${\cal X}$ be a commuting manifold, the quantization of $\Omega({\cal X}\times\C^{2n})\equiv \Omega({\cal X})\otimes\Omega(\C^{2n})$ in the presence of the boundary conditions on $\C^{2n}$ yields the DGA $\Omega[{\cal X}\times\C^{2n}]\equiv\Omega({\cal X})\otimes\Omega[\C^{2n}]$ with total differential
\begin{align}
  \OpTDiff:=d+\Op{Q}
  \,,\label{3.97}
\end{align}
where $d$ is the de Rham differential on ${\cal X}$, and the signs are controlled using Koszul signs governed by the total degree.
Thus, letting $f^i\in \Omega({\cal X})$ and $\Op{f}_i\in \Omega(\C^{2n})$, one has
\begin{align}
  (d+\widehat{Q})\sum_i f^i\Op{f}_i
  =
  \sum_i ((df^i) \Op{f}_i+ (-1)^{{\rm deg} f^i} f^i (\widehat Q \widehat f_i))
  \,.
\end{align}
Introducing a Wigner map $[\cdot]_M:\Omega[{\cal X}\times\C^{2n}]\to \Omega({\cal X}\times \C^{2n})$, where $M$ is a function on ${\cal X}$, it follows that
\begin{align}
  d_M\MoSymOp{M}f
  :=
  \MoSym{M}{d\Op{f}}
  =
  \left(
  dx^{\mu}\,\partial_{x^{\mu}}-\frac{i}{2}\partial_XdM\partial_X
  \right)\MoSymOp{M}f
  \,, \label{3.98}
\end{align}
which is indeed compatible with the $\star_M$-product \eqref{eq:starM exp}.
Applying Eq. \eqref{3.98} to Eq, \eqref{eq:MoSpH} yields 
\begin{align}
  \label{eq:dOpSpH}
  d\OpSpH{U}{K}
  =
  \left(
  dK\partial_K+\frac{i}{4}\left(\partial_K-iK\right)dU\frac{1}{U}\left(\partial_K-iK\right)
  \right)\OpSpH{U}{K}
  \,.
\end{align}
The total differential $\OpTDiff$ can be covariantized by means of a connection $\Op{A}$ that transforms as
\begin{align}
  \Op{A}
  \quad\overset{\Op{G}}{\longrightarrow}\quad \Op{A}^{(\widehat G)}:=
  \Op{G}^{-1}\OpTDiff\Op{G}
  +\Op{G}^{-1}\Op{A}\Op{G}
  \,.
\end{align}
Assuming that $\Op{A}$ is flat and starting from the gauge where $\Op{A}=0$, an $Mp(2n;\C)$ gauge transformation yields 
\begin{align}
  \Op{A}^{(U)}
  :&=
  \left.
  \OpSpH{U}{0}^{-1}\OpTDiff\OpSpH{U}{K}
  \right\vert_{K=0}
  \nonumber\\&=
  \left.
  \left(
  \frac{i}{4}\left(\partial_K-iK\right)dU\frac{1}{U}\left(\partial_K-iK\right)
  +
  \frac12\dX\frac{1-U}{U}\partial_K
  \right)\OpSpH{1}{KU}
  \right\vert_{K=0}
  \nonumber\\&=
  -\frac{i}{4}\OpX U^{-1}dU\OpX
  +\frac{i}{2}\dX(1-U)\OpX
  \,,
\end{align}
where the last step was performed using Eq. \eqref{eq:X=dPW}.
The corresponding flat adjoint covariant derivative
\begin{align}
  \OpTDiff^{(U)}\Op{f}
  :&=
  \OpSpH{U}{0}^{-1}\OpTDiff\left(
  \OpSpH{U}{0}\Op{f}\,\OpSpH{U}{0}^{-1}
  \right)\OpSpH{U}{0}
  =
  \OpTDiff\Op{f}+\left[\Op{A}^{(U)},\,\Op{f}\right]
  \,,
\end{align}
is represented in $M$-order by
\begin{align}
  \MoTDiff{M}^{(U)}
  =
  d+XU^{-1}dU\partial_X
  -\frac{i}{2}\partial_X\left(dM+\left[U^{-1}dU,M\right]\right)\partial_X
  +dXU\partial_X
  \,.
\end{align}

\subsection{Homotopy integration}
\label{sec:homotopy}

Using symbol calculus, the system 
\begin{align}
  \label{eq:Qf=J op}
  \Op{Q}\Op{f}=\Op{J}
  \,,\qquad \Op{Q}\Op{J}=0\ ,
\end{align}
which serves as a model for Eq. \eqref{eq:qf=g}, can be integrated using homotopy contraction methods from commutative geometry.
In $M$-order, the equation reads
\begin{align}
  \label{eq:Qf=J}
  Q\MoSymOp{M}{f}=\MoSymOp{M}{J}
  \,,
\end{align}
where $Q$ is the de Rham differential on the holomorphic symplectic $\C^{2n}$.
A particular solution to this equation is given by the standard homotopy trick
\begin{align}
  \label{eq:rhoMM0}
  \MoSymOp{M}{f}
  =
  Q^{\ast[M,0]}_M\left(\MoSymOpArg{M}{J}{X,\,\dX}\right)
  :=
  \int_0^1 \frac{dt}{t}
  X^I\frac{\partial}{\partial(\dX^I)}
  \MoSymOpArg{M}{J}{tX,\,t\dX}
  \,.
\end{align}
Obviously, any solution to Eq. \eqref{eq:Qf=J} will also solve the same equations in terms of shifted variable $X^{[\Xi]}:=X-\Xi$.
Conversely, one can get a solution to Eq. \eqref{eq:Qf=J} by using the homotopy trick on the shifted variables $X^{[\Xi]}$ after converting $X$ to $X^{[\Xi]}+\Xi$.
This defines a \emph{shifted resolution}
\begin{align}
  \label{eq:rhoMMXi}
  Q_M^{\ast[M,\Xi]}\left(\MoSymOpArg{M}{J}{X,\,\dX}\right)
  :=
  \int_0^1 \frac{dt}{t}
  (X^I-\Xi^I)\frac{\partial}{\partial(\dX^I)}
  \MoSymOpArg{M}{J}{tX+(1-t)\Xi,\,t\dX}
  \,.
\end{align}
As explained in Sec. \ref{sec:VE pert}, that it indeed provides a solution is a consequence of the fact that the object
\begin{align}
  \Op\Pi^{[M,\Xi]}
  =
  1-\Op{Q}^{\ast[M,\Xi]}\Op{Q}-\Op{Q}\Op{Q}^{\ast[M,\Xi]}
  \,,
\end{align}
projects on $H(\Op{Q})$.
Indeed, one can compute in $M$-order:
\begin{align}
  \left(QQ_M^{\ast[M,\Xi]}+Q_M^{\ast[M,\Xi]}Q\right)&
  \MoSymOpArg{M}{f}{X^{[\Xi]}+\Xi,dX}
  \nonumber\\ &=
  \int_0^1\frac{dt}{t}\left[%
  dX\partial_{X^{[\Xi]}},\,X^{[\Xi]}\frac{\partial}{\partial(dX)}
  \right]
  \MoSymOpArg{M}{f}{tX^{[\Xi]}+\Xi,tdX}
  \nonumber\\&=
  \int_0^1\frac{dt}{t}\left(%
  dX\frac{\partial}{\partial(dX)}
  +X^{[\Xi]}\partial_{X^{[\Xi]}}
  \right)
  \MoSymOpArg{M}{f}{tX^{[\Xi]}+\Xi,tdX}
  \nonumber\\&=
  \int_0^1\frac{dt}{t}\,t\frac{d}{dt}
  \MoSymOpArg{M}{f}{tX^{[\Xi]}+\Xi,tdX}
  \nonumber\\&=
  \MoSymOpArg{M}{f}{X^{[\Xi]}+\Xi,dX}
  -\MoSymOpArg{M}{f}{\Xi,0}
  \,.
\end{align}
This proves
\begin{align}
  \label{eq:PXi}
  \Op\Pi^{[M,\Xi]}\Op{f}
  =
  \Pi^{[M,\Xi]}_M\MoSymOp{M}{f}
  =
  \MoSymOpArg{M}{f}{\Xi,0}
  \,,
\end{align}
which generalizes the unshifted case \cite{Didenko:2015cwv}.

Using the $M$-ordering symbol \eqref{eq:MoSpH}, one can derive the lemma
\begin{align}
  \label{eq:rhoOpMXi}
  &
  \Op{Q}^{\ast[M,\Xi]}\left(j(\dX)\;\OpSpH{U}{K}\right)
  \\&=\nonumber
  \int_0^1\frac{dt}{t^{2n+2}}
  \Big(
  -\tfrac{i}{2}\partial_K\left(1+U-(1-U)M\right)
  +\tfrac12K\left(1-U-(1+U)M\right)
  -\Xi
  \Big)^I\frac{\partial}{\partial(\dX)^I}j(t\dX)
  \\\nonumber&\quad\times
  \frac{\sqdet{1-U_t}}{\sqdet{1-U}}
  \exp\left(
  \tfrac{i}{2}K\tfrac{1+U}{1-U}K
  -\tfrac{i}{2}K_t\tfrac{1+U_t}{1-U_t}K_t
  \right)
  \OpSpH{U_t}{K_t}
  \,,
\end{align}
with
\begin{align}
  U_t
  :&=
  \frac%
  {\tfrac{1}{t^2}\tfrac{1+U}{1-U}-\tfrac{1-t^2}{t^2}M-1}%
  {\tfrac{1}{t^2}\tfrac{1+U}{1-U}-\tfrac{1-t^2}{t^2}M+1}
  \,,&
  K_t
  :&=
  \left(
  \frac{1}{t}K\frac{2U}{1-U}
  +\frac{1-t}{t}\Xi
  \right)
  \frac{1-U_t}{2U_t}
  \,,
\end{align}
whose parameters $M$ and $\Xi$ will be used to construct a number resolution operators in what follows.

\section{Orderings, perturbative solutions and Fronsdal fields}
\label{sec:solve Z}

In this Section, we shall apply the formalism above developed to integrating the $\Op Z$-space equations with various choices of homotopy: first, homotopies enconding two types of holomorphic gauges, which simplify the perturbative integration to the point that it will be possible to obtain full, not just linearized, solutions --- failing however to activate non-trivial Fronsdal fields (Sec. \ref{sec:hol g}); then, with a homotopy that no longer factorizes the dependence on $\Op Y$ and $\Op Z$ and corresponding to a rotated Vasiliev gauge (Sec. \ref{sec:rotV g}). 
With the latter homotopy we shall show how it is possible to glue, at first order, the Weyl zero-form to the spacetime connection, thereby activating spacetime gauge fields, independently of the ordering, i.e., in operator language. 
The so-obtained gluing equation is then presented in Weyl ordering for $\OpY$ oscillators, to show that it is identical to the usual formulation of the COMST (\ref{eq:DW=S},\,\ref{eq:S=eeC}), encoding the Fronsdal equations in unfolded form.

For convenience, we shall start from an expansion of the adjoint integration constant $\Op{\Psi}'(\Op Y)$ on Heisenberg group elements. The symbol calculus adapted to $\Op Y$ and $\Op Z$ is described in Sec. \ref{sec:Wo(Y)} and in Secs. \ref{sec:hol g}, respectively, while in Sec. \ref{sec:No(YZ)} we show the choice of homotopy corresponding to the rotated Vasiliev gauge entails a class of orderings that non-trivially entangle $Y$ and $Z$ variables,
and in particular include the one-parameter family of orderings considered in \cite{Didenko:2019xzz}, which interpolates between the entangled normal ordering in which the Vasiliev equations are usually formulated, Weyl ordering, and the limiting spin-local case.
In addition to the formulation in terms of operators of the aforementioned well-known results,
the originality of the present approach resides in the presence of a family of gauges, generalizing Vasiliev's gauge, labelled by spacetime-depedent matrices.
Sec. \ref{sec:Lorentz} shows the relation between this newly found freedom and the local Lorentz symmetry of the free higher-spin equations, 
and contains a discussion about possible procedures to identify the deformed higher-spin transformation (\ref{eq:dgt W},\,\ref{eq:dgt C}) 
inside the larger gauge group \eqref{eq:VE fgt}.

In Sec. \ref{sec:hol g} we construct in particular two families of exact solutions to the operator equations (\ref{eq:VE A},\,\ref{eq:VE Phi}): the gaussian holomorphic gauge dicussed in \cite{2011,2017,COMST}, and a new one, the axial holomorphic gauge, in which the linearized moduli take a form which we expect to facilitate the future analysis of interactions preserving a given set of physical constraints, such as boundary conditions \cite{COMST}.
The gauge function connecting the latter to the rotated Vasiliev gauge at first order is built in Sec. \ref{sec:lin gf}. 
Finally, in Sec. \ref{sec:OWL} we make use of the simplicity of the axial holomorphic gauge to compute Wilson line observables, checking their independence from any twistor-space auxiliary construct of the solutions up to the third order in the integration constant $\Op{\Psi}'$. 
This generalises the first order result of \cite{COMST}.

\subsection{Fiber symbols}
\label{sec:Wo(Y)}

The formalism of Sec. \ref{sec:starprod} can be used to describe $\mathcal{Y}_4$ by  replacing
\begin{align}
  \widehat X_{I}\to \widehat Y_\au
  \,,&&
  K^{I}\to\Lambda^\au
  \,,&&
  \Theta_{IJ}\to \epsilon_{\underline{\a\b}}
  \,,
\end{align}
and defining
\begin{align}
  \OpSpHY{L}{\Lambda}
  :&=\widehat{\cal U}[L, \Lambda]|_{ \widehat X \to \widehat Y }\equiv
  \exp(i\Lambda\OpY)\exp\left(\tfrac{i}{2}\OpY\ell\OpY\right)
  \,,\qquad L:= e^{-2\ell}\ .
\end{align}
As explained in Sec. \ref{sec:OpY}, the fluctuations fields are expanded over totally symmetric monomials
\begin{align}
  \Op{J}_{\au_1\cdots\au_{2(s-1)}}
  &=
  \left.
  (-)^{s-1}
  \frac{\partial}{\partial\Lambda^{\au_1}}
  \cdots
  \frac{\partial}{\partial\Lambda^{\au_{2(s-1)}}}
  e^{i\Lambda\Op{Y}}
  \right\vert_{\Lambda=0}= \Op{Y}_{(\au_1}
  \cdots
  \Op{Y}_{\au_{2(s-1)})}
  \,.
\end{align}
Hence they read in Weyl ordering
\begin{align}
  \MoSym{0}{\Op{J}_{\au_1\cdots\au_{2(s-1)}}}
  =
  \left.
  (-)^{s-1}
  \frac{\partial}{\partial\Lambda^{\au_1}}
  \cdots
  \frac{\partial}{\partial\Lambda^{\au_{2(s-1)}}}
  e^{i\Lambda Y}
  \right\vert_{\Lambda=0}
  =
  Y_{\au_1}
  \cdots
  Y_{\au_{2(s-1)}}
  \,.
\end{align}
Hence, expanding over the totally symmetric generators is equivalent to Taylor-expanding in Weyl order around $Y=0$.
This means the tensorial interpretation (\ref{eq:W Taylor},\,\ref{eq:C Taylor}) of $\Op{W}$ and $\Op{C}$ requires their Weyl ordered symbols to be analytic in $Y$.

The perturbative expansion starts with the vacuum \eqref{eq:AdS W}, and in particular with its gauge function \eqref{eq:AdS gf Op} that can be encoded in a symmetric matrix $L_{\underline{\a\b}}$ as
\begin{align}
  \label{eq:AdS(Y)}
  \Op{L}
  :&=
  \OpSpHY{L}{0}
  \,,&
  \Op{U}^{(0)}
  &=
  \Op{L}^{-1}\;d\Op{L}
  =
  -\frac{i}{4}
  \Op{Y}L^{-1}dL\Op{Y}
  \,.
\end{align}
For the stereographic coordinates \eqref{eq:stereo coord}
\begin{equation}
\MoSym0{\Op{L}}
=
\exp_\star\left(4i\xi x^aP_a\right)
=
\frac{2h}{1+h}
\exp\left(
-\frac{i}{1+h}yx\yb
\right)
\,.
\label{eq:stereo gf WO}
\end{equation}

The starting point of the perturbative expansion (\ref{eq:qPhin}-\ref{eq:DUn}) around the background \eqref{eq:VE BG} is
the first order equations for the zero-form field
\begin{align}
 \Op{q}\Op\Phi^{(1)}
 &=
 0
 \,,&
 \Op{D}_\pi\Op\Phi^{(1)}
 &=
 0
 \,.
\end{align}
The solution is a covariantly constant construct of $\Op{Y}$, that is given by \eqref{eq:Psi=Ck}
in terms of an adjoint intial datum $\Op\Psi$ that we expand for convenience as
\begin{align}
  \label{eq:FT Psi}
  \Op\Psi
  &=
  \int\frac{d^4\Lambda}{(2\pi)^2}
  \Psi_\Lambda\;\OpSpHY{1}{\Lambda}
  \,,&
  D_\Lambda\Psi_\Lambda
  :&=
  \left(d+\Lambda L^{-1}dL\partial_\Lambda\right)\Psi_\Lambda
  =0
  \,.
\end{align}
Assuming appropriate fall-off conditions, such that boundary terms arising via integration by parts on $\Lambda$ vanish, the latter condition on $\Psi_\Lambda$ translates the covariant constancy of $\Op\Phi^{(1)}$.
The above expansion for $\Psi$ follows from an identical expansion of the integration constant in \eqref{eq:Psi},
\begin{align}
  \label{eq:FT PsiLprime}
  \Op\Psi^\prime
  &=
  \int\frac{d^4\Lambda}{(2\pi)^2}
  \Psi^\prime_{\Lambda}\;
  \OpSpHY{1}{\Lambda}
  \,,&
  d\Psi^\prime_{\Lambda}=0
  \ ,
\end{align}
via an application of \eqref{eq:Sp rot}, with $U=L$, followed by a change of integration variable, thereby identifying
\begin{align}
  \label{eq:FT PsiL}
  \Psi_\Lambda
  &=
  \Psi^\prime_{\Lambda L^{-1}}
  =
  \Psi^\prime_{L\Lambda}
   \ .
\end{align}
Let us recall that in addition to its explicit equation,
the Fourier transform $\Psi_\Lambda$ should be such that
\begin{align}
  \label{eq:PhiW}
  \MoSym{0}{\Op\Phi^{(1)}}
  &=
  \MoSym0{\Op\Psi\;\OpSpHy{-1}{0}}
  =
  \int\frac{d^4\Lambda}{(2\pi)^2}
  \Psi_\Lambda\;\MoSpHy0{-1}{\lambda}\;\MoSpHyb0{1}{\bar\lambda}
  =
  \int\frac{d^2\bar\lambda}{2\pi}
  \Psi_{y,\,\bar\lambda}\;
  e^{i\bar\lambda\yb}
  \,,
\end{align}
is analytic in $Y$.
This is the initial datum in terms of which one solves the $\Opz$-dependent equations.

This expansion for $\Op\Psi$ will induce an expansion of the other linearised fields
on which the adjoint covariant derivative defined in \eqref{eq:def acd} will act as
\begin{align}
  \Op{D}\left(
  \int\frac{d^4\Lambda}{(2\pi)^2}
  \Psi_\Lambda\;\Op{f}\right)
  =
  \int\frac{d^4\Lambda}{(2\pi)^2}
  \Psi_\Lambda\;D_\Lambda\Op{f}
  \end{align}
by virtue of \eqref{eq:FT Psi}. Note that the operator $D_\Lambda$ defined by this equation
is still nilpotent and compatible with the operator product, in particular
\begin{align}
  \label{eq:d(PsiUF)=PsiUdF}
  D_\Lambda\OpSpHY{1}{\Lambda}
  &=0
  \,,&
  \Op{D}\left(
  \int\frac{d^4\Lambda}{(2\pi)^2}\,
  \Psi_\Lambda\;\OpSpHY{1}{\Lambda}\;\Op{f}
  \right)
  &=
  \int\frac{d^4\Lambda}{(2\pi)^2}\,
  \Psi_\Lambda\;\OpSpHY{1}{\Lambda}
  \left(D_\Lambda\Op{f}\right)
  \,.
\end{align}

\subsection{Holomorphic gauges}
\label{sec:hol g}
The integration tools of Sec. \ref{sec:homotopy} can be used in the noncommutative holomorphic base space $\mathcal{Z}_2$ to build holomorphic solutions \eqref{eq:fact sol}.
Because of the sign in the commutation relations \eqref{eq:[Z,Z][Y,Z]}
there is a sign subtlety to take care of in the identifications
\begin{align}
  \widehat X_{I}\to \Opz_\ag
  \,,&&
  K^{I}\to\mu^\ag
  \,,&&
  \Theta_{IJ}\to-\epsilon_{\ag\bg}
  \,,\\
  \widehat X^{I}\to -\Opz^\ag
  \,,&&
  K_{J}\to-\mu_\ag
  \,,&&
\end{align}
as well as a sign convention in
\begin{align}
  \label{eq:OpSpHz}
  \OpSpHz{e^{-2g}}{\mu}
  :&=\widehat{\cal U}[e^{-2g},\mu]\rvert_{\widehat X_I\to \Opz_\alpha, \widehat X^I\to -\Opz^\alpha}\equiv
  \exp(i\mu\Opz)\exp\left(-\tfrac{i}{2}\Opz g\Opz\right)
  \,.
\end{align}
Unlike the generator $g_{IJ}$, the reference matrix $R_{IJ}$ of App. \ref{app:X+-} is mapped in the following way:
\begin{align}
  \label{eq:RtoD}
  R_{IJ}
  &\to
  \mathcal{D}_{\ag\bg}
  \,,&
  R_{I}{}^{J}
  &\to
  -\mathcal{D}_{\ag}{}^{\bg}
  \,,&
  R^{I}{}_{J}
  &\to
  -\mathcal{D}^{\ag}{}_{\bg}
  \,,&
  R^{IJ}
  &\to
  \mathcal{D}^{\ag\bg}
  \,.
\end{align}
This fact ensures that the sign is preserved in the classical relations
\begin{align}
  \int d^2z\exp\left(\tfrac{i}{2}z\mathcal{D}z\right)
  &=
  2\pi
  \,,&
  \lim_{\epsilon\to0}
  \frac{1}{\epsilon}
  \exp\left(\tfrac{i}{2\epsilon}z\mathcal{D}z\right)
  &=
  2\pi\delta^2(z)
  \,,
\end{align}
respectively analogous to Eqs. (\ref{startingpt}) and \eqref{eq:deltaS gauss} (with $S=1$, and $\d^2_1(z)\equiv \d^2(z)$).
These in turn allow to choose the reference Kleinian as in Eqs. (\ref{eq:U(-1) seq},\,\ref{eq:def KR}):
\begin{align}
  \label{eq:kappaz from Sp}
  \OpSpHz{-1}0
  :&=
  \lim_{\theta\to\pi}\OpSpHz{e^{i\theta \mathcal{D}}}0
  \,,&
  \OpKz
  &=
  -i\;\OpSpHz{-1}0
  \,.
\end{align}
This convention induces the following analogous to Eqs. (\ref{eq:Tr SpH},\,\ref{eq:U(-1,K) deltaseq})
\begin{align}
  \label{eq:Tr SpHz}
  \Tr\left(
  \OpSpHz{u}{\mu}
  \right)
  &=
  \frac{(2i\pi)^n}{\sqdet{\tfrac{1-u}{2}}}
  \exp\left(-\tfrac{i}{2}\mu\tfrac{1+u}{1-u}\mu\right)
  \,,\\
  \label{eq:Uz(-1,K) deltaseq}
  \lim_{\epsilon\to0}
  \frac{1}{\epsilon^{2}}
  \OpSpHz{-1}{\tfrac{1}{\epsilon}\mu}
  &=
  2i\pi\delta^{2}(\mu)
  \,.
\end{align}

Now, given spacetime independent spinor $\zeta$ and matrix $m$,
the homotopy \eqref{eq:rhoOpMXi}
\begin{align}
  &
  \Op{q}^{\ast[m,\zeta]}\left(j(\dz)\;\OpSpHz{u}{\mu}\right)
  \nonumber\\&=
  \label{eq:q*[m,zeta]}
  -
  \int_0^1\frac{dt}{t^4}
  \left(
  \tfrac{i}{2}\partial_\mu\left(1+u-(1-u)m\right)
  +\tfrac12\mu\left(1-u-(1+u)m\right)
  +\zeta
  \right)^\ag\frac{\partial}{\partial(\dz)^\ag}j(t\dz)
  \nonumber\\&\quad\times
  \frac{\sqdet{1-u_t}}{\sqdet{1-u}}
  \exp\left(
  -\tfrac{i}{2}\mu\tfrac{1+u}{1-u}\mu
  +\tfrac{i}{2}\mu_t\tfrac{1+u_t}{1-u_t}\mu_t
  \right)
  \OpSpH{u_t}{\mu_t}
  \,,
\end{align}
with
\begin{align}
  u_t
  :&=
  \frac%
  {\tfrac{1}{t^2}\tfrac{1+u}{1-u}-\tfrac{1-t^2}{t^2}m-1}%
  {\tfrac{1}{t^2}\tfrac{1+u}{1-u}-\tfrac{1-t^2}{t^2}m+1}
  \,,&
  \mu_t
  :&=
  \left(
  \frac{1}{t}\mu\frac{2u}{1-u}
  -\frac{1-t}{t}\zeta
  \right)
  \frac{1-u_t}{2u_t}
  \,,
\end{align}
indeed obeys (\ref{eq:rhoF},\,\ref{eq:rhoFhol}) and can hence be used to build solutions in holomorphic gauges.
The first step in building such solutions is \eqref{eq:vz1}
\begin{align}
  &
  \Op{v}^{[m,\zeta]}_{1}
  :=
  \frac{e^{i\theta}}{4}\Op{q}^{\ast[m,\zeta]}\left(\OpSpHz{-1}{0}\;\dz^\ag\dz_\ag\right)
  \label{eq:v1[m,zeta]}\\\nonumber&=
  \left.
  \frac{e^{i\theta}}{2}
  \int_0^1\frac{dt}{t^2}\,
  \dz\left(
  \tfrac{i}2\tfrac{t^2}{1-t^2}\tfrac{1+u_t}{u_t}\partial_{\rho_t}
  +\zeta
  \right)
  \sqdet{\tfrac{1-u_t}{2}}
  \exp\left(
  \tfrac{i}{2}\rho_t\tfrac{1+u_t}{1-u_t}\rho_t
  \right)
  \OpSpHz{u_t}{\rho_t}
  \right\vert_{\rho_t
  =
  -\tfrac{1-t}{2t}\zeta
  \frac{1-u_t}{u_t}}
  \,.
\end{align}
The problem with this solution is that it is not bosonic in the sense of \eqref{eq:BP holG} .
It still can be used to build the following bosonic solution
\begin{align}
  \label{eq:BP[m,zeta]}
  \Op{v}_{1}^{[m,\zeta,+]}
  :&=
  \frac{1}{2}\left(
  \Op{v}_{1}^{[m,\zeta]}+\Op{v}_{1}^{[m,-\zeta]}
  \right)
  \,.
\end{align}
To promote this to an exact solution of Vasiliev's equations,
one needs to find a way to systematize the products that appear in \eqref{eq:fact sol}.
While this can be hard in general, let us address two particular cases.

\paragraph{Gaussian holomorphic gauge.}
First, let us consider the case of a non-trivially ordered unshifted homotopy $\Op{q}^{\ast[m,0]}$.
In order to unambiguously define the prefactors, let us specialize to the case where $m=a\mathcal{D}$,
where $a>0$ and where $\mathcal{D}$ is the metric chosen in Eq. \eqref{eq:RtoD}.
As proven in App. \ref{app:g hol g},
the exact holomorphic solution \eqref{eq:fact sol} will be in this case,
independently of $a$,
\begin{align}
  \label{eq:vn(D)}
  \Op{v}_n^{[a\mathcal{D},0]}
  =
  \left.
  \frac12\dz\partial_\rho
  \int_{-1}^{1}\frac{ds}{\sqrt{s}}\,f_n(s)\;
  \OpSpHz{U_s^{(\mathcal{D})}}{\tfrac{1}{4s}\rho(1+s-(1-s)\mathcal{D})}
  \right\vert_{\rho=0}
  \,,
\end{align}
in terms of a parametric function
\begin{align}
  \label{eq:fn(s)}
  f_n(s)
  =
  -\frac{e^{i\theta}}{2}
  \frac{(2n)!}{(n!)^2(n+1)!}\left(
  \frac{e^{i\theta}}{8}
  \log\left(\frac{1}{s^2}\right)\right)^{n-1}
  \,,
\end{align}
and of the abelian subgroup
\begin{align}
  \left\{
  U_s^{(\mathcal{D})}:=
  \frac{1+s^2}{2s}+\frac{1-s^2}{2s}\mathcal{D}
  \;\Big\vert\;
  s\in\left[-1,\,0\right)\cup\left(0,\,1\right]
  \right\}
  \,,
\end{align}
of $Sp(2,\,\R)$.
Note that the conventional overall sign in the definition \eqref{eq:kappaz from Sp} was chosen so that it is the principal square root that appear in the result \eqref{eq:vn(D)} and in the following application of the lemma \eqref{eq:WoSpH}.

This solution was previously built and studied in Weyl order \cite{2011,2017,COMST}, in which it reads
\begin{align}
  \MoSym0{\Op{v}_n^{[a\mathcal{D},0]}}
  =
  \left.
  \dz\partial_\rho
  \int_{-1}^1\frac{ds}{1+s}f_n(s)
  \exp\left(\tfrac{i}{2}\tfrac{1-s}{1+s}z\mathcal{D}z
  +\tfrac{i}{1+s}\rho z\right)
  \right\vert_{\rho=0}
  \,,
\end{align}
and, together with the projectors and twisted projectors (\ref{eq:Pn},\,\ref{eq:Ptn}) on which $\Psi$ was expanded, represented the basic building block of the spherically symmetric higher spin black-hole exact solutions \cite{2011}, as well as of the black-hole plus massless scalar exact solution in Gaussian holomorphic gauge \cite{2017}.

\paragraph{Axial holomorphic gauge.}
The other case of interest is the shifted Weyl-ordered homotopy, yielding, according to \eqref{eq:v1[m,zeta]},
\begin{align}
  \label{eq:v1zeta}
  \Op{v}_{1}^{[0,\zeta]}
  &=
  \frac{e^{i\theta}}{2}(\dz\,\zeta)\int_0^1\frac{dt}{t^2}
  \;
  \OpSpHz{-1}{\tfrac{1-t}{t}\zeta}
  =
  \frac{e^{i\theta}}{2}(\dz\,\zeta)
  \int_0^{+\infty}du
  \;
  \OpSpHz{-1}{u\zeta}
  \,.
\end{align}
The corresponding bosonic solution \eqref{eq:BP[m,zeta]} reads
\begin{align}
  \label{eq:[0,zeta,+]}
  \Op{v}_{1}^{[0,\zeta,+]}
  &=
  \frac12\left(\Op{v}_{1}^{[0,\zeta]}+\Op{v}_{1}^{[0,-\zeta]}\right)
  =
  \frac{e^{i\theta}}{4}(\dz\,\zeta)
  \int_{-\infty}^{+\infty}du
  \left(\theta(u)-\theta(-u)\right)
  \OpSpHz{-1}{u\zeta}
  \,,
\end{align}
in terms of the Heaviside theta function.
As it is proportional to a constant one form, this first order solution is already exact (as well as its bosonized version).
The corresponding exact solutions (\ref{eq:CI LHC},\,\ref{eq:pert part sol}) are said to be in \emph{axial holomorphic gauge} if $\Op{L}\;\Op{H}=1$ or in \emph{axial factorised gauge} if $\Op{H}=1$.
Despite the Gaussian holomorphic gauge being more studied in the literature,
the simplicity of the associated subgroup
\begin{align}
  \{\sigma_1\OpSpHz{\sigma_2}{\mu}\vert\sigma_1=\pm1,\,\sigma_2=\pm1,\,\mu\in\C^2\}
\end{align}
 makes the axial holomorphic gauge a useful choice.

 To the contraction $\Op{q}^{\ast[0,\zeta]}$ used to build this solution is associated a projector \eqref{eq:PXi}
\begin{align}
  \Op{\mathcal{P}}^{[0,\zeta]}\Op{f}
  =
  \left.\MoSymOp{0}{f}\right\vert_{\substack{z\to\zeta\\\dz\to0}}
  \,.
\end{align}
 In Sec. \ref{sec:lin gf}, we will make use of the following particular case :
 \begin{align}
  \label{eq:P[0,zeta]U(-1)}
  \Op{\mathcal{P}}^{[0,\zeta]}\OpSpHz{-1}\mu
  =
  2i\pi
  \delta^2(\mu+\zeta)
  \,.
\end{align}

\subsection{Wilson Line observables in holomorphic gauge}
\label{sec:OWL}
As previously mentioned, an
advantage of building solutions in holomorphic gauge, rather than directly in a gauge where the integration constants obey Eqs. (\ref{eq:deom C},\,\ref{eq:deom W}), is the form (\ref{eq:OWL fact}) that the gauge invariant observables take.
Furthermore, the axial holomorphic gauge \eqref{eq:[0,zeta,+]} allows to determine coefficients $\alpha_{n,p;N,P}$ that go higher in perturbation theory than the first subleading ones (\ref{eq:OWL alpha1 1},\,\ref{eq:OWL alpha1 2}) that were computed in \cite{COMST} using the Gaussian holomorphic gauge \eqref{eq:v1[aD,0]}.

By plugging the coefficients $\beta_{p,k}$ and $\bar\beta_{p,\bar{k}}$ computed in App. \ref{app:OWL} into the definition \eqref{eq:OWL alpha},
one finds that the non-trivial $\alpha_{n,p;N,P}$ coefficients are given by:
\begin{align}
  \label{eq:OWL alpha answer 1}
  \alpha_{2m,1;2M,1}
  &=
  \delta_{m,M}
  \,,&
  \alpha_{2m,0;2M,1}
  &=
  \delta_{m+1,M}
  \,,\\
  \alpha_{2m+1,1;2M+2,1}
  &=
  e^{i\theta}\delta_{m,M}
  \,,&
  \alpha_{2m+1,0;2M+2,1}
  &=
  e^{-i\theta}\delta_{m,M}
  \,,\\
  \alpha_{2m+1,0;2M+1,0}
  &=
  \bar\beta_{0,2(M-m)}
  \,,&
  \alpha_{2m+1,1;2M+1,1}
  &=
  \beta_{0,2(M-m)}
  \,,\\
  \alpha_{2m,0;2M+1,0}
  &=
  e^{i\theta}\bar\beta_{0,2(M-m)}
  \,,&
  \alpha_{2m,0;2M+1,1}
  &=
  e^{-i\theta}\beta_{0,2(M-m)}
  \,,\\
  \label{eq:OWL alpha answer 5}
  \alpha_{2m,0;2M,0}
  &=
  \sum_{\ell=0}^{M-m}\beta_{0,2(M-m-\ell)}\bar\beta_{0,2\ell}
  \,,&&
\end{align}
where $\beta_{0,2\ell}$ and $\bar\beta_{0,2\ell}$ vanish by definition when $\ell=-1$, are
\begin{align}
  \beta_{0,0}
  &=
  8\pi\delta^2(\mu)
  \,,&
  \bar\beta_{0,0}
  &=
  8\pi\delta^2(\bar\mu)
  \,,
\end{align}
when $\ell=0$ and read
\begin{align}
  \beta_{0,2\ell}
  &=
  2\pi
  \left(-\frac{ie^{i\theta}}{2}\right)^{2\ell}
  \int d^{2\ell-1}\tau\int d^{2\ell-1}u
  \nonumber\\&\qquad\times
  \exp\left(%
  -i\sum_{i=1}^{2\ell-1}(-1)^iu_i\left(%
  \sum_{j<i}(-1)^j\tau_j-\sum_{j>i}(-1)^j\tau_i-\tfrac{1}{2}
  \right)(\mu\zeta)
  \right)
  \,,
\end{align}
(and analogously its complex conjugate) otherwise.
The latter integral is defined in terms of the measures:
\begin{align}
  \int d^{2\ell-1}\tau
  :&=
  \int_0^1d\tau_{2\ell}\,\cdots
  \int_0^{\tau_2}d{\tau_1}\,
  \delta\left(\tfrac{1}{2}-\sum_i(-1)^i\tau_i\right)
  \,,\\
  \int d^{2\ell-1}u
  :&=
  \prod_{j=1}^{2\ell}\left(
  \int_{-\infty}^{+\infty}du_j
  \left(\theta(u_j)-\theta(-u_j)\right)
  \right)
  \delta\left(\sum_i(-1)^iu_i\right)
  \,.
\end{align}
Its result is 0 for $\ell=1$ and remains unknown for higher $\ell$.
This means that the $\OpZ$-space contribution to the Wilson Line observables is presently known
up to the third non-trivial order $N$ (i.e. for all $N\leq n+3$) in the expansion \eqref{eq:OWL fact}.

While the observables are constructed to be invariant under all transformations of the form \eqref{eq:VE fgt},
one may suspect that the involved parametric integrals be non-trivially intertwined with the twistor space operations,
so as to compromise that invariance.
However, all the computations that were done so far tend to confirm their gauge independence.
At leading order, the result (\ref{eq:OWL alpha0 1},\,\ref{eq:OWL alpha0 2}) was known already \cite{Bonezzi:2017vha} to be gauge invariant in full generality.
The gauge independence of the first subleading contribution (\ref{eq:OWL alpha1 1},\,\ref{eq:OWL alpha1 2}) computed in $[m,\zeta,+]$ gauge
generalizes their independence from $m$ when they are computed \cite{COMST} in $[m,0]$ gauge.
Finally, an additional evidence is given by the independence of the coefficients (\ref{eq:OWL alpha answer 1}--\ref{eq:OWL alpha answer 1}) from the shift $\zeta$.

\subsection{Rotated Vasiliev gauge and Central On-Mass-Shell Theorem}
\label{sec:rotV g}
As discussed in Sec. \ref{sec:factor}, factorised solutions have trivial spacetime connection.
One way to unfactorise the homotopy $\Op{q}^{\ast[0,\zeta]}$ in such a way as to solve the field content of $\Op{\Phi}$ as gauge field curvatures is to start from the expansion \eqref{eq:FT Psi}
and replace the constant shift $\zeta$ with a momentum-dependent shift $\xi(\Lambda)$.
For a reason that will be explained in Sec. \ref{sec:No(YZ)}, the solution obtained from this construction is said to be in \emph{rotated Vasiliev gauge}.
With the initial datum \eqref{eq:FT Psi}, the holomorphic contribution to the first order equation \eqref{eq:qVn} reads
\begin{align}
  \Op{q}\Op{V}^{(1)}_{z}
  =
  \frac{e^{i\theta}}{4}\dz^\ag\dz_\ag
  \int\frac{d^4\Lambda}{(2\pi)^2}
  \Psi_\Lambda\;\OpSpHY{1}{\Lambda}\;\OpSpHz{-1}{0}
  \,,
\end{align}
or in other words
\begin{align}
  \label{eq:V1hol}
  \Op{V}^{(1)}_{z}
  &=
  \int\frac{d^4\Lambda}{(2\pi)^2}\;
  \Psi_\Lambda\;
  \OpSpHY{1}{\Lambda}\;
  \Op{v}_{1\Lambda}
  \,,&
  \Op{q}\Op{v}_{1\Lambda}
  =
  \frac{e^{i\theta}}{4}\dz^\ag\dz_\ag\;
  \OpSpHz{-1}{0}
  \,.
\end{align}
We take the solution $\Op{v}_{1\Lambda}^{[0,\xi(\Lambda)]}$ defined by Eq. \eqref{eq:v1zeta}.
After integration, the obtained connection corresponds to an operator shift
\begin{align}\label{eq:V1shift}
  \Op{V}^{(1)[0,\xi(-\tfrac12\text{ad}(\OpY))]}_{z}
  &=
  \int\frac{d^4\Lambda}{(2\pi)^2}\;
  \Psi_\Lambda\;
  \OpSpHY{1}{\Lambda}\;
  \Op{v}_{1\Lambda}^{[0,\xi(\Lambda)]}
  \,,
\end{align}
rather than to a constant one.

The next step is to find the spacetime connection at first order from \eqref{eq:qUn}, through the application of $\Op{D}$ to \eqref{eq:V1shift},
or more precisely of $D_\Lambda$ that is defined in Eq. \eqref{eq:FT Psi}.
In fact, since the spacetime and momentum dependence is completely encoded in the shifting spinor $\xi$,
the latter derivative acts like a de Rham differential in the two-dimensional space coordinatized by $\xi$.
Using this and \eqref{eq:dOpSpH}, one shows that the relevant component
\begin{align}
  \label{eq:U(PsiL)}
  \Op{U}^{(1)[0,\xi]}_{z}
  &=
  \int\frac{d^4\Lambda}{(2\pi)^2}\;
  \Psi_\Lambda\;
  \OpSpHY{1}{\Lambda}\;
  \Op{u}_{1\Lambda}^{[0,\xi]}
  \,,
\end{align}
of the spacetime connection
\begin{align}
  \label{eq:U=U+U+W}
  \Op{U}^{(1)[0,\xi,\bar\xi]}
  &=:
  \Op{W}^{(1)[0,\xi,\bar\xi]}
  +\Op{U}^{(1)[0,\xi]}_{z}
  +\Op{U}^{(1)[0,\bar\xi]}_{\zb}
  \,,&
  \Op{q}\Op{W}^{(1)[0,\xi,\bar\xi]}
  &=0
  \,,
\end{align}
satisfies
\begin{align}
  \Op{q}\Op{u}_{1\Lambda}^{[0,\xi]}
  =
  -D_\Lambda\Op{v}_{1\Lambda}^{[0,\xi]}
  =
  \frac{e^{i\theta}}{2}\int_0^1\frac{dt}{t^2}\;
  \dz\left(
  D_\Lambda\xi+\xi((D_\Lambda\xi)\;\partial_\xi)
  \right)
  \OpSpHz{-1}{\tfrac{1-t}{t}\xi}
  \,.
\end{align}
One can solve the latter equation with the same homotopy and get
\begin{align}
  \label{eq:U=rhodV}
  \Op{u}_{1\Lambda}^{[0,\xi]}
  &=
  -\frac{e^{i\theta}}{2}\int_0^1\frac{dt}{t^2}\;
  \left(
  D_\Lambda\xi+\xi(D_\Lambda\xi\;\partial_\xi)
  \right)_\ag
  \Op{q}^{\ast[0,\xi]}\left(
  \dz^\ag\;
  \OpSpHz{-1}{\tfrac{1-t}{t}\xi}
  \right)
  \nonumber\\&=
  -\frac{e^{i\theta}}{2}
  \int_0^1\frac{dt}{t^2}\;
  \int_0^1\frac{ds}{s^3}\;
  (\xi\,D_\Lambda\xi)\;
  \OpSpHz{-1}{(\tfrac{1-t}{st}+\tfrac{1-s}{s})\xi}
  \nonumber\\&=
  -\frac{e^{i\theta}}{2}
  \int_0^1\frac{d\tau}{\tau^3}(1-\tau)\;
  (\xi\,D_\Lambda\xi)\;
  \OpSpHz{-1}{\tfrac{1-\tau}{\tau}\xi}
  \,.
\end{align}
Finally, using \eqref{eq:U=rhodV} in \eqref{eq:DUn}, the cohomological 1-form
\begin{align}
  \label{eq:W(PsiL)}
  \Op{W}^{(1)[0,\xi,\bar\xi]}
  &=:
  \int\frac{d^4\Lambda}{(2\pi)^2}\;
  \Psi_\Lambda\;
  \OpSpHY{1}{\Lambda}\;
  \Op{w}_{1\Lambda}^{[0,\xi,\bar\xi]}
  \end{align}
is hence submitted to the equation
\begin{align}
  D_\Lambda\Op{w}_{1\Lambda}^{[0,\xi,\bar\xi]}
  &=
  -D_\Lambda
  \Op{u}_{1\Lambda}^{[0,\xi]}
  -\text{h.c.}
  \nonumber\\&=
  \frac{e^{i\theta}}{2}
  \int_0^1\frac{d\tau}{\tau^3}(1-\tau)\;
  \left(
  (D_\Lambda\xi\,D_\Lambda\xi)
  +(D_\Lambda\xi\,\xi)(D_\Lambda\xi\,\partial_\xi)
  \right)
  \OpSpHz{-1}{\tfrac{1-\tau}{\tau}\xi}
  -\text{h.c.}
  \nonumber\\&=
  \frac{e^{i\theta}}{2}
  \int_0^1\frac{d\tau}{\tau^3}(1-\tau)\;
  (D_\Lambda\xi\,D_\Lambda\xi)
  \left(
  1+\tfrac{1}{2}(\xi\,\partial_\xi)
  \right)
  \OpSpHz{-1}{\tfrac{1-\tau}{\tau}\xi}
  -\text{h.c.}
  \nonumber\\&=
  \frac{e^{i\theta}}{2}
  (D_\Lambda\xi\,D_\Lambda\xi)
  \int_0^1d\tau\;
  \left(
  \tfrac{1-\tau}{\tau^3}-\tfrac{(1-\tau)^2}{2\tau^2}\partial_\tau
  \right)
  \OpSpHz{-1}{\tfrac{1-\tau}{\tau}\xi}
  -\text{h.c.}
  \nonumber\\&=
  \frac{e^{i\theta}}{2}
  (D_\Lambda\xi\,D_\Lambda\xi)
  \lim_{\tau\to0}
  \left(
  \tfrac{(1-\tau)^2}{2\tau^2}
  \;
  \OpSpHz{-1}{\tfrac{1-\tau}{\tau}\xi}
  \right)
  -\text{h.c.}
  \,,
\end{align}
where the third line comes from a two-dimensional Fierz identity.
Due to the consistency of the Vasiliev equations, the right hand side has to be independent of $\Opz$,
which is indeed the case because of the sequence \eqref{eq:Uz(-1,K) deltaseq}.
Hence the equation on \eqref{eq:W(PsiL)} becomes
\begin{align}\label{eq:gluing}
  \Op{D}\Op{W}_{1}^{[0,\xi,\bar\xi]}
  &=
  \frac{i\pi e^{i\theta}}{2}
  \int\frac{d^4\Lambda}{(2\pi)^2}\;
  \delta^2(\xi)\;
  (D_\Lambda\xi\,D_\Lambda\xi)\;
  \Psi_\Lambda\;
  \OpSpHY{1}{\Lambda}
  -\text{h.c.}
\end{align}
Choosing
\begin{align}
  \label{eq:xi=bl}
  \xi(\Lambda)=b\lambda=:-\lambda b^t\ ,
\end{align}
in terms of an invertible $2\times2$ matrix $b$, one has
\begin{align}
  D_\Lambda\xi
  =
  -\lambda (db^t+\omega b^t)
  -\bar\lambda e^t b^t
  =
  (db-b\omega)\lambda-be\bar\lambda
  \,,&&
  L^{-1}\,dL
  =
  \begin{pmatrix}
    \omega&e\\e^t&\bar\omega
  \end{pmatrix}
  \,,
\end{align}
and the equation becomes, in Weyl order,
\begin{align}
  \label{eq:COMST W VE}
  \MoSym0{\Op{D}\Op{W}_{1}^{[0,\xi,\bar\xi]}}
  &=
  -\frac{ie^{i\theta}}{4\det(b)}
  \int\frac{d^2\bar\lambda}{2\pi}\;
  (\bar\lambda\,e^t\,b^t\,b\,e\,\bar\lambda)\;
  \Psi_{0,\bar\lambda}\;
  e^{i\bar\lambda\yb}
  -\text{h.c.}
  \nonumber\\&=
  \frac{ie^{i\theta}}{4\det(b)}
  (\partial_\yb\,e^t\,b^t\,b\,e\,\partial_\yb)\;
  \MoSym{0}{\Op\Phi^{(1)}}\vert_{y=0}
  -\text{h.c.}
  \nonumber\\&=
  \frac{ie^{i\theta}}{4}
  (\partial_\yb\,e^t\,e\,\partial_\yb)\;
  \MoSym{0}{\Op\Phi^{(1)}}\vert_{y=0}
  -\text{h.c.} \ .
\end{align}
This is indeed equivalent to Eqs. (\ref{eq:DW=S},\,\ref{eq:S=eeC}), i.e., we retrieve the COMST. 
Let us emphasise the fact that,
while we did interpret the final equation in the Weyl-ordered basis for the $\Op Y$ Weyl algebra \eqref{eq:[Y,Y]} in order to present it in the standard basis in which Fronsdal fields are extracted,
no choice of ordering for the $\Opz$ variables was required to achieve the gluing \eqref{eq:gluing} of the Weyl zero-form local datum \eqref{eq:FT Psi} to the gauge field module. Only the group algebra properties of the elements \eqref{eq:OpSpHz} were used.
In particular, the Heisenberg group expansion \eqref{eq:FT Psi} of the initial datum allows to keep their contribution formally factorised,
as would be done for example by presenting the result in total Weyl ordering on $\Op Y$ and $\Op Z$.

\subsection{Normal-ordered homotopies}
\label{sec:No(YZ)}
While the procedure of Sec. \ref{sec:rotV g} properly gives rise to Fronsdal fields as linearised configurations,
there seems to be no natural reason to contract sources that are Fourier-expanded in $\OpY$ (as induced from the expansion \eqref{eq:FT Psi} of the adjoint free field)
along vectors of the form $(Z^\au+\Lambda^\bu (B^t)_\bu\,^\au)\vec{\partial}_{Z^\au}$ as
\begin{align}
  \label{eq:rhoBL}
  &
  \Op{q}^{\ast[0,B\Lambda]}\left(j(\dZ)\;
  \OpSpHY{1}{\Lambda}\;\OpSpHZ{U}{\mathcal{M}}\right)
  =
  \Op{q}^{\ast[0,-\Lambda B^t]}\left(j(\dZ)\;
  \OpSpHY{1}{\Lambda}\;\OpSpHZ{U}{\mathcal{M}}\right)
  \nonumber\\&=
  -
  \int_0^1\frac{dt}{t^4}
  \left(
  \tfrac{i}{2}\partial_{\mathcal{M}}\left(1+U\right)
  +\tfrac12{\mathcal{M}}\left(1-U\right)
  -\Lambda B^t
  \right)^\au\frac{\partial}{\partial(\dZ)^\au}j(t\dZ)
  \nonumber\\&\quad\times
  \frac{\sqdet{1-U_t}}{\sqdet{1-U}}
  \exp\left(
  -\tfrac{i}{2}{\mathcal{M}}\tfrac{1+U}{1-U}{\mathcal{M}}
  +\tfrac{i}{2}{\mathcal{M}}_t\tfrac{1+U_t}{1-U_t}{\mathcal{M}}_t
  \right)
  \OpSpHY{1}{\Lambda}\;\OpSpHZ{U_t}{{\mathcal{M}}_t}
  \,,
\end{align}
with
\begin{align}
  U_t
  :&=
  \frac%
  {(1+U)-t^2(1-U)}%
  {(1+U)+t^2(1-U)}
  \,,&
  {\mathcal{M}}_t
  :&=
  \left(
  \frac{1}{t}{\mathcal{M}}\frac{2U}{1-U}
  +\frac{1-t}{t}\Lambda B^t
  \right)
  \frac{1-U_t}{2U_t}
  \,,
\end{align}
of which $\Op{q}^{\ast[0,b\lambda]}$ is a particular case.
In fact, those shifts have an alternative interpretation in a symbol calculus that involves both kinds of twistor variables together:
\begin{align}
  \OpX_I
  &\to
  \begin{pmatrix}
    \OpY_\au\\\OpZ_\au
  \end{pmatrix}
  \,,&
  K^I
  &\to
  \begin{pmatrix}
    \Lambda^\au&\mathcal{M}^\au
  \end{pmatrix}
  \,,&
  \Theta_{IJ}&\to
  \begin{pmatrix}
    \epsilon_{\au\bu}&0\\0&-\epsilon_{\au\bu}
  \end{pmatrix}
  \,,\\
  \OpX^I
  &\to
  \begin{pmatrix}
    \OpY^\au&-\OpZ^\au
  \end{pmatrix}
  \,,&
  K_I
  &\to
  \begin{pmatrix}
    \Lambda_\au\\-\mathcal{M}_\au
  \end{pmatrix}
  \,.
\end{align}
Indeed, one may use the map \eqref{eq:MoSpH} to represent Eq. \eqref{eq:rhoBL} in an ordering
\begin{align}
  \label{eq:B-order}
  M_{IJ}
  \to
  \begin{pmatrix}
    0&(B^t)_{\au\bu}\\B_{\au\bu}&0
  \end{pmatrix}
  =
  \begin{pmatrix}
    0&B_{\bu\au}\\B_{\au\bu}&0
  \end{pmatrix}
  \,,
\end{align}
where Eq. \eqref{eq:MoSpH} allows to write the source as
\begin{align}
  &
  \MoSymArg{M}{%
  \OpSpHY{1}{\Lambda}\;
  \OpSpHZ{U}{\mathcal{M}}
  }{Y,\,Z}
  =
  \MoSymArg{0}{%
  \OpSpHY{1}{\Lambda}\;
  \OpSpHZ{U}{\mathcal{M}}
  }{Y,\,Z-B\Lambda}
  \,;
\end{align}
and notice that contracting as \eqref{eq:rhoMM0} along the vector $Z^{\au}\vec{\partial}_{Z^\au}$ in this $M$-ordering
corresponds to contracting along $(Z^\au+\Lambda^\bu (B^t)_\bu\,^\au)\vec{\partial}_{Z^\au}$ in Weyl order.
In other words, the shift used in Eqs. (\ref{eq:V1shift},\,\ref{eq:U=rhodV},\,\ref{eq:xi=bl}) corresponds to a standard, unshifted resolution \eqref{eq:rhoMM0} in the particular case of the ordering \eqref{eq:B-order} where the matrix $B$ is block diagonal.

Let us turn our attention to the more specific cases where the matrix $B$ is scalar:
\begin{align}
  B_{\au\bu}
  &=
  -{\check\beta}\,\epsilon_{\au\bu}
  \,,&
  \check\beta\in\R
  \,.
\end{align}
In this case the star product reads
\begin{align}
  \label{eq:star eps}
  &
  \left[\Op{f}\;\Op{g}\right]_{{\check\beta}}
  (Y,Z)
  \\\nonumber
  &=
  \int\frac{d^4Ud^4\widetilde{U}d^4Vd^4\widetilde{V}}{(2\pi)^8}
  \frac{e^{\tfrac{i}{1-{\check\beta}^2}\left(
  VU-\widetilde{V}\widetilde{U}+{\check\beta}\widetilde{V}U-{\check\beta} V\widetilde{U}
  \right)}}{(1-{\check\beta}^2)^4}
  \big[\Op{f}\big]_{M_{\check\beta}}(Y+U,\,Z+\widetilde{U})
  \big[\Op{g}\big]_{M_{\check\beta}}(Y+V,\,Z+\widetilde{V})
  \,,
\end{align}
and the (anti-)holomorphic Kleinians
\begin{align}
  \label{eq:kappa eps}
  \left[\Op{\kappa}\right]_{{\check\beta}}
  &=
  \frac{1}{{\check\beta}^2}
  \exp\left(
  \tfrac{i}{{\check\beta}}yz
  \right)
  \,,&
  \left[\Op{\bar\kappa}\right]_{{\check\beta}}
  &=
  \frac{1}{{\check\beta}^2}
  \exp\left(
  -\tfrac{i}{{\check\beta}}\yb\zb
  \right)
  \,.
\end{align}
Usually, the Vasiliev equations are presented (and the homotopies performed) in normal order,
corresponding to the limit ${\check\beta}\to1$,
or equivalently \cite{Didenko:2019xzz} to $\beta:=(1-{\check\beta})\to0$,
where the star product \eqref{eq:star eps} becomes
\begin{align}
  \label{eq:starM1}
  \MoSym1{\Op{f}\Op{g}}
  (Y,Z)
  =
  \int\frac{d^4Ud^4V}{(2\pi)^4}
  \MoSym1{\Op{f}}(Y+U,\,Z+U)
  \MoSym1{\Op{g}}(Y+V,\,Z-V)
  \exp\left(iVU\right)
  \,.
\end{align}
An advantage of working specifically in that ordering is that the linearised version of the r.h.s. of Eq. \eqref{eq:qVn} is guaranteed by the lemma
\begin{align}
  \MoSym1{\Op\Phi^{(1)}\;\OpK}
  =
  \left.
  \MoSym0{\Op\Phi^{(1)}}
  \right\vert_{y\to-z}
  \,\exp\left(iyz\right)
  \,,
\end{align}
to be analytic in $Y$ and $Z$.
This propagates to the $Z$-dependent part of all linearized master fields \cite{properties}.
This property however still requires extra assumptions in order to be extended beyond the result \eqref{eq:COMST W VE}
(which, as can be seen from Sec. \ref{sec:rotV g}, does not require resorting to it), as it:
\begin{enumerate}[label=\roman*)]
  \item says nothing about the analyticity in $Y$ of the integration constant $W(Y)$, which has to be separately assumed;
  \item is not preserved by star products, hence irrelevant at higher order in perturbation theory.
\end{enumerate}
In Sec. \ref{sec:rotV g}, we have shown not only that the ordering in which the fields are presented is irrelevant,
but also that this result is reached with a wider family of contractions that includes the one in this normal order.
The gauge obtained with the normal ordered contractions $\xi(\Lambda)=-\Lambda$ is usually called the \emph{Vasiliev gauge},
and hence we refer to the solution obtained through $\Op{q}^{\ast[0,b\lambda]}$ as being in \emph{rotated Vasiliev gauge}.
Notice that since the same shifted homotopy is used to solve for $\Op{V}^{(1)}$ and $\Op{U}^{(1)}$,
this is not the same as the \emph{relaxed Vasiliev gauge condition} proposed in \cite{COMST} that allows for the $\Op{V}$ connection to carry an extra piece
which do not contribute to the field equations
if $\Op{U}^{(1)}$ is still built using $\Op{q}^{\ast[0,-\lambda]}$.
In fact, one can relax the rotated Vasiliev gauge condition by adding to $V^{(1)[0,b\lambda]}$ a piece that is $O(Z^2)$ in $(-B)$-ordering.

The linear analysis of Sec. \ref{sec:rotV g} admits three degenerate limiting cases.
The first one is the Weyl ordering ($\check\beta=0$),
where the star product \eqref{eq:star eps} is well defined but where the Kleinians \eqref{eq:kappa eps} are delta functions.
As mentioned in Sec. \ref{sec:hol g}, this implies that the linearized connection \eqref{eq:V1hol} is ill defined in that limit.
In the other two cases, $\check\beta\to\pm\infty$,
the algebra itself is not defined,
as the Kleinians \eqref{eq:kappa eps} and also all star products \eqref{eq:star eps} vanish.
One can however imagine to work with finite values of $\check\beta$ and taking the limit to one of those special values once all symol calculus operations have been performed.
In fact, the $\check\beta\to+\infty$ limit (combined with field-dependent shifts) was useful in figuring out a proposal for a generalized notion of locality of interaction vertices, referred to as spin-locality \cite{Didenko:2019xzz,Gelfond:2019tac,Didenko:2020bxd,Gelfond:2021two}.
It would be interesting to understand what effect the limit to Weyl order may have on interactions.

\subsection{Linearised gauge function}
\label{sec:lin gf}
By virtue of Eq. \eqref{eq:CI LHC}, any solution to the linearized field equations can be written in the form
\begin{align}
  \Op{A}^{(1)}
  &=
  \Op{L}^{-1}\;
  \Op{\mathcal{V}}^{(1)[0,\zeta,+]}\left[\Op{C}^{\prime(1)}\right]\;
  \Op{L}
  -\Op{H}^{(1)[0,\zeta,+]}\Op{L}^{-1}\;\Op\Delta\Op{L}
  +\Op{L}^{-1}\;\Op\Delta(\Op{L}\;\Op{H}^{(1)[0,\zeta,+]})
  -\text{h.c.}
  \nonumber\\&=
  \Op{V}^{(1)[0,\zeta,+]}_z+\Op\Delta\Op{H}^{(1)[0,\zeta,+]}
  +\left[\Op\Omega,\,\Op{H}^{(1)[0,\zeta,+]}\right]
  -\text{h.c.}
  \,,
\end{align}
in terms of the bosonic axial factorised gauge solution \eqref{eq:[0,zeta,+]}.
The purpose of this Section is to find the form of a gauge function $\Op{H}^{(1)[0,\zeta,+]}$ 
that allows to identify inside $\Op{A}^{(1)}$ a connection $\Op{W}^{(1)}$ that carries unfolded Fronsdal fields in the sense that it
\begin{enumerate}[label=(\roman*)]
  \item formally satisfies the propagating equation (\ref{eq:DC=0},\,\ref{eq:DW=S});
  \item is analytic in $Y$ in Weyl order, so as to allow for an expansion of the form \eqref{eq:W Taylor}.
\end{enumerate}
Given an invertible matrix $b$ (and its complex conjugate $\bar{b}$), we know from the procedure of Sec. \ref{sec:rotV g} that if the auxiliary connection $V^{(1)}$ is in the rotated Vasiliev gauge characterized by $\xi=b\lambda$ (and $\bar\xi=\bar{b}\bar\lambda$),
then $\Op{W}^{(1)[0,\xi,\bar\xi]}=\Op{\mathcal{P}}^{[0,\xi]}\Op{A}^{(1)}-\text{h.c.}$ will satisfy the first of those requirement.
We hence require that the auxiliary part of $\Op{A}^{(1)}$ be in such a gauge,
or in other words that it can be decomposed as
\begin{align}
  \Op{A}^{(1)}
  =
  \Op{V}^{(1)[0,\xi]}_z
  +\Op{\bar{V}}^{(1)[0,\bar\xi]}_\zb
  +\Op{U}^{(1)}
  \,,
\end{align}
where in particular
\begin{align}
  \Op{V}^{(1)[0,\xi]}_z
  =
  \Op{V}^{(1)[0,\zeta,+]}_z+\Op{q}\Op{H}^{(1)[0,\zeta,+]}
  \,.
\end{align}
The gauge function $\Op{H}^{(1)[0,\zeta,+]}$ can hence be determined up to a part $\Op{H}^{(1)[0,\zeta,+]}_0$ that commutes with $\OpZ$ by using Eq.  \eqref{eq:h,c(q*)} as
\begin{align}
  \Op{H}^{(1)[0,\zeta,+]}
  &=
  \int\frac{d^4\Lambda}{2(2\pi)^2}\;
  \Psi_\Lambda\;
  \OpSpHY{1}{\Lambda}\;\left(
  \Op{h}_{1\Lambda}^{[\xi,\zeta]}+\Op{h}_{1\Lambda}^{[\xi,-\zeta]}\right)
  +
  \Op{H}^{(1)[0,\zeta,+]}_0
  \,,\\
  \label{eq:h1[xi,zeta]}
  \Op{h}_{1\Lambda}^{[\xi,\zeta]}
  :&=
  \Op{q}^{\ast[0,\zeta]}\Op{v}_{1}^{[0,\xi]}
  =
  \frac{e^{i\theta}}{2} (\xi\zeta)
  \int_0^1\frac{ds}{s^3}\int_0^1\frac{dt}{t^2}\,
  \OpSpHz{-1}{\tfrac{1-t}{st}\xi+\tfrac{1-s}{s}\zeta}
  \nonumber\\&=
  \frac{e^{i\theta}}{2} (\xi\zeta)
  \int_0^{+\infty}dr
  \int_0^{+\infty}dv\,(1+v)\;
  \OpSpHz{-1}{(1+v)r\xi+v\zeta}
  \,,\\
  \Op{H}^{(1)[0,\zeta,+]}_0
  &=
  \int\frac{d^4\Lambda}{2(2\pi)^2}\;
  \Psi_\Lambda\;
  \OpSpHY{1}{\Lambda}\;\left(
  \Op{h}_{01\Lambda}^{[\xi,\zeta]}+\Op{h}_{01\Lambda}^{[\xi,-\zeta]}\right)
\end{align}
where $h_{01\Lambda}^{[\xi,\pm\zeta]}$ are scalar functions of spacetime coordinates.
The first term contributes to the connection $\Op{U}^{(1)}$ as \eqref{eq:CI Un},
\begin{align}
  &
  D_\Lambda\Op{h}_{1\Lambda}^{[\xi,\zeta]}
  \nonumber\\&=
  \frac{e^{i\theta}}{2}
  \int_0^{+\infty}dr
  \int_0^{+\infty}dv\,(1+v)
  \left[
  (D_\Lambda\xi\,\zeta)
  +
  (\xi\zeta)(D_\Lambda\xi\,\partial_\xi)
  \right]
  \OpSpHz{-1}{(1+v)r\xi+v\zeta}
  \nonumber\\&=
  \frac{e^{i\theta}}{2}
  \int_0^{+\infty}dr
  \int_0^{+\infty}dv\,(1+v)
  \left[
  (D_\Lambda\xi\,\zeta)
  +
  (D_\Lambda\xi\,\zeta)(\xi\partial_\xi)
  -
  (D_\Lambda\xi\,\xi)(\zeta\partial_\xi)
  \right]
  \OpSpHz{-1}{(1+v)r\xi+v\zeta}
  \nonumber\\&=
  \frac{e^{i\theta}}{2}
  \int_0^{+\infty}dr
  \int_0^{+\infty}dv\,(1+v)
  \left[
  (D_\Lambda\xi\,\zeta)(1+r\partial_r)
  -
  (D_\Lambda\xi\,\xi)\left(
  r(1+v)\partial_v-r^2\partial_r\right)
  \right]
  \nonumber\\&\qquad\times
  \OpSpHz{-1}{(1+v)r\xi+v\zeta}
  \,,
\end{align}
where a Schouten identity was used in the second line.
The corresponding term in the spacetime gauge field is extracted in rotated Vasiliev gauge as \eqref{eq:c(q*)} using the projection \eqref{eq:P[0,zeta]U(-1)}
\begin{align}
  \label{eq:PDh}
  &
  w^{(1)[\xi,\zeta]}_\Lambda
  =
  \Op{\mathcal{P}}^{[0,\xi]}
  D_\Lambda\Op{h}_{1\Lambda}^{[\xi,\zeta]}
  \nonumber\\&=
  i\pi e^{i\theta}
  \int_0^{+\infty}dr
  \int_0^{+\infty}dv\,(1+v)
  \left[
  (D_\Lambda\xi\,\zeta)(1+r\partial_r)
  -
  (D_\Lambda\xi\,\xi)\left(
  r(1+v)\partial_v-r^2\partial_r\right)
  \right]
  \nonumber\\&\qquad\times
  \delta^2\left((1+r+rv)\xi+v\zeta\right)
  \nonumber\\&=
  i\pi e^{i\theta}
  (D_\Lambda\xi\,\zeta)
  \int_0^{+\infty}dv\,(1+v)
  \left[r\delta^2\left((1+r+rv)\xi+v\zeta\right)\right]%
  _{r=0}^{+\infty}
  \nonumber\\&\quad
  -i\pi e^{i\theta}
  (D_\Lambda\xi\,\xi)
  \int_0^{+\infty}dr\,r
  \left[(1+v)^2\delta^2\left((1+r+rv)\xi+v\zeta\right)\right]%
  _{v=0}^{+\infty}
  \nonumber\\&\quad
  +i\pi e^{i\theta}
  (D_\Lambda\xi\,\xi)
  \int_0^{+\infty}dv\,(1+v)
  \left[r^2\delta^2\left((1+r+rv)\xi+v\zeta\right)\right]%
  _{r=0}^{+\infty}
  \nonumber\\&=
  -i\pi e^{i\theta}
  (D_\Lambda\xi\,\xi)
  \int_0^{+\infty}dr\,r
  \delta^2\left(r\xi+\zeta\right)
  +i\pi e^{i\theta}
  (D_\Lambda\xi\,\xi)
  \int_0^{+\infty}dr\,r
  \delta^2\left((1+r)\xi\right)
  \nonumber\\&\quad
  +i\pi e^{i\theta}
  (D_\Lambda\xi\,\xi)
  \int_0^{+\infty}dv\,(1+v)
  \delta^2\left((1+v)\xi\right)
  \nonumber\\&=
  i\pi e^{i\theta}
  (D_\Lambda\xi\,\zeta)
  \int_0^{+\infty}dw\,
  \delta^2\left(\xi+w\zeta\right)
  \,.
\end{align}
The last two terms were eliminated despite their logarithmically divergent prefactor.
In fact, it is worth to note that this apparent indetermination is the same that appears when applying $\Op{\mathcal{P}}^{[0,\xi]}$ on the expression \eqref{eq:U=rhodV}, where it can be resolved as
\begin{align}
  \lim_{z\to\xi}(D_\Lambda\xi\,\xi)
  \int_0^{+\infty}dr\,r
  \delta^2\left(z+r\xi\right)
  &=
  -\lim_{z\to\xi}(D_\Lambda\xi\,z)
  \int_0^{+\infty}dr
  \delta^2\left(z+r\xi\right)
  \nonumber\\&=
  -(D_\Lambda\xi\,\xi)
  \delta^2\left(\xi\right)
  \int_0^{+\infty}\frac{dr}{(1+r)^2}
  =
  0
  \,,
\end{align}
as assumed in the procedure of Sec. \ref{sec:VE pert}.
When combined back with the Fourier-transformed initial datum \eqref{eq:FT Psi},
the expression \eqref{eq:PDh} contributes to the Weyl-ordered generating function for gauge connections as
\begin{align}
  \label{eq:W^(xi,zeta) Fourier}
  &
  \MoSym0{\Op{W}^{(1)[\xi,\zeta]}}
  =
  i e^{i\theta}
  \int\frac{d^4\Lambda}{8\pi}\;
  \Psi_\Lambda\;
  \MoSpHY0{1}{\Lambda}
  (D_\Lambda\xi\,\zeta)
  \int_0^{+\infty}dw\;\delta^2(\xi+w\zeta)
  \\&=
  \label{eq:W^(xi,zeta) WO}
  -\frac{i e^{i\theta}}{4\det b}
  \int_0^{+\infty}dw\left\{
  w\,\zeta (b^t)^{-1}(db^t+\omega b^t)\zeta
  +i\zeta be\partial_{\yb}
  \right\}
  \MoSymArg0{\Op\Phi^{(1)}}{-wb^{-1}\zeta,\yb}
  e^{iw\zeta(b^t)^{-1}y}
  \,.
\end{align}
This expression may diverge if the Weyl zero-form does not fall off appropriately.
In this case, given that its curvature \eqref{eq:COMST W VE} is regular in $Y$,
the irregular part may be eliminated by a suitable choice of $h_{01\Lambda}^{[\xi,\zeta]}$.
If a regular potential in known in the form \eqref{eq:U(PsiL)},
where $\Op{D}$ acts as a de Rham differential for the rotated momentum $\xi$,
the relevant gauge transformation may be found by applying the homotopy trick \eqref{eq:rhoMM0} in the variable $\xi$.
As will be shown in Sec. \ref{sec:particle}, it is not always guaranteed to work that way.
The improvement that this Section brings with respect to its counterpart in \cite{COMST} is that it does not require to introduce a spacetime vector along which to contract.

\subsection{Sp(4) gauge transformations}
\label{sec:Lorentz}

As mentioned in Sec. \ref{sec:VE}, identifying the gauge transformations (\ref{eq:dgt C},\,\ref{eq:dgt W}) inside the larger group \eqref{eq:VE fgt} is a non-trivial task to be adressed when making the higher order perturbative scheme more precise.
The most intuitive way to do so would be to define a gauge-invariant resolution scheme for the perturbatively defined equations (\ref{eq:qPhin}-\ref{eq:DUn}),
and a perturbative deformation of the parameters that would preserve said scheme.
This is similar in spirit\footnote{%
The main differences is that here the noncommutative space is auxiliary,
and hence the gauge equivalence need not be preserved by the map outside the physical slice.
} to the Seiberg-Witten map \cite{Seiberg:1999vs} used in noncommutative field theory.
An alternative approach would be to keep the definition of the gauge parameters as the $\OpZ$ independent ones,
and to define a covariantization of the scheme with respect to those transformations.
The advantage of the former approach is that the observables \eqref{eq:def OWL} are automatically physically gauge invariant.
The advantage of the latter is that setting it up would also enable one to define perturbation theory around arbitrary backgrounds that are flat in the sense of Eq. \eqref{eq:VEx}.
At the level of the first order in the perturbative expansion, to which this paper is dedicated,
this question can be asked only for $Sp(4,\C)$ transformations.
We will discuss it in both approaches introduced above.

\paragraph{Deformation approach.}
Let us restrict our study of this approach to the Lorentz subalgebra $\msl(2,\mathbb{C})\oplus\msl(2,\mathbb{C})$ under which the COMST \eqref{eq:COMST W VE} is invariant.

It is useful to examine another definition of the Lorentz transformations as the ones
\begin{align}
  \label{eq:fullLorentz}
  \delta_{\lambda,\,\bar\lambda} \Op{F}
  :=&
  \lambda^{\ag\bg}\left(
  \frac{1}{4i}\left[\Opy_\ag\,\Opy_\bg,\,\Op{F}\right]
  -\frac{1}{4i}\left[\Opz_\ag\,\Opz_\bg,\,\Op{F}\right]
  +\frac12\left(\dz_\ag\frac{\partial}{\partial \dz^\bg}+\dz_\bg\frac{\partial}{\partial \dz^\ag}\right)\Op{F}
  \right)
  \nonumber\\&+
  {\bar\lambda}^{\ad\bd}\left(
  \frac{1}{4i}\left[\Opyb_\ad\,\Opyb_\bd,\,\Op{F}\right]
  -\frac{1}{4i}\left[\Opzb_\ad\,\Opzb_\bd,\,\Op{F}\right]
  +\frac12\left(\dzb_\ad\frac{\partial}{\partial \dzb^\bd}+\dzb_\bd\frac{\partial}{\partial \dzb^\ad}\right)\Op{F}
  \right)
  \,,
\end{align}
that act on all the spinor indices appearing in Vasiliev's equations (\ref{eq:[S,S]},\,\ref{eq:[S,Phi]},\,\ref{eq:dS}).
The equations can be written \cite{properties,2011,Sezgin:2011hq} in a manifestly covariant form under these transformations modulo the introduction of the deformed generators
\begin{align}
  \label{eq:defLorentz}
  {\Op{g}}_{\lambda,\,\bar\lambda}
  :=&
  \lambda^{\ag\bg}\left(
  \frac{1}{4i}\Opy_\ag\;\Opy_\bg
  -\frac{1}{4i}\Opz_\ag\,\Opz_\bg
  +\frac{1}{4i}\Op{S}_\ag\,\Op{S}_\bg
  \right)
  +\text{h.c.}
  \,,
\end{align}
that, as can be seen from the vacuum solution \eqref{eq:VE BG},
are indeed a dynamical deformation of $\OpY$ bilinears.

Whether these transformations corresponds to Lorentz transformations of the cohomological fields \eqref{eq:C(Phi) W(U)} is a non-trivial question in general.
It is true in the original perturbative scheme \cite{more,review99,Sezgin:2002ru} where the normal-ordered contraction $\Op{q}^{\ast[0,-\Lambda]}$
is used to solve all equations of the form \eqref{eq:qf=g},
and hence where the physical slice is defined as the $Z=0$ surface in the normal order \eqref{eq:starM1}.
It is not known at the moment whether this is still the case when other homotopies are used.

\paragraph{Covariantization approach.}
This approach is simple to address in the $Sp(4,\C)$ case,
where the consequence of the rotation formula \eqref{eq:FT PsiL} for the generic solution \eqref{eq:sol qf=g} to Eq. \eqref{eq:qf=g} is
\begin{align}
  \label{eq:Lrot sol qf=g}
  &
  \int d^4\Lambda\;
  \OpSpHY{L^{-1}}{0}\left(%
  \Op{q}^{\ast[0,B\Lambda]}\Op{g} + \Op{q}\Op{h} + \Op{c}
  \right)\OpSpHY{1}{\Lambda}\OpSpHY{L}{0}
  \nonumber\\&=
  \int d^4\Lambda\left(%
  \Op{q}^{\ast[0,BL\Lambda]}\Op{g} + \Op{q}\Op{h}^{(L)} + \Op{c}^{(L)}
  \right)\OpSpHY{1}{\Lambda}
  \,,
\end{align}
where $L\in Sp(4;\mathbb{C})$.
As this relation respects the decomposition as a sum of the particular solution, the gauge function and the cohomological element,
it is clear that the transformation of the cohomological part stays undeformed if the matrix $B$ defining the shift undergoes a transformation $B\to BL$.
In particular, decomposing the master field $\Op{U}$ as \eqref{eq:U=U+U+W}, this means the appropriate transformation of $B$ draws the correspondence between a transformation \eqref{eq:VE fgt} with $\Op{G}=\OpSpHY{L}{0}$
and a background gauge transformation \eqref{eq:BG gt} of the cohomological part $\Op{W}^{0,\xi,\bar\xi}$.
Within this approach, the resolution scheme
should be encoded in an adjoint master field
\begin{align}
  \Op{B}^\au
  &=
  B^{\au\bu}\OpY_\bu
  \,,
\end{align}
that would transform under background gauge transformations as
\begin{align}
  \Op{B}^\au\to\OpSpHY{L}{0}^{-1}\;\Op{B}^\au\;\OpSpHY{L}{0}
  =
  B^{\au\bu}L_\bu\,^\cu\OpY_\cu
  \,,
\end{align}
via the prescription that all equations of the form \eqref{eq:qf=g} be resolved with the homotopy contraction $\Op{q}^{\ast[0,\xi(-\tfrac12\text{ad}(\Op{B}))]}$.
General higher-spin backgrounds would then be obtained through general adjoint transformations of the field $\Op{B}$.
Of course, if the resolution scheme entails using different contractions of the above form,
such a $\Op{B}$ should be introduced for each of them.

Independently of the question of non-auxiliary gauge transformations addressed in this section,
the formula \eqref{eq:Lrot sol qf=g} can be used in the context of the vacuum gauge dressing \eqref{eq:AdS(Y)},
where it allows to map resolution operators between the space-time dependent and independent gauges.
Concretely, if one knows what are the resolution operators corresponding to certain boundary conditions,
one can directly apply the appropriately rotated operators to the spacetime-independent local data \eqref{eq:Psi'=C'k} to obtain a solution respecting those boundary conditions.

\section{Massless particle and black-hole-like local data}
\label{sec:PTBH}

In this Section, we apply the method of Sec. \ref{sec:solve Z} for computing the generating function $W$ for unfolded Fronsdal fields to two especially relevant types of local data: AdS massless particles and higher spin black hole states \cite{2011,2017,COMST,Iazeolla:2020jee}, respectively encoded into singleton state projectors and dittos twisted by $\Op{\k}_y$.
First, it will be shown that the singleton projectors are (rescaled) limits of $Sp(4,\mathbb{C})$ elements, in the sense described in Sec. \ref{sec:Sp}. 
Then, starting from the axial holomorphic gauge, where $W$ vanishes, we use Eq. \eqref{eq:W^(xi,zeta) WO} to compute $W$ in rotated Vasiliev gauge for black holes, and show that, as expected, it is analytic in $Y$ in Weyl order.
After that, we show that when applied to the case of a massless scalar particle, this method correctly produces a trivial $W$; in particular, we explicitly construct the gauge function $h^{[\xi,\zeta]}_{01\Lambda}$ that removes the singular yet cohomologically trivial part inherited from the axial holomorphic gauge.
Our treatment here of the latter case constitutes an improvement of the result of \cite{COMST} in that it does not require the introduction of a spacetime vector field along which to perform the homotopy contraction in \eqref{eq:rhoMM0}. 

We leave for future work the construction of $W$ for spin $s>0$ massless particle modes using the above method.
To this end, the remaining non-trivial step would be the construction of the residual gauge function $h^{[\xi,\zeta]}_{01\Lambda}$.

\subsection{Projectors as limits of group sequences}

Massless particle states can be encoded in projectors on singleton states, realized as Gaussian elements in $Y$ that have definite eigenvalue under the action of the energy generator $\Op{E}$ and one spin, e.g. $\Op{M}_{12}$ \cite{fibre,2017,COMST}. The simplest particle states are the ones corresponding to rotationally-invariant scalar modes, encoded into purely $E$-dependent projectors. Spherically symmetric black hole states can be obtained by twisting such projectors by means of a product with the Klein operator \cite{2011,2017,COMST}.

In order to show how the latter correspond to limits of group sequences, let us begin from recalling the matrix representation \eqref{eq:M(Gamma)} of the AdS$_4$ energy operator:
\begin{align}
  E:=M_{0'0}=\frac{i}{2}\Gamma_{0'0}\,.
\end{align}
It generates a one-parameter subgroup of $Sp(4,\C)$,
\begin{align}
  \label{eq:exp(E) matrix}
  \exp\left(-2\theta E\right)
  =
  \exp\left(-i\theta\Gamma_{0'0}\right)
  =
  \begin{pmatrix}
    \cosh(\theta)&i\sinh(\theta)\sigma_0\\
    i\sinh(\theta)\bar\sigma_0&\cosh(\theta)
  \end{pmatrix}
  \,.
\end{align}
The associated $M$-ordering symbols read (see \eqref{eq:stargauss})
\begin{align}
  \label{eq:proj in Mo}
  \MoSpHY{M}{e^{-2\theta E}}{0}
  =
  \frac{1}{\sqdet{\cosh\tfrac{\theta}{2}-iM\Gamma_{0'0}\sinh\tfrac{\theta}{2}}}
  \exp\left(
  \tfrac{i}{2}Y\left(
  i(\tanh\tfrac{\theta}{2})^{-1}\Gamma_{0'0}-M
  \right)^{-1}Y\right)
  \,,
\end{align}
and in particular, in Weyl order,
\begin{align}
  \MoSpHY{0}{e^{-2\theta E}}{0}
  =
  \frac{1}{(\cosh\tfrac{\theta}{2})^2}
  \exp\left(
  -\tfrac{1}{2}\tanh\tfrac{\theta}{2}Y\Gamma_{0'0} Y\right)
  \,.
\end{align}
Clearly, multiplying two matrices of the form \eqref{eq:exp(E) matrix} is equivalent to adding their corresponding angles, viz.
\begin{align}
  \OpSpH{e^{-2\theta_1E}}{0}\OpSpH{e^{-2\theta_2E}}{0}
  =
  \OpSpH{e^{-2(\theta_1+\theta_2)E}}{0}
  \,.
\end{align}

It is interesting to note that the limiting cases $\theta=\pm\infty$ of the matrices \eqref{eq:exp(E) matrix} behave as projectors,
and that their mutual product is ill-defined.
This conclusion is less straightforward in the context of the oscillator realisation,
where their symbols \eqref{eq:proj in Mo} in general Gaussian orderings vanish\footnote{%
This is also the case in the critical order \eqref{eq:Sp crit order}
where the diverging prefactor contributes to the delta function
without compensating the vanishing one.}.
One can however define the finite algebra elements
\begin{align}
  \label{eq:P=e^oo}
  \Op{\mathcal{P}}_{\pm1}
  :=
  \lim_{\theta\to\pm\infty}
  e^{\pm\theta}
  \OpSpHY{e^{-2\theta E}}{0}
  \,.
\end{align}
The Weyl ordering symbol of these objects is
\begin{align}
  \MoSym{0}{\Op{\mathcal{P}}_{\pm1}}
  =
  \lim_{\theta\to\pm\infty}
  \frac{e^{\pm\theta}}{(\cosh\tfrac{\theta}{2})^2}
  \exp\left(\mp\tfrac{1}{2}\tanh\tfrac{\theta}{2}\, Y\Gamma_{0'0}Y\right)
  =
  4\exp\left(\mp4\MoSymOp{0}{E}\right)
  \,,
\end{align}
which shows that they are exactly the projectors studied in \cite{fibre,2011,2017,COMST}.
The normalisation that ensure their finiteness is in fact responsible for the eigenvalue of $E$ that they carry:
\begin{align}
  \Op{\mathcal{P}}_{\epsilon}\;
  \OpSpHY{e^{-2\theta E}}{0}
  =
  e^{-\epsilon\theta}
  \Op{\mathcal{P}}_{\epsilon}
  =
  \OpSpHY{e^{-2\theta E}}{0}\;
  \Op{\mathcal{P}}_{\epsilon}
  \,,
\end{align}
or infinitesimally
\begin{align}
  \Op{E}\;
  \Op{\mathcal{P}}_{\epsilon}
  =
  \frac{\epsilon}{2}\,\Op{\mathcal{P}}_{\epsilon}
  =
  \Op{\mathcal{P}}_{\epsilon}\;
   \Op{E}
  \ ,
\end{align}
where $\epsilon=\pm1$. In other words, $\Op{\mathcal{P}}_{\epsilon}$ behaves as the scalar (anti-)singleton ground state projector, which at the same time, from the point of view of the twisted-adjoint action of the $\msp(4,\mathbb{R})$ algebra, corresponds to an enveloping algebra realisation of the ground state of a massless AdS$_4$ (anti-)scalar with Neumann boundary conditions \cite{fibre,2017,COMST}. Adding an appropriate prefactor and taking the limit gives
\begin{align}
  \Op{\mathcal{P}}_{\epsilon_1}\;
  \Op{\mathcal{P}}_{\epsilon_2}
  =
  \lim_{\theta\to\epsilon_1\infty}
  e^{(\epsilon_1-\epsilon_2)\theta}
  \Op{\mathcal{P}}_{\epsilon_2}
  \ ,
\end{align}
which shows that
\begin{itemize}
  \item $\Op{\mathcal{P}}_{\epsilon}$ are indeed projectors;
  \item Their mutual product $\Op{\mathcal{P}}_{1}\Op{\mathcal{P}}_{-1}$ is infinite.
\end{itemize}

One way to regularise the latter product, thereby achieving orthogonality between singleton and anti-singleton states, 
is with the help of the following family of complex group elements \cite{fibre,2011,2017,COMST},
\begin{align}\label{Ueta}
  U_\eta
  =
  \exp\left(%
  2\log\left(\tfrac{1-\eta}{1+\eta}\right)
  E\right)
  =
  \frac{1+\eta^2}{1-\eta^2}
  -
  \frac{2i\eta}{1-\eta^2}\Gamma_{0'0}
  =
  \begin{pmatrix}
    \tfrac{1+\eta^2}{1-\eta^2}
    &
    \tfrac{2i\eta}{1-\eta^2}\sigma_0
    \\
    \tfrac{2i\eta}{1-\eta^2}\bar\sigma_0
    &
    \tfrac{1+\eta^2}{1-\eta^2}
  \end{pmatrix}
  \,,
\end{align}
to represent the projectors as
\begin{align}
  \Op{\mathcal{P}}_{\epsilon}
  =
  -2\epsilon\oint_{C(\epsilon)}
  \frac{d\eta}{2\pi i}\frac{1}{(\eta-\epsilon)^2}
  \,\OpSpHY{U_\eta}{0}
  \,,
\end{align}
where the countour $C(\epsilon)$ encircles $\epsilon$ anticlockwise in complex plane.
The ground-state projectors $ \Op{\mathcal{P}}_{\epsilon}$ are in fact embedded into a family labelled by non-zero positive integers $n$
\begin{align}
  \label{eq:Pn}
  \Op{\mathcal{P}}_{\epsilon n}
  =
  2\epsilon(-)^n\oint_{C(\epsilon)}
  \frac{d\eta}{2\pi i}\frac{(\eta+\epsilon)^{n-1}}{(\eta-\epsilon)^{n+1}}
  \,\OpSpHY{U_\eta}{0}
  \ ,
\end{align}
corresponding to rank-$n$ rotationally-invariant combination of projectors onto energy level $\frac{\epsilon n}2$ (anti-)supersingleton states, or,
from the twisted-adjoint point of view, to energy level $\epsilon n$ rotationally-invariant massless scalar (anti-)particle modes \cite{fibre,2017,COMST}. The multiplication of two such group elements gives
\begin{align}
  \log\left(\tfrac{1-\eta_1}{1+\eta_1}\right)
  +
  \log\left(\tfrac{1-\eta_2}{1+\eta_2}\right)
  &=
  \log\left(\tfrac{1-\eta_3}{1+\eta_3}\right)
  \,,&
  \eta_3:=\frac{\eta_1+\eta_2}{1+\eta_1\eta_2}
  \,.
\end{align}
The contour integral presentation is not a simple expansion of the projectors, but rather the integral itself is influenced by the product of its integrand.
When multiplying two projectors one assumes that one of the parameters, w.l.o.g. $\eta_2$, is much closer to its base point than the other.
Then one changes variable from $\eta_2$ to $\eta_3$, which will also run a small contour around the same base point according to the previous assumption.
The change of variable gives
\begin{align}
  d\eta_2\frac{(\eta_2+\epsilon_2)^{n_2-1}}{(\eta_2-\epsilon_2)^{n_2+1}}
  =
  (-1)^{n_2}
  d\eta_3
  \frac{(\eta_3+\epsilon_2)^{n_2-1}}{(\eta_3-\epsilon_2)^{n_2+1}}
  \frac{(\eta_1-\epsilon_2)^{n_2}}{(\eta_1+\epsilon_2)^{n_2}}
  \,,
\end{align}
which allows in particular to rebuild $\Op{\mathcal{P}}_{\epsilon_2 n_2}$ out of the $\eta_3$ integral.
With the help of the following consequence of the residue theorem
\begin{align}
  \oint_{C(\epsilon)}\frac{d\eta_1}{2\pi i}
  \frac{(\eta_1+\epsilon)^{a-1}}{(\eta_1-\epsilon)^{a+1}}
  &=
  \frac{1}{2\epsilon}\begin{pmatrix}a-1\\a\end{pmatrix}
  \,,&
  \begin{pmatrix}a-1\\a\end{pmatrix}=\delta_{a0}
  \quad\text{if}\quad a\in\Z\,,
\end{align}
one can perform the $\eta_1$ integral and find the projector algebra \cite{2011}
\begin{align}
  \Op{\mathcal{P}}_{\epsilon_1 n_1}\Op{\mathcal{P}}_{\epsilon_2 n_2}
  =
  \delta_{\epsilon_1\epsilon_2}\delta_{n_1n_2}
  \Op{\mathcal{P}}_{\epsilon_1 n_1}
  \,.
\end{align}

\subsection{Black-hole-like solution}

The twisted-adjoint initial datum $\Op\Phi'_{\rm bh}$ for spherically symmetric higher spin black-hole-like solutions \cite{Didenko:2009td,2011,2017,COMST} was shown in \cite{2011} to correspond to twisted rotationally-invariant supersingleton projectors $\Op{\mathcal{P}}_{\epsilon n}\Op{\kappa}_y$. Equivalently,
\begin{align}
  \Op\Psi^\prime_{\rm bh}
  =
  \sum_{\epsilon=\pm1}\sum_{n=1}^{+\infty}\nu_{\epsilon n}
  \Op{\mathcal{P}}_{\epsilon n}
  \,.
\end{align}
Fourier transforming as in \eqref{eq:FT OpSpH}, one obtains
\begin{align}
  \Psi^\prime_{\Lambda\,\rm{bh}}
  =
  2
  \sum_{\epsilon=\pm1}\sum_{n=1}^{+\infty}
  \epsilon(-1)^n\nu_{\epsilon n}
  \oint_{C(\epsilon)}\frac{d\eta}{2\pi i\eta^2}
  \left(\frac{\eta+\epsilon}{\eta-\epsilon}\right)^{n}
  \exp\left(-\tfrac{1}{2\eta}\Lambda\Gamma_{0'0}\Lambda\right)
  \,.
\end{align}
This can be recast in the more compact form
\begin{align}
  \Psi^\prime_{\Lambda\,\rm{bh}}
  &=
  2\Sene\nu_{\epsilon n}
  \frac{1}{\eta^2}
  \exp\left(-\tfrac{1}{2\eta}\Lambda\Gamma_{0'0}\Lambda\right)
  \,,\\
  \label{eq:Sene}
  \Sene f(\epsilon,n,\eta)
  :&=
  \sum_{\epsilon=\pm1}\sum_{n=1}^{+\infty}
  \epsilon(-1)^n
  \oint_{C(\epsilon)}\frac{d\eta}{2\pi i}
  \left(\frac{\eta+\epsilon}{\eta-\epsilon}\right)^{n}
  f(\epsilon,n,\eta)\,,
\end{align}
so the spacetime-dependent adjoint zero-form in $L$-gauge, by using \eqref{eq:FT PsiL}, reads
\begin{align}
  \Psi_{\Lambda\,\rm{bh}}
  =
  \Sene2\nu_{\epsilon n}
  \exp\left(-\tfrac{1}{2\eta}\Lambda L^{-1}\Gamma_{0'0}L\Lambda\right)
  \,.
\end{align}
In terms of stereographic gauge function \eqref{eq:stereo gf WO},
one has
\begin{align}
  -L^{-1}\Gamma_{0'0}L
  =
  \frac{1}{1-x^2}
  \begin{pmatrix}
    \sigma_0\xb-x\bar\sigma_0
    &
    \sigma_0-x\bar\sigma_0x
    \\
    \bar\sigma_0-\xb\sigma_0\xb
    &
    \bar\sigma_0x-\xb\sigma_0
  \end{pmatrix}
  =:
  \begin{pmatrix}
    \vark^L&v^L\\\bar{v}^L&\bar\vark^L
  \end{pmatrix}
  \,.
\end{align}
Note the properties
\bea
&(\vark^L)^2=(\bar\vark^L)^2=r^2
\,,\qquad
v^L\bar{v}^L=\bar{v}^Lv^L=-(1+r^2)&
\,,\\
&\vark^Lv^L=-v^L\bar\vark^L
\,,\qquad
\bar{v}^L\vark^L=-\bar\vark^L\bar{v}^L&
\,,\\
&\det \vark^L_{\a\b} = \det\bar\vark^L_{\ad\bd} = -r^2\,,&
\eea
where we recall that,
as in the rest of the paper, the products and squares are taken according to the NW-SE contraction rule
and where $r$ is the radial coordinate of the global $AdS_4$ spherical chart
\begin{equation}
r=\frac{2\sqrt{(x_0)^2+x^2}}{(1-x^2)}
\,.
\end{equation}
The resulting Weyl tensor generating function (via \eqref{eq:PhiW}) is thus
\begin{align}\label{Phibh}
  \MoSym{0}{\Op\Phi^{(1)}_{\rm{bh}}}
  =
  \frac{2i}{r}
  \Sene\frac{1}{\eta}\nu_{\epsilon n}
  \exp\left(%
  -\tfrac{1}{2\eta}y(\vark^L)^{-1}y
  +iy(\vark^L)^{-1}v^L\yb
  -\tfrac{\eta}{2}\yb(\bar\vark^L)^{-1}\yb
  \right)
\end{align}
(of course coinciding with the one found in \cite{2011} without using the plane-wave expansion in $\Lambda$). A corresponding gauge connection is obtained by plugging this expression into \eqref{eq:W^(xi,zeta) WO}.
This involve computing an incomplete Gaussian integral, for which we use the lemma
\begin{align}
  &
  \int_0^{\infty}dw(\ag w + \bg)\exp\left(
  -\cg w^2+\dg w
  \right)
  \nonumber
  \\&=
  \frac{\ag}{2\cg}
  +
  \frac{\sqrt{\pi}}{4\sqrt\cg}
  \left(2\bg+\frac{\ag\dg}{\cg}\right)
   \exp\left(\frac{\dg^{2}}{4 \cg}\right)
  \left(1+\operatorname{erf}{\left(\frac{\dg}{2 \sqrt{\cg}} \right)}\right) 
  \,,
\end{align}
that is valid for $\Re(\cg)>0$.
Hence the relevant part of the connection is
\begin{align}
  \MoSym{0}{\Op{W}^{(1)[\xi,\zeta]}}
  &=
  -\frac{e^{i\theta}}{2r\det b}\Sene\nu_{\epsilon n}
  \exp\left(-\tfrac{\eta}{2}\yb(\bar\vark^L)^{-1}\yb\right)
  \left[-\frac{\zeta(db-b\omega+be\bar{v}^L(\vark^L)^{-1})b^{-1}\zeta}{\zeta(b^t)^{-1}(\vark^L)^{-1}b^{-1}\zeta}\right.
  \nonumber\\&\quad
  +i\left(%
  \zeta be(\bar\vark^L)^{-1}\yb
  -
  (y\vark^L+\yb\bar{v}^L)(\vark^L)^{-1}b^{-1}\zeta
  \frac{\zeta(db-b\omega+be\bar{v}^L(\vark^L)^{-1})b^{-1}\zeta}{\zeta(b^t)^{-1}(\vark^L)^{-1}b^{-1}\zeta}
  \right)
  \nonumber\\&\qquad\times
  \sqrt{-\frac{\pi\eta}{2\zeta(b^t)^{-1}(\vark^L)^{-1}b^{-1}\zeta}}
  \exp\left(%
  \tfrac{\eta}{2}\tfrac%
  {\left((y\vark^L+\yb v^L)(\vark^L)^{-1}b^{-1}\zeta\right)^2}%
  {\zeta(b^t)^{-1}(\vark^L)^{-1}b^{-1}\zeta}\right)
  \nonumber\\&\qquad\times\left.
  \left(%
  1-\erf\left(i
  \sqrt{-\tfrac{\eta}{2\zeta(b^t)^{-1}(\vark^L)^{-1}b^{-1}\zeta}}
  (y\vark^L+\yb v^L)(\vark^L)^{-1}b^{-1}\zeta
  \right)
  \right)
  \right]
  \,.
\end{align}
To obtain this result, we need to assume the negativity of the real part of the spinor contraction $\tfrac{1}{\eta}\zeta(b^t)^{-1}(\vark^L)^{-1}b^{-1}\zeta$.
In the absence of concurrent constraints, this condition can be reached by controlling locally the phase of $b$, 
which we recall has no influence on the result \eqref{eq:COMST W VE}.
One may argue that the integral in \eqref{eq:W^(xi,zeta) WO} is purely formal, and in fact its convergence has no incidence on it being a potential for the cocycle $\Op\Sigma$ built from the Weyl zero-form \eqref{Phibh},
as the terms regularised to zero obviously do not contribute to this check.
One could also consider analytically continue that lemma, in a similar fashion to what is done in App. \ref{app:anal},
but the embedding of half-lines in the square root Riemann surface is not trivial to define unambiguously.
In any case, it would be puzzling that an incompatible regularisation be needed higher in perturbation theory.

The result is clearly analytic in $Y$,
hence allowing for an expansion of the form \eqref{eq:W Taylor}.
Note also that, the error function being odd,
the symmetrization \eqref{eq:[0,zeta,+]} will cause this connection to be bosonic.

\subsection{Scalar particle modes}
\label{sec:particle}

Within $Sp(4,\mathbb{C})$ (identifying $\Op{X}$ with $\Op{Y}$) the holomorphic Kleinian $\OpKy$ is realized as
\begin{align}
  \OpKy
  &=
  -i\;\OpSpH{K}{0}
  \,,&
  K
  &=
  \begin{pmatrix}-\mathbb{I}&0\\0&\mathbb{I}\end{pmatrix}
  \,.
\end{align}
Recalling the group element \eqref{Ueta} used to build regularised projectors \eqref{eq:Pn}, clearly the group element
\begin{align}
  \label{eq:UhK}
  U_\eta\,K
  =
  \begin{pmatrix}
    -\tfrac{1+\eta^2}{1-\eta^2}
    &
    \tfrac{2i\eta}{1-\eta^2}\sigma_0
    \\
    -\tfrac{2i\eta}{1-\eta^2}\bar\sigma_0
    &
    \tfrac{1+\eta^2}{1-\eta^2}
  \end{pmatrix}
  \,.
\end{align}
is instrumental to obtaining the \emph{twisted} regularised projector \cite{2017,COMST}
\begin{align}
  \label{eq:Ptn}
  \Op{\widetilde{\mathcal{P}}}_{\epsilon n}
  &=
  2\epsilon(-)^n\oint_{C(\epsilon)}
  \frac{d\eta}{2\pi i}\frac{(\eta+\epsilon)^{n-1}}{(\eta-\epsilon)^{n+1}}
  \OpSpHY{U_\eta}{0}\OpKy
  \nonumber\\&=
  2i\epsilon(-)^{n+1}\oint_{C(\epsilon)}
  \frac{d\eta}{2\pi i}\frac{(\eta+\epsilon)^{n-1}}{(\eta-\epsilon)^{n+1}}
  \OpSpHY{U_\eta K}{0}
  \,.
\end{align}
Clearly
\begin{align}
  KU_\eta K=U_{-\eta}\,,
\end{align}
hence, with an appropriate change of variable in \eqref{eq:Pn}
\begin{align}
  \OpKy\Op{\mathcal{P}}_{\epsilon n}\OpKy=\Op{\mathcal{P}}_{-\epsilon n}
  \,,
\end{align}
and the generalized projectors (\ref{eq:Pn},\,\ref{eq:Ptn}) satisfy the algebra \cite{2017}
\begin{align}
  \Op{\mathcal{P}}_{\epsilon_1 n_1}
  \Op{\mathcal{P}}_{\epsilon_2 n_2}
  &=
  \delta_{\epsilon_1\epsilon_2}\delta_{n_1n_2}
  \Op{\mathcal{P}}_{\epsilon_1 n_1}
  \,,&
  \Op{\mathcal{P}}_{\epsilon_1 n_1}
  \Op{\widetilde{\mathcal{P}}}_{\epsilon_2 n_2}
  &=
  \delta_{\epsilon_1\epsilon_2}\delta_{n_1n_2}
  \Op{\widetilde{\mathcal{P}}}_{\epsilon_1 n_1}
  \,,\\
  \Op{\widetilde{\mathcal{P}}}_{\epsilon_1 n_1}
  \Op{\mathcal{P}}_{\epsilon_2 n_2}
  &=
  \delta_{\epsilon_1,-\epsilon_2}\delta_{n_1n_2}
  \Op{\widetilde{\mathcal{P}}}_{\epsilon_1 n_1}
  \,,&
  \Op{\widetilde{\mathcal{P}}}_{\epsilon_1 n_1}
  \Op{\widetilde{\mathcal{P}}}_{\epsilon_2 n_2}
  &=
  \delta_{\epsilon_1,-\epsilon_2}\delta_{n_1n_2}
  \Op{\mathcal{P}}_{\epsilon_1 n_1}
  \,.
\end{align}

The massless scalar particle initial datum is
\begin{align}
  \Op\Psi^\prime_{\rm pt}
  &=
  \sum_{\epsilon=\pm1}\sum_{n=1}^{+\infty}
  \widetilde\nu_{\epsilon n}
   \Op{\widetilde{\mathcal{P}}}_{\epsilon n}
  =
  -2i\Sene 
  \widetilde\nu_{\epsilon n}\frac{1}{\eta^2-1}
  \OpSpHY{U_\eta K}{0}
  \,,
\end{align}
where we have used the notation \eqref{eq:Sene}.
Since the matrix \eqref{eq:UhK} squares to 1 (independently of the value of $\eta$), the Fourier transform \eqref{eq:FT OpSpH} is a limiting case.
In terms of a 2-dimensional symmetric matrix $D$
, one can define
\begin{align}
  \label{eq:Ke}
  K_\varepsilon
  :&=
  \begin{pmatrix}
    -e^{\varepsilon D}&0\\0&1
  \end{pmatrix}
  \,,&
  K
  &=
  \lim_{\varepsilon\to0}K_\varepsilon
  \,.
\end{align}
Hence the previous group matrix is the $\varepsilon\to0$ limit of
\begin{align}
  U_\eta\,K_\varepsilon
  =
  \begin{pmatrix}
    -\tfrac{1+\eta^2}{1-\eta^2}e^{\varepsilon D}
    &
    \tfrac{2i\eta}{1-\eta^2}\sigma_0
    \\
    -\tfrac{2i\eta}{1-\eta^2}\bar\sigma_0e^{\varepsilon D}
    &
    \tfrac{1+\eta^2}{1-\eta^2}
  \end{pmatrix}
  \,.
\end{align}
Using the sequence \eqref{eq:delta gauss}, the plane wave representation \eqref{eq:FT OpSpH} of the limit element \eqref{eq:Ke} is
\begin{align}
  \FTSpH{U_\eta\,K}{0}{\Lambda}
  &=
  \lim_{\varepsilon\to0}
  \FTSpH{U_\eta\,K_\varepsilon}{0}{\Lambda}
  \nonumber\\&=
  \lim_{\varepsilon\to0}
  \frac{1-\eta^2}{\eta^2}
  \frac{1}{\sqdet{\varepsilon D}}
  (1+\mathcal{O}(\varepsilon^2))
  \nonumber\\&\qquad\times
  \exp\left(%
  \tfrac{i}{\varepsilon}
  (\lambda+\tfrac{i}{\eta}\bar\lambda\bar\sigma_0)
  D^{-1}
  (\lambda-\tfrac{i}{\eta}\sigma_0\bar\lambda)
  +
  \eta\lambda\sigma_0\bar\lambda
  +
  \mathcal{O}(\varepsilon)
  \right)
  \nonumber\\&=
  -i\pi
  \frac{1-\eta^2}{\eta^2}
  \delta^2\left(\lambda-\tfrac{i}{\eta}\sigma_0\bar\lambda\right)
  \,.
\end{align}
Hence, the spacetime-independent momentum-space initial datum is
\begin{align}
  \Psi^\prime_{\Lambda\,\rm{pt}}
  &=
  2\pi\Sene\widetilde\nu_{\epsilon n}\frac{1}{\eta^2}
  \delta^2\left(\lambda-\tfrac{i}{\eta}\sigma_0\bar\lambda\right)
  \,,
\end{align}
and the spacetime-dependent adjoint zero-form is
\begin{align}
  \Psi_{\Lambda\,\rm{pt}}
  &=
  2\pi\Sene\widetilde\nu_{\epsilon n}
  \frac{1-x^2}{1-2i\eta x_0+\eta^2x^2}
  \delta^2\left(\bar\lambda-\lambda N\right)
  \,,&
 N
  :&=
  \left(%
  1-\tfrac{i}{\eta}x\bar\sigma_0
  \right)\left(%
  \bar{x}-\tfrac{i}{\eta}\bar\sigma_0
  \right)
  \,.
\end{align}
A simple integration \eqref{eq:PhiW} gives the Weyl zero-form
\begin{align}\label{Phipt}
  \MoSym0{\Op\Phi^{(1)}_{\rm pt}}
  =
  \Sene\widetilde\nu_{\epsilon n}
  \frac{1-x^2}{1-2i\eta x_0+\eta^2x^2}
  \exp\left(iy N\yb\right)
\end{align}
which indeed coincides with the expression found in \cite{2017,COMST} without using the Fourier expansion.
Because of the vanishing difference between the powers in $y$ and $\yb$, the encoded configuration consists in a scalar field and its tower of derivatives,
but no curvature that would contribute to the right hand side of the free equation \eqref{eq:COMST W VE}.
This means that the associated connection \eqref{eq:W^(xi,zeta) WO} can be trivialized via a $\OpZ$-independent linear gauge transformation \eqref{eq:VE igt}.
This gauge function $\Op{H}^{(1)[\xi,\zeta]}_0$ must be non-trivial as in this case the Weyl-ordered connection
\begin{align}
  &
  \MoSym0{\Op{W}^{(1)[\xi,\zeta]}}
  \nonumber\\&=
  \label{eq:Wpt(xi)}
  -\Sene\widetilde\nu_{\epsilon n}
  \frac{1-x^2}{1-2i\eta x_0+\eta^2x^2}
  \int d^4\Lambda\;\MoSpH{0}{1}{\Lambda}
  \delta^2\left(\bar\lambda-\lambda N\right)
  (D_\Lambda\xi\;\zeta)
  \int_0^{+\infty}dw\,\delta^2\left(\xi+w\zeta\right)
  \\&=
  \frac{1}{\det b}
  \Sene\widetilde\nu_{\epsilon n}
  \frac{1-x^2}{1-2i\eta x_0+\eta^2x^2}
  \frac{%
  \zeta(b^t)^{-1}(db^t+\omega b^t+Ne^t b^t)\zeta
  }{%
  (\zeta(b^t)^{-1}(y+N\yb))^2
  }
\end{align}
is itself non-trivial, and it is not even analytic in $Y$, thereby blurring the identification \eqref{eq:W Taylor} of the gauge fields.

Since on the momentum integral in \eqref{eq:W^(xi,zeta) Fourier}
the effective covariant derivative $D_\Lambda$ act according to the following particular form of \eqref{eq:d(PsiUF)=PsiUdF}
\begin{align}
  \Op{D}\left(
  \int\frac{d^4\Lambda}{(2\pi)^2}\,
  \Psi_\Lambda\;\OpSpHY{1}{\Lambda}\;f(\xi)
  \right)
  &=
  \int\frac{d^4\Lambda}{(2\pi)^2}\,
  \Psi_\Lambda\;\OpSpHY{1}{\Lambda}
  \left(D_\Lambda\xi\,\partial_\xi\right)f(\xi)
  \,,
\end{align}
one is tempted to use a resolution operator \eqref{eq:rhoMMXi} in the spinorial variable $\xi$ to integrate the connection.
However, in this case, the integrability condition in the counterpart to Eq. \eqref{eq:qq*g=g}
\begin{align}
  \Op{D} D_\Lambda^{\ast[\mu]} \MoSym0{\Op{W}^{(1)[\xi,\zeta]}}
  =
  \left(1-D_\Lambda^{\ast[\mu]}\Op{D}-\Op{\mathcal\pi}^{[\mu]}\right)
  \MoSym0{\Op{W}^{(1)[\xi,\zeta]}}
  \,,
\end{align}
is blurred by the momentum dependence of $\xi$:
\begin{align}
  D_\Lambda^{\ast[\mu]}\left(\Op{D}\MoSym0{\Op{W}^{(1)[\xi,\zeta]}}\right)
  &\propto
  D_\Lambda^{\ast[\mu]}\left(%
  \int d^4\Lambda\;\MoSpH{0}{1}{\Lambda}
  \delta^2\left(\bar\lambda-\lambda N\right)
  (D_\Lambda\xi\;D_\Lambda\xi)
  \delta^2(\xi)
  \right)
  \nonumber\\&=
  2
  \int d^4\Lambda\;\MoSpH{0}{1}{\Lambda}
  \delta^2\left(\bar\lambda-\lambda N\right)
  \int_0^1dt\,t\,
  ((\xi-\mu)\;D_\Lambda\xi)
  \delta^2(t\xi+(1-t)\mu)
  \nonumber\\&=
  \frac2{\det b}\;
  \mu\left(db-b\omega+beN^t\right)b^{-1}\mu
  \int_0^1dt\,\frac{1-t}{t^3}\,
  \exp\left(%
  i\tfrac{1-t}{t}\mu b^{-1}(y+N\yb)\right)
  \nonumber\\&=
  \frac2{\det b}\;
  \frac{%
  \mu\left(db-b\omega+beN^t\right)b^{-1}\mu
  }{
  \left(\mu b^{-1}(y+N\yb)\right)^2
  }
  \neq0
  \,.
\end{align}
To find a proper potential for $\MoSym0{\Op{W}^{(1)[\xi,\zeta]}}$, one needs a homotopy contraction that is compatible with the vanishing of that integral,
i.e. that does not interfere with the variable $\Lambda$.
One solution is to notice that after performing the replacement $\xi=b\lambda=:B\Lambda$, the whole spacetime dependence is encoded in the matrix $b$ and $D_\Lambda$ still acts as an exterior differential, now on the 8-dimensional variable $B_{\ag\bu}$, viz.
\begin{align}
  \label{eq:DB}
  D_\Lambda\;f(\Lambda B^t)
  &=
  \Lambda^\au\left(%
  dB^\bg{}_{\au}+(L^{-1}\,dL)_\au{}^{\cu}B^\bg{}_{\cu}\right)
  \frac{\partial}{\partial(\Lambda B^t)^\bg}\;f(\Lambda B^t)
  \nonumber\\&
  =
  \left(%
  dB^{\bg\au}+(L^{-1}\,dL)^{\au\cu}B^\bg{}_{\cu}\right)
  \frac{\partial}{\partial B^{\bg\au}}\;f(\Lambda B^t)
  =:
  DB^{\bg\au}\frac{\partial}{\partial B^{\bg\au}}\;f(\Lambda B^t)
  \,.
\end{align}
Hence the relevant component of the connection \eqref{eq:W0} can be annihilated by the following residual gauge fixing:
\begin{align}
  h^{[\xi,\zeta,C]}_{10\Lambda}
  &=
  -D^{\ast[C]}_\Lambda
  w^{[\xi,\zeta]}_{1\Lambda}
  \nonumber\\&=
  i\pi e^{i\theta}\,
  \zeta\left((b-c)\lambda-g\bar\lambda\right)
  \int_0^1dt\int_0^{+\infty}dw\;
  \delta^2\left(%
  \left(tb+(1-t)c\right)\lambda+(1-t)g\bar\lambda+w\zeta
  \right)
  \,,
\end{align}
in terms of a shift $C_{\ag\bu}=(c_{\ag\bg},\,g_{\ag\bd})$
and of the partial inverse $D^{\ast[C]}_\Lambda$ that it defines for the differential \eqref{eq:FT Psi}.
Note that, in order for this homotopy contractor to actually provide a solution to the equation $D_\Lambda h^{[\xi,\zeta,C]}_{10\Lambda}=-w^{[\xi,\zeta]}_{1\Lambda}$,
the shift $C_{\ag\bu}$ is assumed to:
\begin{itemize}
  \item not vanish, since, because of the aforementioned condition, the homotopy has to be well defined on the object $(D_\Lambda\xi\;D_\Lambda\xi)
  \delta^2(\xi)$
  (which is of order 0 in $B$);
  \item be covariantly constant in the sense of \eqref{eq:DB}, in particular it has a non-vanishing $g_{\ag\bd}$ part.
\end{itemize}
Modulo those assumptions, one finds back $\Op{D} \Op{H}^{(1)[\xi,\zeta,C]}_0=-\Op{W}^{(1)[\xi,\zeta]}$, hence the connection is trivial as expected.

Note that, because of the triviality of their gauge part, scalar solutions satisfy \eqref{eq:COMST W VE} in significantly many more gauges with respect to generic solutions.
However, the purpose of this Section is to test the resolution scheme based on the axial holomorphic gauge \eqref{eq:v1zeta} and rotated Vasiliev gauge (\ref{eq:V1shift},\,\ref{eq:U=rhodV}) on simple solutions, which we believe will be particularly useful to perturbatively imposing asymptotically AdS boundary conditions following the scheme presented in \cite{COMST} (see also \cite{Iazeolla:2020jee}).
Ultimately, the idea will thus be to impose the rotated gauge condition on more general solutions where scalar particles will appear as one term among others in the expansion of the initial datum.

\section{Conclusions}

Thinking of Vasiliev's equations as describing a family of noncommutative twistor spaces fibered over a spacetime manifold, and using properties of the inhomogenous metaplectic groups over the complex as well as real numbers, we integrate the system in operator form and in a variety of gauges without making reference to one specific ordering.
More precisely, we have used a family of orderings in which the contraction of twistor space coordinates above a given spacetime point is given by a constant matrix (with fixed anti-symmetric part), which can be used to interpolate between normal, Weyl and other orderings.
This method allows us to construct a number of interesting exact solutions, to compute Wilson lines in twistor space and to set up a scheme for switching on a gauge function that activates Fronsdal fields in ALAdS regions order by order in perturbation theory, as we have demonstrated explicitly at first order.

The methods developed here may facilitate the further investigation of a number of interesting problems in the theory, such as the computation of holographic multi-point functions by perturbative expansion of the second Chern class around various ALAdS backgrounds \cite{COMST}. In particular, it would be interesting to examine the mixing of black-hole states and particle states in correlation functions, thereby extending the work of \cite{neiman}.
Another interesting problem concerns the activation of the Wigner deformation parameter in the four-dimensional context by giving an adjoint vacuum expectation value to $\widehat B$, thereby breaking $Mp(4;\R)$ down to $Mp(2;\R)\times Mp(2;\R)$ such that the full master fields, valued in the full enveloping algebra, describe fractional spin fluctuations around domain walls, providing a natural generalization of \cite{Boulanger:2013naa}.

Finally, recasting the Vasiliev system in operator form may make analogies with the underlying quantum mechanical conformal particle system more manifest, allowing for some transfer of techniques as well as facilitating the embedding of Vasiliev's theory into the multi-parton gauge theory  \cite{Engquist:2005yt,Vasiliev:2012tv,Gaberdiel:2021qbb} that has been proposed to describe the tensionless limit of string theory in anti-de Sitter spacetime.

\paragraph{Acknowledgements.} We have benefited from discussions with C. Arias, R. Aros, P. Bieliavsky, N. Boulanger, A. Canazas, F. Diaz, V. E. Didenko, Y. Neiman, C. Reyes, E. Sezgin, A. Sharapov, E. Skvortsov, M. Valenzuela, B. Vallilo, M. A. Vasiliev. We are grateful to the anonymous Referee for useful suggestions that improved the presentation.
D.D.F. was a Research Fellow at the F.R.S.-FNRS (Belgium) during most of the preparation of this work.
P.S. acknowledges the DCF at UNAB for his academic freedom and his work is partially supported by Fondecyt Regular grants 1140296 and 1151107. He also acknowledges the support of the Centro de Ciencias Exactas at Universidad del Bio-Bio during the last stages of this work.

\appendix

\section{Conventions}
\label{app:cvt}

In this Section we spell out some of the conventions that are used throughout the paper.
Most of the equation written in this paper involve operator-valued differential forms. 
Operators, thought of as elements of a non-abelian algebra, are denoted with hats.
Whenever a product of differential forms appear, it is meant to involve at least an exterior product, although it is typically accompanied with an operator product. 

We use two types of indices: spinorial and vectorial.
On the one hand, we use spinor indices:
\begin{itemize}
  \item $Sp(2)$ (anti-)holomorphic indices denoted by (dotted) greek letters from the beginning of the alphabet;
  \item $Sp(4)$ indices denoted by underlined greek letters from the beginning of the alphabet;
  \item $Sp(2n)$ indices denoted by upper-case letters from the middle of the alphabet.
\end{itemize}
Of course all these cases are included in the last one, to which we turn our attention.
The invariant antisymmetric tensors $\theta^{IJ}$ and $\theta_{IJ}$ are chosen with the convention
\begin{align}
  \theta_{IK}\theta^{JK}=\delta_I^{J}
  \,.
\end{align}
They are implictly used to raise, lower and contract indices following the so-called Northwest-southeast rule:
\begin{align}\label{implicit1}
  V^I
  &=
  \theta^{IJ}V_J
  \,,&
  V_I
  &=
  V^J\theta^{JI}
  \,,&
  VW
  &:=
  V^IW_I
  =
  -WV\,.
\end{align}
This also works for spin-tensors, a more involved contraction reading
\begin{align}\label{implicit2}
  VABW:=
  V^IA_J^{\phantom{J}K}B_K^{\phantom{K}L}W_L
  \,.
\end{align}
The rules also apply to the symplectic antisymmetric tensor itself, in particular
\begin{align}
  \theta^{IK}\theta^{JL}\theta_{KL}
  &=
  \theta^{IJ}
  \,,&
  \theta_I^{\phantom{I}J}
  &=
  \delta_I^J
  =
  -\theta^J_{\phantom{J}I}
  \,.
\end{align}
Because of the latter relation, the indices of a scalar tensor $a\theta_{IJ}$ ($a\in\C$) will often be omitted
by assimilating it with the scalar $a$.
The particular cases of relevant for the four-dimensional Vasiliev system make use of the same convention,
except for the fact that the components of the symplectic invariant matrix are written $\epsilon_{\au\bu}$ for $Sp(4)$ and $\epsilon_{\ag\bg}$ (resp. $\epsilon_{\ad\bd}$) for (anti-)holomorphic two-dimensional spinors.

On the other hand, tensor indices associated to a $N+1$-dimensionl\footnote{%
In particular, for the four-dimensional case to which this paper is dedicated, one needs to set $N=3$.} 
tangent space are indicated by:
\begin{itemize}
  \item Lower-case letters from the beginning of the alphabet for Lorentz indices, that transform under $SO(1,N)$;
  \item Upper-case letters from the beginning of the alphabet for ambient indices, that transform under $SO(2,N)$.
\end{itemize}
The latter take value in the set $\{0',0,\cdots,3\}$ and are raised, lowered and contracted by the constant metric of components $\eta_{AB}$.
The infinitesimal generators of the orthogonal group who operates on them verify the commutation relations
\begin{align}
  \label{eq:SO(2,d) ambient}
  \left[
  \Op{J}_{A,B}
  ,\,
  \Op{J}_{C,D}
  \right]
  =
  i
  \left(
  \eta_{AD}\Op{J}_{B,C}+\eta_{BC}\Op{J}_{A,D}
  -\eta_{AC}\Op{J}_{B,D}-\eta_{BD}\Op{J}_{A,C}
  \right)
  =
  4i
  \eta_{[D[A}\Op{J}_{B],C]}
  \,.
\end{align}
Tensors that transform irreducibly under this algebra are characterized by Young diagrams,
among which we are interested in the two-row rectangular ones
\begin{align}
  \label{eq:YT ambient rect}
  \begin{ytableau}
     A_1&A_2&\none[\dots]&A_{n}\\B_1&B_2&\none[\dots]&B_{n}
  \end{ytableau}
\end{align}
Here we use the symmetric basis for tensors of the orthogonal groups,
that is to say that projecting on the symmetries of a given tableau is done by
removing all traces, anti-symmetrizing over columns, symmetrizing over rows and, finally, normalizing so that this operation constitutes a projection.
We use a notation in which commas separate symmetrized and traceless sets of vector indices appearing on a given tensor, and subindexed indices from the same letter of the alphabet appearing on different tensors are symmetrized.
The covariance of such tensors can be reduced to be manifest only under Lorentz transformations, by highlighting a subspace with a constant metric $\eta_{ab}$ of signature $(-,+,\cdots,+)$.
One defines
\begin{align}
  \Op{M}_{ab}:=\Op{J}_{a,b}
  \,,\qquad 
  \Op{P}_{a}:=\Op{J}_{0',a}
  \,,
\end{align}
in terms of which the $AdS_d$ isometry algebra becomes
\begin{align}
  \label{eq:SO(2,N) Lorentz}
  [\Op{M}_{ab},\Op{M}_{cd}]&=4i\eta_{[d\vert[a}\Op{M}_{b]\vert c]}
  \,,&
  [\Op{M}_{ab},\Op{P}_c]&=2i\eta_{c[b}\Op{P}_{a]}
  \,,&
  [\Op{P}_a,\Op{P}_b]&=i\Op{M}_{ab}
  \,.
\end{align}
The tableau \eqref{eq:YT ambient rect} branches into
\begin{align}
  \label{eq:YT Lorentz}
  \begin{ytableau}
     a_1&a_2&\none[\dots]&a_{t}&a_{t+1}&\none[\dots]&a_{n}\\b_1&b_2&\none[\dots]&b_{t}&\none&\none&\none
  \end{ytableau}
\end{align}
for $t=0,\cdots,n$.

Assuming from now on that the spacetime dimension is equal to 4,
vector and tensor indices are related by special sets of matrices.
In the case of the ambient vectors, the Howe duality $\mso(2,3)\approx\msp(4)$ can be realised in terms of five $4\times4$ antisymmetric matrices $(\Gamma_A)_{A=0',0,\cdots,3}$ that generate a Clifford algebra
\begin{align}
  \label{eq:Gamma alg}
  (\Gamma_A)_\au\,^\cu
  (\Gamma_B)_\cu\,^\bu
  +
  (\Gamma_B)_\au\,^\cu
  (\Gamma_A)_\cu\,^\bu
  =
  2\,\eta_{AB}\,\delta_\au^\bu
  \,.
\end{align}
From them one builds the spin matrices
\begin{align}
  \label{eq:Gamma_AB}
  (\Gamma_{AB})_{\underline{\a\b}}
  :=
  \frac12\left(
  (\Gamma_A)_\au\,^\cu
  (\Gamma_B)_{\cu\bu}
  -
  (\Gamma_B)_\au\,^\cu
  (\Gamma_A)_{\cu\bu}
  \right)
  \,,
\end{align}
and the related matrix representation of the commutators \eqref{eq:SO(2,d) ambient}
\begin{align}
  \label{eq:M(Gamma)}
  M_{AB}
  :&=
  \frac{i}{2}\Gamma_{AB}
  \,.
\end{align}
A field $\phi_{A_1\cdots A_{n},B_1\cdots B_{n}}$ transforming in the representation \eqref{eq:YT ambient rect} is now equivalent to a totally symmetric spin-tensor $\phi_{\au_1\cdots\au_{2n}}$.
The analogous to the Lorentz subalgebra on the spinorial side is the $\msu(2)\oplus\msu(2)$ algebra of purely (anti-)holomorphic generators,
a relation that is realised on the components $v^a$ of a Lorentz vectors as
\begin{equation}
v_{\a\ad}:=v^a(\sigma_a)_{\a\ad}
=:\bar{v}_{\ad\a}
\,,\qquad
v^a=-\frac12(\sigma^a)_{\a\ad}v^{\a\ad}
\,,
\end{equation}
where the Van der Waerden symbols, 
with components $(\sigma_a)_{\ag\ad}=({\bar\sigma}_a)_{\ad\ag}$,
have the following properties
\begin{align}
\sigma_a{\bar{\sigma}}_b
&=
\eta_{ab}+\sigma_{ab}
\,,&
\sigma_{ab}:=\frac12(\sigma_a{\bar{\sigma}}_b-\sigma_b{\bar{\sigma}}_a)
\,,\\
{\bar{\sigma}}_a\sigma_b
&=
\eta_{ab}+{\bar{\sigma}}_{ab}
\,,&
{\bar{\sigma}}_{ab}:=\frac12({\bar{\sigma}}_a\sigma_b-{\bar{\sigma}}_b\sigma_a)
\,,\\
&&
(\sigma^a)_{\a\ad}(\sigma_{a})_{\b\bd}
=-2\epsilon_{\a\b}\epsilon_{\ad\bd}
\,.
\end{align}
Now the data of the Lorentz tensor \eqref{eq:YT Lorentz} is equivalent to the one of a spin-tensor $\phi_{\ag_1\cdots\ag_{n+t},\ad_1\cdots\ad_{n-t}}$ and of its dual  $\phi_{\ag_1\cdots\ag_{n-t},\ad_1\cdots\ad_{n+t}}$,
both symmetric under the exchange of 2 (anti-)holomorphic indices.
The associated block decomposition of the Clifford elements \eqref{eq:Gamma alg} is
\begin{align}
  \label{eq:Gamma(sigma)}
  \Gamma_{0'}
  &=
  \begin{pmatrix}
    i&0\\0&-i
  \end{pmatrix}
  \,,&
  \Gamma_a
  &=
  \begin{pmatrix}
    0&i\sigma_a\\-i\bar\sigma_a&0
  \end{pmatrix}
  \,,&
  \Gamma_{0'a}
  &=
  \begin{pmatrix}
    0&-\sigma_a\\-\bar\sigma_a&0
  \end{pmatrix}
  \,,&
  \Gamma_{ab}
  &=
  \begin{pmatrix}
    \sigma_{ab}&0\\0&\bar\sigma_{ab}
  \end{pmatrix}
  \,.
\end{align}

The Cartan connection of AdS${}_4$ is the pull-back of a Maureer-Cartan form, that is to say it is an algebra-value 1-form field 
\begin{equation}
\Op\Omega
:= 
-\frac{i}{2}\Omega^{AB}\Op{J}_{A,B}
=
-i \left(e^{a}\Op{P}_a +  \tfrac{1}{2}\omega^{ab}\Op{M}_{ab}\right)
\,,
\end{equation} 
satisfying
\begin{align}
  \label{eq:AdS R=0 app}
  d\Op\Omega+\frac12[\Op\Omega,\Op\Omega]=0\,.
\end{align}
The Lorentz components of this equation read 
\begin{equation}
d\omega^{ab} + \omega^{a}_{\phantom{a}c}\omega^{cb} + e^{a} e^{b}=0
\,,\qquad
de^{a} + \omega^{a}_{\phantom{a}b}e^{b}=0
\,,
\end{equation}
thereby allowing to identify $e^{a}$ and $\omega^{ab}$ respectively as the vierbein and spin connection of $AdS_4$,
in units where the cosmological constant is equal to -3.
The (1-form valued) matrix components of the curvatures, defined by the realisation \eqref{eq:M(Gamma)}, satisfy the assocative version of Eq. \eqref{eq:AdS R=0 app}:
\begin{align}
  d\Omega_\au^{\phantom\au\bu}
  +\Omega_\au^{\phantom\au\cu}\Omega_\cu^{\phantom\cu\bu}
  =0\,,
\end{align}
that can be solved by means of a gauge function $L$ as 
\begin{align}
  \label{eq:AdS gf mat}
  \Omega_\au^{\phantom\au\bu}
  =
  (L^{-1})_\au^{\phantom\au\cu}dL_\cu^{\phantom\cu\bu}
  \,.
\end{align}

In most cases in which we need to make the spacetime dependence explicit, we choose to do it with the help of stereographic coordinates,
that can be written in Lorentz covariant fashion
\begin{equation}
\label{eq:stereo coord}
x^a\in\mathbb{R}^4
\,,\qquad
x^2\neq1
\,,\qquad
ds^2=\frac{4}{(1-x^2)^2}dx^2
\,.
\end{equation}
This vectorial form allows to define associated matrix coordinates
\begin{align}
  x_{\ag\ad}:=x^a(\sigma_a)_{\ag\ad}=:{\bar x}_{\ad\ag}\,.
\end{align}Two useful definitions are
\begin{equation}
h:=\sqrt{1-x^2}
\,,\qquad
\xi:=
\vert x^2\vert^{-\tfrac12}\tanh^{-1}\left(\sqrt{\frac{1-h}{1+h}}\right)
\,.
\end{equation}
The block form of the vacuum gauge function \eqref{eq:AdS gf mat} corresponding to stereographic coordinates and of its inverse are then given by \cite{Bolotin:1999fa}
\begin{align}
  L
  &=
  \frac{1}{h}
  \begin{pmatrix}
    1&x\\\bar{x}&1
  \end{pmatrix}
  \,,&
  L^{-1}
  &=
  \frac{1}{h}
  \begin{pmatrix}
    1&-x\\-\bar{x}&1
  \end{pmatrix}
  \,.
\end{align}
We also refer to the global spherical coordinates $(t,r,\theta,\varphi)$, in which the metric reads
\bea  ds^2 \ = \ -(1+r^2)dt^2+\frac{dr^2}{1+ r^2}+r^2(d\theta^2+\sin^2\theta d\varphi^2) \ ,\label{metricglob}\eea
and which provide a single cover of $AdS_4$ for $t\in [0,2\pi)$, $r\in[0,\infty)$, $\theta\in[0,\pi]$ and $\varphi\in[0,2\pi)$. 
The radial coordinate $r$ can be related to the stereographic coordinates via
\begin{equation}
r=\frac{2\sqrt{(x_0)^2+x^2}}{1-x^2}
\,.
\end{equation}

\section{Analytic continuation of Gaussian integrals and delta functions}
\label{app:anal}

As we have recalled in the paper, notable solutions to the Vasiliev equations --- both at linearized and full level, such as massless particles, bulk-to-boundary propagators, higher-spin generalizations of black holes and black branes, FLRW backgrounds, etc. --- have local data $\Phi'$ (or $\Psi'$) expanded over Gaussians and plane waves in oscillators \cite{GiombiYin2,Didenko:2012tv,Iazeolla:2020jee,Didenko:2009td,2011,Sundell:2016mxc,2017,review,COMST,Iazeolla:2020jee,Didenko:2021vui,Iazeolla:2007wt,Iazeolla:2015tca,cosmo}. Their symbols in various orderings, as well as star products involving them, generally bring on square roots of (generally) complex determinants. Preserving the characteristic metaplectic $4\pi$-periodicity of operators at the level of symbols, imposing reality conditions and other physical constraints on relevant master field configurations \cite{2011,2017}, together with demanding a consistent action of $\tau$ on Weyl-ordering symbols and that multi-dimensional integration be unambiguous, force us to analytically continue the square root of all determinants to complex values, preserving their phases rather than extracting the principal square root. 
The same applies to degenerate Gaussian symbols defining delta sequences: the latter are, in fact, far from exotic objects of little physical interest, considering that they naturally appear, working in Weyl order on ${\cal Y}$, as the configuration that the spherically-symmetric higher-spin black holes approach at the origin \cite{Didenko:2009td,2011,2017}, and that basic linearized solutions like massless particles and bulk-to-boundary propagators are obtained from the corresponding local data by considering analytic extensions of the delta function preserving the phase of their argument \cite{2017,COMST,Iazeolla:2020jee}, in a sense to be made precise in the following.  

In this appendix we shall therefore first recall some of the reasons for our analytic continuation of Gaussian integration and delta functions to the complex plane, and then explain the details of this continuation, thereby giving a more precise interpretation to such manipulations with symbols.  As we shall only refer to Weyl-ordering symbols, for notational simplicity we shall drop the label $0$ both from symbols and star product unless otherwise specified.

\subsection{Some motivations}\label{appD:motivations}

Reality conditions $\Phi'^\dagger=\pi(\Phi')$ on the higher-spin black hole Weyl zero-form can be imposed at the level of the corresponding local datum
\be \Phi'_{{\rm bh}} = \sum_{\ve=\pm 1}\sum_{n=1}^{\infty} \n_n \cP_n\star\k_y = \sum_{\ve=\pm 1}\sum_{n=1}^{\infty} \n_n \tcP_n \ee
(where $\cP_n$ are the projectors \eqref{eq:Pn} presented in Weyl order), implying that the deformation parameters $\n_n=i^n\m_n$ where $\m_n\in \R$. Passing to the resulting $x$-dependent master fields gives the Weyl tensor generating function \eqref{Phibh},
%
%
where the $i$ in the prefactor comes from continuing $(\det \vark^L)^{-1/2}=(-r^2)^{-1/2}$ \cite{2011}. Without that prefactor, the scalar field extracted from \eqref{Phibh} would be imaginary, i.e., reality of the scalar field requires that the Gaussian determinant be extracted with its complex phase. 

Considering the realization of the Klein operators \eqref{eq:K=U(-1)} as Gaussian delta sequences in Weyl order, namely 
\be\label{KWO}  [\Op K]_0 = 
 \lim_{M\to 0} \frac{(-i)^n}{\sqdet{-M}}
  \exp\left(
  -\tfrac{i}{2}XM^{-1}X\right)
  \,, \ee
in accordance with \eqref{eq:OpSp U}, the above requirement of phase preservation should also apply to $\k_y$ and $\k_z$, that are therefore to be treated as analytic continuations of the Dirac delta function. This is in accordance with (and further motivated by) the physical interpretation of the twisted projectors \eqref{eq:Ptn} as local data for AdS massless particle states \cite{fibre,2017,COMST}, and of similar Gaussian elements as local data for bulk-to-boundary propagators (see \cite{Iazeolla:2020jee} for their relation).  In such sectors --- that, following the terminology of \cite{Iazeolla:2020jee}, we refer to as regular --- the Weyl zero-form in Weyl ordering reads
\be\Phi_{{\rm reg}}^{(1)}(x,Y)  \ = \ \sum_{\ve=\pm 1}\sum_{n=1}^{\infty} \tn_n \,L^{-1}\star \cP_n\star \pi(L) \ = \ \sum_{\ve=\pm 1}\sum_{n=1}^{\infty} \tn_n \tcP^L_n\star\k_y \ , \label{Phireg}\ee
with
\bea  \tcP_n^L \ := \  4\pi (-)^{n-1}\ve_n\,\oint_{C(\ve_n)} \frac{d\eta}{2\pi i}\,\left(\frac{\eta+1}{\eta-1}\right)^{n}\,\d^2(Ay+B\yb)
\ ,\label{tPE} \eea
%
%
%
where $A$ and $B$ are matrices depending on $x$ and $\eta$. Analytic continuation of the delta function suggests to identify
\be \d^2(Ay+B\yb) = \frac{1}{\det A}\d^2(y+N\yb) \ , \label{anald}\ee
where $N:=A^{-1}B$. Ultimately, as seen explicitly for the scalar particle in \eqref{Phipt}, 
\be  \Phi_{{\rm reg}}^{(1)}(x,Y) \ = \ \Sene\widetilde\nu_{\epsilon n} \frac{1}{\det A}\,\exp\left(iyN\yb\right) \label{lastreg} \ . \ee
Thus, the scalar field is entirely encoded in $(\det A)^{-1}$, and the above analytic continuation of the delta function is crucial to encode the proper phases, corresponding, for the particle case, to the energy eigenvalues. This way, for instance, the Breitenlohner-Freedman scalar with Neumann boundary conditions is correctly reproduced from \eqref{Phipt} for $n=1$ at $y=0=\bar y$ as \cite{2017,COMST,Iazeolla:2020jee}
\be \phi_{1;(0)} = \frac{e^{-it}}{\sqrt{1+r^2}}  \ee
in AdS global spherical coordinates.

Another simple motivation is the analytic structure of the two-point function
\be \left\langle :\varphi^2(\xi)::\varphi^2(0):\right\rangle= \frac{-1}{\xi^2}\,\ee
of the scalar composite operator built from a free conformal scalar field $\varphi(\xi)$ in Minkowski spacetime corresponding to $\phi$ expanded on boundary-to-bulk propagators; for the three-dimensional Green's functions, see for example Eq. (16) in \cite{Zhang:2008jy}.
Indeed, the above two-point function, with its  analytical structure, is reproduced by the zero-form charges of HSG computed using the analytical continuation of Gaussian integrals to be spelled out below \cite{Colombo:2010fu,Colombo:2012jx,Bonezzi:2017vha}.

We therefore analytically continue, when necessary, all expressions, including star products and re-ordering formulas such as \eqref{eq:stargauss}. This implies taking square roots of products of complex numbers, so without a specific prescription the analytic continuation is only defined up to a sign. This ambiguity at the level of symbols can be put in correspondence with the one inherent to the projectivity of the metaplectic representation. Thus, with the help of a specific prescription to extract phases from the square root, it is possible to distinguish the two symbol representatives of a given $Sp(2n,\mathbb{C})$ element: in other words, by virtue of the map \eqref{eq:stargauss}, the two branches of the double covering $Mp(2n,\mathbb{C})$ can be put in correspondence with the two sheets of the Riemann surface ${\cal S}_2$ of the square root. Besides, without a specific phase assignment, integration over oscillator symbols may not be well-defined: for instance, the order of integration on different variables in multi-dimensional integrals in $Y$ or $Z$ cannot be interchanged, in general, as that may result in a sign discrepancy due to the assignment of complex determinants to one or another Riemann sheet at intermediate steps of the calculation. Similarly, when acting with the $\tau$-map \eqref{eq:def tau} on Gaussian functions $f(Y)$, $g(Y)$ of the oscillator with no prescription on how to extract phases from square roots may result in a sign ambiguity, $\tau(f\star g) = \pm\tau(g)\star\tau(f)$. 

In this paper, we will only give an analytic continuation prescription suitable for Weyl-ordering symbols --- which means, by virtue of \eqref{eq:stargauss}, and recalling that Weyl order corresponds to $M=0$, symbols involving the determinant of a single matrix under square root --- leaving the general problem for future work. For the same reason, it remains to be studied whether this prescription is compatible with associativity of the star products of generic symbols, as the resulting integration will in general give rise to determinants of sums and products of matrices under square root, whose treatment go beyond the scope of the present paper.

One way to analytically continue Gaussian integrals involving elements of type \eqref{eq:stargauss} for $M=0$, which will be instrumental in Sec. \ref{sec:fresnel} to distinguish $K$ from $-K$ and which overcomes the ambiguities above described, proceeds as follows: in order to establish the correspondence between the metaplectic double cover and ${\cal S}_2$, given an $\msp(2n,\mathbb{C})$ element $A$, we first establish $\exp\left(\frac{1}2 XiAX \right)$ as a reference Gaussian element, and prescribe that the eigenvalues of the matrix $iA$ are to be taken in the first Riemann sheet; then, we prescribe that the integral 
\be\label{eq:gauss int} \int_{\mathbb{R}^{2n}}\frac{d^{2n}X}{(2\pi)^n}\,\exp\left(\frac{w}2 XiAX \right) = (\det w i A)^{-1/2}=\exp[-\tfrac12(2n\Log w+\Log i\lambda_1+...+\Log i\lambda_n)] \ee
where $\lambda_i$ are the eigenvalues of the complex $2n\times 2n$ matrix $A$, $w$ is a complex factor, and $\Log$ denotes the principal logarithm (i.e., the logarithm whose imaginary part belongs to the interval $(-\pi,\pi]$). This means, in particular, that we interpret the determinant on the ($2n$-dimensional generalization of the) Riemann surface of the square root, projecting the result to the principal branch only at the end of the computation, after all square roots have been evaluated (see Appendices \ref{1dGaussian} and \ref{sec:fresnel}).

For instance, the ambiguity in applying the $\tau$-map on Gaussian function is also resolved by the prescription \eqref{eq:gauss int}: indeed, in view of its action on the oscillators, $\tau$ effectively adds $\pi$ to the phase of every eigenvalue, hence, operating on the reference Gaussian element, $\tau\left(\exp\left(\frac{i}2 XAX\right)\right)=\exp\left(-\frac{i}2 XAX\right)$, and 
\be\label{eq:tauprescr} 
(\det (-i A))^{-1/2} = \exp[-\tfrac12(2n\Log(-1)+\Log i\lambda_1+...+\Log i\lambda_n)] \ . \ee

The prescription described above is consistent with assigning the Weyl ordering symbol $\k_y$  to $[(-1)^{\Op{\mathcal{N}}_y}]_0$, where we use the shorthand notation $\Op{\mathcal{N}_y}:=\widehat{\rm Op}_R[\mathcal{N}_R(Y)]=-\frac{i}{2}\Op{y}^+ \Op{y}^-$ (see Eq. \eqref{eq:variousN}, with $\Op{X}^I\to \Op{y}^\a$), and with the analytic continuation of the delta function first mentioned in Sec. \ref{sec:Sp} (analogous considerations apply to $\k_z$ with the obvious identification $\Op{X}^I\to \Op{z}^\a$). Indeed, $(-1)^{\Op{\mathcal{N}}_y}=e^{ i\pi \Op{\mathcal{N}}_y}$ is distinguished by the properties that \cite{fibre}
\be (-1)^{\Op{\mathcal{N}}_y} (-1)^{\Op{\mathcal{N}}_y} = 1 \ , \label{Nsquared}\ee
and that, as a consequence, it is odd under $\tau$,
\be \tau((-1)^{\Op{\mathcal{N}}_y})= (-1)^{\frac{i}{2}\Op{y}^- \Op{y}^+} =(-1)^{\frac{i}{2}\Op{y}^+ \Op{y}^--1} = -(-1)^{-\Op{\mathcal{N}}_y}=-(-1)^{\Op{\mathcal{N}}_y} \ . \ee
Eq. \eqref{Nsquared} determines the identification of $[(-1)^{\Op{\mathcal{N}}_y}]_0$ with $\k_y$, as\footnote{Note that, instead, defining $\Op{\widetilde w}_y:=\widehat{\rm Op}_0[\mathcal{N}_R(Y)]=-\frac{i}{4}(\Op{y}^+ \Op{y}^-+\Op{y}^- \Op{y}^+)$, $(-1)^{\Op{\widetilde w}_y} (-1)^{\Op{\widetilde w}_y} =-1$, since, as can be seen from Eq.\eqref{eq:variousN}, $\Op{\widetilde w}_y=\Op{\mathcal{N}}_y+\frac12$, hence $(-1)^{\Op{\widetilde w}_y}=i (-1)^{\Op{\mathcal{N}}_y}$.} $\k_y\star\k_y = \pi(1) =1$. 

Consistently, interpreting the holomorphic Kleinian as an analytic continuation of the delta function in the sense already anticipated with \eqref{anald}, the action of $\tau$ on the Weyl-ordering symbol gives
\be \tau(\k_y)=\tau(2\pi\d^2(y))=2\pi\d^2(iy)=-\k_y \ . \label{taudelta}\ee
In turn, the latter equation is consistent with the above-described prescription to extract phases from the Gaussian determinants, in view of the Gaussian delta-sequence corresponding to $\k_y$ in Weyl ordering \eqref{KWO}. Indeed, the star product of two delta-sequences with generic sign $\s$ at the exponent, $\lim_{\e\to 0^+}\frac{1}{\e}\,e^{-\frac{i\s}{\e}w_y}$, where $w_y=y^+ y^- = -\frac12 y R y $, and $R$ is a two-dimensional matrix squaring to the unit matrix (see \eqref{eq:R^2=1}), gives (taking the limits after the star product and using \eqref{eq:tauprescr} in the last step)
\bea \lim_{\e\to 0^+}\frac{1}{\e}\,e^{-\frac{i\s}{\e}w_y}\star\lim_{\e'\to 0^+}\frac{1}{\e'}\,e^{-\frac{i\s'}{\e'}w_y} = \frac{1}{\sqrt{\det (i\s R)}}\frac{1}{\sqrt{\det (i\s' R)}} = \exp(-\Log \s -\Log\s')    \ ,  \label{starseq}\eea
which in particular gives $-1$ when $\s=-\s'$. In this sense, representing $\k_y=\lim_{\e\to 0^+}\frac{1}{\e}\,e^{-\frac{i}{\e}w_y}$, we can interpret
\be \tau(\k_y)=\lim_{\e\to 0^+}\frac{1}{\e}\,e^{\frac{i}{\e}w_y}=-\k_y \ , \label{taukappa}\ee
as $\k_y\star\tau(\k_y)=-1$ from \eqref{starseq}, consistently with \eqref{taudelta}.  Note also that, in the light of this identification, the direction of the limit $\e\to 0$ defining $\k_y$ is immaterial, as $\lim_{\e \to 0^-}\frac{1}{\e}\,e^{-\frac{i}{\e}w_y}=-\lim_{\e \to 0^+}\frac{1}{\e}\,e^{\frac{i}{\e}w_y}=-\tau(\k_y)=\k_y$. Hence, we can simply refer to the delta sequence 
\be \k_y=\lim_{\e \to 0}\frac{1}{\e}\,e^{-\frac{i}{\e}w_y} \ .  \ee
Thus, the phase prescription we propose, while not strictly relevant for the linear analysis that we carry on in the body of the paper, is consistent with a number of physical and mathematical requirements, as shown above.

Summarizing, working within the framework of the Vasiliev equations in various orderings and in operator form, it is natural to consider complex Gaussian symbols in $Y$ and $Z$, and to analytically continue Gaussian integrals and delta functions in such a way as to preserve the phases of any complex scaling of their arguments. The prescription here proposed for such analytic extension --- preserving algebraic properties of operators at the level of their symbols in different orderings, and coherent with the physical interpretation of notable Gaussian solutions to the Vasiliev equations --- involves considering Gaussian determinants as defined on the $2n$-dimensional generalization of the Riemann surface of the square root ${\cal S}_2^{2n}$, as in \eqref{eq:gauss int}, projecting to the first branch only after all square roots have been taken.

Let us now turn to examining the details of this analytic continuation of Gaussian integrals and delta sequences, beginning with the one-dimensional case. As we shall see, a natural framework within which to interpret the above proposed extension of delta functions to complex variable is given by a one-parameter family $\d_\varphi(z)$ of delta densities defined on a bundle of lines through the origin of ${\cal S}_2$.

\subsection{One-dimensional complex Gaussian integrals and analytic delta one-forms}\label{1dGaussian}

\begin{figure}
\centering
\includegraphics[scale=0.9]{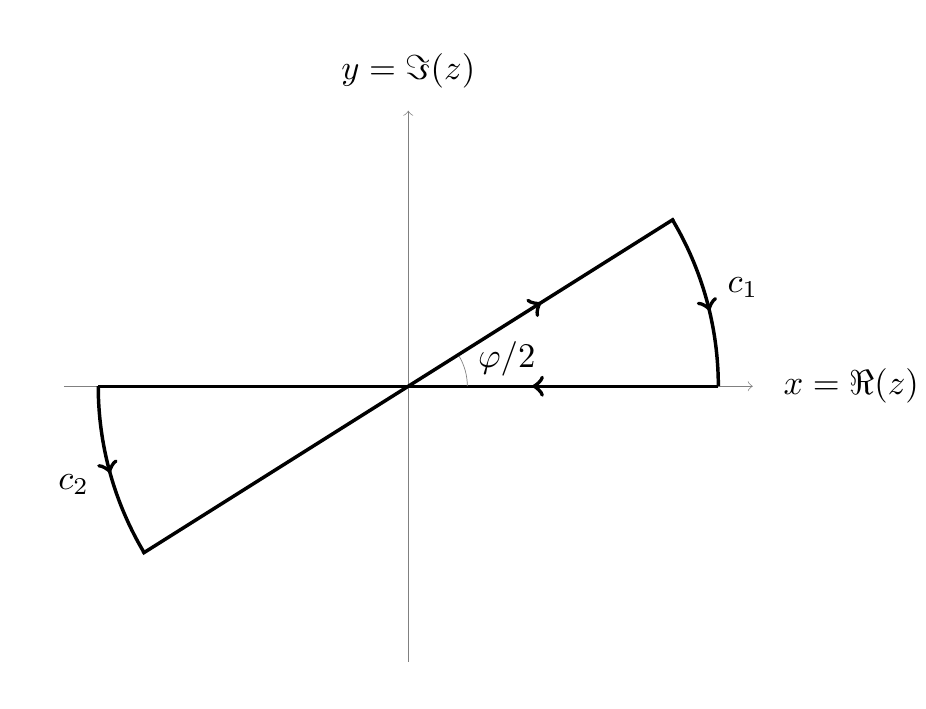}
\caption{The closed contour used in Eq. \eqref{eq:contour}.}\label{Fig:contour}
\end{figure}   

The standard one-dimensional Gaussian integral 
\begin{align}
 I_0 := \int_{-\infty}^{+\infty}dx\;\exp\left(-\tfrac{a}2 x^2\right)
  =
  \sqrt{\frac{2\pi}{a}}\,, \qquad a>0\ ,\label{orig}
\end{align}
can be deformed to the contour integral
\be I_\varphi:=\int_{e^{i\varphi/2}\mathbb{R}} dz\,\exp(-\tfrac{a}2 z^2)  =   \sqrt{\frac{2\pi}{a}}\ ,\qquad \varphi\in (-\tfrac{\pi}2,\tfrac{\pi}2)\ ,\quad {\rm Re}(a)>0\ , \label{Iphi}\ee
along the tilted real axis
\be e^{i\varphi/2}\mathbb{R}:=\{z=e^{i\varphi/2} x\,|\,x\in(-\infty,+\infty)\}\ ,\ee
as follows from Cauchy's theorem applied to the closed contour integral
\be \oint_{C} dz\,\exp(-\tfrac{a}2 z^2)  =  I_{\varphi}+I_{c_1} -I_0+I_{c_2} =0 , \label{eq:contour}\ee
where the integrals along the arcs $c_1$ and $c_2$ (see Figure \ref{Fig:contour}) vanish in the limit of infinite radius, as long as $\varphi\in (-\tfrac{\pi}2,\tfrac{\pi}2)$ and ${\rm Re}(a)>0$.
Likewise, 
\be I_\varphi=   (-1)^{N}\sqrt{\frac{2\pi}{a}}\ ,\qquad \varphi-2N\pi\in (-\tfrac{\pi}2,\tfrac{\pi}2)\ ,\quad {\rm Re}(a)>0\ , \label{Iphibis}\ee
for $N\in \mathbb{Z}$.
Adding a linear term to the exponent of the integrand in \eqref{Iphi} and completing the square, it follows that
\be \int_{e^{i\varphi/2}\mathbb{R}} dz\,\exp(-\tfrac{a}2 z^2+bz)  =  (-1)^N \sqrt{\frac{2\pi}{a}} \exp\left(\tfrac{b^2}{2a}\right)\ ,\qquad b\in \mathbb{C}\ , \label{Iphib}\ee
for $\varphi-2N\pi \in (-\tfrac{\pi}2,\tfrac{\pi}2)$ and ${\rm Re}(a)>0$.

For the reasons collected in App. \ref{appD:motivations}, we define the analytic delta one-form $\delta\in \Omega_{[1,0]}({\cal S}_2)$, where ${\cal S}_2$ is the two-sheeted Riemann surface of the square-root function, by 
\be \int_{e^{i\varphi/2}\mathbb{R}} \delta f =f(0)\ ,\qquad \varphi\in \mathbb{R}\ ,\ee
where $f:{\cal S}_2\to \mathbb{C}$ is a Fourier transformable test function on $e^{i\varphi/2}\mathbb{R}$, that is, if  $\imath_\varphi:\mathbb{R}\to {\cal S}_2$ embeds the oriented real line $\mathbb{R}$ into ${\cal S}_2$ as $e^{i\varphi/2}\mathbb{R}$, then 
\be \imath_\varphi^\ast \delta=dz \,\delta_\varphi(z)\ ,\qquad \int_{e^{i\varphi/2}\mathbb{R}} dz \,\delta_\varphi(z) f(z) =f(0)\ .\ee
The analytic properties of the delta one-form thus amounts to that the delta densities are related by
\be dz\,\d_\varphi(z)|_{z=e^{i\varphi/2}x} = dz'\,\d_{\varphi'}(z')|_{z'=e^{i\varphi'/2}x}\ ,\ee
that is,
\be \delta_\varphi(e^{i\varphi/2}x) = e^{-i(\varphi-\varphi')/2} \delta_{\varphi'}(e^{i\varphi'/2}x) \ ,\qquad \varphi,\varphi'\in\mathbb{R}\ ,\label{covariant1} \ee
while the reflection of the argument of $\delta_\varphi$ with respect to the origin of $e^{i\varphi/2}\mathbb{R}$, which is simply a reparametrization of $e^{i\varphi/2}\mathbb{R}$, yields
\be \d_\varphi(-z)=\d_\varphi(z) \,.\label{reflection} \ee
It follows that
\be  \delta_\varphi(e^{i\varphi/2}x) = e^{-i\varphi/2} \delta_{0}(x) \ ,\label{B27}\ee
where $\delta_{0}$ is thus the delta measure on $\mathbb{R}$; in particular,
\be \delta_{2\pi}(x)=\delta_{2\pi}(-x)=\delta_{2\pi}(e^{i\pi}x)=e^{-i\pi}\d_0(x)=-\d_0(x)\ ,\label{minusdelta}\ee
which is the one-dimensional counterpart of the property \eqref{taudelta} of the inner Klein operator of the Weyl algebra.

As it is evident from \eqref{Iphib} setting $a=\frac{1}{\e}$ and taking the limit $\e\to 0^+$, in the regions of convergence of the Gaussian integral the delta density $\delta_\varphi(z)$ can thus be represented by the delta sequences
\begin{align}\label{deltaseqphi}
 \delta_\varphi(z)|_{z\in e^{i\varphi/2}\mathbb{R}}&= 
  \lim_{\epsilon\to 0^+}
  \frac{(-1)^N}{\sqrt{2\pi \epsilon}}
  \exp\left(-\tfrac{1}{2\epsilon} z^2\right)|_{z\in e^{i\varphi/2}\mathbb{R}}\nonumber\\
& = \lim_{\epsilon\to0^+}
  \frac{(-1)^N}{\sqrt{2\pi \epsilon}}
  \exp\left(-\tfrac{1}{2\epsilon} e^{i\varphi}x^2\right)|_{x\in\mathbb{R}}\ ,
\end{align}
for $\varphi-2N\pi\in (-\tfrac{\pi}2,\tfrac{\pi}2)$; in particular
\be \delta_0(x)=\lim_{\epsilon\to 0^+}
  \frac{1}{\sqrt{2\pi \epsilon}}
  \exp\left(-\tfrac{1}{2\epsilon} x^2\right)|_{x\in \mathbb{R}}\ ,\label{eq:1ddeltaseqR}\ee
and
\be \delta_{2\pi}(x)=\lim_{\epsilon\to 0^+}
  \frac{-1}{\sqrt{2\pi \epsilon}}
  \exp\left(-\tfrac{1}{2\epsilon} (e^{i\pi}x)^2\right)|_{x\in \mathbb{R}}=-\delta_0(x)|_{x\in \mathbb{R}}\ .\ee
Due to property \eqref{covariant1}, in fact, the direction of the limit $\e \to 0$ is actually immaterial, and we may analytically continue the delta sequences, which yields
\begin{align}
  \label{eq:1ddeltaseq2}
 \delta_\varphi(z)= \lim_{\epsilon\to0}
  \frac{1}{\sqrt{2\pi\epsilon}}
  \exp\left(-\tfrac{1}{2\epsilon}z^2\right)\ ,\qquad \epsilon \in \C\setminus\{0\}\ .
\end{align}
Conversely, viewing Eq. \eqref{eq:1ddeltaseq2} as a definition, property \eqref{covariant1} follows upon absorbing phases into the complex dummy parameters, viz. 
\begin{align}  \delta_\varphi(z)|_{z=e^{i\varphi/2}x} &= \lim_{\epsilon\to0}
  \frac{1}{\sqrt{2\pi\epsilon}}
  \exp\left(-\tfrac{1}{2\epsilon}z^2\right)|_{z=e^{i\varphi/2}x}= \lim_{\epsilon\to0}
  \frac{1}{\sqrt{2\pi\epsilon}}
  \exp\left(-\tfrac{1}{2\epsilon}e^{i(\varphi-\varphi')}z^{\prime 2}\right)|_{z'=e^{i\varphi'/2}x}\nonumber\\ &=  \lim_{\epsilon'\to0}e^{-i(\varphi-\varphi')/2}
  \frac{1}{\sqrt{2\pi\epsilon'}}
  \exp\left(-\tfrac{1}{2\epsilon'}z^{\prime 2}\right)|_{z'=e^{i\varphi'/2}x} = e^{-i(\varphi-\varphi')/2} \delta_{\varphi'}(z')|_{z'=e^{i\varphi'/2}x} \ ,\end{align}
where $x\in \mathbb{R}$ and $\epsilon'=\e e^{-i(\varphi-\varphi')}$.
We note that while $I_\varphi$ does not admit any unambigous analytic continuation from its disjoint regions of convergence to $\R$, the delta one-form  $\delta$ is analytic on ${\cal S}_2$ (rather than $\C$) as can be seen from Eq. \eqref{B27}, from which it follows that
\be \imath_\varphi^\ast \delta=\imath_{\varphi'}^\ast \delta\ ,\qquad \varphi-\varphi'\in 4\pi \mathbb{Z}\ ,\ee
that is, $\delta|_{e^{i\varphi/2}\mathbb{R}}$ are distinct on the real lines $e^{i\varphi/2}\mathbb{R}$ with $\varphi\in [0,4\pi)$ passing through the origin, which sweep through the square-root Riemann surface ${\cal S}_2$ with ramification point at the origin.
More precisely, viewing ${\cal S}_2$ as consisting of two Riemann sheets in which $z=|z|e^{i\phi}$ with $\phi\in(-\pi,\pi]$ in the first sheet and $\phi\in(\pi,3\pi]$ in the second, the lines $e^{i\varphi/2}\R$ sweep out the entire ${\cal S}_2$  as follows: lines with $\varphi\in [0,2\pi)$ sweep the first sheet; lines with $\varphi\in [2\pi,4\pi)$ stretch between the two sheets; and lines with $\varphi\in [4\pi,6\pi)$ sweep the second sheet; and so on.

Centering the delta form on a specific point $z'=e^{i\varphi/2}x'$ of $e^{i\varphi/2}\mathbb{R}$, we define
\be \int_{e^{i\varphi/2}\mathbb{R}}dz\,\d_\varphi(z-z')\,f(z)=f(z')\ .\ee
Setting instead $a=\epsilon$ in \eqref{Iphib}, again taking the limit $\e\to 0$ and using \eqref{eq:1ddeltaseq2}, one obtains the (generalized) Fourier representation of any $\d_{\theta}(k)$, $k=e^{i\theta/2}k_0$, $k_0\in\mathbb{R}$:
\bea  & \displaystyle \int_{e^{i\varphi/2}\mathbb{R}}dz\;
  \exp\left(ikz\right)
   =
  \lim_{\epsilon\to0}\int_{e^{i\varphi/2}\mathbb{R}}dz\;
  \exp\left(ik z-\frac{\e}2 z^2\right) = \lim_{\epsilon\to 0}
  \sqrt{\frac{2\pi}{\epsilon}}
  \exp\left(-\tfrac{k^2}{2\epsilon}\right)
  = 2\pi\delta_\theta(k)   \ ,\label{FTDC}
 \eea
for any line $e^{i\varphi/2}\mathbb{R}$. It is important to note that, here and in the body of the paper, we are \emph{defining} the Fourier representation of this complex generalization of the delta function via the above Gaussian dressing, and, as above, we analytically extend the result to any phase angle $\varphi$.
With this definition, the Fourier representation respects the specific properties that characterize the family $\d_\varphi$: for instance, given $k=e^{i\theta/2}k_0$ and $k'=e^{i\theta'/2}k_0$,
\bea  \delta_{\theta}(k) &=&
 \int_{e^{i\varphi/2}\mathbb{R}}\frac{dz}{2\pi}\;
  \exp(ikz) = \int_{e^{i\varphi/2}\mathbb{R}}\frac{dz}{2\pi}\;
  \exp(ie^{i(\theta-\theta')/2}k'z) \nonumber\\
  &=& e^{-i(\theta-\theta')/2}\int_{e^{i\varphi'/2}\mathbb{R}}\frac{dz'}{2\pi}\;
  \exp(ik'z') = e^{-i(\theta-\theta')/2}\d_{\theta'}(k') \ ,
 \eea
where $z'=e^{i\varphi'/2}x$, $\varphi'=\frac{\varphi+\theta-\theta'}{2}$, and the result follows from the independence of the Fourier representation of $\d_\theta(k)$ from the integration line.

\subsection{Multi-dimensional complex Gaussian integrals and analytic delta forms}\label{sec:fresnel}

Armed with the framework of delta forms on ${\cal S}_2$ described above, it is easy to extend the definition of the corresponding delta sequences in the $2n$-dimensional case. Starting from the standard multi-dimensional Gaussian integral,
\begin{align}
  \label{standardgauss}
  \int_{\R^k}d^kX\exp\left(-\tfrac{1}{2}X^{I}A_{IJ}X^{J}\right)
  =
  \frac{2\pi}{\sqdet{A}}
  \,,
\end{align}
where $A$ is a positive-definite symmetric real matrix, an immediate generalization of the one-dimensional results above which is of relevance in the body of the paper is the generalized multi-dimensional Fresnel integral (where we use the implicit indices matrix notation \eqref{implicit1}-\eqref{implicit2})
\begin{align}
 \int_{\R^{2n}} d^{2n}X\;\exp\left(\tfrac{i}{2}XRX\right)
 = \frac{(2\pi)^n}{\sqrt{\det iR}}=(2\pi)^n\exp\left[-\frac12\left(n\Log i + n\Log(-i)\right)\right]=
   (2\pi)^n
  \ ,\label{startingpt}
\end{align}
in terms of a $2n$-dimensional real matrix $R$ that is a square root of the identity in the sense of \eqref{eq:R^2=1}. Again, \eqref{eq:gauss int} was used, with $iR$ chosen as ``reference matrix''. Eq. \eqref{startingpt} is a particular case of the more general Fresnel-type integral
\be I_Q:=\int_{\R^{2n}} d^{2n}X\;\exp\left(-\tfrac{i}{2}X^I Q_{IJ} X^J\right) \ , \ee
where $Q$ is a symmetric real matrix. The evaluation of this integral starting from our one-dimensional results can be carried out by diagonalizing $Q_{IJ}$ via a special orthogonal matrix to $Q_{\rm{diag}}=\left(\begin{array}{ccc}  \s_1 q_1 &\, & \,\\\, & \ddots & \,\\\, &\, &  \s_{2n}q_{2n}\end{array}\right)$, $\s_k=\{1,-1\}$, $q_k\in \mathbb{R}^+$, and by applying the results of App. \ref{1dGaussian} to each of the resulting one-dimensional Gaussian integral factors, thereby obtaining
\begin{align} I_Q & = \prod_{K=1}^{2n}\int_{-\infty}^{+\infty}d X^{\prime K} e^{-\tfrac{i}2(\s_K q_K)(X^{\prime K})^2} = \prod_{K=1}^{2n}\sqrt{\frac{2\pi}{\s_{K}i q_{K}}} \nonumber\\
& =(2\pi)^n\exp\left[-\frac12 \sum_{K=1}^{2n}\Log(\s_K i q_K)\right]=\frac{(2\pi)^n}{\sqrt{|\det Q|}}\,\exp\left(-\tfrac{i\pi}{4}{\rm sgn}(Q)\right)\ ,
\end{align}
where ${\rm sgn}(Q)$ is the signature of $Q$, using the prescription \eqref{eq:gauss int} for extracting the square root\footnote{Indeed, the prescription \eqref{eq:gauss int} is the one usually employed when evaluating generalized Fresnel integrals (see for instance \cite{Guillemin:1990ew}).}. 

Then, we can implement in a $2n$-dimensional set-up the same generalization of the Gaussian integration \eqref{Iphi}, by rotating the integration variables with a $GL(2n,\C)$ transformation, and performing the Gaussian integral on the resulting $2n$-(real)dimensional surface
\begin{align}
  S\R^{2n}
  :&=
  \left\{SX,\;X\in\R^{2n}\right\}
  \,,&
  S\in GL(2n,\C) \ ,
\end{align}
which, extending the one-dimensional case treated above, we view as embedded in the multi-dimensional square-root Riemann surface ${\cal S}_2^{2n}$. Thus, the generalization of \eqref{Iphi} applied to our reference integral \eqref{startingpt} is
\be \int_{S\R^{2n}} d^{2n}X' \exp\left(\frac{i}2 X'RX'\right)=\det S \int_{\R^{2n}} d^{2n}X \exp\left(-\frac{i}2 XS^tRSX\right)=\frac{(2\pi)^n}{\sqrt{\det iR}}=(2\pi)^n \ .\ee
Completing the square at the exponent,
\be \int_{S\R^{2n}} d^{2n}X' \exp\left(\frac{i}2 X'RX'+iKX'\right)=\frac{(2\pi)^n}{\sqrt{\det iR}}\exp\left(\frac{i}2 KR^{-1}K\right)=(2\pi)^n \exp\left(\frac{i}2 KR^{-1}K\right)\ ,\label{GaussianSR}\ee
and, rescaling $R$ with $\frac{1}{\e}$ and taking the limit $\e\to 0$ along any direction one finds that the sequence
\begin{align}
  \label{eq:deltaS gauss}
  \lim_{\epsilon\to0}\frac{1}{\e^n}
  \exp\left(\tfrac{i}{2\epsilon}X'RX'\right)
 =(2\pi)^n \delta^{2n}_S(X')
\end{align}
defines a $2n$-dimensional generalization of the delta density $\d_\varphi(z)$ on the slice $S\mathbb{R}^{2n}$. Analogously, integrating it on $\mathbb{R}^{2n}$ gives
\bea &\displaystyle \lim_{\epsilon\to0}\frac{1}{\e^n} \int_{\R^{2n}} d^{2n}X \exp\left(\tfrac{i}{2\epsilon}X'RX'+iKX \right)  = \lim_{\epsilon\to0}\frac{1}{\e^n}   \int_{\R^{2n}} d^{2n}X \exp\left(-\frac{i}2 XS^tRSX+iKX\right)&\nonumber\\
& \displaystyle \ \ \ \ \ =\frac{(2\pi)^n}{\det S}\,\lim_{\epsilon\to0}\exp\left(-\frac{i\epsilon}2 K(S^{-1})^tR^{-1}S^{-1}K\right)=\frac{(2\pi)^n}{\det S}& \ \,\eea
which means that the relation between $\d^{2n}_S$ and $\d^{2n}_1$, the delta function on $\mathbb{R}^{2n}$, is
\begin{align}
  \label{eq:deltaSgaussR}
 \d^{2n}_S(X')
 \equiv \delta^{2n}_S(SX) =
 \frac{1}{\det S} \,\delta^{2n}_1(X)
  \,,
\end{align}
which is the $2n$-dimensional analogue of the property \eqref{B27}. In other words, \eqref{eq:deltaS gauss} defines a family of delta densities which generalizes the family $\d_\varphi$ of the one-dimensional case and which, together, extend the notion of delta function to the $2n$-dimensional complex case --- more precisely to be thought of as living on ${\cal S}_2^{2n}$. It is in terms of delta $2n$-forms on ${\cal S}_2^{2n}$ and corresponding delta densities that the properties of the Kleinians $\k_y$ and $\k_z$, discussed in App. \ref{appD:motivations} and first introduced in \cite{2011}, can be naturally understood: in particular, the  preservation of the phase of the argument of the Kleinian $\k_y$ seen, e.g., in \eqref{anald} and \eqref{taudelta} (idem for $\k_z$ and their complex conjugates), acquires a natural explanation when thinking of the Kleinians as the extension of the delta function to the $2$-dimensional complex case defined by the family $\d^2_S$ described above, related to each other via \eqref{eq:deltaSgaussR}. 

We can thus think of $\delta^{2n}_S(X')$ as the delta function with respect to $\int_{S\R^{2n}} d^{2n}X'$,
\begin{align}
  \int_{S\R^{2n}}d^{2n}X'\;\delta^{2n}_S(X'-X'_0)\;f(X')
  =
  f(X'_0)
  \,.
\end{align}
Like in the one-dimensional case (see Eqs. \eqref{covariant1}-\eqref{reflection}), property  \eqref{eq:deltaSgaussR} relating the delta functions living on different complex slices $S\mathbb{R}^{2n}$ should be distinguished from rotations and reflections within a given slice, which leave $\d_S^{2n}$ invariant,
\be \d_S^{2n}(O X')=\frac{1}{|\det O|}\d_S^{2n}(X') = \d_S^{2n}(X') \ ,\ee
for any real orthogonal matrix $O$ acting on $X'$. 

The Fourier representation instead arises in the opposite limit of Eq. \eqref{GaussianSR}, rescaling $R$ by a vanishing parameter $\e$. More precisely, for any $K'= \widetilde{S}K$, by inserting a Gaussian factor in the integrand and computing the complex Gaussian integral as above one obtains
\begin{align}
  \int_{S\R^{2n}}d^{2n}X'\;\exp\left(iK'X'\right)
  &=
  \lim_{\epsilon\to0}
  \int_{S\R^{2n}}d^{2n}X'\;\exp\left(iK'X'+\tfrac{i}{2}\e X'RX'\right)
  \nonumber\\
  &=
  \lim_{\epsilon\to0} \frac{(2\pi)^n}{\sqrt{\det i\e R}}\exp\left(\frac{i}{2\e}K'R^{-1}K'\right) =  (2\pi)^n \d_{\widetilde{S}}^{2n}(K')
  \,,
\end{align}
where $R=R^{-1}$ (see \eqref{eq:R^2=1}) was used. In other words, one can expand this analytic delta function in plane waves along any slice of ${\cal S}_2^{2n}$. We stress again that, technically, the above complex generalization of the Fourier representation of the delta function rests on its definition via the above Gaussian dressing.
As a consequence, like in the one-dimensional case, the Fourier representation is compatible with property \eqref{eq:deltaSgaussR} of this family of delta functions, e.g,.
\be \d^{2n}_S(SK) = \int_{\R^{2n}}\frac{d^{2n}X}{(2\pi)^n} \exp\left(-iKS^tX\right) = \frac{1}{\det S} \int_{S\R^{2n}}d^{2n}X'\;\exp\left(iKX'\right)= \frac{1}{\det S}\d^{2n}_1(K) \ .\ee

Thus, throughout the paper, we use the Gaussian integration formula \eqref{eq:gauss int}
and the associated delta-sequence
\begin{align}
  \label{eq:delta gauss}
  \lim_{\epsilon\to0}\frac{1}{\e^n}
  \exp\left(\tfrac{i}{2\epsilon}XAX\right)
 =\frac{(2\pi)^n}{\sqrt{\det iA}} \delta^{2n}(X)
 \,,
\end{align}
where we note that we frequently omit the label $S$ of the slice $S\R^{2n}$ letting it be inferred by the space that the argument belongs to. 

To summarize, it is natural to consider a complex analytic continuation of the delta function to represent the Kleinians $\Op \k_y$ and $\Op \k_z$ (as well as their complex conjugates) in Weyl ordering --- both for continuity with infinitesimally close ordering prescriptions, in which the Kleinians are represented by regular Gaussians, and for consistency with algebraic properties of the operators $(-1)^{\Op {\cal N}}$ on the $y$ or $z$ space as well as with reality conditions on solutions of the Vasiliev equations expanded over Gaussians. The resulting complex extension of the delta function admits a natural formulation in terms of analytic delta $2n$-forms projecting to delta densities defined on ${\cal S}_2^{2n}$, which can be represented via Gaussian delta sequences as well as via generalized Fourier transform. 

\section{One-parameter symplectic subgroups} \label{1psymp}
In this appendix we describe how the subgroup of $Sp(2n,\R)$ generated by an element of its infinitesimal algebra is described in the context of symbol calculus.
The purpose of App. \ref{app:stargauss} is to prove the general formula \eqref{eq:stargauss} for a one-dimensional group defined as the exponential of a symplectic operator.
In App. \ref{app:X+-}, we discuss some of its properties in the cases  of a non-degenerate matrix $G$.
In App. \ref{app:U(1) ex}, we give an explicit example to illustrate the projectivity of the metaplectic representation \eqref{eq:def OpSpH}.

\subsection{$M$-ordered Gaussians}
\label{app:stargauss}

This Subsection is dedicated to deriving the expression of the $M$-ordered symbol of an operator exponential
$\exp\left(\tfrac{i\xi}{2}\OpX G\OpX\right)$, where $\xi$ is a complex number.
Singling out the scalar prefactor $\xi$ in the exponent allows to write the equation
\begin{align}
  \left(\frac{\partial}{\partial\xi}-\frac{i}{2}\OpX G\OpX\right)
  \exp\left(\tfrac{i\xi}{2}\OpX G\OpX\right)
  =0
  \,.
\end{align}
In the language of the star-product \eqref{eq:starM bopp}, one gets the differential equation
\begin{align}
  0
  &=
  \left(\frac{\partial}{\partial\xi}
  -\frac{i}{2}\left(X+i\partial_X(M+1)\right)G\left(X-i(M-1)\partial_X\right)
  \right)
  \MoSym{M}{\exp\left(\tfrac{i\xi}{2}\OpX G\OpX\right)}
  \nonumber\\&=
  \left(\frac{\partial}{\partial\xi}
  -\frac{i}{2}\partial_X(M+1)G(M-1)\partial_X
  -XG(M-1)\partial_X
  -\frac{i}{2}XGX
  -\frac{1}{2}\Tr{(M+1)G}
  \right)
  \MoSym{M}{e^{\tfrac{i\xi}{2}\OpX G\OpX}}
  \,,
\end{align}
to be solved with the initial condition
\begin{align}
  \alpha(0)
  &=
  1
  \,,&
  B(0)
  &=
  0
  \,.
\end{align}
One can make the following ansatz
\begin{align}
  \MoSym{M}{\exp\left(\tfrac{i\xi}{2}\OpX G\OpX\right)}
  =
  \alpha(\xi)\exp\left(\tfrac{i}{2}XB(\xi)X\right)
  \,,
\end{align}
where $\alpha(\xi)$ is a number and $B(\xi)$ is a matrix.
The part of the resulting equation that is bilinear in $\OpX$ gives an equation for $B$:
\begin{align}
  \partial_\xi B
  &=
  \big(1+B(M+1)\big)G\big(1+(M-1)B\big)
  \,.
\end{align}
A solution that works for any generator $G$ (i.e. excluding possible projectors that may preserve the equation for some particular $G$) is
\begin{align}
  B(\xi)=\left((\tanh \xi G)^{-1}-M\right)^{-1}
  \,.
\end{align}
This can then be plugged in the remaining equation that becomes 
\begin{align}
  \partial_\xi\alpha
  &=
  \tfrac{1}{2}\Tr\left((M+1)G\big(1+(M-1)B\big)\right)\alpha
  \nonumber\\&=
  \tfrac{1}{2}\Tr\left((M+1)G
  \left(\cosh(\xi G)-\sinh(\xi G)\right)
  \left(\cosh(\xi G)-M\sinh(\xi G)\right)^{-1}
  \right)\alpha
  \nonumber\\&=
  \tfrac{1}{2}\Tr\left(
  \left(M\cosh(\xi G)-\sinh(\xi G)\right)G
  \left(\cosh(\xi G)-M\sinh(\xi G)\right)^{-1}
  \right)\alpha
  \nonumber\\&=
  -\tfrac{1}{2}\Tr\left(\partial_\xi\log\left(
  \cosh(\xi G)-M\sinh(\xi G)
  \right)\right)\alpha
  \,,
\end{align}
whose solution compatible with the initial condition finally gives
\begin{align}\label{}
  \MoSym{M}{\exp\left(\tfrac{i\xi}{2}\OpX G\OpX\right)}
  =
  \frac{1}{\sqdet{\cosh(\xi G)-M\sinh(\xi G)}}
  \exp\left(
  \tfrac{i}{2}X\left(
  \tanh(\xi G)^{-1}-M
  \right)^{-1}X\right)
  \,.
\end{align}
This concludes the proof of eq. \eqref{eq:stargauss}.

\subsection{Number operators}
\label{app:X+-}
The symplectic coordinates $\Op{X}^I$ can be split into canonical coordinates
\be
  \label{eq:XpmR}
  \Op{X}_{\epsilon R}
  :=
  P_{\epsilon R}\Op{X}\ ,\qquad P_{\epsilon R}
  :=
  \frac{1}{2}\left(1+\epsilon R\right)\ ,\qquad \epsilon=\pm1\ ,
\ee
obeying
\be [\Op{X}_{\epsilon R}^{I},\Op{X}_{\epsilon' R}^{J}]= 2i\epsilon'  \delta_{\epsilon,-\epsilon'} P_{\epsilon R}^{IJ}\ ,\qquad RX_{\epsilon R}=
  \epsilon X_{\epsilon R}
  \,,
\ee
using a complete pair of orthogonal projectors $P_{\pm R}$ formed out of a symmetric square root $R$ of the unit, viz.
\be
  \label{eq:R^2=1}
  R_{IJ}=R_{JI}
  \,,\qquad
  R_I{}^J R_J{}^K=\delta_I^K
  \,.
\ee
From $\Tr R:=R_I{}^I=0$ it follows that $R$ has eigenvalues $\pm 1$ with multiplicity $n$, and hence
\be
  \label{eq:det R}
  \det R=(-1)^n
  \,.
\ee
From the Baker--Campbell--Hausdorff lemma \eqref{eq:BCH}, it follows that the quantization map $\widehat{\rm Op}_R$, given by \eqref{eq:OpM(EK)} with $M=R$, obeys
\be \widehat{\rm Op}_R[e^{iKX}]\equiv e^{iK\OpX+\tfrac{i}2KRK}=e^{iK\OpX_{+R}} e^{iK\OpX_{-R}}\ ,
\ee
such that \eqref{eq:WeylMapM} implies
\be \widehat{\rm Op}_R[f_+(X_{+R})f_-(X_{-R})]
=\widehat{\rm Op}_R[f_+(X_{+R})] \widehat{\rm Op}_R[f_-(X_{-R})]
=\widehat{\rm Op}_0[f_+(X_{+R})] \widehat{\rm Op}_0[f_-(X_{-R})]\ ,\ee
that is, $R$-ordering is a normal ordering.
The associated Hamiltonian function
\be
\mathcal{N}_R := \frac{i}{4} XRX=-\frac{i}2 X^I_{+R} X_{-R,I}\ ,\ee
generates the similarity transformations
\begin{align}\label{U1action}
  \exp\left(i\theta\widehat{\rm Op}_0[\mathcal{N}_R]\right)
  \Op{X}_{\epsilon R}
  \exp\left(-i\theta\widehat{\rm Op}_0[\mathcal{N}_R]\right)
  =
  \exp\left(i\epsilon\theta\right)
  \Op{X}_{\epsilon R}
  \,,
\end{align}
as can be seen from
\be\label{eq:variousN} \widehat{\rm Op}_0[\mathcal{N}_R]=\frac{i}{4} \OpX R\OpX\ ,\qquad \widehat{\rm Op}_R[\mathcal{N}_R]=-\frac{i}2 \OpX^I_{+R} \OpX_{-R,I}\ ,\qquad \widehat{\rm Op}_0[\mathcal{N}_R]=\widehat{\rm Op}_R[\mathcal{N}_R]+\frac{n}2\ ,\ee
which implies
\be \left[\widehat{\rm Op}_0[\mathcal{N}_R],\,\OpX_{\epsilon R}\right]
  =
  \epsilon\OpX_{\epsilon R}
  \,.
\ee
As is well known (see for instance \cite{CSM,Woit:2017vqo} for more details), $\exp\left(i\theta\widehat{\rm Op}_0[\mathcal{N}_R]\right)$ generates a $U(1)$ subgroup (of the double-covering $Mp(2n,\mathbb{R})$ of $Sp(2n,\mathbb{R})$) with noteworthy properties: at $\theta=\tfrac{\pi}2$, it exchanges coordinates and momenta (as it appears from \eqref{U1action}), thereby corresponding (up to a phase) to a Fourier transform; at $\theta=\pi$ it changes sign to $\Op{X}$ via adjoint action;
and, as it appears from the shift in the third equation in \eqref{eq:variousN}, for $\theta=2\pi$ the operator $\exp\left(2\pi i\widehat{\rm Op}_0[\mathcal{N}_R]\right)= (-1)^n \exp\left(2\pi i\widehat{\rm Op}_R[\mathcal{N}_R]\right)$, giving rise to the distinctive sign of the metaplectic representation in its action on a Fock space.

Its $M$-ordering symbol is
\begin{align}
  \label{eq:Mo exp(ithN)}
  \MoSym{M}{\exp\left(i\theta\widehat{\rm Op}_0[\mathcal{N}_R]\right)}
  =
  \frac{\exp\left(
  -\tfrac{i}{2}X\left(
  i\cotan\left(\tfrac{\theta}{2}\right)R+M
  \right)^{-1}X\right)}{\sqdet{
  \cos\left(\tfrac{\theta}{2}\right)
  -i\sin\left(\tfrac{\theta}{2}\right)MR}}
  \,,
\end{align}
in accordance with Eq. \eqref{eq:stargauss}, and the prescription we choose to extract the square root of complex Gaussian symbols as discussed in App. \ref{appD:motivations} and exhibited in Eq. \eqref{eq:gauss int}, is such as to preserve the characteristic behaviour of the corresponding operator as a $2\pi$-periodic or $2\pi$-anti-periodic function of $\theta$ for $n$ even or odd, respectively\footnote{%
For reasons mentioned in Sec. \ref{sec:Sp} and explained in detail in App. \ref{app:anal},  to avoid ambiguities we extract square roots of Gaussian determinants as in Eq. \eqref{eq:gauss int}, projecting the result to the principal branch only at the end of the computation. With this prescription, for $\theta =2\pi$ \eqref{eq:Mo exp(ithN)} gives $(\det \cos\pi)^{-1/2}=((-1)^{2n})^{-1/2}=e^{-i\pi n}=(-1)^n$. Note that a generic different choice, such as, for instance, extracting the principal square root, would not preserve this sign.}.
Note that, because of the singularity that appears when plugging $U=-1$ into Eq. \eqref{eq:OpSp U}, there is no canonical sign associated with the group element $\OpSpH{-1}{0}$.
By convention, we decide to choose the representative
\begin{align}
  \label{eq:U(-1) seqbis}
  \OpSpH{-1}0
  :=
  \lim_{\theta\to\pi}\OpSpH{e^{i\theta R}}0
  =
  \exp\left(-i\pi\widehat{\rm Op}_0[\mathcal{N}_R]\right)
  \,.
\end{align}
From the considerations above it follows in particular that,
\be
\label{eq:def KR}
  \Op{K}_{R}
  :=
  \exp\left(-i\pi\widehat{\rm Op}_R[\mathcal{N}_R]\right)
  =
  (-i)^n\exp\left(-i\pi\widehat{\rm Op}_0[\mathcal{N}_R]\right)
  \equiv
  (-i)^n \;\widehat{\mathcal{U}}[-1,0]
  \,,
\ee
is an inner Klein operator, viz.
\be \OpX\;\Op{K}_{R}
  =
  -\Op{K}_{R}\;\OpX
  \,,\qquad
  (\Op{K}_R)^2
  =1
  \,.
\ee
Clearly, $-\Op{K}_R$ also satisfies these defining properties.
The $M$-ordering symbol of the Kleinian $\Op{K}_R$ is
\be
  \label{eq:OpK_A}
  \MoSym{M}{\Op{K}_R}
  =
  (-i)^n\MoSym{M}{\widehat{\mathcal{U}}[-1,0]}
  =
  \frac1{\sqdet{-iM}}
  \exp\left(
  -\tfrac{i}{2}XM^{-1}X\right)
  \,,
\ee
independent of the polarization.
In particular, its Weyl-ordering symbol is
\be
  \MoSym{0}{\Op{K}_R}
  = (-i)^n\MoSym{0}{\widehat{\mathcal{U}}[-1,0]}=
  (2\pi )^n\delta^{2n}(X)
  \,,
\ee
where the analytical delta function is defined in App. \ref{app:anal}.
Choosing
\be (R_{IJ})^\ast=\sigma R_{IJ}\ ,\qquad \sigma=\pm1\ ,\ee
gives rise to compact and non-compact real forms in which
\be (\Op{X}_{\epsilon R})^\dagger =\Op{X}_{\sigma\epsilon R}\,\qquad (\widehat{\rm Op}_0[\mathcal{N}_R])^\dagger=-\sigma \widehat{\rm Op}_0[\mathcal{N}_R]\ ,\ee
and
\be \exp\left(i\theta\widehat{\rm Op}_0[\mathcal{N}_R]\right)\in \left\{\begin{array}{cc} SO(2)\subset Sp(2n,\mathbb{R})& \mbox{for $\sigma=-1$ and $\theta\in\mathbb{R}$}\ ,\\
SO(1,1)\subset Sp(2n,\mathbb{R})& \mbox{for $\sigma=+1$ and $\theta\in i\mathbb{R}$}\ ,\end{array}\right.\ .\ee

\subsection{The two-dimensional metaplectic groups}\label{app:U(1) ex} 

For $n=1$, and working with Weyl-ordering symbols, it follows from 
\be \det (1+U)=2+{\rm tr}\,U\ ,\ee
that the assignment of patches for $Mp(2;\C)$ and its $Mp(2;\R)$ restriction is controlled by the choice of argument of $2+{\rm tr}\,Pr(g)$, where $g\in Mp(2;\C)$.
For example, in the image of the exponential map it follows from 
\be G(\alpha,\beta,\gamma)_I{}^J=\left[\begin{array}{cc} \alpha&\beta\\ \gamma&-\alpha\end{array}\right]\in \mathfrak{sp}(2;\C)\ ,\qquad \exp(-2G)=\cosh \omega-2G \frac{\sinh \omega}{\omega}\ ,\qquad \omega^2 =-4\det G\ ,\ee
that $\tfrac12 {\rm tr}\, e^{-2G}=\cosh \omega$; hence, 
\be \left[\OpSp{e^{-2G}}\right]_0(X) =\sqrt{\frac{2}{1+\cosh\omega}} \exp\left(\frac{\sinh \omega}{\omega} \frac{iX G X}{1+\cosh\omega}\right)\ .\ee
Restricting to $Sp(2;\R)$, one has $\tfrac12 {\rm tr}\, e^{-2G}|_{Sp(2;\R)}\geqslant -1$.

In particular, the $U(1)$ subgroup elements
\be e^{-2G(0,-\theta/2,\theta/2)}=\left[\begin{array}{cc} \cos \theta&\sin\theta\\ -\sin\theta&\cos\theta\end{array}\right]\ ,\qquad \omega^2=-\theta^2\ ,\label{U1Sp}\ee
are represented by the Weyl-ordering symbol
\begin{align}
  \label{eq:OpSp(e^(-2G(0,-th/2,th/2)))}
  \left[\OpSp{e^{-2G(0,-\theta/2,\theta/2)}}\right]_0(X) &=\sqrt{\frac{2}{1+\cos\theta}} \exp\left(\frac{\sin \theta}{\theta} \frac{iX G(0,-\theta/2,\theta/2) X}{1+\cos\theta}\right)\nonumber\\&=\sqrt{\frac{2}{1+\cos\theta}} \exp\left(\frac{i\sin \theta}{2(1+\cos\theta)} \left((X^1)^2+ (X^2)^2 \right)\right)\ ,
\end{align}
which can be extended across $Mp(2;\R)$ by using the real-analytic extension of $\sqrt{1+\cos\theta}$ defined by 
\be \left. \sqrt{1+\cos\theta}\right|_{\theta=2\pi N}:= \sqrt{2}(-1)^N\ ,\qquad \left.\frac{d}{d\theta} \sqrt{1+\cos\theta}\right|_{\theta=(2N-1)\pi}:= \tfrac1{\sqrt{2}}(-1)^N\ ,\qquad N\in\mathbb{Z}\ ,\ee
that is, by choosing the square-root branch such that $\sqrt{1+\cos\theta}=\sqrt{2} \cos (\tfrac{\theta}2)$.
Thus, letting $\eta:= \pi-\theta$, such that $\sqrt{1+\cos \theta}=\eta/\sqrt{2}+O(\eta^3)$ and $\sin\theta=\eta+O(\eta^3)$, and using the analytic delta sequence \eqref{eq:1ddeltaseqR} (taking into account also \eqref{eq:1ddeltaseq2}), it follows that
\begin{align} \lim_{\theta\to\pi} \left[\OpSp{e^{-2G(0,-\theta/2,\theta/2)}}\right]_0(X)&=\lim_{\eta\to 0}\tfrac{2}{\eta} e^{\frac{i}{\eta}\left((X^1)^2+ (X^2)^2\right)}\nonumber\\&= \lim_{\epsilon\to 0}\tfrac{i}{\epsilon} e^{-\frac{1}{2\epsilon}\left((X^1)^2+ (X^2)^2\right)}=2\pi i \delta_0(X^1)\delta_0(X^2)\ ,\end{align}
in accordance with \eqref{eq:K0}.
Likewise, letting $\eta':= \pi+\theta$, such that $\sqrt{1+\cos \theta}=\eta'/\sqrt{2}+O(\eta'^3)$ and $\sin\theta=-\eta'+O(\eta^3)$, it follows that
\begin{align} \lim_{\theta\to-\pi} \left[\OpSp{e^{-2G(0,-\theta/2,\theta/2)}}\right]_0(X)&=\lim_{\eta'\to 0}\tfrac{2}{\eta'} e^{-\frac{i}{\eta}\left((X^1)^2+ (X^2)^2\right)}\nonumber\\&= -\lim_{\epsilon\to 0}\tfrac{i}{\epsilon} e^{-\frac{1}{2\epsilon}\left((X^1)^2+ (X^2)^2\right)}=-2\pi i \delta_0(X^1)\delta_0(X^2)\ ,\end{align}
also in accordance with \eqref{eq:K0}.

The $Sp(2;\C)$ group element $-I_2$ can also be approached along the $SO(1,1)\subset Sp(2;\R)$ consisting of the elements
\be U(\alpha)=-I_2 e^{-2G(-\alpha/2,0,0)}=\left[\begin{array}{cc} -e^{\alpha}&0\\ 0&-e^{-\alpha}\end{array}\right]\ ,\ee
with $2+{\rm tr}\, U(\alpha)<0$; close to $-I_2$, these elements are represented by
\be [\widehat{\cal U}(U(\alpha)]_0(X)= \frac{2}{\sqrt{-\alpha^2}}e^{\frac{2i}{\alpha} X^1 X^2}(1+O(\alpha))\ ,\ee
with limit 
\be \lim_{\alpha\to 0}[\widehat{\cal U}(U(\alpha)]_0(X)=\pm 2\pi i \delta_0(X^1)\delta_0(X^2)\ ,\ee
depending on the choice of branch for the square root.

Thus, the Weyl-ordering representation map can be extended real-analytically from $U(1)\cup SO(1,1)$ into 
\be Mp(2;\R)\cong Mp(2;\R)_+\cup Mp(2;\R)_-\ ,\ee
by using the two patches defined by
\be {\rm arg} (2+{\rm tr}(Pr(g)))|_{g\in Mp(2;\R)_+}\in \{0,\pi\}\ ,\qquad {\rm arg} (2+{\rm tr}(Pr(g)))|_{g\in Mp(2;\R)_-}\in \{2\pi,3\pi\}\ ,\ee
and then complex-analytically from $Mp(2;\R)$ into
\be Mp(2;\C)\cong Mp(2;\C)_+\cup Mp(2;\C)_-\ ,\ee
by using the two patches defined by
\be {\rm arg} (2+{\rm tr}(Pr(g)))|_{g\in Mp(2;\C)_+}\in (-\pi,\pi]\ ,\qquad {\rm arg} (2+{\rm tr}(Pr(g)))|_{g\in Mp(2;\C)_-}\in (\pi,3\pi]\ ,\ee
where $Pr:Mp(2,\C)\to Sp(2,\C)$.

Finally, as shown in Section \ref{sec:3.1} via the Weyl transform, the real-analytic extension of metaplectic group elements in terms of symbols of operators is analogous to the one more commonly used in the literature, in which elements of $Mp(2n;\R)$ are realized as Gaussian integral operators on functions $\psi(x)\in L^{2}(\R^n)$. For example, letting $x=X^1/\sqrt{2}$, $p=X^2/\sqrt{2}$ and $t=-\theta$, the lift of the $Sp(2;\R)$ matrix \eqref{U1Sp} into $Mp(2;\R)$ is the integral operator 
\be {\cal I}[e^{-2G(0,t/2,-t/2)}]\psi(x) =\int_{-\infty}^\infty {\cal G}(x,x';t)\psi(x') dx' \ ,\label{calI}  \ee
where ${\cal G}(x,x';t)$ is the quantum-mechanical evolution kernel for the harmonic oscillator $H=\frac12(p^2+x^2)$, viz.
\be  {\cal G}(x,x';t):=\langle x|e^{-i\widehat{H}t}|x'\rangle=\frac{1}{\sqrt{2\pi i \sin t}}\exp\left[\frac{i}{2\sin t}\Big(\big(x^2+(x')^2\big)\cos t-2xx'\Big)\right]\ , \ee
viewed as a real-analytic function of $t\in \R$ valued in a space of distributions in $x$ and $x'$. Indeed, \eqref{calI} is precisely the integral operator $W_{\exp(-iHt)}$ as defined in \eqref{Weyltr1} for $n=1$, as it is easy to check directly by recalling (from \eqref{eq:OpSp(e^(-2G(0,-th/2,th/2)))} or from the general expression \eqref{eq:stargauss}) that the Weyl-ordered symbol of the evolution operator of the harmonic oscillator is $[e^{-i\Op H t}]_0(x,p)=\frac{1}{\sqrt{\cos^2 (t/2)}}\exp\left(-i\tan(t/2)(x^2+p^2)\right)$, which gives 
\be \frac{1}{2\pi}\int_{-\infty}^\infty dp\, [e^{-i\Op H t}]_0\left(\frac{x+x'}{2},p\right)e^{i(x-x')p}=\frac{1}{\sqrt{2\pi i \sin t}}\exp\left[\frac{i}{2\sin t}\Big(\big(x^2+(x')^2\big)\cos t-2xx'\Big)\right]\ . \ee
The phase can be extracted by fixing a branch, thus rewriting \cite{deGosson} 
\be  {\cal G}(x,x';t)=\frac{e^{-\tfrac{i\pi}2 [t/\pi]}}{\sqrt{2\pi i |\sin t|}}\exp\left[\frac{i}{2\sin t}\Big(\big(x^2+(x')^2\big)\cos t-2xx'\Big)\right]\ ,\qquad t\neq n\pi\ ,\ee
where $[\xi]$ denotes the integer part of $\xi\in \R$.
The real-analyticity in $t$ is manifest for $t\in \pi\mathbb{Z}$, and 
\be \lim_{t\to n\pi^\pm}{\cal G}(x,x';t) =i^{-n}\delta(x-(-1)^n x')\ ,\ee
independently of whether the limit is taken from below or above.
The integer $n=-[t/\pi]$ is referred to as the \emph{Maslov index} of the operator ${\cal I}[e^{-2G(0,t/2,-t/2)}]$; in general, this index is a topological invariant assigned to Hamiltonian flows along paths in $Sp(2n;\R)$ whereby to each symplectic matrix are associated two Maslov indices modulo $4$, $n$ and $n+2$ determining the sign of the two corresponding operators $Mp(2n;\R)$ associated to it. This is analogous to the choice of ${\rm arg}(\det(1+U))={\rm arg} (2+{\rm tr}(Pr(g)))$, assigning the metaplectic patch to a symbol of an operator.

\section{Star-product details}
\label{app:star}

The repeated application of the Baker-Campbell-Hausdorff formula \eqref{eq:BCH} allows to compute the product of an arbitrary number of delta operators \eqref{eq:DeltaM}:
\begin{align}
  \prod_{i=0}^I\OpDM{M_i}\left(X_i\right)
  &=
  \int\frac{d^{2n(I+1)}K}{(2\pi)^{2n(I+1)}}
  \exp\left(
  \tfrac{i}{2}\sum_{i=1}^I K_i M_i K_i
  +i\sum_{i=0}^{I-1}\sum_{j=i}^{I}K_i K_j
  +i\sum_{i=0}^IK_i\left(\OpX-X_i\right)
  \right)
  \nonumber\\&=
  \int\frac{d^{2n(I+1)}K}{(2\pi)^{2n(I+1)}}
  \exp\left(
  \tfrac{i}{2}\sum_{i=0}^I\sum_{j=0}^I
  K_i\left(\delta_{ij}M_j+\theta_{ij}\right)K_j
  +i\sum_{i=0}^IK_i\left(\OpX-X_i\right)
  \right)
  \,.
\end{align}
Tracing this object using \eqref{eq:TrPW} will produce a Dirac delta distribution
that will allow to integrate out $X_0$.
The remaining $2nI$-dimensional Gaussian integration \eqref{startingpt} yields
\begin{align}
  &
  \Tr\left(\prod_{i=0}^I\OpDM{M_i}\left(X_i\right)\right)
  \\\nonumber&=
  \frac1{(2\pi)^{nI}\sqdeti{(\delta_{ij}M_j+\theta_ij +M_0)}}
  \exp\left(
  \tfrac{i}{2}\sum_{i=0}^I\sum_{j=0}^I
  (X_i-X_0)(\delta_{ij}M_j+\theta_{ij}+M_0)^{-1}(X_j-X_0)
  \right)
  \,,
\end{align}
where we used the abusive notation $(A_{ij})^{-1}:=(A^{-1})_{ij}$.
The following 2 particular cases are useful in Sec. \ref{sec:starprod} :
\begin{align}
  \label{eq:Tr(opD^2)}
  &
  \Tr(\OpDM{M_0}(X_0)\OpDM{M_1}(X_1))
  \\\nonumber&=
  \frac1{(2\pi)^n\sqdeti{(M_0+M_1)}}
  \exp\Big(
  \tfrac{i}{2}(X_0-X_1)(M_0+M_1)^{-1}(X_0-X_1)
  \Big)
  \,,\\
  \label{eq:Tr(opD^3)}
  &
  \Tr(\OpDM{M}(X_0)\OpDM{M}(X_1)\OpDM{M}(X_2))
  \\\nonumber&=
  \frac1{(2\pi)^{2n}\sqdet{(1+3M^2)}}
  \exp\left(
  iX_0\frac{M}{1+3M^2}X_0
  +iX_0\frac{M}{1+3M^2}X_0
  +iX_0\frac{M}{1+3M^2}X_0
  \right)\\\nonumber&\quad\times\exp\left(
  -iX_0\frac{1+M}{1+3M^2}X_0
  -iX_0\frac{1+M}{1+3M^2}X_0
  -iX_0\frac{1+M}{1+3M^2}X_0
  \right)
  \,.
\end{align}

\section{Gaussian holomorphic gauge}
\label{app:g hol g}

The purpose of this appendix is to prove that the solution (\ref{eq:CI LHC},\,\ref{eq:pert part sol},\,\ref{eq:fact sol}) in the holomorphic gauge \eqref{eq:vn(D)}
is the one that is obtained by solving all equations of the form \eqref{eq:qf=g} with the help of the specific homotopy $\Op{q}^{\ast[a\mathcal{D},0]}$.
This solution is based on a family of real symplectic matrices
\begin{align}
  \label{eq:def UsD}
  U_s^{(\mathcal{D})}:&=
  \frac{1+s^2}{2s}+\frac{1-s^2}{2s}\mathcal{D}
  \,,
\end{align}
that interpolate between the identity matrix $1=U_1^{(\mathcal{D})}$ and its opposite $-1=U_{-1}^{(\mathcal{D})}$.
Note that this interpolation in the group is not continuous as the $s=0$ element is not defined.
They have the following properties:
\begin{align}
  \frac{1-U_s^{(\mathcal{D})}}{U_s^{(\mathcal{D})}}
  &=
  \frac{(1-s)^2}{2s}\left(1-\frac{1+s}{1-s}\mathcal{D}\right)
  \,,\\
  \det\left(1-\frac{1+s}{1-s}\mathcal{D}\right)
  &=
  \left(1-\frac{1+s}{1-s}\mathcal{D}\right)\left(1+\frac{1+s}{1-s}\mathcal{D}\right)
  =
  -\frac{4s}{(1-s)^2}
  \,.
\end{align}
that are useful in derive the bahaviour of (the $\Opz$ space Hodge dual of) the operators \eqref{eq:def UsD} under the relevant homotopy contractor.
Indeed, using them after applying the homotopy formula \eqref{eq:v1[m,zeta]} and proceeding to the change of variables
\begin{align}
  t
  &=
  \sqrt{\frac{\tfrac{1+s}{1-s}+a}{\tfrac{1+S}{1-S}+a}}
  \,&
  dt
  &=
  -\frac{t^3}{\tfrac{1+s}{1-s}+a}\frac{dS}{(1-S)^2}
  \,,
\end{align}
enables one to prove the property
\begin{align}
  \label{eq:q*[aD,0]}
  \Op{q}^{\ast[a\mathcal{D},0]}\left(%
  \dz\,\dz\,\OpSpHz{U_s^{(\mathcal{D})}}0\right)
  =
  \left.
  2i\,\frac{\sqrt{s}}{1-s}\,\dz\,\partial_\rho
  \int_{s}^1\frac{dS}{\sqrt{S}}\;
  \OpSpHz{U_S^{(D)}}{\tfrac{1}{4S}\rho(1+S-(1-S)\mathcal{D})}
  \right\vert_{\rho=0}
  \,,
\end{align}
where the square root that appears is the principal one (i.e. defined for this range of values by $\sqrt{-1}=i$),
meaning in particular that one cannot group the factors inside the square root.
The $s=-1$ case allows to specialize the linear auxiliary connection \eqref{eq:v1[m,zeta]} to the case of interest, viz.
\begin{align}
  \Op{v}_1^{(a\mathcal{D},0)}
  &=
  -\frac{e^{i\theta}}{4}
  \Op{q}^{\ast(a\mathcal{D},0)}\left(%
  \dz\dz\,\OpSpHz{U_{-1}^{(\mathcal{D})}}0\right)
  \nonumber\\&=
  \label{eq:v1[aD,0]}
  \left.
  -\frac{e^{i\theta}}4\dz\partial_\rho
  \int_{-1}^{1}\frac{ds}{\sqrt{s}}\;
  \OpSpHz{U_s^{(\mathcal{D})}}{\tfrac{1}{4s}\rho(1+s-(1-s)\mathcal{D})}
  \right\vert_{\rho=0}
  \,.
\end{align}

Using the previous equation as a base case for the recursive proof of the result \eqref{eq:vn(D)}, let us assume that there is an order $N$ up to which the one-form fields $\Op{v}_{n<N}^{[a\mathcal{D},0]}$ are all of the form
\begin{align}
  \Op{v}_n^{[a\mathcal{D},0]}
  =
  \left.
  \frac12\dz\partial_\rho
  \int_{-1}^{1}\frac{ds}{\sqrt{s}}\,f_n(s)\;
  \OpSpHz{U_s^{(\mathcal{D})}}{\tfrac{1}{4s}\rho(1+s-(1-s)\mathcal{D})}
  \right\vert_{\rho=0}
  \,.
\end{align}
To compute the product of two such elements, one must notice that the family of elements \eqref{eq:def UsD} forms in fact a group
\begin{align}
  U_{s_1}^{(\mathcal{D})}\,U_{s_2}^{(\mathcal{D})}
  &=
  U_{s_1s_2}^{(\mathcal{D})}
  \,,
\end{align}
and that, because it consists in two continuous branches respectively connected to $1$ and to $-1$,
its oscillator realisation \eqref{eq:repr Sp} comes with a sign
\begin{align}
  \frac{1}{\sqrt{s_1}}
  \OpSpH{U_{s_1}^{(\mathcal{D})}}{0}
  \frac{1}{\sqrt{s_2}}
  \OpSpH{U_{s_2}^{(\mathcal{D})}}{0}
  &=
  \frac{1}{\sqrt{s_1s_2}}
  \OpSpH{U_{s_1s_2}^{(\mathcal{D})}}{0}
  \,.
\end{align}
The product of interest reads\begin{align}
  &
  \Op{v}_{n_1}^{(a\mathcal{D},0)}
  \Op{v}_{n_2}^{(a\mathcal{D},0)}
  \nonumber\\&=
  \frac{1}{4}\,\dz\partial_{\rho_1}\,\dz\partial_{\rho_2}
  \int_{-1}^1\frac{ds_1}{\sqrt{s_1}}\,f_{n_1}(s_1)
  \int_{-1}^1\frac{ds_2}{\sqrt{s_2}}\,f_{n_2}(s_2)
  \nonumber\\&\quad\times
  \left.
  \OpSpHz{U_{s_1}^{(\mathcal{D})}}{\tfrac{1}{4s_1}\rho_1(1+s_1-(1-s_1)\mathcal{D})}
  \OpSpHz{U_{s_2}^{(\mathcal{D})}}{\tfrac{1}{4s_2}\rho_2(1+s_2-(1-s_2)\mathcal{D})}
  \right\vert_{\rho_1=\rho_2=0}
  \nonumber\\&=
  \left.
  -\frac{i}{32}\dz\dz
  \int_{-1}^1ds_1\,f_{n_1}(s_1)
  \int_{-1}^1ds_2\,f_{n_2}(s_2)
  \frac{1+s_1s_2}{(s_1s_2)^{3/2}}
  \left(1-\frac{i}{2}\frac{1-s_1s_2}{1+s_1s_2}\partial_\mu\mathcal{D}\partial_\mu\right)
  \OpSpHz{U_{s_1s_2}^{(\mathcal{D})}}{\mu}
  \right\vert_{\mu=0}
  \,.
\end{align}
Defining a new variable $s:=s_1s_2$,
one can use the lemma \eqref{eq:dOpSpH} applied to its one-dimensional differential to convert the momentum derivatives into a parametric one.
This change of variable is performed via the introduction of $1=\int_{-1}^{1}ds\,\delta(s-s_1s_2)$.
In this context, it is useful to define the functional operation
\begin{align}
  f_{n_1}\circ f_{n_2}(s)
  :=
  \int_{-1}^1ds_1\,
  \int_{-1}^1ds_2\,
  f_{n_1}(s_1)f_{n_2}(s_2)
  \delta(s-s_1s_2)
  \,.
\end{align}
Note that the parametric integrals are treated as intrinsically real.
In particular, a change of sign is treated as a transformation of the real segment itself rather than a global phase rotation of it.
The product becomes
\begin{align}
  \Op{v}_{n_1}^{(a\mathcal{D},0)}
  \Op{v}_{n_2}^{(a\mathcal{D},0)}
  &=
  -\frac{i}{32}\dz\dz
  \int_{-1}^1ds\,f_{n_1}\circ f_{n_2}(s)
  \frac{1+s}{s^{3/2}}
  \left(1-2s\frac{1-s}{1+s}\frac{d}{ds}\right)
  \OpSpHz{U_{s}^{(\mathcal{D})}}{0}
  \,.
\end{align}

Plugging the result of this product and the resolution \eqref{eq:q*[aD,0]} into the formula \eqref{eq:fact sol} gives
\begin{align}
  &
  \Op{v}_{n}^{[a\mathcal{D},0]}
  =
  -\Op{q}^{\ast[a\mathcal{D},0]}\left(\sum_{k=1}^{n-1}
  \Op{v}_{n-k}^{(a\mathcal{D},0)}
  \Op{v}_{k}^{(a\mathcal{D},0)}\right)
  \nonumber\\&=
  -\frac{1}{16}\dz\partial_\rho
  \int_{-1}^1ds\left(\sum_{k=1}^{n-1}f_{n-k}\circ f_{k}(s)\right)
  \frac{1+s}{s^{3/2}}
  \left(1-2s\frac{1-s}{1+s}\frac{d}{ds}\right)
  \frac{\sqrt{s}}{1-s}\nonumber\\&\qquad\qquad\quad  \left.
  \int_{s}^1\frac{dS}{\sqrt{S}}\;
  \OpSpHz{u_S^{(D)}}{\tfrac{1}{4S}\rho(1+S+(1-S)\mathcal{D})}
  \right\vert_{\rho=0}
  \nonumber\\&=
  -\frac{1}{16}\dz\partial_\rho
  \int_{-1}^1ds\left(\sum_{k=1}^{n-1}f_{n-k}\circ f_{k}(s)\right)
  \frac{1+s}{s^{3/2}}
  \frac{\sqrt{s}}{1-s}
  \left(-2s\frac{1-s}{1+s}\frac{d}{ds}\right)\nonumber\\&\qquad\qquad\quad  \left.
  \int_{s}^1\frac{dS}{\sqrt{S}}\;
  \OpSpHz{u_S^{(D)}}{\tfrac{1}{4S}\rho(1+S+(1-S)\mathcal{D})}
  \right\vert_{\rho=0}
  \nonumber\\&=
  \left.
  -\frac{1}{8}\dz\partial_\rho
  \int_{-1}^{-1}\frac{ds}{\sqrt{s}}
  \left(\sum_{k=1}^{n-1}f_{n-k}\circ f_{k}(s)\right)
  \OpSpHz{U_s^{(\mathcal{D})}}{\tfrac{1}{4s}\rho(1+s-(1-s)\mathcal{D})}
  \right\vert_{\rho=0}
  \,.
\end{align}
This converts the operator recursion \eqref{eq:fact sol} with its base case \eqref{eq:v1[aD,0]} into a parametric recursion
\begin{align}
  f_n(s)
  &=
  -\frac{1}{4}\sum_{k=1}^{n-1}f_{n-k}\circ f_{k}(s)
  \,,&
  f_1(s)
  &=
  -\frac{e^{i\theta}}{2}
  \,.
\end{align}
The homogeneity of the first equation, related to the one of the pertubative expansion \eqref{eq:pert part sol} it comes from,
allows to rewrite it as
\begin{align}
  f_n(s)
  &=
  -4\sum_nC_{n-1}\left(-\frac{1}{4}f_1(s)\right)^{\circ n}
  \,,
\end{align}
where $C_n$ are the Catalan numbers:
\begin{align}
  C_0:&=1
  \,,&
  C_{n+1}:&=\sum_{k=0}^n C_kC_{n-k}
  \,,&
  C_n&=\frac{(2n)!}{n!(n+1)!}
  \,.
\end{align}
The $\circ$-power can be established recursively to be
\begin{align}
  \left(-\frac{1}{4}f_1(s)\right)^{\circ n}
  =
  \frac{1}{(k-1)!}
  \left(\frac{e^{i\theta}}{8}\right)^n
  \left(\log\left(\tfrac{1}{s^2}\right)\right)^{k-1}
  \,.
\end{align}
Indeed
\begin{align}
  \left(-\frac{1}{4}f_1(s)\right)^{\circ n}\circ\left(-\frac{1}{4}f_1(s)\right)
  &=
  \frac{1}{(k-1)!}
  \left(\frac{e^{i\theta}}{8}\right)^{n+1}
  \int_{-1}^{1}ds_1\int_{-1}^{1}ds_2
  \delta(s-s_1s_2)
  \left(\log\left(\tfrac{1}{s_1^2}\right)\right)^{k-1}
  \nonumber\\&=
  \frac{2}{(k-1)!}
  \left(\frac{e^{i\theta}}{8}\right)^{n+1}
  \int_{\left\vert s_2\right\vert}^{1}ds_1
  \left(\log\left(\tfrac{1}{s_1^2}\right)\right)^{k-1}
  \nonumber\\&=
  \frac{1}{(k-1)!}
  \left(\frac{e^{i\theta}}{8}\right)^{n+1}
  \left(\log\left(\tfrac{1}{s^2}\right)\right)^{k}
  \,.
\end{align}
Combining these results gives
\begin{align}
  \label{eq:fn(s)bis}
  f_n(s)
  =
  -\frac{e^{i\theta}}{2}\left(\frac{e^{i\theta}}{8}\right)^{n-1}
  \frac{(2n)!}{(n!)^2(n+1)!}\left(\log\left(\frac{1}{s^2}\right)\right)^{n-1}
  \,,
\end{align}
concluding the proof.

\paragraph{Limit to Weyl order.}
In Weyl order, the linearised solution \eqref{eq:v1[aD,0]} built in this appendix corresponds to identifying 
\begin{align}
  \label{eq:vn(D)0}
  \Op{q}^{\ast[a\mathcal{D},0]}_0\Big(\delta^2(z)\dz\dz\Big)
  =
  \left.
  \dz\partial_\rho
  \int_{-1}^1\frac{ds}{1+s}
  \exp\left(\tfrac{i}{2}\tfrac{1-s}{1+s}z\mathcal{D}z
  +\tfrac{i}{1+s}\rho z\right)
  \right\vert_{\rho=0}
  \,.
\end{align}
This result is to be contrasted with the fact that the straight application of the standard homotopy $q^{\ast[0,0]}_0$ to a delta-function source is ill-defined \cite{COMST}. In this respect, the fact that the r.h.s. of \eqref{eq:vn(D)0} is independent from the parameter $a$ can be interpreted as a regularization of the latter resolution, i.e.,
\begin{align}
  \left[q^{\ast[ 0,0]}_0\Big(\delta^2(z)\dz\dz\Big)\right]_{q-{\rm reg}}
  =
  \lim_{a\to0}
  {q}^{\ast[a\mathcal{D},0]}_0\Big(\delta^2(z)\dz\dz\Big)
  \,,
\end{align}
obtained by flowing to $q^{\ast[0,0]}_0$ from the resolution operator ${q}^{\ast[a\mathcal{D},0]}_0$ corresponding to the Gaussian holomorphic gauge. What is interesting to observe is that the solution so obtained is identical to the one obtained in \cite{COMST} by sticking to $q^{\ast[0,0]}_0$ and regularizing the delta-function source as a Gaussian delta sequence, 
\begin{align}
  \left[q^{\ast[0,0]}_0\Big(\delta^2(z)\dz\dz\Big)\right]_{\d-{\rm reg}}
  =
  \lim_{a\to 0}
  {q}^{\ast[0,0]}_0\Big(\frac{1}{2\pi}\MoSym{a\mathcal{D}}{\OpKz}\dz\dz\Big)
  \,, 
\end{align}
i.e., 
\be \lim_{a\to0}
  {q}^{\ast[a\mathcal{D},0]}_0\Big(\delta^2(z)\dz\dz\Big) = \lim_{a\to0}
  {q}^{\ast[0,0]}_0\Big(\frac{1}{2\pi}\MoSym{a\mathcal{D}}{\OpKz}\dz\dz\Big) \ .  \ee
In other words, we find here a correspondence between two regularizations: one that entails regularizing the source itself and one that consists in acting directly on the singular source with a reordered homotopy. This is one explicit example that shows how the freedom in choosing $M$ in \eqref{eq:rhoOpMXi} can encode the freedom of realizing the source term on different functional classes. It would be interesting to understand whether this principle can be extended to solve more general equations involving singular sources. 

Of course, the $\mathcal{D}$-dependence of the obtained solution shows that the limit to Weyl order may not be uniquely defined. However, the results collected in \cite{COMST} and in Sec. \ref{sec:OWL} (via the computations in App. \ref{app:OWL}) show that the Wilson line observables are in fact independent from $\mathcal{D}$ or similar spinor constructs used to define the regularizations, at least in the first few perturbative orders.

\section{Observable computations in holomorphic gauge}
\label{app:OWL}
In this appendix, we compute the coefficients in Eqs. \eqref{eq:OWL beta} and \eqref{eq:OWL bb} that are relevant for establishing the results of Sec. \ref{sec:OWL}.

\subsection{Background terms and protection}
\label{sec:OWL0}
Their leading-order contribution only depends on the background,
viz.
\begin{align}
  \beta_{p,0}
  =
  \frac{(-i)^p}{2\pi}
  \Tr_{\Opz}\left(
  \OpSpHz{(-1)^p}{\tfrac{1}{2}\mu}\right)
  \,,
\end{align}
that is, using the lemma \eqref{eq:Tr SpHz},
\begin{align}
  \label{eq:beta(0,0)}
  \beta_{0,0}
  &=
  8\pi\delta^2(\mu)
  \,,&
  \bar\beta_{0,0}
  &=
  8\pi\delta^2(\bar\mu)
  \,,&
  \beta_{1,0}
  =
  \bar\beta_{1,0}
  &=
  1
  \,.
\end{align}
Plugging this into Eq. \eqref{eq:OWL alpha}, one sees that the non-trivial leading coefficients \eqref{eq:OWL alpha} read
\begin{align}
  \label{eq:OWL alpha0 1}
  \alpha_{2m,0;2m;0}
  &=
  (8\pi)^2\delta^4({\cal M})
  \,,&
  \alpha_{2m,1;2m;1}
  &=
  1
  \,,\\
  \label{eq:OWL alpha0 2}
  \alpha_{2m+1,0;2m+1;0}
  &=
  8\pi\delta^2(\bar\mu)
  \,,&
  \alpha_{2m+1,1;2m+1;1}
  &=
  8\pi\delta^2(\mu)
  \,,
\end{align}
consistently with the literature \cite{Colombo:2012jx,Bonezzi:2017vha}.
As shown in \cite{COMST}, half of the coefficients $\beta_{p,k}$ and $\bar\beta_{p,k}$ are protected,
namely $\beta_{1,k}$ and $\bar\beta_{1,k}$,
that is to say the ones that are well defined at $\mu=0$.
Indeed, they can then be expanded in powers of $\mu$ as
\begin{align}
  \beta_{1,k}
  =
  \frac{1}{2\pi}
  \frac{\left(\partial_\nu\right)^k}{k!}
  \sum_{\ell=0}^{+\infty}\frac{1}{\ell!}
  \left.\STr_{\Opz}\left(\left(
  \frac{i}{2}
  \mu^\ag\,\left(
  \Opz_\ag-2i
  \sum_{j=1}^\infty \nu^{j}\;\Op{v}^{[F]}_{\ag\,j}\right)\right)^\ell
  \right)\right\vert_{\nu=0}.
\end{align}
The vanishing of the $\ell\neq0$ terms
is ensured by the graded cyclicity of the supertrace \eqref{eq:def STr} and by the bosonic projection \eqref{eq:BP holG}.
This in turn gives
\begin{align}
  \label{eq:OWL protected}
  \beta_{1,k}&=\delta_{k,0}
  \,,&
  \bar\beta_{1,\bar{k}}&=\delta_{\bar{k},0}
  \,.
\end{align}

\subsection{Linear terms in $[m,\zeta,+]$ gauge}
\label{sec:OWL1}
To compute the subleading terms, one needs to evaluate the potential $\Op{V}$ along the line, which is done using the formula \eqref{eq:SpH on f(X)}.
The first subleading contribution to the coefficient \eqref{eq:OWL beta} is given in $[m,\zeta]$ gauge by
\begin{align}
  \beta_{p,1}
  &=
  \frac{(-i)^p}{2\pi}
  \Tr_{\Opz}\left(
  \int_0^1d\tau\,\mu^\ag\;
  \OpSpHz{1}{\tfrac{\tau}{2}\mu}
  \Op{v}_{\ag\,1}^{[m,\zeta]}
  \OpSpHz{1}{-\tfrac{\tau}{2}\mu}
  \OpSpHz{(-1)^p}{\tfrac{1}{2}\mu}
  \right)
  \nonumber\\&=
  -\frac{(-i)^pe^{i\theta}}{4\pi}
  \int_0^1d\tau
  \int_0^1\frac{dt}{t^2}\,
  \sqdet{\tfrac{1-u_t}{2}}
  \mu\left(
  \tfrac{i}2\tfrac{t^2}{1-t^2}\tfrac{1+u_t}{u_t}\partial_{\rho_t}
  +\zeta
  \right)
  \exp\left(
  \tfrac{i}{2}\rho_t\tfrac{1+u_t}{1-u_t}\rho_t
  \right)
  \nonumber\\&\qquad\times\left.
  \Tr_{\Opz}\left(
  \OpSpHz{(-1)^p}{\tfrac{1-\tau+(-1)^p\tau}{2}\mu}
  \OpSpHz{u_t}{\rho_t}
  \right)
  \right\vert_{\rho_t
  =
  -\tfrac{1-t}{2t}\zeta
  \frac{1-u_t}{u_t}}
  \nonumber\\&=
  \frac{i^{-p+1}e^{i\theta}}{2}
  \int_0^1d\tau
  \int_0^1\frac{dt}{t^2}\,
  \frac{\sqdet{1-u_t}}{\sqdet{1-(-1)^pu_t}}
  \mu\left(
  \tfrac{i}2\tfrac{t^2}{1-t^2}\tfrac{1+u_t}{u_t}\partial_{\rho_t}
  +\zeta
  \right)
  \nonumber\\&\qquad\times
  \exp\left(
  i(1-(-1)^p)
  \rho_t\tfrac{u_t}{(1-u_t)(1-(-1)^pu_t)}\rho_t
  -i(1-\tau+(-1)^p\tau)
  \rho_t\tfrac{u_t}{1-(-1)^pu_t}\mu
  \right)
  \nonumber\\&\qquad\times\left.
  \exp\left(
  \tfrac{i}{8}(1-\tau+(-1)^p\tau)^2
  \mu\tfrac{1+(-1)^pu_t}{1-(-1)^pu_t}\mu
  \right)
  \right\vert_{\rho_t
  =
  -\tfrac{1-t}{2t}\zeta
  \frac{1-u_t}{u_t}}
\end{align}
In the first case, the expression can be evaluated directly as
\begin{align}
  \beta_{0,1}
  &=
  \frac{ie^{i\theta}}{2}
  \int_0^1\frac{dt}{t^2}\,
  \left(
  \mu\zeta
  -\tfrac{1}{2t}\mu m\mu
  \right)
  \exp\left(
  -\tfrac{i}{2}\tfrac{1-t}{t}\mu\zeta
  +\tfrac{i}{8}\tfrac{1-t^2}{t^2}\mu m\mu
  \right)
  \nonumber\\&=
  e^{i\theta}
  \left[
  \exp\left(
  -\tfrac{i}{2}\tfrac{1-t}{t}\mu\zeta
  +\tfrac{i}{8}\tfrac{1-t^2}{t^2}\mu m\mu
  \right)\right]_{t=0}^1
  =
  e^{i\theta}\,,
\end{align}
where we used the lemma
\be \lim_{\epsilon\to 0}e^{\frac{i}{\epsilon}\mu m \mu}=0 \ ,\ee
which follows from the delta sequence $\d^2(\mu)=\lim_{\epsilon\to 0}\frac{1}{\epsilon} e^{\frac{i}{\epsilon}\mu m \mu} $ (see Sec. \ref{sec:Sp} and App. \ref{app:anal}).
The observable vanishes in the other case due to the protection argument \eqref{eq:OWL protected}.
Plugging these results and the ones from App. \ref{sec:OWL0} into the defintion \eqref{eq:OWL alpha} reproduces the results \cite{COMST}
\begin{align}
  \label{eq:OWL alpha1 1}
  \a_{2m+1,1;2m+2,1}
  &=
  e^{i\theta}
\,,&
\a_{2m+1,0;2m+2,1}
&=
e^{-i\theta}
\,,\\
\label{eq:OWL alpha1 2}
\a_{2m,0;2m+1,0}
&=
8\pi e^{i\theta}\delta^2(\bar\mu)
\,,&
\a_{2m,0;2m+1,1}
&=
8\pi e^{-i\theta}\delta^2(\mu)
\,.
\end{align}

\subsection{Higher orders in $[0,\zeta,+]$ gauge}
\label{sec:OWL2}
In the gauge \eqref{eq:v1zeta}, the residual gauge invariance of the observables (i.e. their independence from the spinor parameter $\zeta$) can be shown to all orders in perturbation theory.
At order $n$, the holomorphic contribution reads
\begin{align}
  &
  \beta_{p,n}
  \nonumber\\&=
  \frac{(-i)^p}{2\pi}
  \Tr_{\Opz}\left(
  \int_0^1d\tau_n\,\cdots
  \int_0^{\tau_2}d{\tau_1}\,
  \prod_{i=1}^n\left(
  \mu^\ag\;
  \OpSpHz{1}{\tfrac{\tau_i}{2}\mu}
  \Op{v}_{\ag\,1}^{[0,\zeta]}
  \OpSpHz{1}{-\tfrac{\tau_i}{2}\mu}
  \right)
  \OpSpHz{(-1)^p}{\tfrac{1}{2}\mu}
  \right)
  \,,
\end{align}
When applying the product formula \eqref{eq:repr SpH},
the sign is the one
\begin{align}
  {\OpSpHz{-1}0}^{n+p}
  =
  (-1)^{\left\lfloor{\tfrac{n}2}\right\rfloor}
  \OpSpHz{(-1)^{n+p}}0
  \,,
\end{align}
that can be deduced from the projectivity of the metaplectic representation.
One finds
\begin{align}
  \beta_{p,n}
  &=
  \frac{(-i)^p}{2\pi}
  \left(\frac{e^{i\theta}}{2}(\mu\zeta)\right)^n
  (-1)^{\left\lfloor\tfrac{n+p}{2}\right\rfloor}
  \int d^n\tau\int d^nu
  \Tr_{\Opz}\left(
  \OpSpHz{(-1)^{n+p}}{%
  \tfrac{(-1)^n}{2}\mu
  -\sum_{i=1}^n(-1)^i\left(u_i\zeta+\tau_i\mu\right)
  }
  \right)
  \nonumber\\&\qquad\times
  \exp\left(%
  -i\sum_{i=1}^n(-1)^iu_i\left(%
  \sum_{j<i}(-1)^j\tau_j-\sum_{j>i}(-1)^j\tau_i+\tfrac{(-1)^n}{2}
  \right)(\mu\zeta)
  \right)
  \,,
  \label{eq:beta(p,n)[zeta]}
\end{align}
where the theta functions present in \eqref{eq:[0,zeta,+]} are implicitly encoded in the measure.
In the case where $n+p$ is odd, one can apply the lemma \eqref{eq:Tr SpHz}, then integrate all the homotopy variables $u_i$ and get
\begin{align}
  \beta_{(n+1\mod2),n}
  =
  (-1)^{\left\lfloor\tfrac{n}{2}\right\rfloor}
  \left(\frac{e^{i\theta}}{2}\right)^n
  \int\prod_{i=1}^{n}\frac{d\tau_i}{%
  \sum_{j<i}(-1)^j\tau_j-\sum_{j>i}(-1)^j\tau_i+\tfrac{(-1)^n}{2}}
  \,,
\end{align}
which is manifestly $\zeta$-independent.
When $n$ is odd, one combination of the $\tau$ parameters can be naturally integrated out to give
\begin{align}
  \beta_{0,2\ell+1}
  =
  e^{i\theta}\beta_{1,2\ell}
  =
  e^{i\theta}\delta_{\ell,0}
  \,.
\end{align}
On the other hand, examining \eqref{eq:beta(p,n)[zeta]} when $n+p$ is even gives
\begin{align}
  \beta_{(p\equiv n\mod2),n}
  &=
  2\pi i^{2p-n}
  \left(\frac{e^{i\theta}}{2}(\mu\zeta)\right)^n
  \int d^n\tau\int d^nu
  \;\delta^2\left(%
  \tfrac{(-1)^n}{2}\mu
  -\sum_{i=1}^n(-1)^i\left(u_i\zeta+\tau_i\mu\right)
  \right)
  \exp\left(...\right)
\end{align}
A change of variable  allows to convert this twistor delta function to a parametric one,
modulo taking the jacobian $(\mu\zeta)$.
Because of the analytic continuation discussed in App. \ref{app:anal}, it does not come under an absolute value. 
One finds
\begin{align}
  &
  \beta_{(p\equiv n\mod2),n}
  \nonumber\\&=
  2\pi i^{2p-n}
  \left(\frac{e^{i\theta}}{2}\right)^n
  (\mu\zeta)^{n-1}
  \int d^n\tau\int d^nu
  \;\delta\left(\sum_i(-1)^iu_i\right)
  \;\delta\left(\tfrac{(-1)^n}{2}-\sum_i(-1)^i\tau_i\right)
  \exp\left(...\right)
  \nonumber\\&=
  \label{eq:OWL n+p even}
  2\pi i^{2p-n}
  \left(\frac{e^{i\theta}}{2}\right)^n
  (\mu\zeta)^{n-1}
  \int d^{n-1}\tau\int d^{n-1}u
  \nonumber\\&\qquad\times
  \exp\left(%
  -i\sum_{i=1}^{n-1}(-1)^iu_i\left(%
  \sum_{j<i}(-1)^j\tau_j-\sum_{j>i}(-1)^j\tau_i+\tfrac{(-1)^{n-1}}{2}
  \right)(\mu\zeta)
  \right)
  \,,
\end{align}
While the exact boundaries of the $u$ integrals are quite involved in this latest expression,
they are still defined only in terms of each other, 0 and $\infty$.
This means that one can rescale those $n-1$ integrals by a factor of $(\mu\zeta)$ to make the dependence on the latter completely disappear.
While the general evaluation of this integral is left for future work, the $n=2$ case is more straightforward:
\begin{align}
  \beta_{0,2}
  =
  -\frac{i\pi}{2}e^{2i\theta}
  \int_{0}^{\tfrac12}d\tau
  \int_{-\infty}^{+\infty}du
  \exp\left(-\tfrac{i}{2}u\right)
  =
  -\frac{i\pi}{4}e^{2i\theta}
  \delta\left(-\tfrac{1}{2}\right)
  =
  0
  \,.
\end{align}
Hence, among the coefficients \eqref{eq:OWL beta}, the ones that remain unknown are $\beta_{0,2\ell}$ for $\ell\geq2$.

\providecommand{\href}[2]{#2}\begingroup\raggedright\endgroup



\begin{thebibliography}{10}

\bibitem{barnich}
G.~Barnich and M.~Henneaux, ``{Consistent couplings between fields with a gauge
  freedom and deformations of the master equation},''
  \href{http://dx.doi.org/10.1016/0370-2693(93)90544-R}{{\em Phys. Lett. B}
  {\bf 311} (1993)  123--129}, \href{http://arxiv.org/abs/hep-th/9304057}{{\tt
  arXiv:hep-th/9304057}}.

\bibitem{Sleight:2017pcz}
C.~Sleight and M.~Taronna, ``{Higher-Spin Gauge Theories and Bulk Locality},''
  \href{http://dx.doi.org/10.1103/PhysRevLett.121.171604}{{\em Phys. Rev.
  Lett.} {\bf 121} (2018) no.~17, 171604},
\href{http://arxiv.org/abs/1704.07859}{{\tt arXiv:1704.07859 [hep-th]}}.

\bibitem{Vasiliev:1990en}
M.~A. Vasiliev, ``{Consistent equation for interacting gauge fields of all
  spins in (3+1)-dimensions},''
\href{http://dx.doi.org/10.1016/0370-2693(90)91400-6}{{\em Phys. Lett.} {\bf
  B243} (1990)  378--382}.

\bibitem{properties}
M.~A. Vasiliev, ``{Properties of equations of motion of interacting gauge
  fields of all spins in (3+1)-dimensions},''
\href{http://dx.doi.org/10.1088/0264-9381/8/7/014}{{\em Class. Quant. Grav.}
  {\bf 8} (1991)  1387--1417}.

\bibitem{more}
M.~A. Vasiliev, ``{More on equations of motion for interacting massless fields
  of all spins in (3+1)-dimensions},''
\href{http://dx.doi.org/10.1016/0370-2693(92)91457-K}{{\em Phys. Lett.} {\bf
  B285} (1992)  225--234}.

\bibitem{review99}
M.~A. Vasiliev, ``{Higher spin gauge theories: Star product and AdS space},''
\href{http://arxiv.org/abs/hep-th/9910096}{{\tt arXiv:hep-th/9910096
  [hep-th]}}.

\bibitem{Vasiliev:2003ev}
M.~A. Vasiliev, ``{Nonlinear equations for symmetric massless higher spin
  fields in (A)dS(d)},''
  \href{http://dx.doi.org/10.1016/S0370-2693(03)00872-4}{{\em Phys. Lett.} {\bf
  B567} (2003)  139--151},
\href{http://arxiv.org/abs/hep-th/0304049}{{\tt arXiv:hep-th/0304049
  [hep-th]}}.

\bibitem{Bekaert:2005vh}
X.~Bekaert, S.~Cnockaert, C.~Iazeolla, and M.~A. Vasiliev, ``{Nonlinear higher
  spin theories in various dimensions},'' in {\em {Higher spin gauge theories:
  Proceedings, 1st Solvay Workshop: Brussels, Belgium, 12-14 May, 2004}},
  pp.~132--197.
\newblock 2004.
\newblock
\href{http://arxiv.org/abs/hep-th/0503128}{{\tt arXiv:hep-th/0503128
  [hep-th]}}.
\newblock

\bibitem{Didenko:2014dwa}
V.~E. Didenko and E.~D. Skvortsov, ``{Elements of Vasiliev theory},''
\href{http://arxiv.org/abs/1401.2975}{{\tt arXiv:1401.2975 [hep-th]}}.

\bibitem{Vasiliev:2017cae}
M.~A. Vasiliev, ``{On the Local Frame in Nonlinear Higher-Spin Equations},''
  \href{http://dx.doi.org/10.1007/JHEP01(2018)062}{{\em JHEP} {\bf 01} (2018)
  062},
\href{http://arxiv.org/abs/1707.03735}{{\tt arXiv:1707.03735 [hep-th]}}.

\bibitem{Didenko:2018fgx}
V.~E. Didenko, O.~A. Gelfond, A.~V. Korybut, and M.~A. Vasiliev, ``{Homotopy
  Properties and Lower-Order Vertices in Higher-Spin Equations},''
  \href{http://dx.doi.org/10.1088/1751-8121/aae5e1}{{\em J. Phys.} {\bf A51}
  (2018) no.~46, 465202},
\href{http://arxiv.org/abs/1807.00001}{{\tt arXiv:1807.00001 [hep-th]}}.

\bibitem{Didenko:2019xzz}
V.~E. Didenko, O.~A. Gelfond, A.~V. Korybut, and M.~A. Vasiliev, ``{Limiting
  Shifted Homotopy in Higher-Spin Theory and Spin-Locality},''
  \href{http://dx.doi.org/10.1007/JHEP12(2019)086}{{\em JHEP} {\bf 12} (2019)
  086}, \href{http://arxiv.org/abs/1909.04876}{{\tt arXiv:1909.04876
  [hep-th]}}.

\bibitem{Gelfond:2019tac}
O.~A. Gelfond and M.~A. Vasiliev, ``{Spin-Locality of Higher-Spin Theories and
  Star-Product Functional Classes},''
  \href{http://dx.doi.org/10.1007/JHEP03(2020)002}{{\em JHEP} {\bf 03} (2020)
  002}, \href{http://arxiv.org/abs/1910.00487}{{\tt arXiv:1910.00487
  [hep-th]}}.

\bibitem{Didenko:2020bxd}
V.~E. Didenko, O.~A. Gelfond, A.~V. Korybut, and M.~A. Vasiliev,
  ``{Spin-locality of $\eta^{2}$ and $ {\overline{\eta}}^2 $ quartic
  higher-spin vertices},''
  \href{http://dx.doi.org/10.1007/JHEP12(2020)184}{{\em JHEP} {\bf 12} (2020)
  184}, \href{http://arxiv.org/abs/2009.02811}{{\tt arXiv:2009.02811
  [hep-th]}}.

\bibitem{Gelfond:2021two}
O.~A. Gelfond and A.~V. Korybut, ``{Manifest form of the spin-local higher-spin
  vertex $\varUpsilon ^{\eta \eta }_{\omega CCC}$},''
  \href{http://dx.doi.org/10.1140/epjc/s10052-021-09401-4}{{\em Eur. Phys. J.
  C} {\bf 81} (2021) no.~7, 605}, \href{http://arxiv.org/abs/2101.01683}{{\tt
  arXiv:2101.01683 [hep-th]}}.

\bibitem{Vasiliev:2012vf}
M.~A. Vasiliev, ``{Holography, Unfolding and Higher-Spin Theory},''
  \href{http://dx.doi.org/10.1088/1751-8113/46/21/214013}{{\em J. Phys. A} {\bf
  46} (2013)  214013}, \href{http://arxiv.org/abs/1203.5554}{{\tt
  arXiv:1203.5554 [hep-th]}}.

\bibitem{neiman}
Y.~Neiman, ``{The holographic dual of the Penrose transform},''
  \href{http://dx.doi.org/10.1007/JHEP01(2018)100}{{\em JHEP} {\bf 01} (2018)
  100}, \href{http://arxiv.org/abs/1709.08050}{{\tt arXiv:1709.08050
  [hep-th]}}.
  
\bibitem{Pierre Bieliavsky} Pierre Bieliavsky, private communication.

\bibitem{2011}
C.~Iazeolla and P.~Sundell, ``{Families of exact solutions to Vasiliev's 4D
  equations with spherical, cylindrical and biaxial symmetry},''
  \href{http://dx.doi.org/10.1007/JHEP12(2011)084}{{\em JHEP} {\bf 12} (2011)
  084},
\href{http://arxiv.org/abs/1107.1217}{{\tt arXiv:1107.1217 [hep-th]}}.

\bibitem{2017}
C.~Iazeolla and P.~Sundell, ``{4D Higher Spin Black Holes with Nonlinear Scalar
  Fluctuations},'' \href{http://dx.doi.org/10.1007/JHEP10(2017)130}{{\em JHEP}
  {\bf 10} (2017)  130},
\href{http://arxiv.org/abs/1705.06713}{{\tt arXiv:1705.06713 [hep-th]}}.

\bibitem{review}
C.~Iazeolla, E.~Sezgin, and P.~Sundell, ``{On Exact Solutions and Perturbative
  Schemes in Higher Spin Theory},''
  \href{http://dx.doi.org/10.3390/universe4010005}{{\em Universe} {\bf 4}
  (2018) no.~1, 5},
\href{http://arxiv.org/abs/1711.03550}{{\tt arXiv:1711.03550 [hep-th]}}.

\bibitem{COMST}
D.~De~Filippi, C.~Iazeolla, and P.~Sundell, ``{Fronsdal fields from gauge
  functions in Vasiliev\textquoteright{}s higher spin gravity},''
  \href{http://dx.doi.org/10.1007/JHEP10(2019)215}{{\em JHEP} {\bf 10} (2019)
  215}, \href{http://arxiv.org/abs/1905.06325}{{\tt arXiv:1905.06325
  [hep-th]}}.

\bibitem{cosmo}
R.~Aros, C.~Iazeolla, J.~Noreña, E.~Sezgin, P.~Sundell, and Y.~Yin, ``{FRW and
  domain walls in higher spin gravity},''
  \href{http://dx.doi.org/10.1007/JHEP03(2018)153}{{\em JHEP} {\bf 03} (2018)
  153},
\href{http://arxiv.org/abs/1712.02401}{{\tt arXiv:1712.02401 [hep-th]}}.

\bibitem{Boulanger:2015ova}
N.~Boulanger, P.~Kessel, E.~D. Skvortsov, and M.~Taronna, ``{Higher spin
  interactions in four-dimensions: Vasiliev versus Fronsdal},''
  \href{http://dx.doi.org/10.1088/1751-8113/49/9/095402}{{\em J. Phys.} {\bf
  A49} (2016) no.~9, 095402},
\href{http://arxiv.org/abs/1508.04139}{{\tt arXiv:1508.04139 [hep-th]}}.

\bibitem{GiombiYin1}
S.~Giombi and X.~Yin, ``{Higher Spin Gauge Theory and Holography: The
  Three-Point Functions},''
  \href{http://dx.doi.org/10.1007/JHEP09(2010)115}{{\em JHEP} {\bf 09} (2010)
  115}, \href{http://arxiv.org/abs/0912.3462}{{\tt arXiv:0912.3462 [hep-th]}}.

\bibitem{Colombo:2010fu}
N.~Colombo and P.~Sundell, ``{Twistor space observables and quasi-amplitudes in
  4D higher spin gravity},''
  \href{http://dx.doi.org/10.1007/JHEP11(2011)042}{{\em JHEP} {\bf 11} (2011)
  042},
\href{http://arxiv.org/abs/1012.0813}{{\tt arXiv:1012.0813 [hep-th]}}.

\bibitem{Sezgin:2011hq}
E.~Sezgin and P.~Sundell, ``{Geometry and Observables in Vasiliev's Higher Spin
  Gravity},'' \href{http://dx.doi.org/10.1007/JHEP07(2012)121}{{\em JHEP} {\bf
  07} (2012)  121},
\href{http://arxiv.org/abs/1103.2360}{{\tt arXiv:1103.2360 [hep-th]}}.

\bibitem{Colombo:2012jx}
N.~Colombo and P.~Sundell, ``{Higher Spin Gravity Amplitudes From Zero-form
  Charges},''
\href{http://arxiv.org/abs/1208.3880}{{\tt arXiv:1208.3880 [hep-th]}}.

\bibitem{Didenko:2012tv}
V.~E. Didenko and E.~D. Skvortsov, ``{Exact higher-spin symmetry in CFT: all
  correlators in unbroken Vasiliev theory},''
  \href{http://dx.doi.org/10.1007/JHEP04(2013)158}{{\em JHEP} {\bf 04} (2013)
  158},
\href{http://arxiv.org/abs/1210.7963}{{\tt arXiv:1210.7963 [hep-th]}}.

\bibitem{Bonezzi:2017vha}
R.~Bonezzi, N.~Boulanger, D.~De~Filippi, and P.~Sundell, ``{Noncommutative
  Wilson lines in higher-spin theory and correlation functions of conserved
  currents for free conformal fields},''
  \href{http://dx.doi.org/10.1088/1751-8121/aa8efa}{{\em J. Phys.} {\bf A50}
  (2017) no.~47, 475401},
\href{http://arxiv.org/abs/1705.03928}{{\tt arXiv:1705.03928 [hep-th]}}.

\bibitem{Sharapov:2020quq}
A.~Sharapov and E.~Skvortsov, ``{Characteristic Cohomology and Observables in
  Higher Spin Gravity},'' \href{http://dx.doi.org/10.1007/JHEP12(2020)190}{{\em
  JHEP} {\bf 12} (2020)  190}, \href{http://arxiv.org/abs/2006.13986}{{\tt
  arXiv:2006.13986 [hep-th]}}.

\bibitem{Kontsevich:1997vb}
M.~Kontsevich, ``{Deformation quantization of Poisson manifolds. 1.},''
  \href{http://dx.doi.org/10.1023/B:MATH.0000027508.00421.bf}{{\em Lett. Math.
  Phys.} {\bf 66} (2003)  157--216},
  \href{http://arxiv.org/abs/q-alg/9709040}{{\tt arXiv:q-alg/9709040}}.

\bibitem{fibre}
C.~Iazeolla and P.~Sundell, ``{A Fiber Approach to Harmonic Analysis of
  Unfolded Higher-Spin Field Equations},''
  \href{http://dx.doi.org/10.1088/1126-6708/2008/10/022}{{\em JHEP} {\bf 10}
  (2008)  022},
\href{http://arxiv.org/abs/0806.1942}{{\tt arXiv:0806.1942 [hep-th]}}.

\bibitem{GiombiYin2}
S.~Giombi and X.~Yin, ``{Higher Spins in AdS and Twistorial Holography},''
  \href{http://dx.doi.org/10.1007/JHEP04(2011)086}{{\em JHEP} {\bf 04} (2011)
  086}, \href{http://arxiv.org/abs/1004.3736}{{\tt arXiv:1004.3736 [hep-th]}}.

\bibitem{Iazeolla:2020jee}
C.~Iazeolla, ``{On boundary conditions and spacetime/fibre duality in
  Vasiliev's higher-spin gravity},''
  \href{http://dx.doi.org/10.22323/1.376.0181}{{\em PoS} {\bf CORFU2019} (2020)
   181}, \href{http://arxiv.org/abs/2004.14903}{{\tt arXiv:2004.14903
  [hep-th]}}.

\bibitem{Didenko:2009td}
V.~E. Didenko and M.~A. Vasiliev, ``{Static BPS black hole in 4d higher-spin
  gauge theory},'' \href{http://dx.doi.org/10.1016/j.physletb.2013.04.021,
  10.1016/j.physletb.2009.11.023}{{\em Phys. Lett.} {\bf B682} (2009)
  305--315}, \href{http://arxiv.org/abs/0906.3898}{{\tt arXiv:0906.3898
  [hep-th]}}.
[Erratum: Phys. Lett.B722,389(2013)].

\bibitem{Sundell:2016mxc}
P.~Sundell and Y.~Yin, ``{New classes of bi-axially symmetric solutions to
  four-dimensional Vasiliev higher spin gravity},''
  \href{http://dx.doi.org/10.1007/JHEP01(2017)043}{{\em JHEP} {\bf 01} (2017)
  043},
\href{http://arxiv.org/abs/1610.03449}{{\tt arXiv:1610.03449 [hep-th]}}.

\bibitem{Didenko:2021vui}
V.~E. Didenko and A.~V. Korybut, ``{Planar solutions of higher-spin theory.
  Part I. Free field level},''
  \href{http://dx.doi.org/10.1007/JHEP08(2021)144}{{\em JHEP} {\bf 08} (2021)
  144}, \href{http://arxiv.org/abs/2105.09021}{{\tt arXiv:2105.09021
  [hep-th]}}.

\bibitem{Didenko:2021vdb}
V.~E. Didenko and A.~V. Korybut, ``{Planar solutions of higher-spin theory.
  Part II. Nonlinear corrections},''
  \href{http://arxiv.org/abs/2110.02256}{{\tt arXiv:2110.02256 [hep-th]}}.

\bibitem{Iazeolla:2007wt}
C.~Iazeolla, E.~Sezgin, and P.~Sundell, ``{Real forms of complex higher spin
  field equations and new exact solutions},''
  \href{http://dx.doi.org/10.1016/j.nuclphysb.2007.08.002}{{\em Nucl. Phys.}
  {\bf B791} (2008)  231--264},
\href{http://arxiv.org/abs/0706.2983}{{\tt arXiv:0706.2983 [hep-th]}}.

\bibitem{Iazeolla:2015tca}
C.~Iazeolla and J.~Raeymaekers, ``{On big crunch solutions in
  Prokushkin-Vasiliev theory},''
  \href{http://dx.doi.org/10.1007/JHEP01(2016)177}{{\em JHEP} {\bf 01} (2016)
  177},
\href{http://arxiv.org/abs/1510.08835}{{\tt arXiv:1510.08835 [hep-th]}}.

\bibitem{CSM}
R.~Carter, G.~Segal, and I.~MacDonald,
  \href{http://dx.doi.org/https://doi.org/10.1017/CBO9781139172882}{{\em
  {Lectures on Lie groups and Lie algebras}}}.
\newblock Cambridge University Press, 1995.

\bibitem{Folland}
G.~Folland, {\em Harmonic Analysis in Phase Space}.
\newblock Annals of Mathematics Studies. Princeton University Press, 1989.

\bibitem{Guillemin:1990ew}
V.~Guillemin and S.~Sternberg, {\em Symplectic Techniques in Physics}.
\newblock Cambridge University Press, 1990.

\bibitem{Woit:2017vqo}
P.~Woit, \href{http://dx.doi.org/10.1007/978-3-319-64612-1}{{\em {Quantum
  Theory, Groups and Representations}}}.
\newblock Springer, 2017.

\bibitem{deGosson}
M.~A.~de Gosson, {\em The Principles of Newtonian and Quantum Mechanics}.
\newblock Imperial College Press, 2001.

\bibitem{Vasiliev:1988sa}
M.~A. Vasiliev, ``{Consistent Equations for Interacting Massless Fields of All
  Spins in the First Order in Curvatures},''
  \href{http://dx.doi.org/10.1016/0003-4916(89)90261-3}{{\em Annals Phys.} {\bf
  190} (1989)  59--106}.

\bibitem{Didenko:2015cwv}
V.~E. Didenko, N.~G. Misuna, and M.~A. Vasiliev, ``{Perturbative analysis in
  higher-spin theories},''
  \href{http://dx.doi.org/10.1007/JHEP07(2016)146}{{\em JHEP} {\bf 07} (2016)
  146},
\href{http://arxiv.org/abs/1512.04405}{{\tt arXiv:1512.04405 [hep-th]}}.

\bibitem{Bychkov:2021zvd}
A.~S. Bychkov, K.~A. Ushakov, and M.~A. Vasiliev, ``{The
  \ensuremath{\sigma}\ensuremath{-} Cohomology Analysis for Symmetric
  Higher-Spin Fields},'' \href{http://dx.doi.org/10.3390/sym13081498}{{\em
  Symmetry} {\bf 13} (2021) no.~8, 1498},
  \href{http://arxiv.org/abs/2107.01736}{{\tt arXiv:2107.01736 [hep-th]}}.

\bibitem{Engquist:2005yt}
J.~Engquist and P.~Sundell, ``{Brane partons and singleton strings},''
  \href{http://dx.doi.org/10.1016/j.nuclphysb.2006.06.040}{{\em Nucl. Phys.}
  {\bf B752} (2006)  206--279},
\href{http://arxiv.org/abs/hep-th/0508124}{{\tt arXiv:hep-th/0508124
  [hep-th]}}.

\bibitem{Sezgin:2005pv}
E.~Sezgin and P.~Sundell, ``{An Exact solution of 4-D higher-spin gauge
  theory},'' \href{http://dx.doi.org/10.1016/j.nuclphysb.2006.06.038}{{\em
  Nucl. Phys.} {\bf B762} (2007)  1--37},
\href{http://arxiv.org/abs/hep-th/0508158}{{\tt arXiv:hep-th/0508158
  [hep-th]}}.

\bibitem{BTZ}
R.~Aros, C.~Iazeolla, P.~Sundell, and Y.~Yin, ``{Higher spin fluctuations on
  spinless 4D BTZ black hole},''
  \href{http://dx.doi.org/10.1007/JHEP08(2019)171}{{\em JHEP} {\bf 08} (2019)
  171},
\href{http://arxiv.org/abs/1903.01399}{{\tt arXiv:1903.01399 [hep-th]}}.

\bibitem{Eastwood:2002su}
M.~G. Eastwood, ``{Higher symmetries of the Laplacian},''
  \href{http://dx.doi.org/10.4007/annals.2005.161.1645}{{\em Annals Math.} {\bf
  161} (2005)  1645--1665}, \href{http://arxiv.org/abs/hep-th/0206233}{{\tt
  arXiv:hep-th/0206233}}.

\bibitem{Sharapov:2019vyd}
A.~Sharapov and E.~Skvortsov, ``{Formal Higher Spin Gravities},''
  \href{http://dx.doi.org/10.1016/j.nuclphysb.2019.02.011}{{\em Nucl. Phys. B}
  {\bf 941} (2019)  838--860}, \href{http://arxiv.org/abs/1901.01426}{{\tt
  arXiv:1901.01426 [hep-th]}}.

\bibitem{Bekaert:2009pt}
X.~Bekaert, M.~Rausch~de Traubenberg, and M.~Valenzuela, ``{An infinite
  supermultiplet of massive higher-spin fields},''
  \href{http://dx.doi.org/10.1088/1126-6708/2009/05/118}{{\em JHEP} {\bf 05}
  (2009)  118}, \href{http://arxiv.org/abs/0904.2533}{{\tt arXiv:0904.2533
  [hep-th]}}.

\bibitem{Bars:2000qm}
I.~Bars, ``{Survey of two time physics},''
  \href{http://dx.doi.org/10.1088/0264-9381/18/16/303}{{\em Class. Quant.
  Grav.} {\bf 18} (2001)  3113--3130},
  \href{http://arxiv.org/abs/hep-th/0008164}{{\tt arXiv:hep-th/0008164}}.

\bibitem{Engquist:2007pr}
J.~Engquist, P.~Sundell, and L.~Tamassia, ``{On Singleton Composites in
  Non-compact WZW Models},''
  \href{http://dx.doi.org/10.1088/1126-6708/2007/02/097}{{\em JHEP} {\bf 02}
  (2007)  097},
\href{http://arxiv.org/abs/hep-th/0701051}{{\tt arXiv:hep-th/0701051
  [hep-th]}}.

\bibitem{Gaberdiel:2021qbb}
M.~R. Gaberdiel and R.~Gopakumar, ``{String Dual to Free N=4 Supersymmetric
  Yang-Mills Theory},''
  \href{http://dx.doi.org/10.1103/PhysRevLett.127.131601}{{\em Phys. Rev.
  Lett.} {\bf 127} (2021) no.~13, 131601},
  \href{http://arxiv.org/abs/2104.08263}{{\tt arXiv:2104.08263 [hep-th]}}.

\bibitem{Vasiliev:2012tv}
M.~A. Vasiliev, ``{Multiparticle extension of the higher-spin algebra},''
  \href{http://dx.doi.org/10.1088/0264-9381/30/10/104006}{{\em Class. Quant.
  Grav.} {\bf 30} (2013)  104006}, \href{http://arxiv.org/abs/1212.6071}{{\tt
  arXiv:1212.6071 [hep-th]}}.

\bibitem{Flato:1978qz}
M.~Flato and C.~Fronsdal, ``{One Massless Particle Equals Two Dirac Singletons:
  Elementary Particles in a Curved Space. 6.},''
\href{http://dx.doi.org/10.1007/BF00400170}{{\em Lett. Math. Phys.} {\bf 2}
  (1978)  421--426}.

\bibitem{Sezgin:2001zs}
E.~Sezgin and P.~Sundell, ``{Doubletons and 5-D higher spin gauge theory},''
  \href{http://dx.doi.org/10.1088/1126-6708/2001/09/036}{{\em JHEP} {\bf 09}
  (2001)  036}, \href{http://arxiv.org/abs/hep-th/0105001}{{\tt
  arXiv:hep-th/0105001}}.

\bibitem{Sezgin:2001ij}
E.~Sezgin and P.~Sundell, ``{7-D bosonic higher spin theory: Symmetry algebra
  and linearized constraints},''
  \href{http://dx.doi.org/10.1016/S0550-3213(02)00299-7}{{\em Nucl. Phys. B}
  {\bf 634} (2002)  120--140}, \href{http://arxiv.org/abs/hep-th/0112100}{{\tt
  arXiv:hep-th/0112100}}.

\bibitem{Vasiliev:2004cm}
M.~A. Vasiliev, ``{Higher spin superalgebras in any dimension and their
  representations},''
  \href{http://dx.doi.org/10.1088/1126-6708/2004/12/046}{{\em JHEP} {\bf 12}
  (2004)  046}, \href{http://arxiv.org/abs/hep-th/0404124}{{\tt
  arXiv:hep-th/0404124}}.

\bibitem{Basile:2018dzi}
T.~Basile, X.~Bekaert, and E.~Joung, ``{Twisted Flato-Fronsdal Theorem for
  Higher-Spin Algebras},''
  \href{http://dx.doi.org/10.1007/JHEP07(2018)009}{{\em JHEP} {\bf 07} (2018)
  009}, \href{http://arxiv.org/abs/1802.03232}{{\tt arXiv:1802.03232
  [hep-th]}}.

\bibitem{DixmierBook}
J.~Dixmier, {\em Enveloping Algebras}.
\newblock North-Holland mathematical library. North-Holland Publishing Company,
  1977.

\bibitem{Gelfond:2008td}
O.~A. Gelfond and M.~A. Vasiliev, ``{Sp(8) invariant higher spin theory,
  twistors and geometric BRST formulation of unfolded field equations},''
  \href{http://dx.doi.org/10.1088/1126-6708/2009/12/021}{{\em JHEP} {\bf 12}
  (2009)  021}, \href{http://arxiv.org/abs/0901.2176}{{\tt arXiv:0901.2176
  [hep-th]}}.

\bibitem{Wigner:50}
E.~P. Wigner, ``Do the equations of motion determine the quantum mechanical
  commutation relations?,''
  \href{http://dx.doi.org/10.1103/physrev.77.711}{{\em Physical Review} {\bf
  77} (1950) no.~5, 711--712}.

\bibitem{Yang:51}
L.~M. Yang, ``A note on the quantum rule of the harmonic oscillator,''
  \href{http://dx.doi.org/10.1103/physrev.84.788}{{\em Physical Review} {\bf
  84} (1951) no.~4, 788--790}.

\bibitem{Vasiliev:1989re}
M.~A. Vasiliev, ``{Higher Spin Algebras and Quantization on the Sphere and
  Hyperboloid},'' \href{http://dx.doi.org/10.1142/S0217751X91000605}{{\em Int.
  J. Mod. Phys. A} {\bf 6} (1991)  1115--1135}.

\bibitem{Vasiliev:1989qh}
M.~A. Vasiliev, ``{Quantization on sphere and high spin superalgebras},'' {\em
  JETP Lett.} {\bf 50} (1989)  374--377.

\bibitem{Ferrari:2005kx}
A.~F. Ferrari, M.~Gomes, A.~Y. Petrov, and A.~J. da~Silva, ``{Supersymmetric
  non-Abelian noncommutative Chern-Simons theory},''
  \href{http://dx.doi.org/10.1016/j.physletb.2006.05.031}{{\em Phys. Lett. B}
  {\bf 638} (2006)  275--282}, \href{http://arxiv.org/abs/hep-th/0511059}{{\tt
  arXiv:hep-th/0511059}}.

\bibitem{Wu:1990ci}
S.-y. Wu, ``{Topological Quantum Field Theories on Manifolds With a
  Boundary},'' \href{http://dx.doi.org/10.1007/BF02096795}{{\em Commun. Math.
  Phys.} {\bf 136} (1991)  157--168}.
  
\bibitem{Boulanger:2012bj}
N.~Boulanger, N.~Colombo and P.~Sundell,
``A minimal BV action for Vasiliev's four-dimensional higher spin gravity,''
JHEP \textbf{10} (2012), 043
doi:10.1007/JHEP10(2012)043
[arXiv:1205.3339 [hep-th]].


\bibitem{Boulanger:2015kfa}
N.~Boulanger, E.~Sezgin and P.~Sundell,
``4D Higher Spin Gravity with Dynamical Two-Form as a Frobenius-Chern-Simons Gauge Theory,''
[arXiv:1505.04957 [hep-th]].



\bibitem{Kubo}
R.~Kubo, ``{Wigner Representation of Quantum Operators and Its Applications to
  Electrons in a Magnetic Field},''
  \href{http://dx.doi.org/10.1143/JPSJ.19.2127}{{\em Journal of the Physical
  Society of Japan} {\bf 19} (1964)  2127–--2139}.

\bibitem{Cohen66}
L.~Cohen, ``{Generalized Phase-Space Distribution Functions},''
  \href{http://dx.doi.org/10.1063/1.1931206}{{\em Journal of Mathematical
  Physics} {\bf 7} (1966)  781--786}.

\bibitem{Cohenbook}
L.~Cohen, {\em The Weyl Operator and its Generalization}.
\newblock Pseudo-Differential Operators. Springer Basel, 2012.

\bibitem{Seiberg:1999vs}
N.~Seiberg and E.~Witten, ``{String theory and noncommutative geometry},''
  \href{http://dx.doi.org/10.1088/1126-6708/1999/09/032}{{\em JHEP} {\bf 09}
  (1999)  032}, \href{http://arxiv.org/abs/hep-th/9908142}{{\tt
  arXiv:hep-th/9908142}}.

\bibitem{Sezgin:2002ru}
E.~Sezgin and P.~Sundell, ``{Analysis of higher spin field equations in four
  dimensions},'' {\em JHEP} {\bf 07} (2002)  055,
\href{http://arxiv.org/abs/hep-th/0205132}{{\tt arXiv:hep-th/0205132}}.

\bibitem{Boulanger:2013naa}
N.~Boulanger, P.~Sundell, and M.~Valenzuela, ``{Three-dimensional
  fractional-spin gravity},''
  \href{http://dx.doi.org/10.1007/JHEP02(2014)052}{{\em JHEP} {\bf 02} (2014)
  052}, \href{http://arxiv.org/abs/1312.5700}{{\tt arXiv:1312.5700 [hep-th]}}.
  [Erratum: JHEP 03, 076 (2016)].
  

\bibitem{Boulanger:2015uha}
N.~Boulanger, P.~Sundell and M.~Valenzuela,
``Gravitational and gauge couplings in Chern-Simons fractional spin gravity,''
JHEP \textbf{01} (2016), 173
[erratum: JHEP \textbf{03} (2016), 075]
doi:10.1007/JHEP01(2016)173
[arXiv:1504.04286 [hep-th]].

  \bibitem{Iazeolla:2022dal}
C.~Iazeolla and P.~Sundell,
``Unfolding, higher spins, metaplectic groups and resolution of classical singularities,''
[arXiv:2205.00296 [hep-th]].

\bibitem{Bolotin:1999fa}
K.~Bolotin and M.~A. Vasiliev, ``{Star product and massless free field dynamics
  in AdS(4)},'' \href{http://dx.doi.org/10.1016/S0370-2693(00)00307-5}{{\em
  Phys.Lett.} {\bf B479} (2000)  421--428},
\href{http://arxiv.org/abs/hep-th/0001031}{{\tt arXiv:hep-th/0001031
  [hep-th]}}.

\bibitem{Zhang:2008jy}
H.-H. Zhang, K.-X. Feng, S.-W. Qiu, A.~Zhao, and X.-S. Li, ``{On analytic
  formulas of Feynman propagators in position space},''
  \href{http://dx.doi.org/10.1088/1674-1137/34/10/005}{{\em Chin. Phys. C} {\bf
  34} (2010)  1576--1582}, \href{http://arxiv.org/abs/0811.1261}{{\tt
  arXiv:0811.1261 [math-ph]}}.
  
\bibitem{Wong1998}
M.~W. Wong, \href{http://dx.doi.org/10.1007/0-387-22778-4_1}{{\em Weyl
  Transforms}}.
\newblock Springer New York, New York, NY, 1998.
\newblock \url{https://doi.org/10.1007/0-387-22778-4_1}.

\end{thebibliography}
\end{document}